\documentclass{aa} 

\usepackage[dvipsnames]{xcolor}

\newcommand{\appropto}{\mathrel{\vcenter{
  \offinterlineskip\halign{\hfil$##$\cr
    \propto\cr\noalign{\kern2pt}\sim\cr\noalign{\kern-2pt}}}}}
\usepackage{graphicx,booktabs,array}
\usepackage{txfonts}
\usepackage{hyperref}
\usepackage{amssymb}
\usepackage{amsmath}
\usepackage{mathtools}
\usepackage{bm}
\begin{document} 

   \title{Towards Nebular Spectral Modelling of Magnetar-Powered Supernovae}

   \author{C. M. B. Omand
          \inst{1}
          \and
          A. Jerkstrand\inst{1}
          }

   \institute{The Oskar Klein Centre, Department of Astronomy, Stockholm University, Albanova 10691, Stockholm, Sweden\\
              \email{conor.omand@astro.su.se}
             }

   \date{}

 
  \abstract
   {Many energetic supernovae (SNe) are thought to be powered by the rotational-energy of a highly-magnetized, rapidly-rotating neutron star.  The emission from the associated luminous pulsar wind nebula (PWN) can photoionize the supernova ejecta, leading to a nebular spectrum of the ejecta with signatures possibly revealing the PWN. 
   SN 2012au is hypothesized to be one such supernova.}
   {We investigate the impact of different ejecta and PWN parameters on the supernova nebular spectrum, and test if any photoionization models are consistent with SN 2012au. We study how constraints from the nebular phase can be linked into modelling of the diffusion phase and the radio emission of the magnetar.}
   {We present a suite of late-time (1-6y) spectral simulations of SN ejecta powered by an inner PWN. Over a large grid of 1-zone models, we study the behaviour of the SN physical state and line emission as PWN luminosity ($L_{\rm PWN}$), injection spectral energy distribution (SED) temperature ($T_{\rm PWN}$), ejecta mass ($M_{\rm ej}$), and composition (pure O or realistic) vary. We discuss the resulting emission in the context of the observed behaviour of SN 2012au, a strong candidate for a PWN-powered SN.  We use optical light curve models and broadband PWN models to make predictions about possible radio emission from SN 2012au.}
   {The supernova nebular spectrum varies as $T_{\rm PWN}$ varies, as the ejecta become less ionized as $T_{\rm PWN}$ increases.  Low ejecta mass models, at high PWN power, obtain runaway ionization for O I and, in extreme cases, also O II, causing a sharp decrease in their ion fraction over a small change in the parameter space.  Certain models can reproduce the oxygen lines luminosities of SN 2012au reasonably well at individual epochs,
   but we find no model that fits over the whole time evolution; this is likely due to uncertainties and simplifications in the model setup. 
   Using our derived constraints from the nebular phase, we predict that the magnetar powering SN 2012au had an initial rotation period $\sim$ 15 ms, and should be a strong radio source ($F > 100$ $\mu$Jy) for decades.}
   {}

   \keywords{radiative transfer -- 
                stars: magnetars --
                supernovae: general --
                supernovae: individual: SN 2012au
               }

   \maketitle
%

\section{Introduction} \label{sec:intro}

Recent optical transient surveys have revealed several classes of core-collapse supernovae (CCSNe) more luminous and energetic than predicted by standard supernova models. These include hydrogen-poor superluminous supernovae (SLSNe), which are the brightest known optical transient with $\sim$ 100 times more energy radiated than a typical supernova \citep{Gal-Yam2012, Nicholl2021}, and broad-line Type Ic supernovae (SNe Ic-BL), which are slightly overluminous and have higher inferred kinetic energies that the canonical supernova explosion energy of 10$^{51}$ erg \citep[e.g.][]{Taddia2019}. Some SNe Ic-BL have been associated with long-duration gamma-ray bursts (LGRBs) \citep{Cano2017}, while a small number of SLSNe have been associated with ultra-long gamma-ray bursts (ULGRBS) \citep{Gendre2013, Nakauchi2013, Levan2014}, although strong radio and X-ray constraints show that jet formation in SLSNe is rare \citep{Coppejans2018, Margutti2018}.  SLSNe and SNe Ic-BL tend to also have similar spectral features both at early \citep{Pastorello2010, Inserra2013, Nicholl2013, Blanchard2019} and late \citep{Milisavljevic2013,Jerkstrand2017,Nicholl2017} phases, and similar host-galaxies \citep{Lunnan2014, Leloudas2015b, Angus2016, Schulze2018, Orum2020}.

Because of these similarities, many studies have suggested that these two supernova classes 
could be powered by the same mechanism 
\citep{Metzger2015, Ioka2016, Kashiyama2016, Margalit2018, Suzuki2021}.  SLSNe have been suggested to arise from pair-instability or pulsational pair instability explosions \citep{Barkat1967, Heger2002, Gal-Yam2009} or by interaction with circumstellar medium (CSM) \citep{Chatzopoulos2012, Ginzburg2012}, but neither model scenario has been demonstrated to be able to produce both light curves and spectra.
Another suggested powering mechanism is fallback onto a newly-formed black hole \citep{Dexter2013}, but the accretion rate needed to produce a SLSN is unphysically high in most cases \citep{Moriya2018}.  Another model that seems consistent with both SLSNe and SNe Ic-BL light curves is the magnetar-driven model \citep{Kasen2010, Woosley2010}, where the rotational-energy of a highly-magnetized, rapidly-rotating neutron star powers a luminous pulsar wind nebula (PWN), which both exerts a pressure on the ejecta, accelerating it, and produces broadband emission which is thermalized in the ejecta, increasing the temperature and ionization of the gas and the luminosity of the supernova.

The magnetar engine model makes several multiwavelength predictions that can be used to test it and characterize the magnetar further.  The PWN is expected to be bright in both radio and X-rays at late times \citep{Murase2015, Kashiyama2016, Omand2018}, and a few observations have been performed \citep[e.g.][]{Margutti2018, Eftekhari2021, Murase2021}; so far only one promising PWN candidate has been detected in radio, PTF10hgi \citep{Eftekhari2019, Law2019, Mondal2020, Hatsukade2021}.  An infrared excess was predicted by \cite{Omand2019} due to the PWN emission heating newly-formed dust in the ejecta, which has been detected in four SLSNe \citep{Chen2021, Sun2022}, although can also be illuminated by the supernova in an IR echo \citep{Bode1980, Dwek1983} or heated through shock interaction \citep{Smith2008, Fox2010, Sarangi2018}, and collapsar models also predict an infrared excess due to r-process nucleosynthesis in the accretion disk \citep{Barnes2022, Anand2023}.  Polarization has also been observed in several SLSNe, which could either be from a central engine \citep{Inserra2016, Saito2020} or CSM interaction \citep{Pursiainen2022}, but other SLSNe are unpolarized \citep{Leloudas2015, Lee2019, Lee2020, Poidevin2022b, Poidevin2022}, so this signal is not ubiquitous. 

At later times SN ejecta become more transparent, and emission from the metal-rich core can reveal important information about the origin and explosion of the star. This phase is often referred to the nebular phase, even though radiative transfer effects can still operate for many years or decades.
There are only limited spectroscopic observations of SLSNe and Ic-BL SNe in their nebular phase due to their relative rarity and often significant distances.
A few relatively nearby events have been well studied \citep{Gal-Yam2009, Chen2015, Yan2015, Yan2017, Nicholl2016, Nicholl2018, Lunnan2016, Lunnan2018, Jerkstrand2017, Quimby2018, Blanchard2021, West2023} leading to some small-scale statistical analysis \citep{Nicholl2019}. Both SLSNe and SNe Ic-BL show strong [O I] $\lambda \lambda$6300, 6364 and [Ca II] $\lambda \lambda$7291, 7323 emission, with SLSNe also showing strong O I $\lambda$7774 emission and an elevated flux around 5000 \AA, where [Fe II] is thought to dominate \citep{Nicholl2019} - although this has also occurred in some SNe Ic-BL, such as SN 1998bw \citep{Mazzali2001}. Some SLSNe also show [O III] lines \citep{Lunnan2016, Jerkstrand2017, Yan2017} and possibly [O II] as well.  

Efforts to model the nebular spectra of these exotic supernovae are so far limited.  Observed spectra are inconsistent with pair-instability spectral models \citep{Dessart2013, Jerkstrand2016}, 
as these massive, $^{56}$Ni-powered supernovae tend to produce cool and neutral ejecta, with red, strongly line-blanked spectra with narrow lines of species such as Fe I and Si I. This conflicts even qualitatively with observed spectra of SLSN such as SN 2007bi and PTF12dam \citep{Gal-Yam2009, Chen2015} which are blue with broad lines.  Other studies have tried to mimic the effects of additional late-time energy deposition, e.g. by a magnetar, into normal CCSN ejecta \citep{Dessart2012, Jerkstrand2017, Dessart2018, Dessart2019}. 
Such models are more successful, mostly due to the lower ejecta masses and lower amount of line-blanketing iron. They treat the magnetar power, however, as either a purely thermal on-the-spot energy injection or in the same manner as radioactive decay.  
This approach may miss effects of increased photoionization and does not probe the PWN spectral energy distribution (SED), which is not yet well constrained and has strong implications for multiwavelength studies.

Nebular spectra can also probe clumping in the ejecta, and studies have inferred high amounts of clumping for both normal stripped envelope SNe \citep{Taubenberger2009} and SLSNe \citep{Jerkstrand2017,Nicholl2019, Dessart2019}.

In this paper, we develop the \texttt{SUMO} spectral synthesis code \citep{Jerkstrand2011, Jerkstrand2012} to be able to handle powering by high-energy radiation from a PWN, and use this code to investigate spectral formation over a parameter space relevant for SLSNe and Ic-BL SNe. We study the physical processes occurring in the ejecta and search for distinct predictions for emergent spectra that could reveal the presence of a central magnetar. In particular, we compare the models to observations of SN 2012au, a peculiar luminous stripped-envelope (Type Ib) SN with spectra taken at $\sim$ 1 year and at $\sim$ 6 years \citep{Milisavljevic2013, Takaki2013, Milisavljevic2018, Pandey2021}.  

The paper is structured as follows. Section \ref{sec:mod} summarizes the numerical model and model setup.  Section \ref{sec:olines} presents some general theory regarding the formation of oxygen lines.  In Section \ref{sec:12au}, we overview previous observations of SN 2012au.   Results are presented in Section \ref{sec:res} and their implications are discussed in Section \ref{sec:dis}.  Finally, a summary is given in Section \ref{sec:conc}.

\section{Modelling Overview} \label{sec:mod}

\subsection{\texttt{SUMO}}

Non-local thermodynamic equilibrium (NLTE) spectral synthesis is performed using the \texttt{SUMO} code \citep{Jerkstrand2011, Jerkstrand2012, Jerkstrand2017hb}, which takes an input composition and density as a function of ejecta velocity; solves for the temperature, ionic abundances, atomic level populations, and the radiation field; and uses these to compute the emergent supernova spectrum. \texttt{SUMO} is specialized to the late, NLTE phase of the supernova. It can compute the physical state of the ejecta either in steady-state or time-dependently \citep{Pognan2022}. The code currently treats 22 elements between hydrogen and nickel, several r-process elements \citep{Pognan2022b}, and several molecules \citep{Liljegren2020, Liljegren2022}.  An arbitrary number of ionization stages can now be treated‚ in this study the first four stages were allowed for.
Code updates for this study are given in Appendix \ref{app:code}.

\subsection{Model Setup} \label{sec:modsetup}

We explore two types of compositions. The most abundant element in virtually all stripped-envelope CCSN models is oxygen. Oxygen provides the the dominant lines in the 6y spectrum of SN 2012au, with strong emission from all its first three ionization stages. The first composition is therefore pure oxygen.  

The second composition is taken from a stellar evolution/explosion model of a $M_{\rm ZAMS} = 25~M_\odot$ progenitor core-collapse supernova from \cite{Woosley2007}. This progenitor has a CO core of 6.8 $M_\odot$, of which 1.8 $M_\odot$ forms a compact object and 5.0 $M_\odot$ of nuclear processed material is ejected in the supernova. For this model we assume all the material to become homogeneously mixed. Whereas complete microscopic mixing is not indicated in normal SNe \citep{Fransson1989,Jerkstrand2017hb}, stronger mixing may be induced when a powerful PWN is present \citep[see e.g. simulations by][]{Chen2020,Suzuki2021}. Since ejecta mass $M_{\rm ej}$ can vary in this study, the mass fractions are held constant and element masses are scaled with ejecta mass. The mass fractions for this model are listed in Table \ref{tbl:hegermf}.

The short mean-free path of X-rays and UV radiation means that morphology details are likely important for SNe powered by PWNe. One-dimensional photoionization models show that the ejecta develop regions of different ionization states \citep[e.g.][]{Chevalier1992}. This is in contrast to a radioactivity-powered SNe, where gamma rays have a mean-free path similar to the ejecta scale in the nebular phase and so the gamma-ray field strength has small or moderate spatial variation. 

On the other hand, the strong Rayleigh-Taylor mixing seen in multi-dimensional simulations of engine-driven supernovae \citep{Chen2020, Suzuki2021} calls into question the applicability of stratified multi-zone models in 1D - the onion-shell structure is probably such a poor representation of the real morphology that gains in accuracy are not obvious. As illustrated to some extent by the images of the Crab Nebula, the synchrotron generating regions may become mixed in velocity space with clumps and filaments of ejecta. If such mixing is strong enough, one may end up in a situation closer to a 1-zone setup rather than a stratified 1D ejecta.

As a starting point, we mostly explore 1-zone models for our two compositions. In Section \ref{sec:multiz}, we however also explore the effects of zoning (in 1D) using one of the models. That analysis shows that allowing for 1D ionization stratification, differences of a factor of a few can be expected the main oxygen emission lines. We then use this information to assess which part of parameter space of 1-zone models are viable for SN 2012au.


\begin{table}
\centering
\begin{tabular}{c|c}
   Element & Mass Fraction \\ \hline
   $^{56}$Ni & 1.1 $\times$ 10$^{-2}$ \\
   $^{57}$Co & 4.4 $\times$ 10$^{-4}$ \\  
   $^{44}$Ti & 5.1 $\times$ 10$^{-6}$ \\  \hline  
   He & 1.5 $\times$ 10$^{-3}$ \\
   C & 6.3 $\times$ 10$^{-2}$ \\
   N & 2.7 $\times$ 10$^{-5}$ \\
   O & 0.69 \\
   Ne & 8.9 $\times$ 10$^{-2}$ \\
   Na & 1.1 $\times$ 10$^{-3}$ \\
   Mg & 4.3 $\times$ 10$^{-2}$ \\
   Al & 3.4 $\times$ 10$^{-3}$ \\
   Si & 5.7 $\times$ 10$^{-2}$ \\
   S & 3.3 $\times$ 10$^{-2}$ \\
   Ar & 4.8 $\times$ 10$^{-3}$ \\
   Ca & 1.9 $\times$ 10$^{-3}$ \\
   Sc & 1.3 $\times$ 10$^{-6}$ \\
   Ti & 4.4 $\times$ 10$^{-5}$ \\
   V & 6.9 $\times$ 10$^{-6}$ \\
   Cr & 2.1 $\times$ 10$^{-4}$ \\
   Mn & 3.5 $\times$ 10$^{-5}$ \\
   Fe & 2.5 $\times$ 10$^{-3}$ \\
   Co & 1.4 $\times$ 10$^{-4}$ \\
   Ni & 1.2 $\times$ 10$^{-3}$ \\
\end{tabular}
\caption{Composition of the realistic Ic model.}
\label{tbl:hegermf}
\end{table}

The shell in our model expands homologously between 2000 and 3000 km s$^{-1}$, with the inner boundary representing the contact discontinuity between the ejecta and the PWN. The inner boundary velocity is set from the low velocity emission lines detected around one year in SN 2012au \citep{Milisavljevic2013} by assuming these lines are produced close to the contact discontinuity, while the outer boundary is chosen so the single-zone density is similar to that of the inner ejecta region with a $\rho \propto v^{-6}$ density profile motivated by multi-dimensional numerical simulations of engine driven supernovae \citep{Suzuki2017, Suzuki2019}; this density profile is expected for SNe with a total engine energy greater than the supernova explosion energy, injected over a timescale less than the diffusion time, and without significant early radiative losses dampening the efficiency of energy transport from the central ejecta to the outer layers \citep{Suzuki2021}. 
The shell has a constant density and a filling factor of 1 - clumping is not considered in this study. 

The SED of the photons injected at the inner boundary, which represents the emission from the pulsar-wind nebula, is taken as a blackbody with a temperature $T_{\rm PWN}$ and luminosity $L_{\rm PWN}$; the number of ionizing photons injected will be proportional to $L_{\rm PWN}T_{\rm PWN}^{-1}$, as long as $T_{\rm PWN}$ is high enough for most of the photons to be ionizing.  While a more realistic treatment would use a power-law or broken power-law to represent the broadband synchrotron emission, this would involve 2 or 3 parameters for the SED shape. By instead using a blackbody SED, we limit the grid to introduction of just a single further parameter. Furthermore, one of our goals is to understand the basic physical mechanisms at play, and to this end using blackbodies of different temperatures allow us to study the connection between characteristic photon energies and resulting gas properties.  

More complicated PWN models, such as those which calculate synchrotron emission and self absorption, pair cascades, Compton and inverse Compton scattering, adiabatic cooling, and both internal and external attenuation \citep{Murase2015, Murase2016} mostly still find that the SED resembles a broken power-law over the infrared to X-ray range (see Figure 3 from \citet{Omand2018}, for example), since these additional processes mostly affect the spectrum at radio and hard X-ray/gamma ray wavelengths.  \citet{Vurm2021} showed that to reproduce the late-time light curve decay in SN 2015bn and SN 2017egm, the PWN spectrum must be inverse Compton-dominated, and emit mostly in X-rays and gamma rays; however, this scenario would predict extremely faint radio emission, which conflicts with radio observations of PTF10hgi \citep{Eftekhari2019, Law2019, Mondal2020, Hatsukade2021}.



At $t$ = 6 years, we compute a model grid with

\begin{enumerate}
    \item $M_{\rm ej}$ = 1.0, 1.5, 2.5, 4.0, 6.0, and 10.0 $M_\odot$
    \item $L_{\rm PWN}$ = $1 \times 10^{38}$, $2 \times 10^{38}$, $5 \times 10^{38}$, $1 \times 10^{39}$, $2 \times 10^{39}$, $5 \times 10^{39}$, and $1 \times 10^{40}$ erg s$^{-1}$
    \item $T_{\rm PWN}$ = $1 \times 10^{5}$, $3 \times 10^{5}$, and $1 \times 10^{6}$ K
    \item Composition: Pure O and realistic SN Ic (as described above)
\end{enumerate}
At $t$ = 320 days we compute a smaller grid motivated by the results at $t$ = 6 years.  The models are named by time (approximate in years), composition, $M_{\rm ej}$, $L_{\rm PWN}$, and $T_{\rm PWN}$: an example would be 6Ic-4-1e40-1e5.

We investigate a range of ejecta masses $1-10$ $M_\odot$, motivated by both the mass range estimated for SN 2012au in previous studies \citep{Takaki2013, Milisavljevic2013, Pandey2021} and the mass range estimated for the Ic-BL population in previous sample studies \citep{Taddia2019, Corsi2022, Anand2023}.  This mass range and the velocities chosen give kinetic energies of $\sim$ $10^{50-51}$ erg, less than that expected for a typical supernova. A lower kinetic energy will cause a higher ejecta density, which may cause the temperature and ionization to be lower \citep{Dessart2019}.  When the difference is large, one may interpret the models to represent the line forming region; the inner ejecta lying close to the edge of the PWN and receiving its input.  

The luminosity of a PWN at late times is roughly 

\begin{equation}
    L_{\rm PWN} (t) \approx L_{\rm 0, PWN} \left(\frac{t}{t_{\rm SD}}\right)^{-2},
    \label{eqn:lpwnt}
\end{equation}
where 

\begin{equation}
    L_{\rm 0, PWN}\approx \frac{E_{\rm rot}}{t_{\rm SD}}
    \label{eqn:lpwno}
\end{equation}
is the initial PWN luminosity and $t_{\rm SD}$ is the pulsar spin-down time.  The pulsar rotational energy $E_{\rm rot}$ and spin-down time are related to the initial pulsar spin period $P_0$ and pulsar magnetic field $B$ (assumed constant in time) by 

\begin{align}
    E_{\rm rot} \approx & \, 2.5 \times 10^{52} \text{ erg } \left( \frac{P_0}{\text{1 ms}}\right)^{-2} \label{eqn:erot}, \\
    t_{\rm SD} \approx & \, 8 \text{ hours } \left( \frac{P_0}{\text{1 ms}}\right)^{2} \left( \frac{B}{\text{10$^{14}$ G}}\right)^{-2}. \label{eqn:tsd}
\end{align}
Combining these gives a PWN luminosity of

\begin{equation}
    L_{\rm PWN} (t) \approx 7 \times 10^{41} \text{ erg s$^{-1}$ } \left( \frac{B}{\text{10$^{14}$ G}}\right)^{-2} \left( \frac{t}{\text{1 year}}\right)^{-2}.
    \label{eqn:lpwntnum}
\end{equation}
The typical magnetic field of a rapidly-rotating pulsar that reproduces a Ic-BL SN light curve if the diffusion phase is magnetar-powered is estimated to be $\sim 5 \times 10^{14}$ G \citep{Kashiyama2016}, which gives a PWN luminosity at 6 years of $\sim$ $10^{39}$ erg s$^{-1}$, which motivates the range of PWN luminosities examined.   

The characteristic synchrotron energy $\nu_b$ in a PWN, at which $\nu F_\nu$ peaks, is given by \citep{Murase2021}

\begin{multline}
     \nu_b \approx 5 \times 10^8 \text{ GHz } \left( \frac{\gamma_b}{10^5}\right)^{2} \left( \frac{\epsilon_B}{0.003}\right)^{1/2}
     \\ \times \left( \frac{P_0}{\text{1 ms}}\right)^{1/2} \left( \frac{M_{\rm ej}}{\text{10 $M_\odot$}}\right)^{3/4} \left( \frac{t}{\text{1 year}}\right)^{-3/2}
    \label{eqn:nub}   
\end{multline}
where $\epsilon_B$ is the fraction of spin down energy carried by nebular magnetic fields and $\gamma_b$ is the average electron injection Lorentz factor.  The values shown in Equation \ref{eqn:nub} are typical of those found from modelling Galactic PWNe, including the Crab Nebula \citep{Kennel1984, Tanaka2010, Tanaka2013}; however, these values, as well as those for spectral indices, are not well constrained for highly luminous newborn PWNe, but there is some evidence that the value differs from that of Galactic PWNe \citep{Law2019, Eftekhari2021, Murase2021, Vurm2021}.  For these default parameters, this gives a characteristic photon energy of $\sim$ 130 eV, which corresponds to the peak of a blackbody with temperature $\sim 5 \times 10^5$ K, which motivates the range of injection SED temperatures examined. We allow three values for $T_{\rm PWN}$: $1 \times 10^{5}$, $3 \times 10^{5}$, and $1 \times 10^{6}$ K, corresponding to characteristic photon energies ($2.8 kT$) of 24, 72 and 240 eV, respectively; these energies are all high enough for the majority of photons to be able to photoionize oxygen.

There are a few pieces of physics relevant to an engine-driven supernova that \texttt{SUMO} does not yet treat correctly.  The most prominent is inner-shell ionizations. While electron and photon impact ionization partially considers contribution from inner shells, their contribution to ionization rates are approximate because we do not follow Auger and fluorescence processes in detail. 
In this study, we therefore avoid hard X-ray illumination of the ejecta to avoid these inner shell processes. This restricts us to using $T_{\rm PWN} \lesssim 10^6$ K. Once these processes are better treated, we will be able to explore higher SED temperatures, more realistic broadband injection spectra, and Compton-dominated spectra \citep{Vurm2021}.  We also assume that photoelectrons are fully thermalized, ignoring possibilities for them to ionize or excite the gas. This approximation should be reasonable for this study due to the lack of hard X-rays injected, because photoelectrons will typically not have the energy to further ionize the ejecta. The other processes may however be important for a more realistic PWN spectrum.

\section{Formation of Oxygen Emission Lines} \label{sec:olines}

The strength of the oxygen emission lines we study will depend on the parameters of the atomic transitions that correspond to the observed lines.  The emissivity of a line is given, in general, by

\begin{equation}
    j = \frac{1}{4\pi}n_u A \beta_S h\nu,
    \label{eqn:lline}
\end{equation}
where $n_u$ is the population of the excited state for the line transition, $A$ is the spontaneous radiative decay rate, and $\beta_S$ is the Sobolev escape probability (which we take to unity for the analysis in this section, corresponding to optically thin lines).  The rate equilibrium equation for a 2-level system with only thermal collisional and spontaneous radiative decay terms is

\begin{equation}
    n_u(A + C_{\rm down}) = n_g C_{\rm up} ,
    \label{eqn:transrate}
\end{equation}
where $C_{\rm up} = c_1 \Upsilon(T) e^{-T_{\rm exc}/T} n_e$ is the collisional excitation rate; $C_{\rm down} = c_1 \frac{g_{\rm down}}{g_{\rm up}} \Upsilon(T) n_e$ is the collisional deexcitation rate; $n_g \approx n_{\rm ion}$ is the ground state population; $n_e$ is the electron density; $g_{\rm down}$ and $g_{\rm up}$ are the multiplicities of the ground and excited states, respectively; $\Upsilon(T)$ is the effective collision strength, which is dependent on the cross section function for the transition but normally has a weak temperature dependence \citep{Jerkstrand2017hb}, and so we take it as a constant; and $c_1$ is also a constant.  The excitation temperature $T_{\rm exc} = E_{\rm exc}/k_{\rm B}$ is the temperature that corresponds to the excitation energy for the transition.  Solving for $n_u A$ gives

\begin{equation}
    n_u A = n_{\rm ion} e^{-T_{\rm exc}/T} \frac{c_1 \Upsilon n_e}{1 + \frac{g_{\rm down}c_1 \Upsilon n_e}{g_{\rm up} A}}.
    \label{eqn:nAfromrate}
\end{equation}
From this equation, we find that the line luminosity

\begin{figure}
    \centering
    \includegraphics[width=0.5\textwidth]{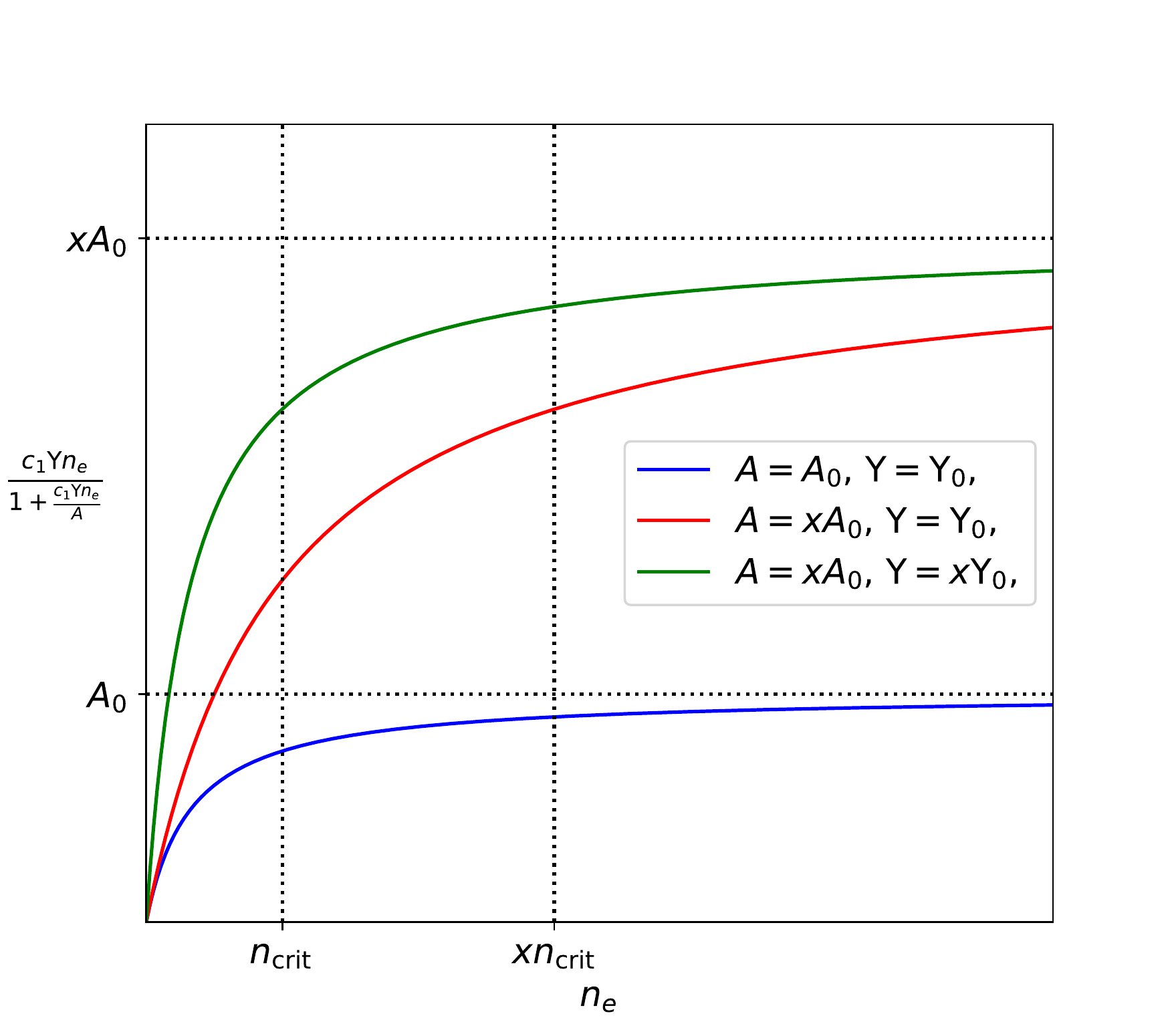}
    \caption{A schematic showing the dependence of the third line luminosity factor on $n_e$, $A$, and $\Upsilon$.  Lines with the same $A$ values (red and green) will approach the same luminosity in the high $n_e$ (LTE) limit, lines with the same $\Upsilon$ values (blue and red) show the same luminosity in the low $n_e$ limit, and lines with the same $A/\Upsilon$ (blue and green) show the same value of $n_{\rm crit}$, with $n_{\rm crit}$ for the red line being a factor of $x$ larger.}
    \label{fig:linelum}
\end{figure}

\begin{enumerate}
    \item is proportional to the ion density.
    \item decreases exponentially when the temperature is below the excitation temperature. 
    \item depends on the atomic parameters $\Upsilon$ and $A$, and, at low electron densities, $n_e$.
\end{enumerate}
The quantity $A/\left(c_1\Upsilon\right)$ is known as the critical density $n_{\rm crit}$, which is the cutoff between local thermodynamic equilibrium (LTE) and non-LTE (NLTE).  If $n_e \gg  n_{\rm crit}$, the system is in LTE and the line emissivity is

\begin{equation}
   j = \frac{1}{4\pi}n_{\rm ion} h \nu e^{-T_{\rm exc}/T} A,
    \label{eqn:ltellum}
\end{equation}
proportional to $A$, while if $n_e \ll n_{\rm crit}$, then the line emissivity is

\begin{equation}
   j = \frac{1}{4\pi}n_{\rm ion} h \nu e^{-T_{\rm exc}/T} c_1 \Upsilon n_e,
    \label{eqn:nltellum}
\end{equation}
proportional to $\Upsilon n_e$. A schematic dependence of the third line luminosity factor in Equation \ref{eqn:nAfromrate} (putting $g_{\rm down}/g_{\rm up} = 1$ for simplicity, so $c_1 \Upsilon n_e / \left(1 + c_1 \Upsilon n_e/A\right)$) on $n_e$, $A$, and $\Upsilon$ is shown in Figure \ref{fig:linelum}.  For the same $T_{\rm exc}$, lines with the same $A$ values will show the same luminosity in the high $n_e$ (LTE) limit, lines with the same $\Upsilon$ values will show the same luminosity in the low $n_e$ limit, and lines with the same $A/\Upsilon$ value will have the same value of $n_{\rm crit}$.  The values of these properties for the [O I], [O II] and [O III] forbidden lines examined in this study are shown in Table \ref{tbl:olines}. 

\begin{table*}
\centering
\begin{tabular}{cccccc}
   Line & Transition & $\lambda$ ($\AA$) & $T_{\rm exc}$ (10$^3$ K) & $A$ (s$^{-1}$) & $\Upsilon$   \\ \hline
   $\left[\rm O~I\right]$ & 2p$^4$($^1$D$_2$) $\rightarrow$ 2p$^4$($^3$P$_2$) & 6300 & 22.8 & 5.63 $\times$ 10$^{-3}$ & 6.47 $\times$ 10$^{-2}$  \\
   & 2p$^4$($^1$D$_2$) $\rightarrow$ 2p$^4$($^3$P$_1$) & 6364 & 22.8 & 1.82 $\times$ 10$^{-3}$ & 3.88 $\times$ 10$^{-2}$ \\
   $\left[\rm O~II\right]$ & 2p$^3$($^2$P$^{\rm o}_{3/2}$) $\rightarrow$ 2p$^3$($^2$D$^{\rm o}_{5/2}$) & 7320 & 58.2 & 9.07 $\times$ 10$^{-2}$ & 0.72 \\
   & 2p$^3$($^2$P$^{\rm o}_{3/2}$) $\rightarrow$ 2p$^3$($^2$D$^{\rm o}_{3/2}$) & 7331 & 58.2 & 3.85 $\times$ 10$^{-2}$ & 0.40 \\
   & 2p$^3$($^2$P$^{\rm o}_{1/2}$) $\rightarrow$ 2p$^3$($^2$D$^{\rm o}_{5/2}$) & 7319 & 58.2 & 5.19 $\times$ 10$^{-2}$ & 0.29 \\
   & 2p$^3$($^2$P$^{\rm o}_{1/2}$) $\rightarrow$ 2p$^3$($^2$D$^{\rm o}_{3/2}$) & 7330 & 58.2 & 7.74 $\times$ 10$^{-2}$ & 0.27 \\
   $\left[\rm O~III\right]$ & 2p$^2$($^1$D$_2$) $\rightarrow$ 2p$^2$($^3$P$_1$) & 4959 & 29.2 & 6.21 $\times$ 10$^{-3}$ & 0.710 \\
   & 2p$^2$($^1$D$_2$) $\rightarrow$ 2p$^2$($^3$P$_0$) & 5007 & 29.2 & 1.81 $\times$ 10$^{-2}$ & 1.183
\end{tabular}
\caption{Transition properties for the forbidden lines examined in this study.  The effective collision strengths $\Upsilon$ are for $T$ = 5000 K \citep{Jerkstrand2017hb}.}
\label{tbl:olines}
\end{table*}

\paragraph{[O I] $\lambda \lambda$ 6300, 6364.} The [O I] $\lambda \lambda$ 6300, 6364 line has the lowest excitation temperature of any of the lines studied, meaning that the transition can occur at  lower ejecta temperatures than [O II] and [O III]. However, it has lower $A$ and $\Upsilon$ values, which means that apart from the exponential temperature term, its intrinsic emissivity is weaker than [O II] and [O III] both in LTE and NLTE. With its 
$A/\Upsilon \sim 0.1$, it has a critical density of $n_{\rm crit} = 3.8 \times 10^6$ cm$^{-3}$. At 6y, the electron density of all our models is below this, meaning that the line will form in NLTE.   At 1 year, however, the ejecta electron density is typically higher than $n_{\rm crit}$, so [O I] will then be in LTE.  
The line will be dominant over [O II] and [O III] either when the gas is mostly neutral, and/or when temperatures are low or moderate. For equal ion abundances ($n_{\rm O I} = n_{\rm O II} = n_{\rm O III}$), it becomes stronger than [O II] for $T \lesssim 15,000$ K, and stronger than [O III] for $T \lesssim 3,000$ K.

\paragraph{[O II] $\lambda \lambda$ 7319, 7330.} The [O II] $\lambda \lambda$ 7319, 7330 line has the highest excitation temperature of any of the lines studied, meaning that the transition will be strong only at high ejecta temperatures.  Both its $A$ value and its $\Upsilon$ value are a factor $\sim$ 10 higher than for [O I]: its critical density is therefore about the same, and it will be in LTE at 1y and NLTE at 6y. 
For equal ion abundances it is stronger than [O III] for $T \gtrsim 15,000$ K in LTE, due to its 10 times higher A-value. As its collision strength is similar to [O III], it is always weaker in NLTE.

\paragraph{[O III] $\lambda \lambda$ 4959, 5007.} The [O III] $\lambda \lambda$ 4959, 5007 line has an excitation temperature a little higher than [O I], and the transition has the highest value of $\Upsilon$. With its $A/\Upsilon\sim 0.01$ it has a critical density about a factor 10 lower than [O I] and [O II], therefore staying the longest in LTE.  [O III] will be in LTE at 1 year and just transitioning from LTE to NLTE around 6 years; the critical density is reached in models with $\gtrsim$ 3 $M_{\odot}$ of ejecta. For equal ion abundances, [O III] easily becomes strong in NLTE due to its large value of $\Upsilon$, but still quite low excitation temperature. It would also be comparable to [O I] in LTE due to its somewhat larger $A$ value.

\section{Summary of SN 2012au Observations} \label{sec:12au}

SN 2012au is a luminous SN Ib that was discovered on March 14 2012 \citep{Howerton2012} in NGC 4790 at 23.5 $\pm$ 0.5 Mpc (z = 0.0054 $\pm$ 0.0001).  \citet{Milisavljevic2013}, \citet{Takaki2013}, and \citet{Pandey2021} presented a series of photometric and spectroscopic observations from pre-peak to 1 year post-maximum. These studies estimate a quasi-bolometric peak luminosity of $6-7$ $\times$ 10$^{42}$ erg/s, a factor few larger than a typical SN Ib/c. Arnett-type modelling of the diffusion phase gave best-fitting kinetic energies of $\left(5-10\right)\times 10^{51}$ erg \citep{Milisavljevic2013,Takaki2013,Pandey2021}. 
The studies all estimate a $M_{\rm Ni}$ of $\approx$ 0.3 $M_\odot$ (assuming the diffusion phase is $^{56}$Co-powered) and ejecta masses in the range $3-8$ $M_\odot$. 
\cite{Milisavljevic2013} and \cite{Kamble2014} both find the metallicity of the supernova environment to be $Z$ $\sim 1-2$ $Z_{\odot}$.

SN 2012au showed unusual spectroscopic evolution during its first year.  Spectra taken around peak closely resemble SNe of Type Ib, such as SN 2008D \citep{Soderberg2008}, with prominent He I, Fe II, Si II, Ca II, Na I, and O I absorption features, although with velocities of $\sim$ 2 $\times$ 10$^4$ km s$^{-1}$, much higher than normal SNe Ib.  In the late photospheric phase (about 50 days post-peak), the spectrum more closely resembles a SN Ic such as SN 1997dq \citep{Matheson2001}, although still with higher observed velocities ($\sim$ 7000 km s$^{-1}$) than typical SN Ib/c at this epoch.  At about 110 days post peak, the [O I] and [Ca II] nebular emission lines appear \citep{Takaki2013}, and between then and $\sim$ 250 days post-peak, other emission lines, such as Mg I], Ca II H$\&$K, Na I D, and O I, appear \citep{Milisavljevic2013}. The nebular spectra also show persistent Fe II P-Cyg absorptions at $\lesssim$ 2000 km s$^{-1}$ and an iron peak plateau between 4000 and 5600 \AA, and the widths of the emission lines indicate two distinct emission regions, with Mg I] $\lambda$ 4571; Na I $\lambda \lambda$ 5890, 5896; [O I] $\lambda \lambda$ 6300, 6364; and [Ca II] $\lambda \lambda$ 7291, 7324 all showing full-width half-maximum (FWHM) velocities $\gtrsim$ 4500 km s$^{-1}$, and O I $\lambda$ 7774, O I 1.317$\mu$m, and Mg I 1.503$\mu$m all showing FWHM velocities $\approx$ 2000 km s$^{-1}$. These features are more comparable to SLSN spectra at similar epochs, such as SN 2007bi \citep{Gal-Yam2009}.  Large imaging polarization values indicate asymmetry in the ejecta \citep{Pandey2021}. 

\cite{Kamble2014} performed radio observations with the Jansky Very Large Array (VLA) between 5-37 GHz and X-ray observations with the \textit{Swift} X-ray Telescope (XRT) between 0.3-10 keV within the first two months post-explosion.  Emission was detected by both instruments, with the bright radio emission having a flux between 10-100 mJy and and a radio energy budget of $\sim$ 10$^{47}$ erg, intermediate between SN 1998bw \citep{Kulkarni1998, Li1999} and SN 2002ap \citep{Berger2002}, two similarly energetic supernovae \citep{Iwamoto1998, Mazzali2002}.  The shock wave was estimated to have a velocity of 0.2$c$ and a mass-loss rate of $\dot{M} = 3.6 \times 10^{-6} M_\odot$ yr$^{-1}$ for the wind velocity of 1000 km s$^{-1}$ that was inferred for the progenitor.  The lack of sharp features or discontinuities in the radio light curves suggest relatively smooth mass loss in the final few decades before the supernova.

\cite{Milisavljevic2018} were able to obtain another optical spectrum at $\sim$ 6 years post-explosion; which showed only [O I]; [Ca II]/[O II]; [O III]; and [S III] $\lambda \lambda$ 9069, 9531; with a possible detection of a broad but weak O I $\lambda$ 7774 line. From 1y to 6y, the [O I] and [Ca II]/[O II] lines both narrowed from $\gtrsim$ 4500 km s$^{-1}$ to $\sim$ 2300 km s$^{-1}$, while maintaining roughly the same peak intensity ratio, while the O I line broadened significantly from $\sim$ 2000 km s$^{-1}$ while decreasing in peak intensity more significantly compared to [O I] and [Ca II]/[O II].  \cite{Milisavljevic2018} argue that this spectrum is inconsistent with those predicted by either radioactivity or SN-CSM interaction \citep[e.g.][]{Chugai2006, Milisavljevic2015, Mauerhan2018}, but can be consistent with PWN powering models \citep[e.g.][]{Chevalier1992}.  
\cite{Milisavljevic2018} also obtained X-ray upper limits of $L_X < 2 \times 10^{38}$ erg s$^{-1}$ between 0.5-10 keV using the Chandra X-ray Observatory (CXO). 

\cite{Stroh2021} were able to detect radio emission from SN 2012au at $\sim$ 7 years post-explosion with the VLA Sky Survey (VLASS).  The signal was in the 2-4 GHz band and had a flux of 4.5 $\pm$ 0.3 mJy.  They considered emission from CSM interaction, off-axis jets, and emergent PWNe, and concluded that the emission was likely from a PWN, similar to \cite{Milisavljevic2018}.

Other supernovae have had spectra taken at very-late times as well \citep[for a review, see][]{Milisavljevic2012}, but most of them, with the exception of the Type IIb SN 1993J \citep{Filippenko1993} and the Type II-pec SN 1987A \citep{Menzies1987, Chugai1996} are Type IIn or Type IIL SNe \citep[for a review of their spectral evolution, see][]{Chevalier1994}, and all of them show clear signs of interaction with circumstellar material.  SN 2012au is unique as the only stripped envelope supernova with spectra taken at very late times, as well as the only one without clear signs of strong CSM interaction.

\section{Results} \label{sec:res}

We present grids of one-zone models for the pure oxygen and realistic compositions, as well as a multi-zone model for one of our highest optical depth models.  The overall goodness of fit for our models is evaluated with a model score

\begin{equation}
\text{Score} = \sum_i \left(\log\left[\frac{L_{\rm i, model}}{L_{\rm i, obs}}\right]\right)^2   
\label{eqn:modscore}
\end{equation}
where $L_i$ is the line luminosity, with $i$ running over all the lines used to generate the score.  We use a logarithm so that brighter lines do not overwhelm the score.  A perfect match for a line will give a score of 0, and factors of 2, 5, and 10 difference in luminosity will give scores of $\sim$ 0.1, $\sim$ 0.5, and 1, respectively. We study line luminosities as opposed to line ratios, which are insensitive to distance uncertainties, because certain ejecta properties do not scale simply with certain parameters; for example, increasing $L_{\rm PWN}$ to increase the luminosity of two lines would also change the ionization structure of the ejecta \citep[e.g][]{Jerkstrand2017}.

The line luminosity is calculated for both observations and models using \citep{Jerkstrand2017}

\begin{equation}
    L_{\rm line} = I_1 - \frac{I_2 - I_1}{d\lambda_2/d\lambda_1 -1},
    \label{eqn:linelum}
\end{equation}
where $I_1$ and $I_2$ are defined as the integral of the flux

\begin{equation}
    I_n = \int^{\lambda_0+d\lambda_n}_{\lambda_0-d\lambda_n}F_\lambda d\lambda
\end{equation}
with $\lambda_0$ and $d\lambda_1$ being chosen to enclose as much line flux as possible without contamination from other lines in both the model and observation, and we use $d\lambda_2 = 1.25d\lambda_1$.  Values of $\lambda_0$ and $d\lambda_1$ are unique to each line and each epoch, but shared between model and observation. The values of $d\lambda_1$ are given in Table \ref{tbl:dlam}.

\begin{table}
\centering
\begin{tabular}{ccc}
   Line & $d\lambda_{\rm 1, 6y}$ ($\AA$)& $d\lambda_{\rm 1, 1y}$ ($\AA$)  \\ \hline
   $\left[\rm O~I\right]$ & 130 & 300 \\
   $\left[\rm O~II\right]$ & 130 & 300 \\
   $\left[\rm O~III\right]$ & 130 & 100 \\
   O I & 180 & 100 \\
\end{tabular}
\caption{The values of $d\lambda_1$ for each line and epoch used to determine the observed and model line luminosities.}
\label{tbl:dlam}
\end{table}

The model spectra at 6 years are compared to the spectrum from \cite{Milisavljevic2018}, which has clear detections of forbidden oxygen transitions [O I] $\lambda \lambda$ 6300, 6364 (hereafter referred to as [O I]), either [O II] $\lambda \lambda$ 7319, 7330 or [Ca II] $\lambda \lambda$ 7291, 7323 (hereafter referred to as [O II] and [Ca II]), and [O III] $\lambda \lambda$ 4959, 5007 (hereafter referred to as [O III]), and a possible weak detection of O I $\lambda$ 7774 (hereafter referred to as O I).  
The model spectra is not corrected dust extinction along the line of sight (which has $E(B - V ) < 0.1$ mag \citep{Milisavljevic2013}); however, the model spectra are adjusted for extinction from dust formed within the supernova by dividing the line luminosities by factors of 3.31, 3.37, and 2.10 for [O I], [O II], and [O III] respectively \citep{Niculescu-Duvaz2022}; the flux of the O I line was too low for strong constraints to be placed on dust formation, so we assume this line is unaffected by dust.  The spectra at 1 year are compared to observations from \cite{Milisavljevic2013}, focusing mostly on the [O I] and O I lines, since [O III] is not clearly detected and [Ca II] probably contributes to the [O II] line.

\subsection{Pure Oxygen Composition}

\subsubsection{6 Years}

\begin{figure*}
\newcolumntype{D}{>{\centering\arraybackslash} m{6cm}}
\noindent
\makebox[\textwidth]{
\begin{tabular}{m{0.8cm} DDD}
& \boldsymbol{$T_{\rm PWN} = 10^5$} \textbf{ K} & \boldsymbol{$T_{\rm PWN} = 3 \times 10^5$} \textbf{ K} & \boldsymbol{$T_{\rm PWN} = 10^6$} \textbf{ K}\\
\textbf{O I} &
\includegraphics[width=1.1\linewidth]{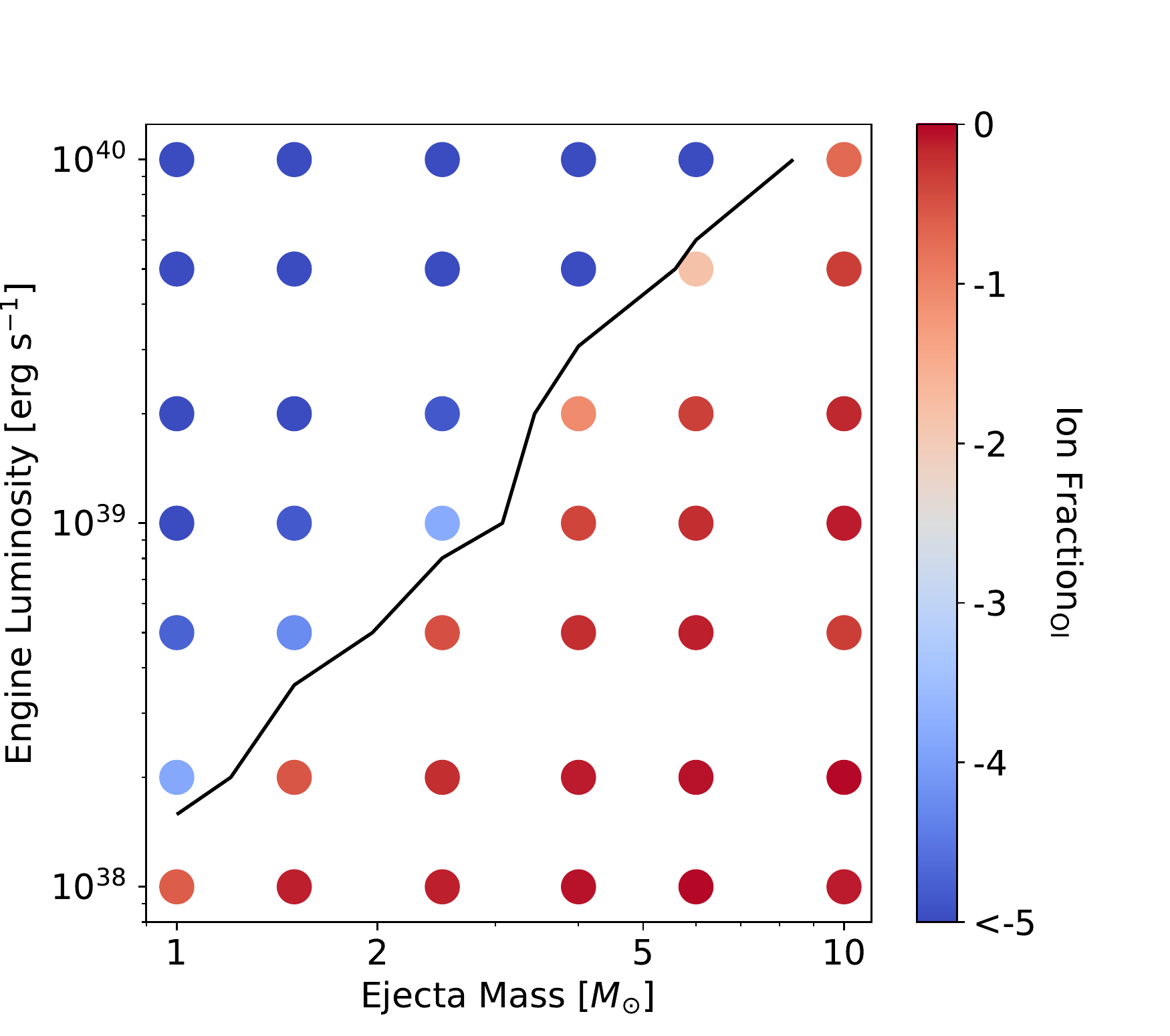}&
\includegraphics[width=1.1\linewidth]{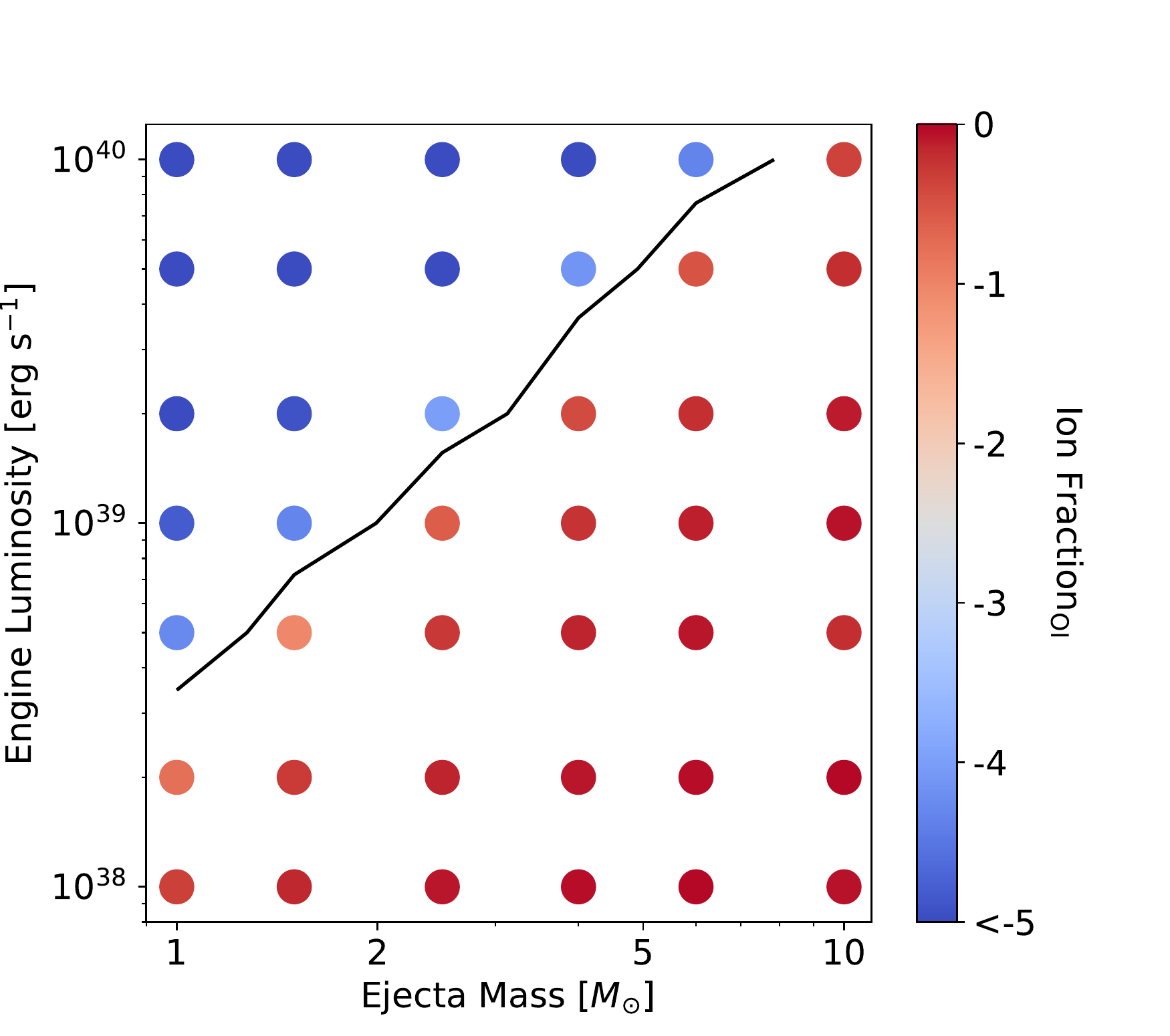}&
\includegraphics[width=1.1\linewidth]{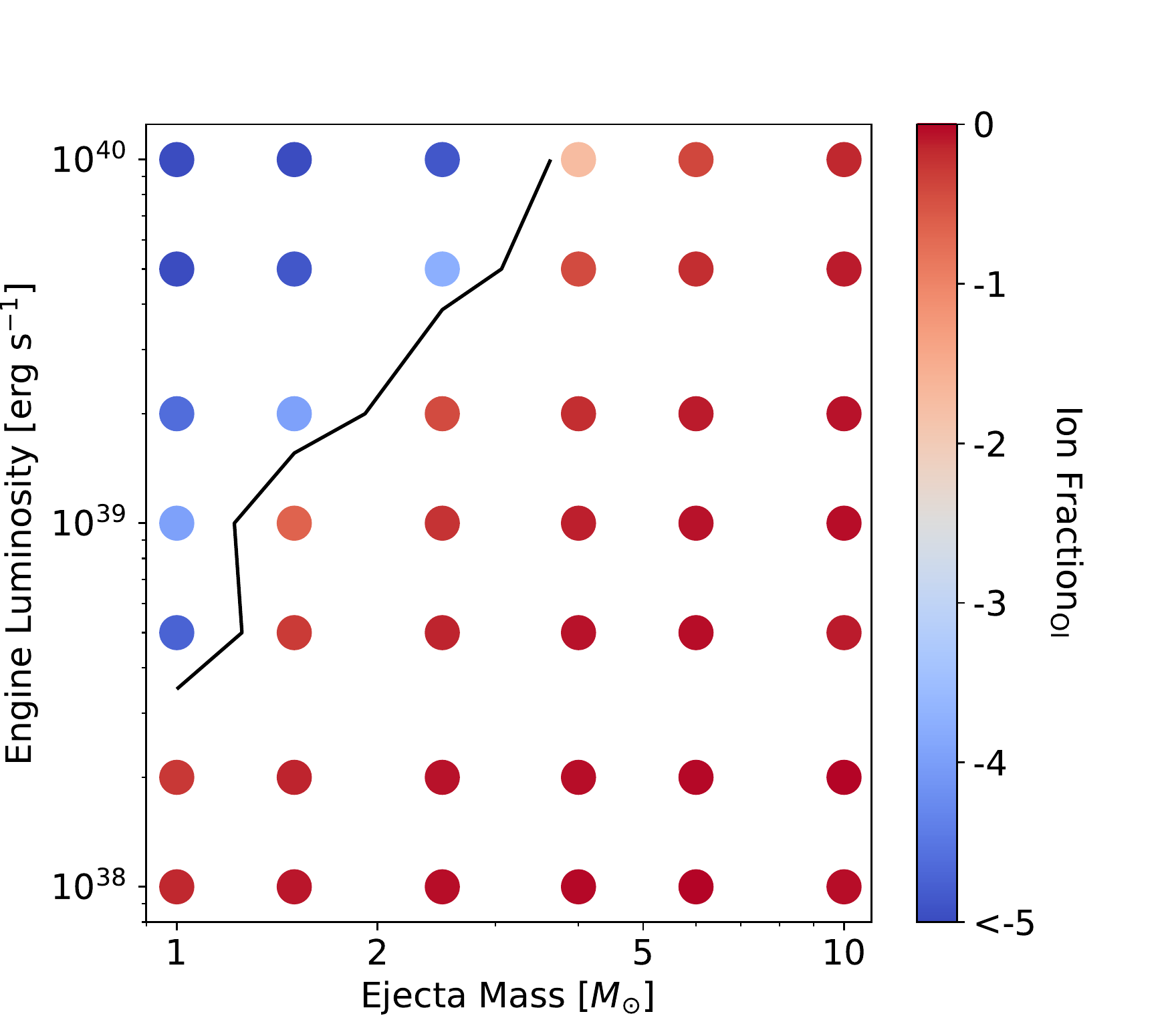}\\[-1.5ex]
\textbf{O II}&
\includegraphics[width=1.1\linewidth]{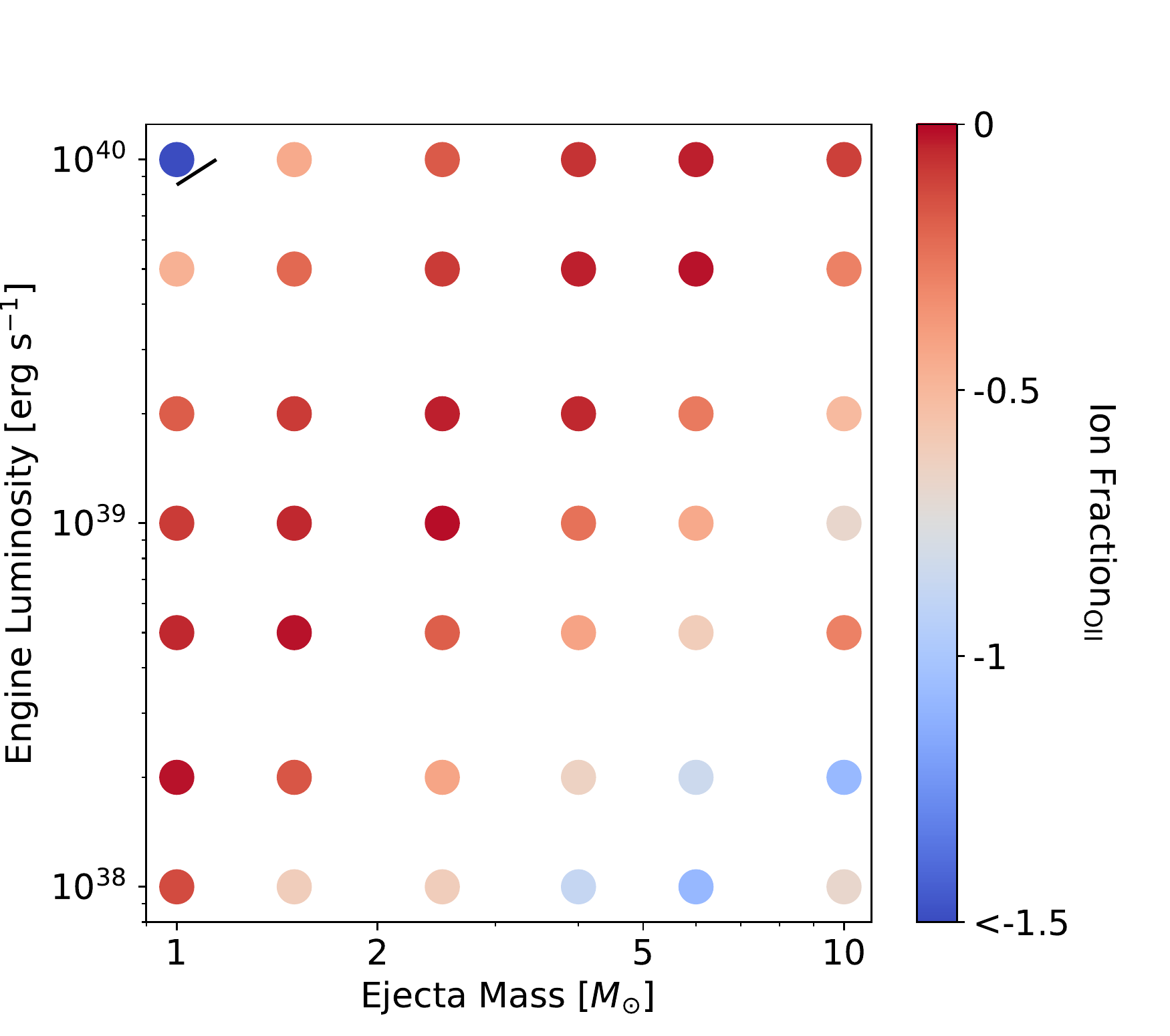}&
\includegraphics[width=1.1\linewidth]{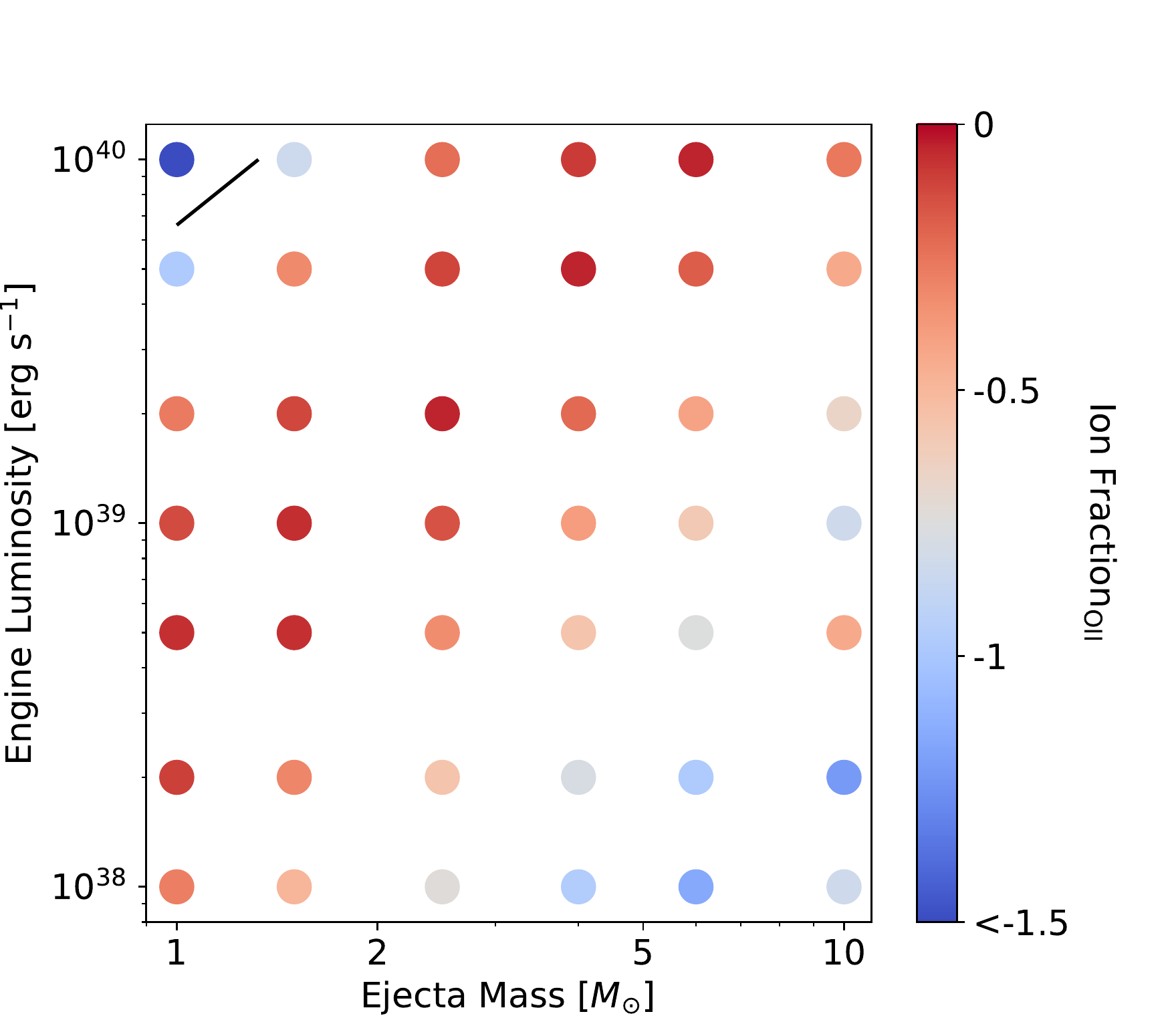}&
\includegraphics[width=1.1\linewidth]{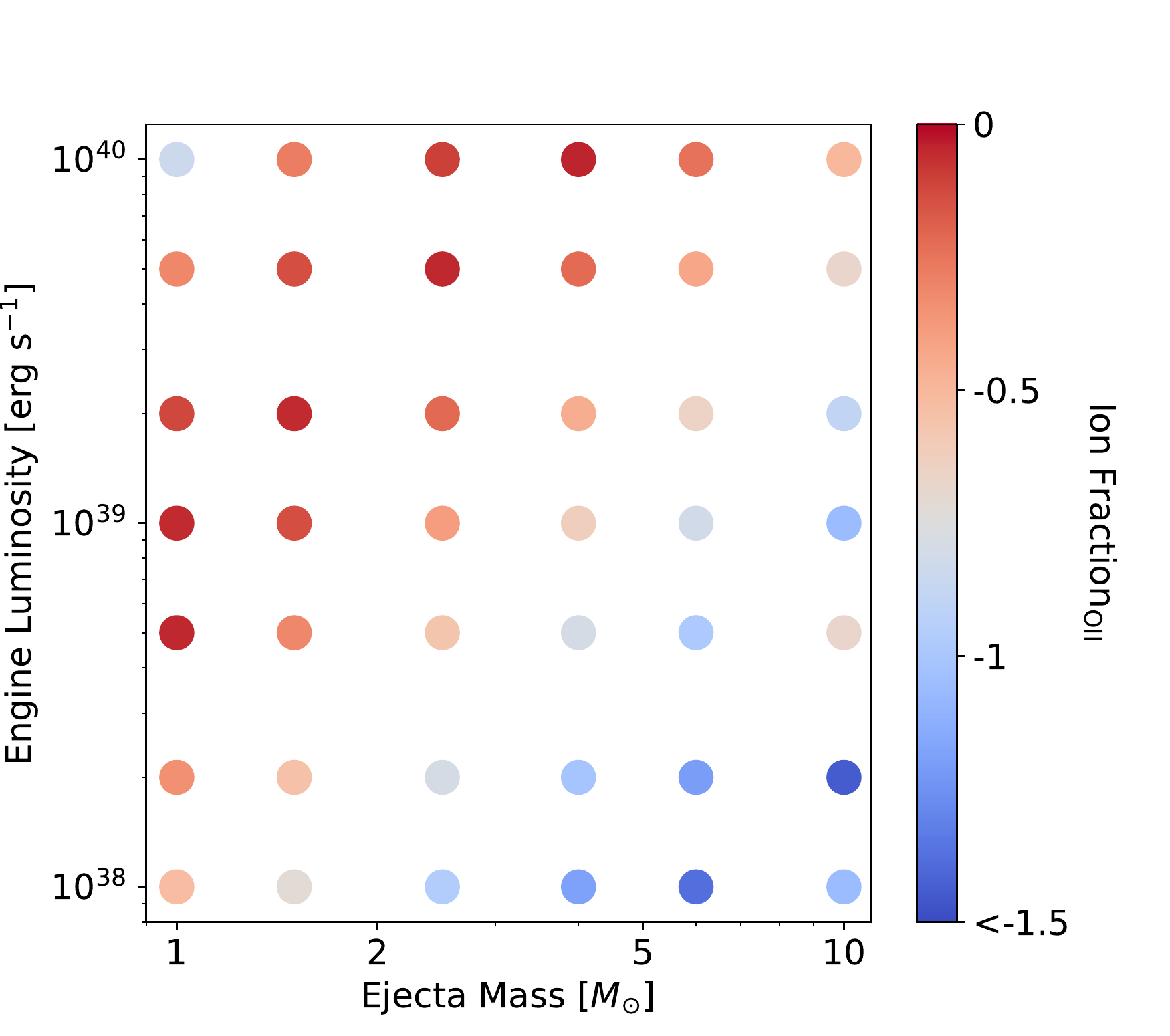}\\[-1.5ex]
\textbf{O III}&
\includegraphics[width=1.1\linewidth]{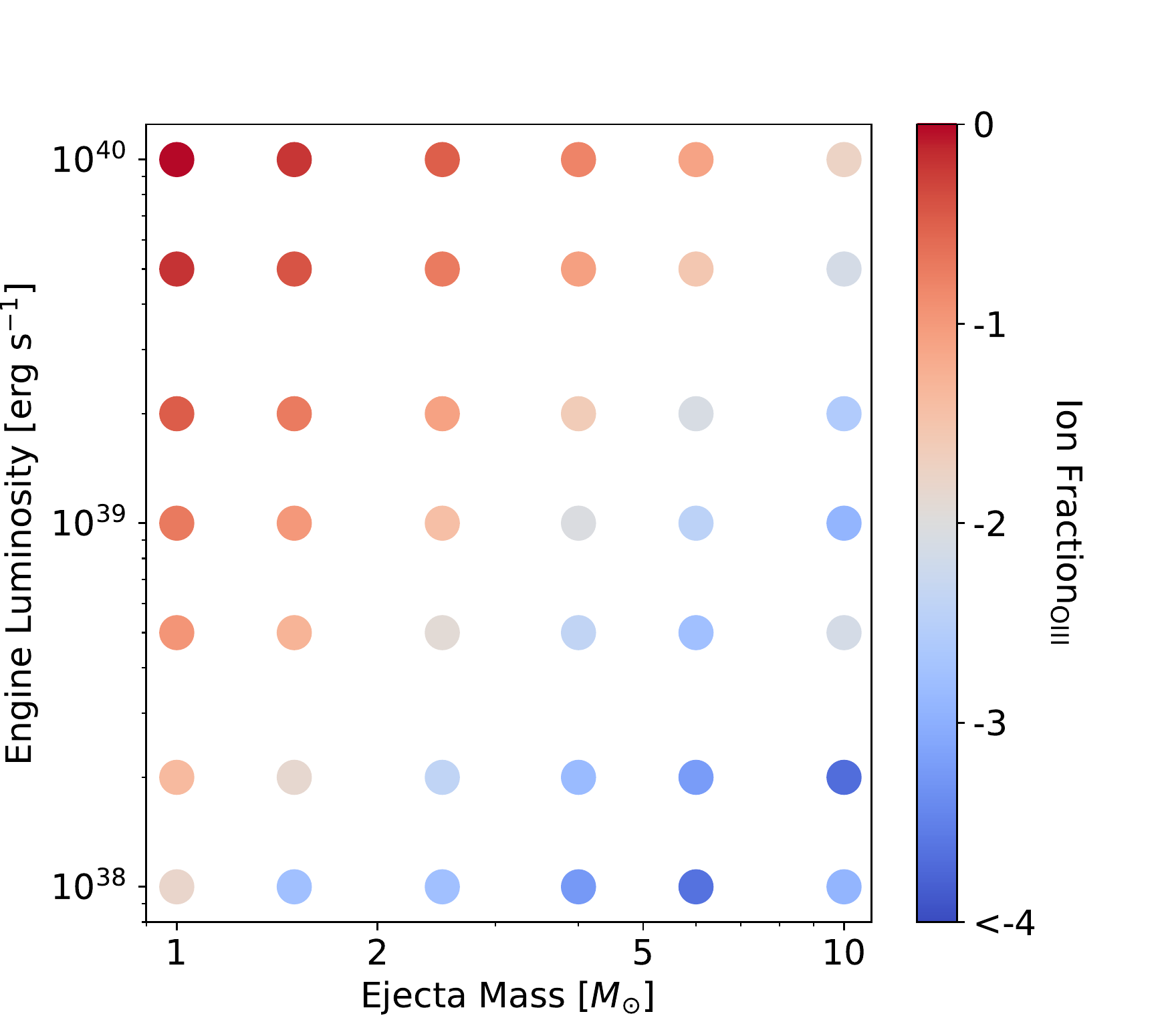}&
\includegraphics[width=1.1\linewidth]{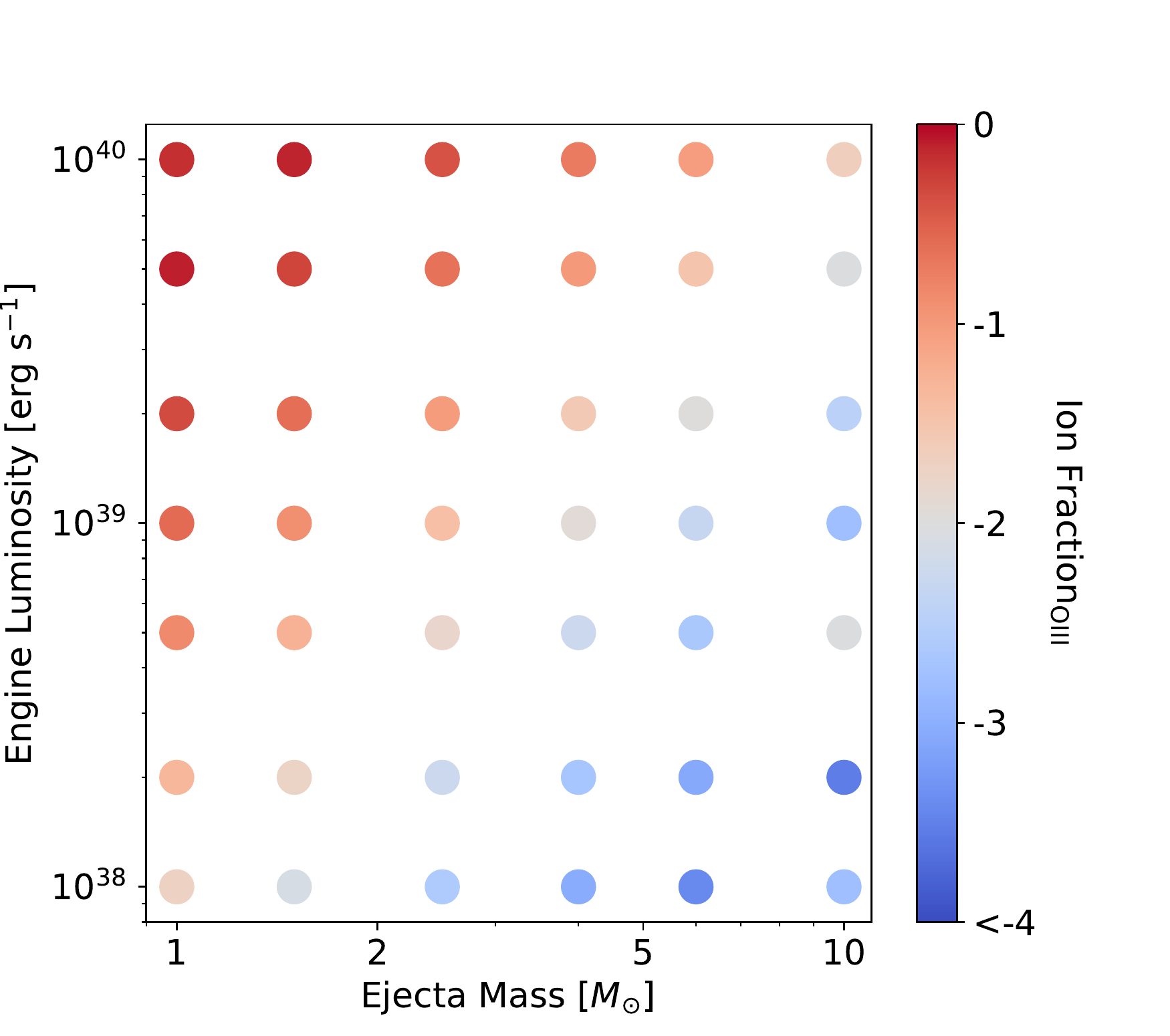}&
\includegraphics[width=1.1\linewidth]{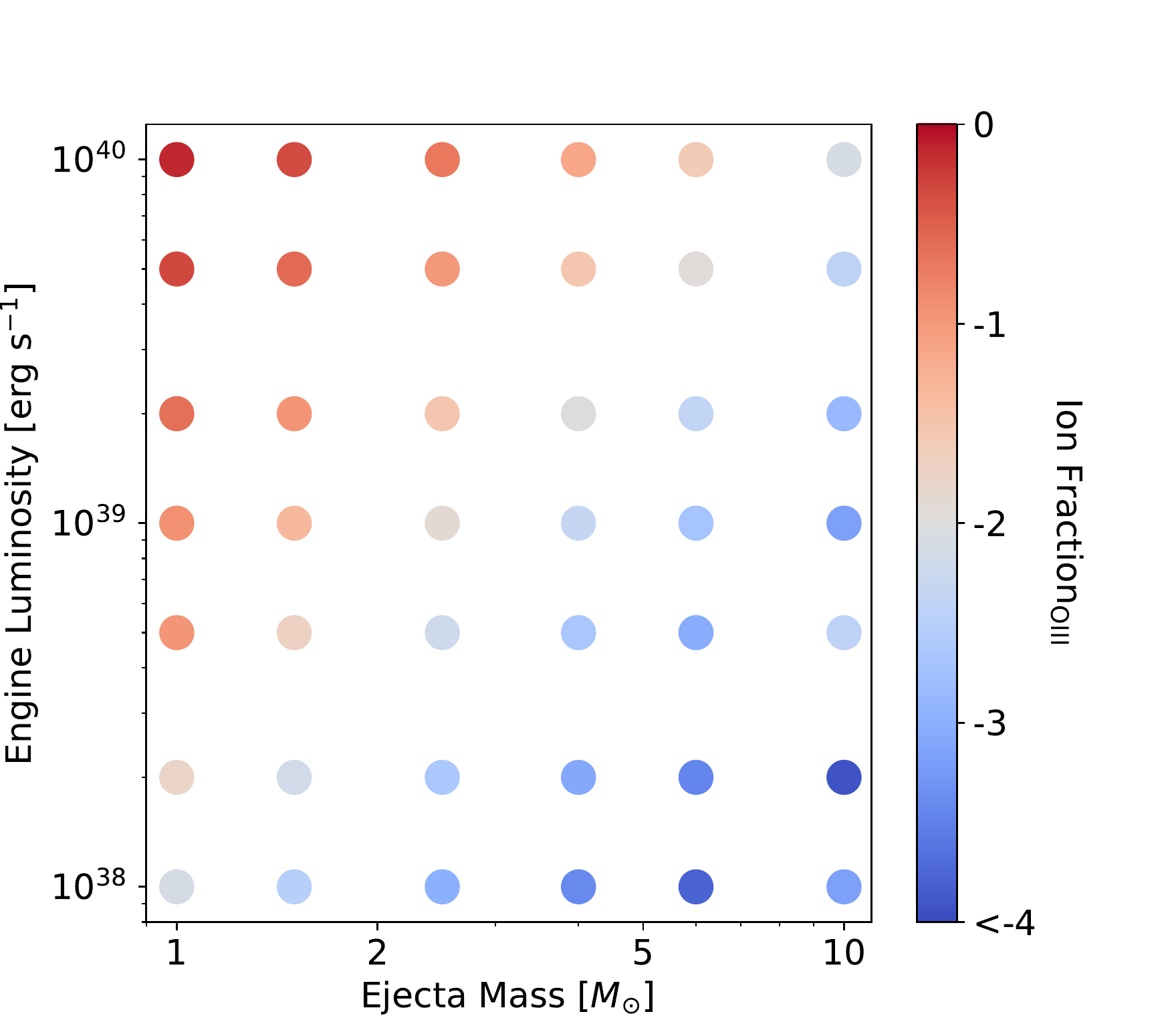}\\[-1.5ex]
\boldsymbol{$T_{\rm ej}$}&
\includegraphics[width=1.1\linewidth]{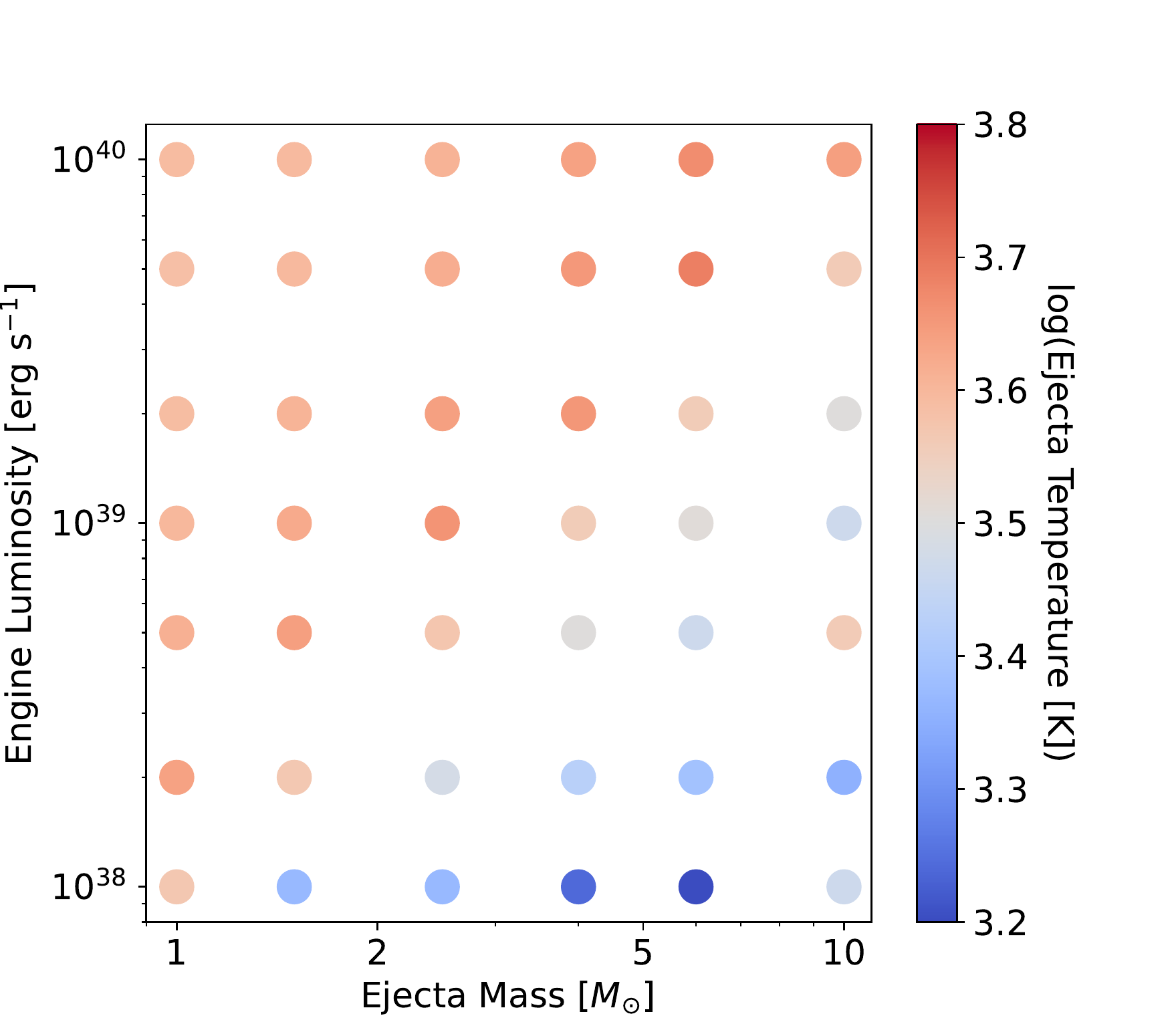}&
\includegraphics[width=1.1\linewidth]{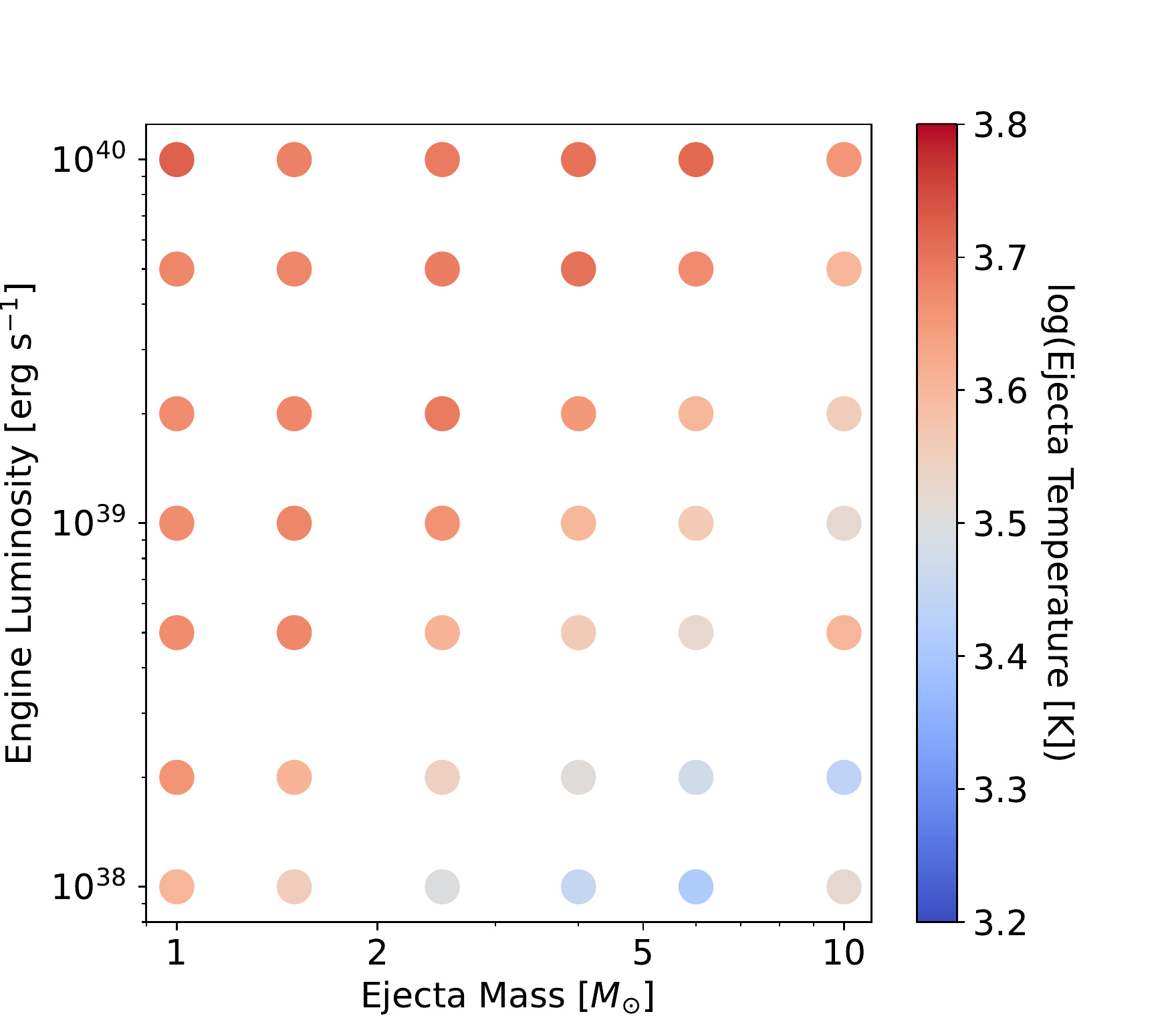}&
\includegraphics[width=1.1\linewidth]{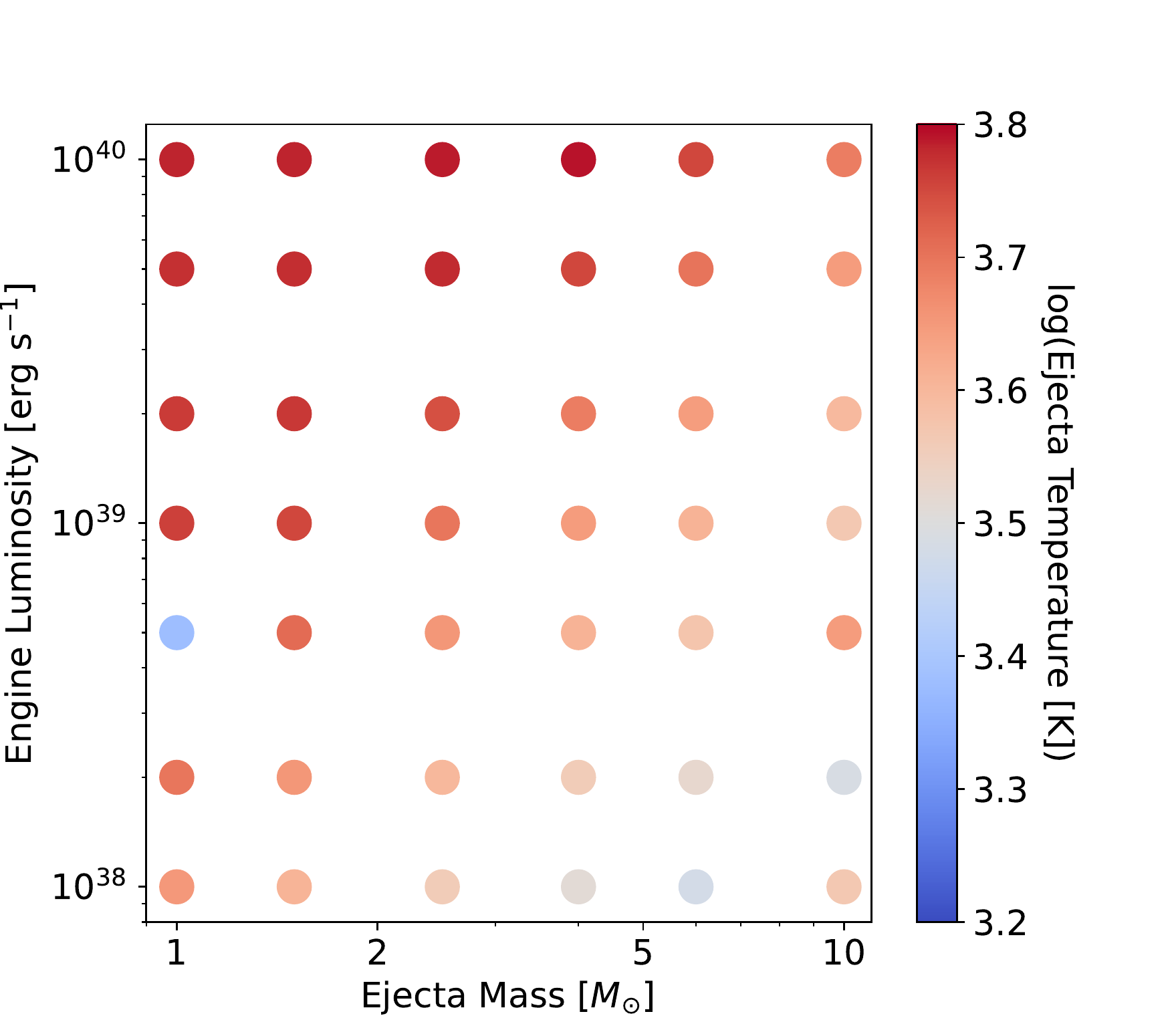}\\[-1.5ex]
\end{tabular}}
\caption{The ion fractions of O I (top), O II (second row), and O III (third row), and the ejecta temperature $T_{\rm ej}$ (bottom) in the simulations at 6 years for a pure oxygen composition at three different values of $T_{\rm PWN}$.  The black contour denotes the low ejecta mass, high engine luminosity regime where runaway ionization can occur for both O I and O II.}%
\label{fig:o6y_ionfrac}
\end{figure*}

The ejecta temperature and ion fractions for O I, O II, and O III over the grid are shown in Figure \ref{fig:o6y_ionfrac}.  Models with large ejecta mass and low engine luminosity are dominated by O I, and increasing luminosity and decreasing the ejecta mass shift the ionization balance more towards O II, which is more prominent in the middle of the grid, and O III, which is most prominent at low ejecta masses and high engine luminosities.  

There is a sharp decrease in O I fraction as engine luminosity increases and ejecta mass decreases; this is due to runaway ionization, a sharp transition in the ionization balance solution as the radiation field strength exceeds a critical value \citep{Jerkstrand2017}; this is related but not identical to ionization breakout \citep{Metzger2014}, which refers to when the whole ejecta become optically thin at some wavelength due to ionization. 
Runaway ionization for O II can also be seen at the highest engine luminosity and lowest mass in the model grid, leading to an ejecta that is dominated by O III.  Since the photons at low $T_{\rm PWN}$ are already energetic enough to ionize both O I and O II (a $10^5$ K blackbody peaks at 24 eV, and the PWN will have enough photons more energetic than their ionization potentials of 13.6 and 35 eV, respectively), the decrease in ionizing photon count as $T_{\rm PWN}$ increases causes runaway ionization to occur only in models with comparatively higher engine luminosities and lower ejecta masses.

\begin{figure*}
\newcolumntype{D}{>{\centering\arraybackslash} m{6cm}}
\noindent
\makebox[\textwidth]{
\begin{tabular}{m{1cm} DDD}
& \boldsymbol{$T_{\rm PWN} = 10^5$} \textbf{ K} & \boldsymbol{$T_{\rm PWN} = 3 \times 10^5$} \textbf{ K} & \boldsymbol{$T_{\rm PWN} = 10^6$} \textbf{ K}\\
\textbf{[O I]}&
\includegraphics[width=1.1\linewidth]{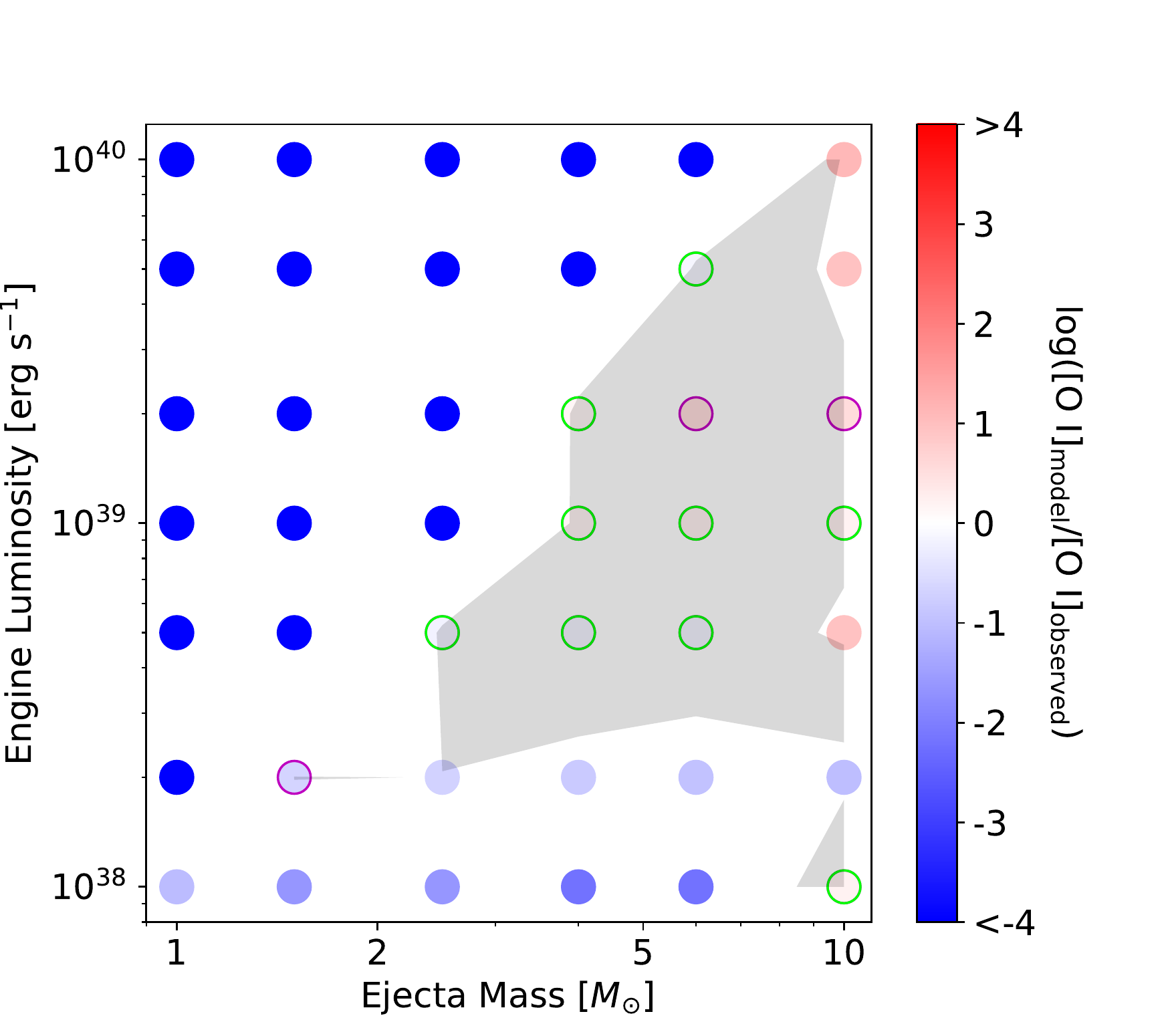}&
\includegraphics[width=1.1\linewidth]{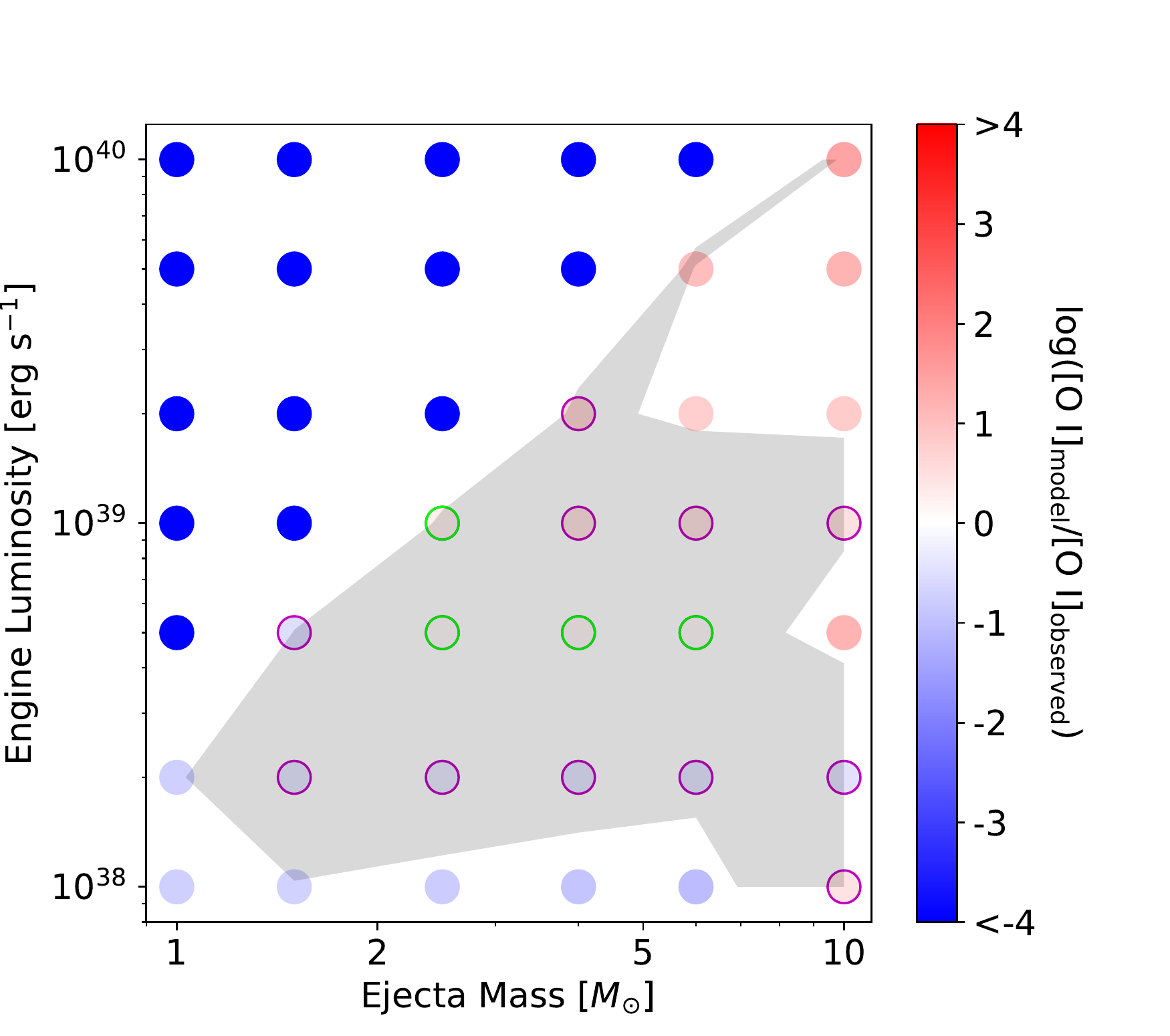}&
\includegraphics[width=1.1\linewidth]{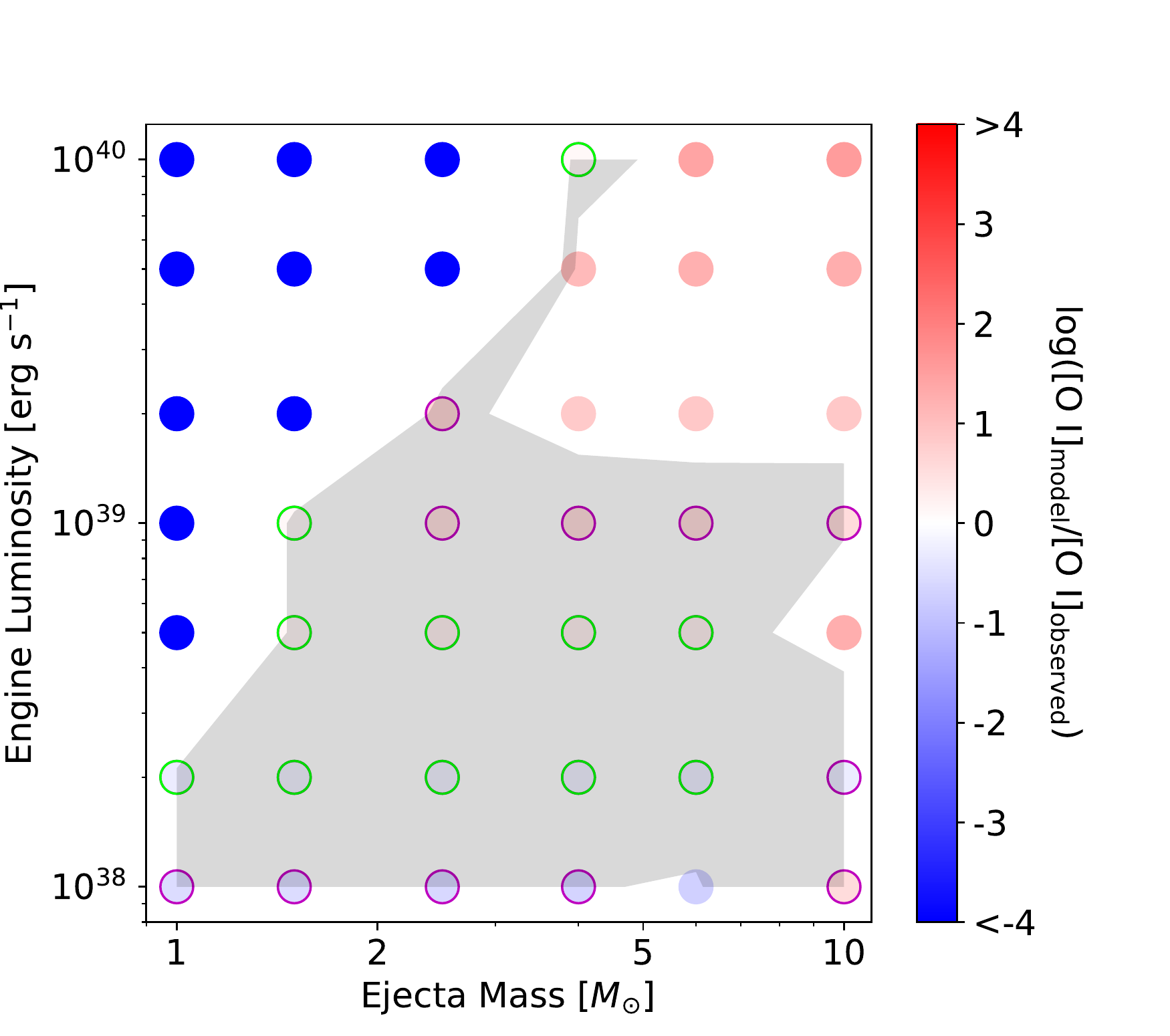}\\[-1.5ex]
\textbf{[O II]}&
\includegraphics[width=1.1\linewidth]{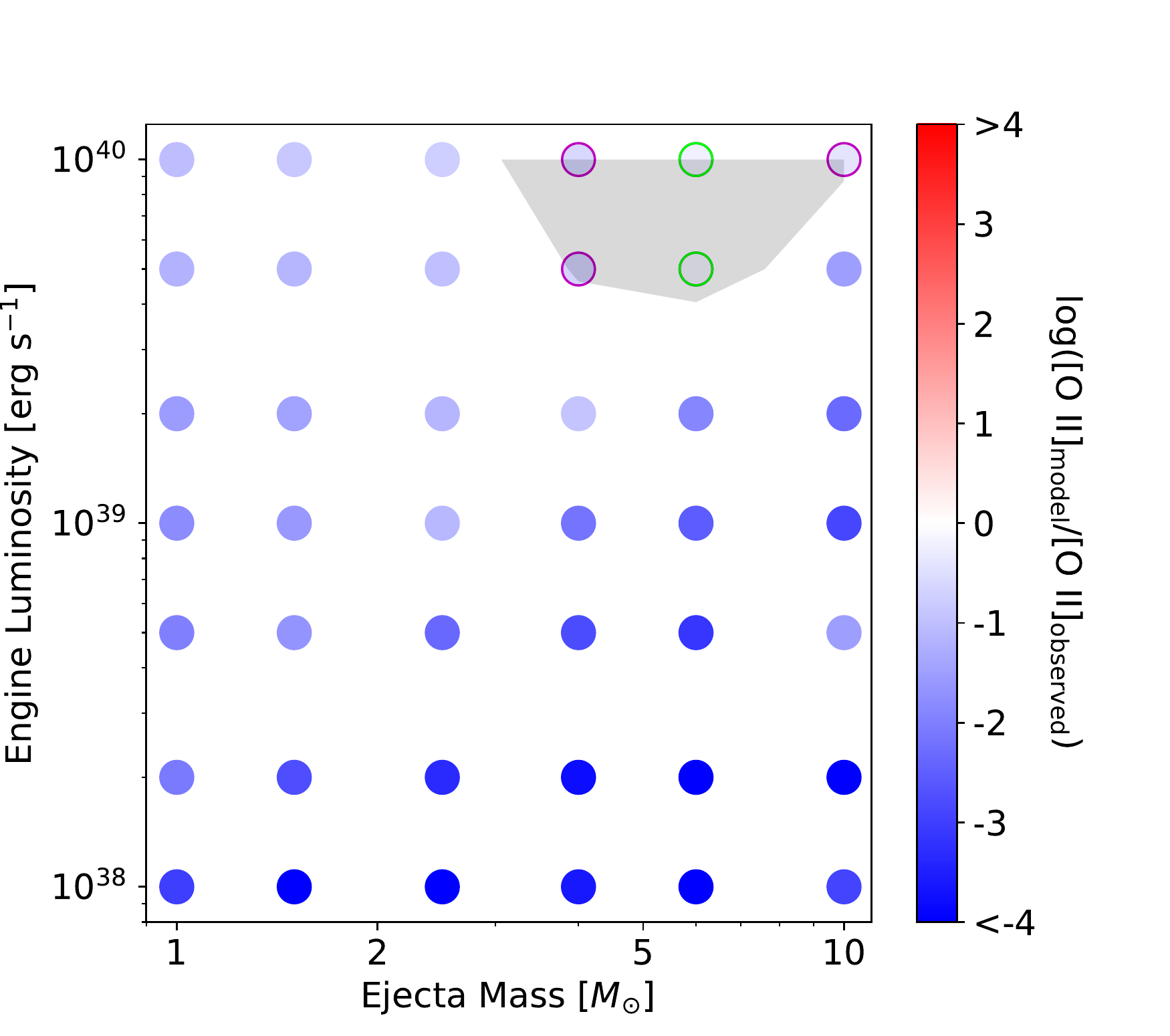}&
\includegraphics[width=1.1\linewidth]{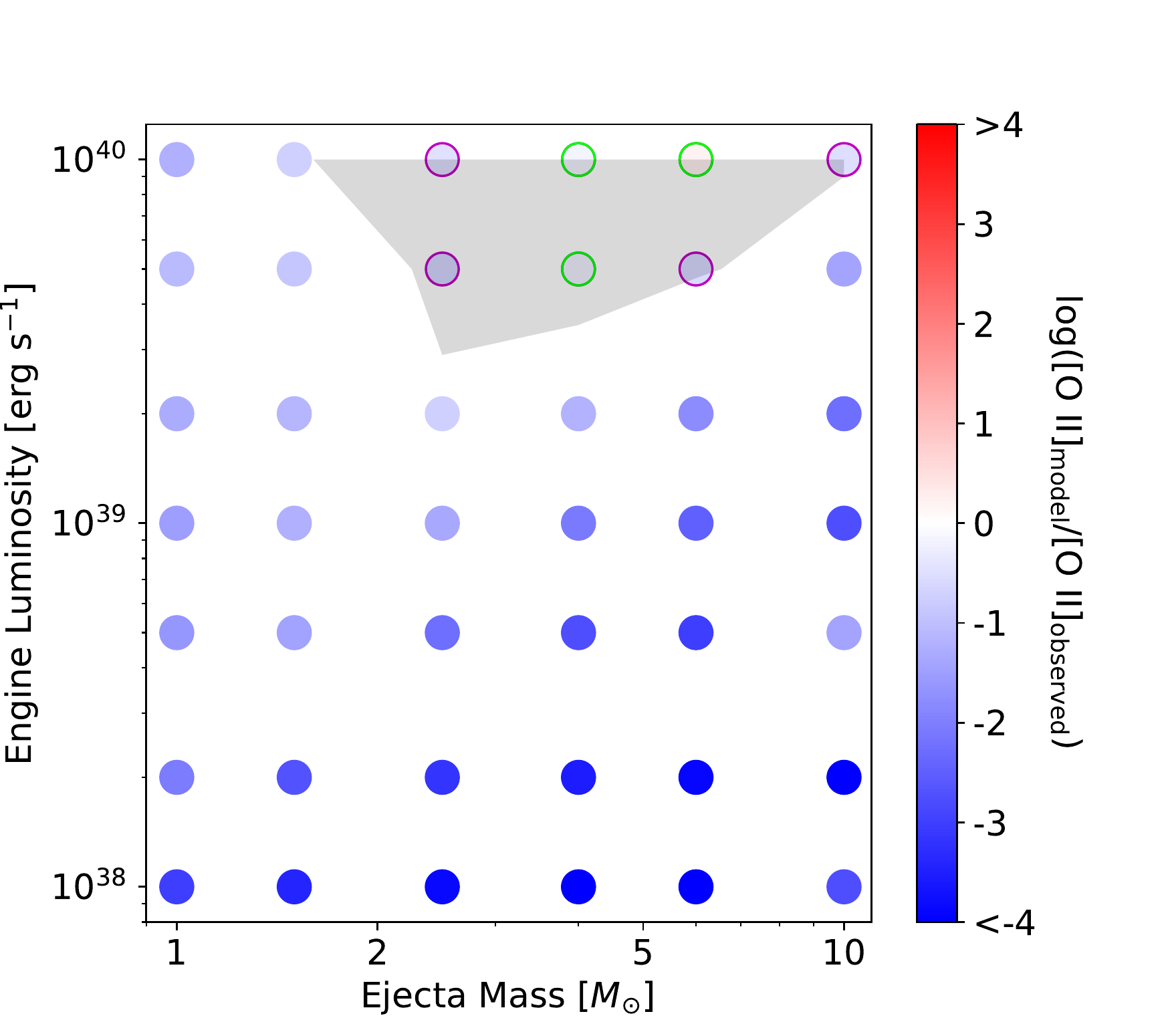}&
\includegraphics[width=1.1\linewidth]{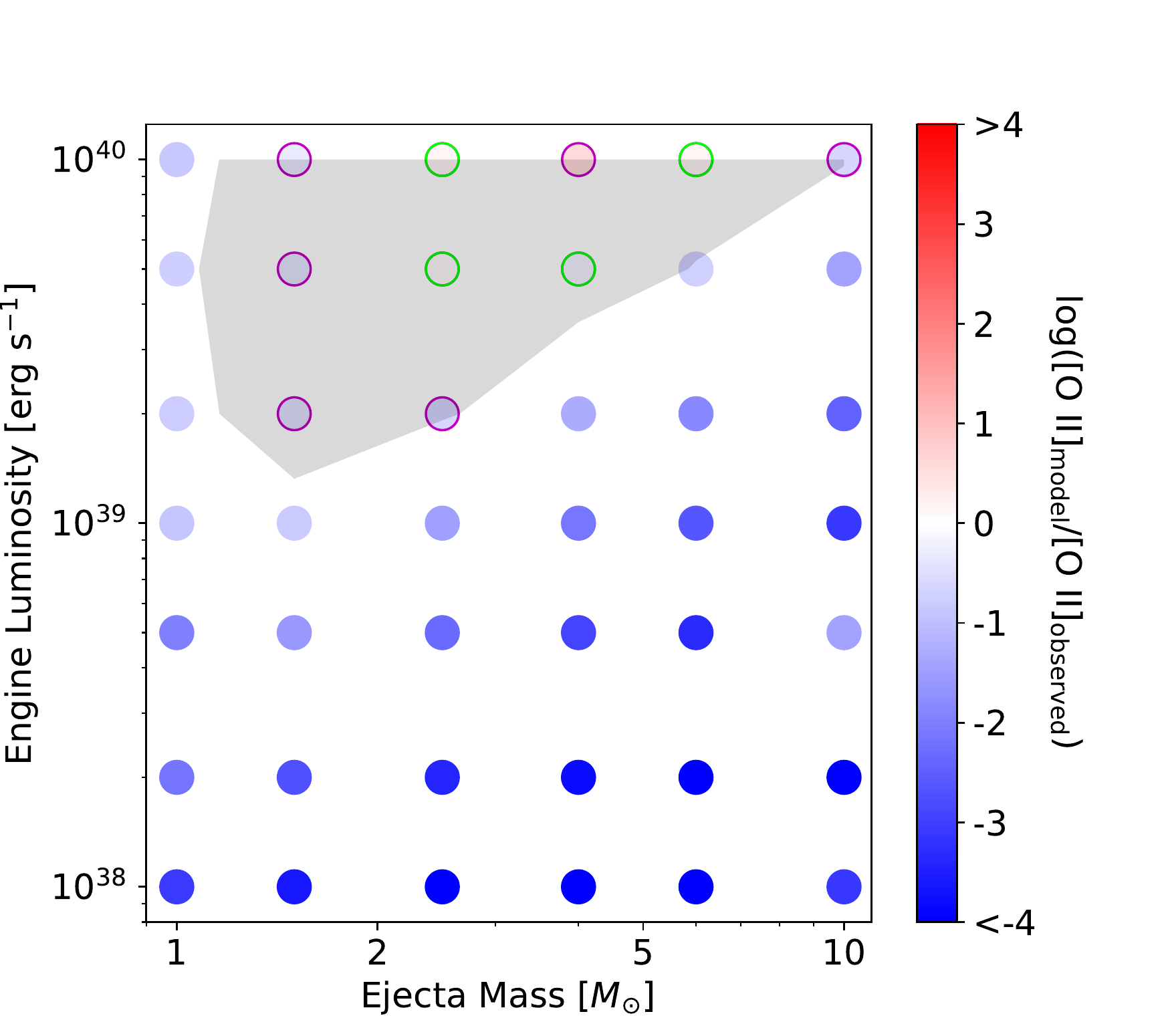}\\[-1.5ex]
\textbf{[O III]}&
\includegraphics[width=1.1\linewidth]{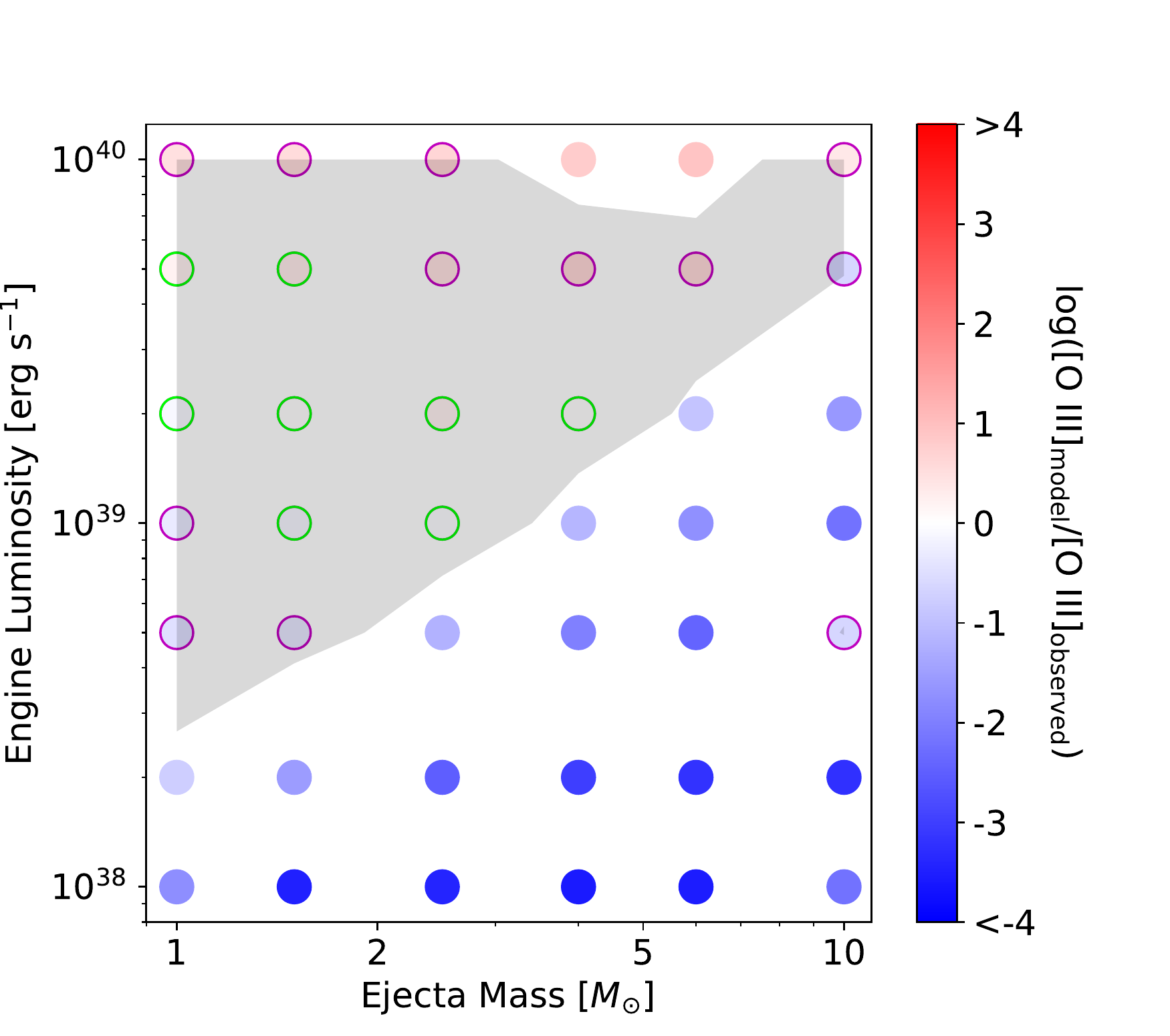}&
\includegraphics[width=1.1\linewidth]{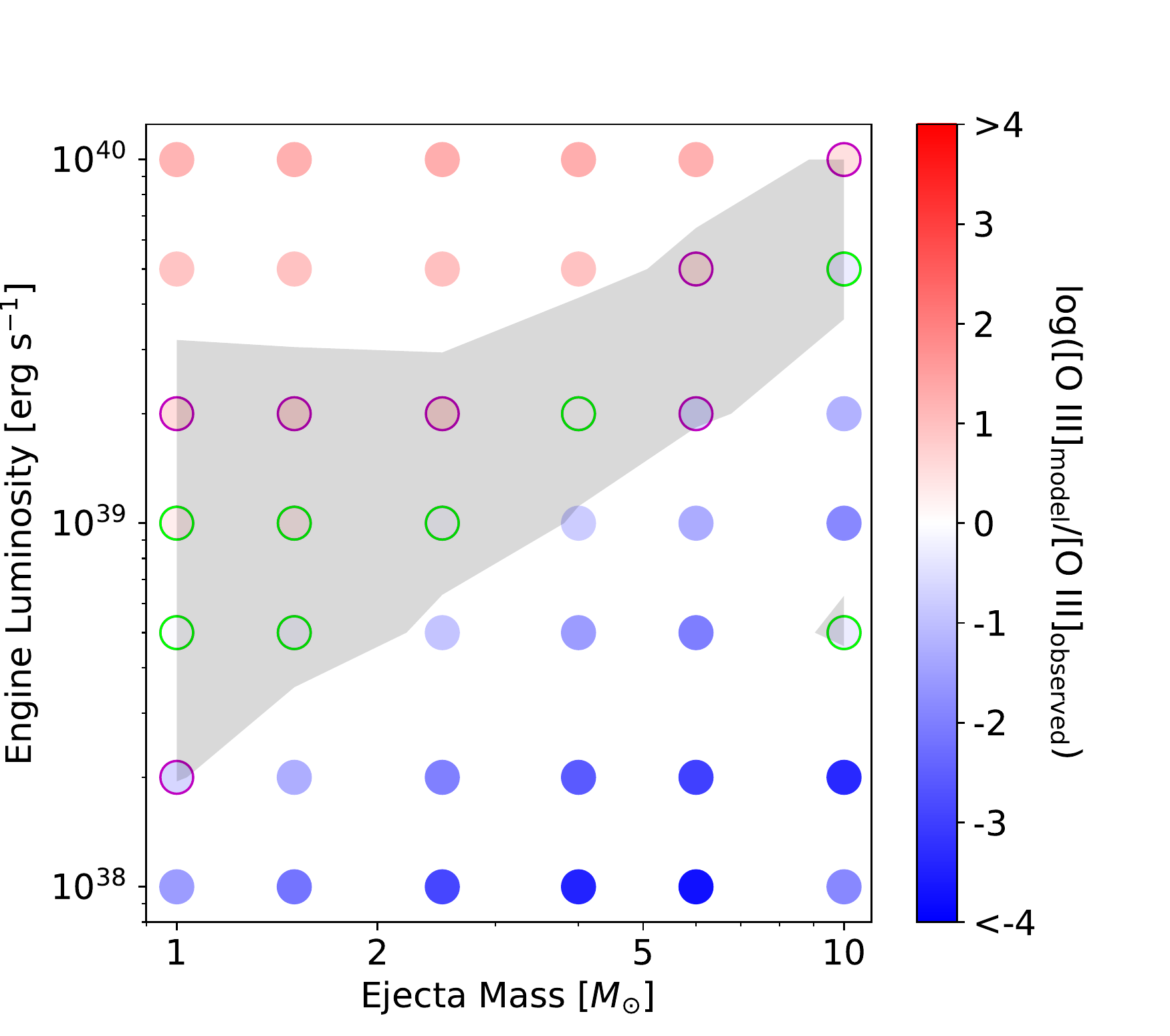}&
\includegraphics[width=1.1\linewidth]{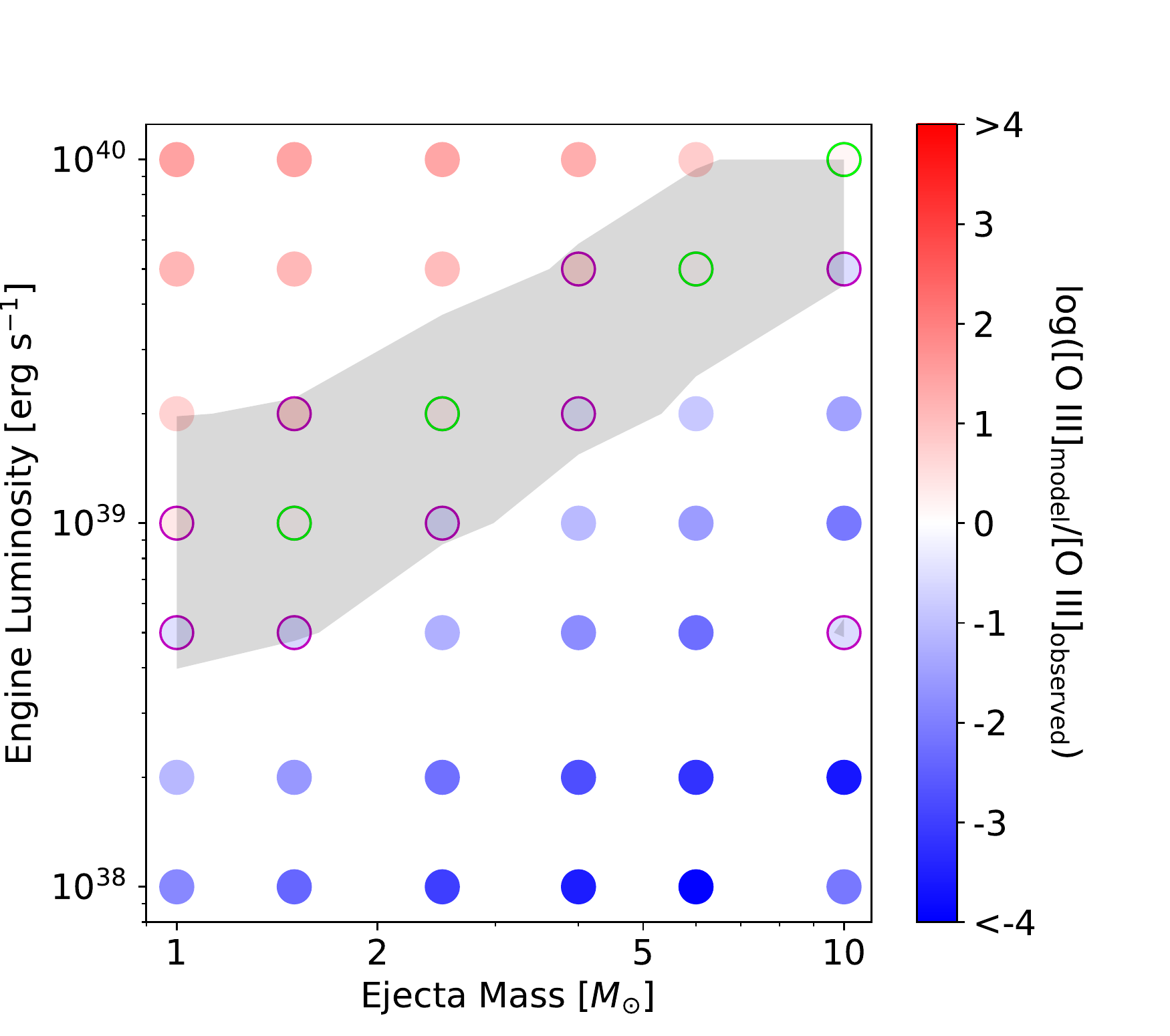}\\[-1.5ex]
\textbf{O I}&
\includegraphics[width=1.1\linewidth]{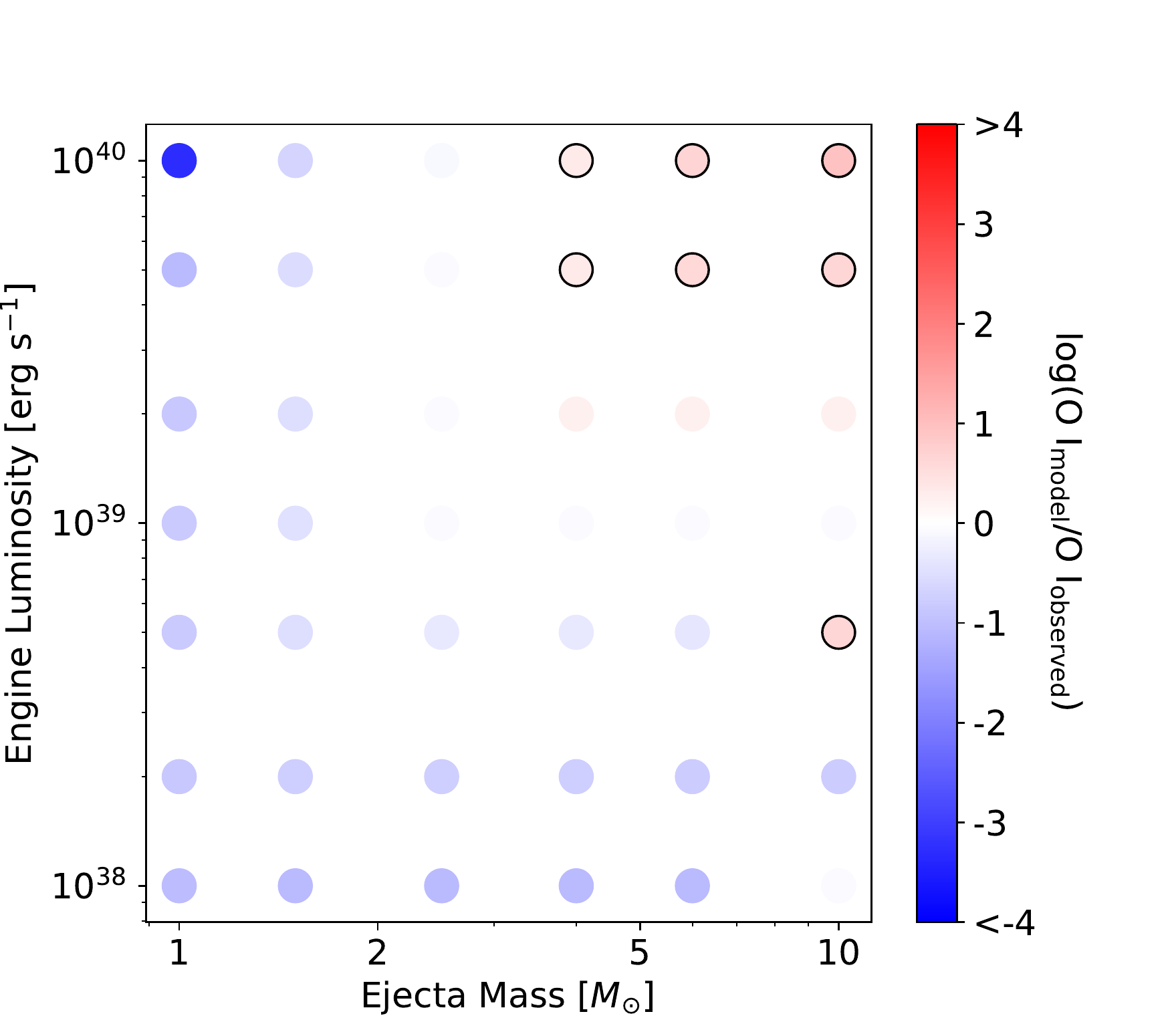}&
\includegraphics[width=1.1\linewidth]{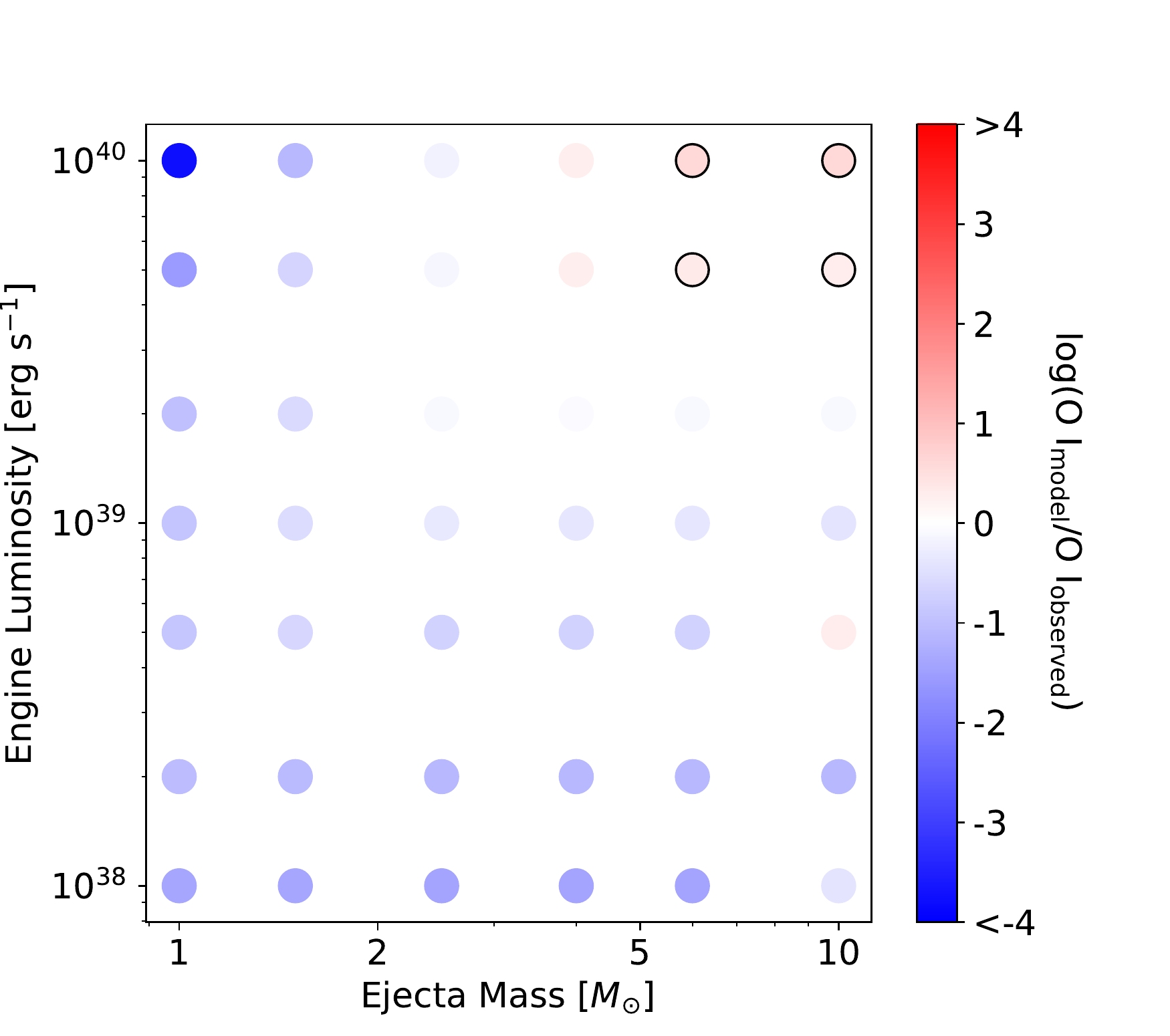}&
\includegraphics[width=1.1\linewidth]{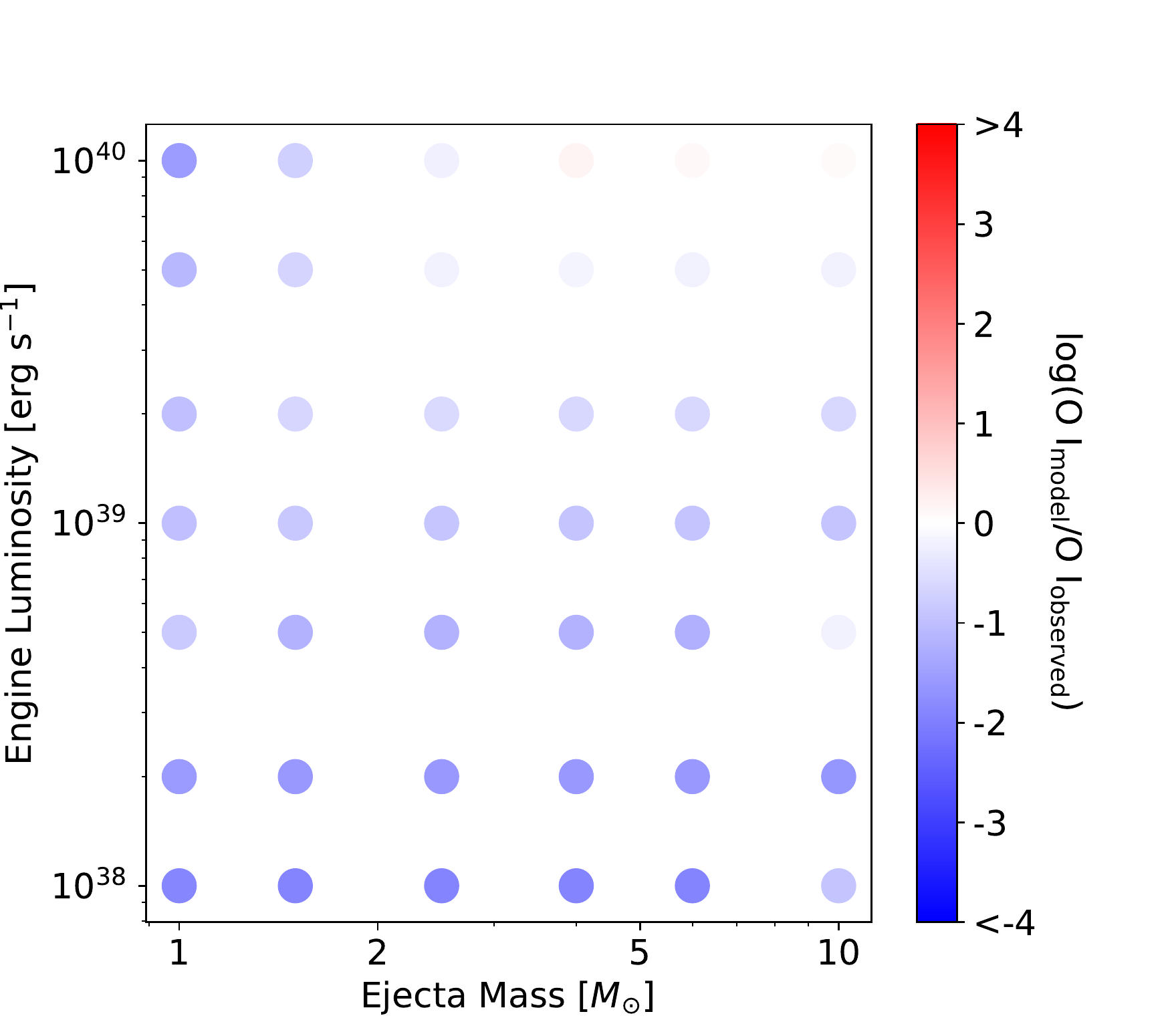}\\[-1.5ex]
\end{tabular}}
\caption{The luminosity of the model [O I] (top), [O II] (second row), [O III] (third row), and O I (bottom) lines in units of the observed line luminosities and limits for SN 2012au at 6 years for a pure oxygen composition at three different values of $T_{\rm PWN}$.  The green circled points represent where the model and observed values are within a factor of 2, the purple circles within a factor of 5, and the grey shaded region also within a factor of 5 to aid in the visualization.  For O I (bottom panel), the black circled points represent where the model luminosity is more than a factor 2 larger than the observational upper limit.}%
\label{fig:o6y_linecomp}
\end{figure*}

The ejecta temperature increases when the ejecta mass decreases, the engine luminosity increases, or the injection SED temperature increases. The temperature covers the range $1,600 - 6,300$ K. 

The dust-corrected model line luminosities of [O I], [O II], and [O III]; normalized to the observed line luminosities of 7.3 $\times$ 10$^{37}$, 1.0 $\times$ 10$^{38}$, and 1.4 $\times$ 10$^{38}$ erg s$^{-1}$, respectively; and the O I line luminosity normalized to the observational limit of 3.5 $\times$ 10$^{37}$ erg s$^{-1}$ \citep{Milisavljevic2018}; are shown in Figure \ref{fig:o6y_linecomp}.  For all four lines, the PWN SED temperature has a clear effect on the line luminosities, as the [O I], [O II], and [O III] luminosities increase with increasing $T_{\rm PWN}$, while the O I luminosity decreases.  The O I luminosity is only too high in a small region of the parameter space with high $L_{\rm PWN}$, high $M_{\rm ej}$, and low $T_{\rm PWN}$, and this can likely exclude any models with $T_{\rm PWN} < 10^5$ K due to an overluminous O I line.  

Although some features are easily relatable to the ion fractions from Figure \ref{fig:o6y_ionfrac}, such as the decrease in [O I] luminosity due to the runaway ionization of O I, the line luminosities are not a simple function of the ion fractions (Section \ref{sec:olines}).  The [O I] line luminosity roughly traces the ionization fraction, but is stronger in the high mass, high engine luminosity regime when the ejecta is hottest while still mostly neutral. The [O II] line luminosity is only strong when the ejecta temperature is high due to the high excitation energy of the [O II] transition. Both the [O I] and [O II] lines also show a strong dependence on ejecta mass, being strongest in high mass ejecta; this is because both lines are in NLTE, and show luminosities $\appropto n_{\rm ion} n_e \appropto M_{\rm ej}^2$.  The [O III] line is transitioning between LTE and NLTE, and doesn't show as strong an ejecta mass dependence as [O I] and [O III]; it is strongest in hot, highly ionized ejecta.

\begin{figure*}
\newcolumntype{D}{>{\centering\arraybackslash} m{6cm}}
\makebox[\textwidth]{
\begin{tabular}{DDD}
\boldsymbol{$T_{\rm PWN} = 10^5$} \textbf{ K} & \boldsymbol{$T_{\rm PWN} = 3 \times 10^5$} \textbf{ K} & \boldsymbol{$T_{\rm PWN} = 10^6$} \textbf{ K}\\
\includegraphics[width=1.1\linewidth]{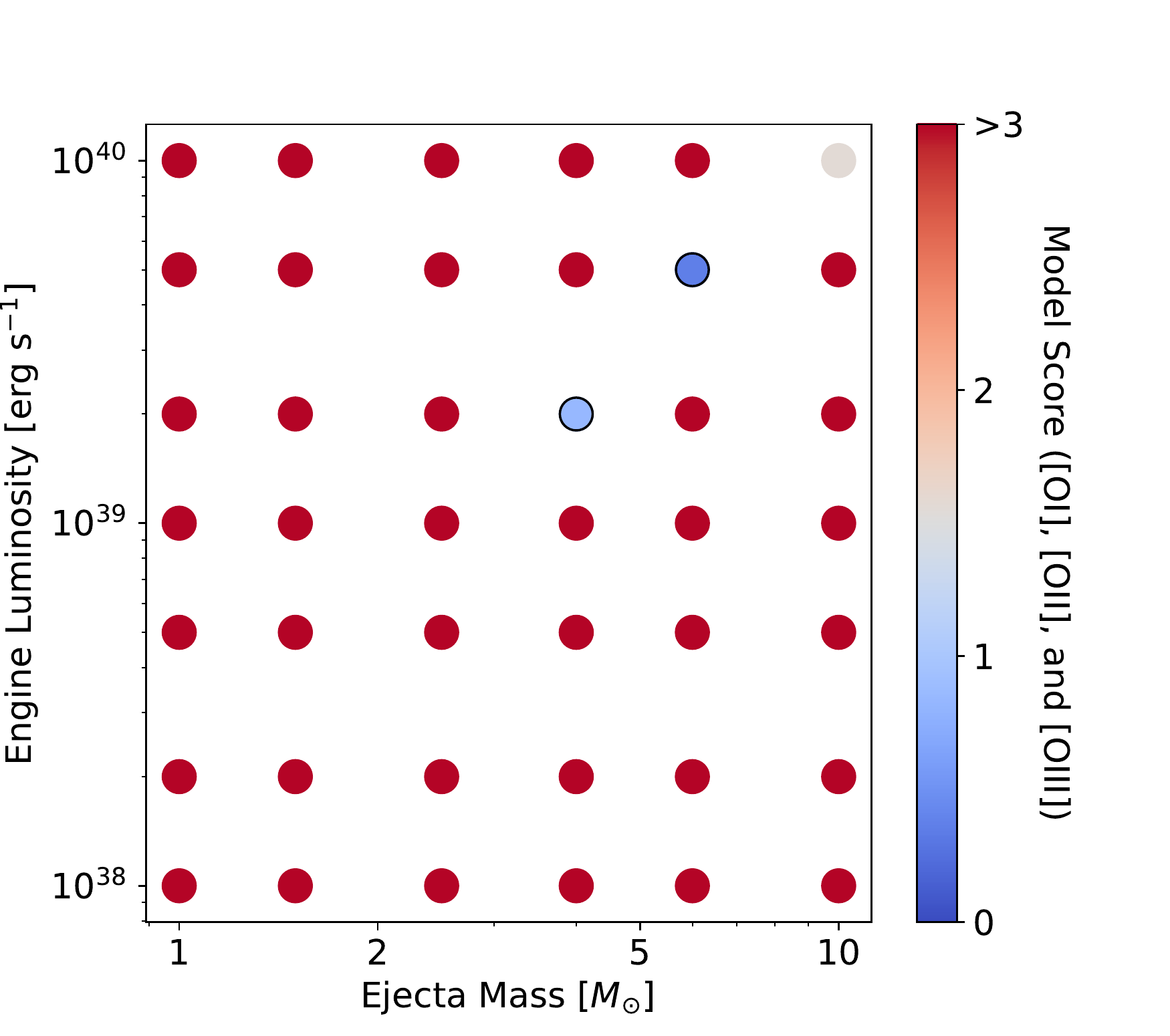}&
\includegraphics[width=1.1\linewidth]{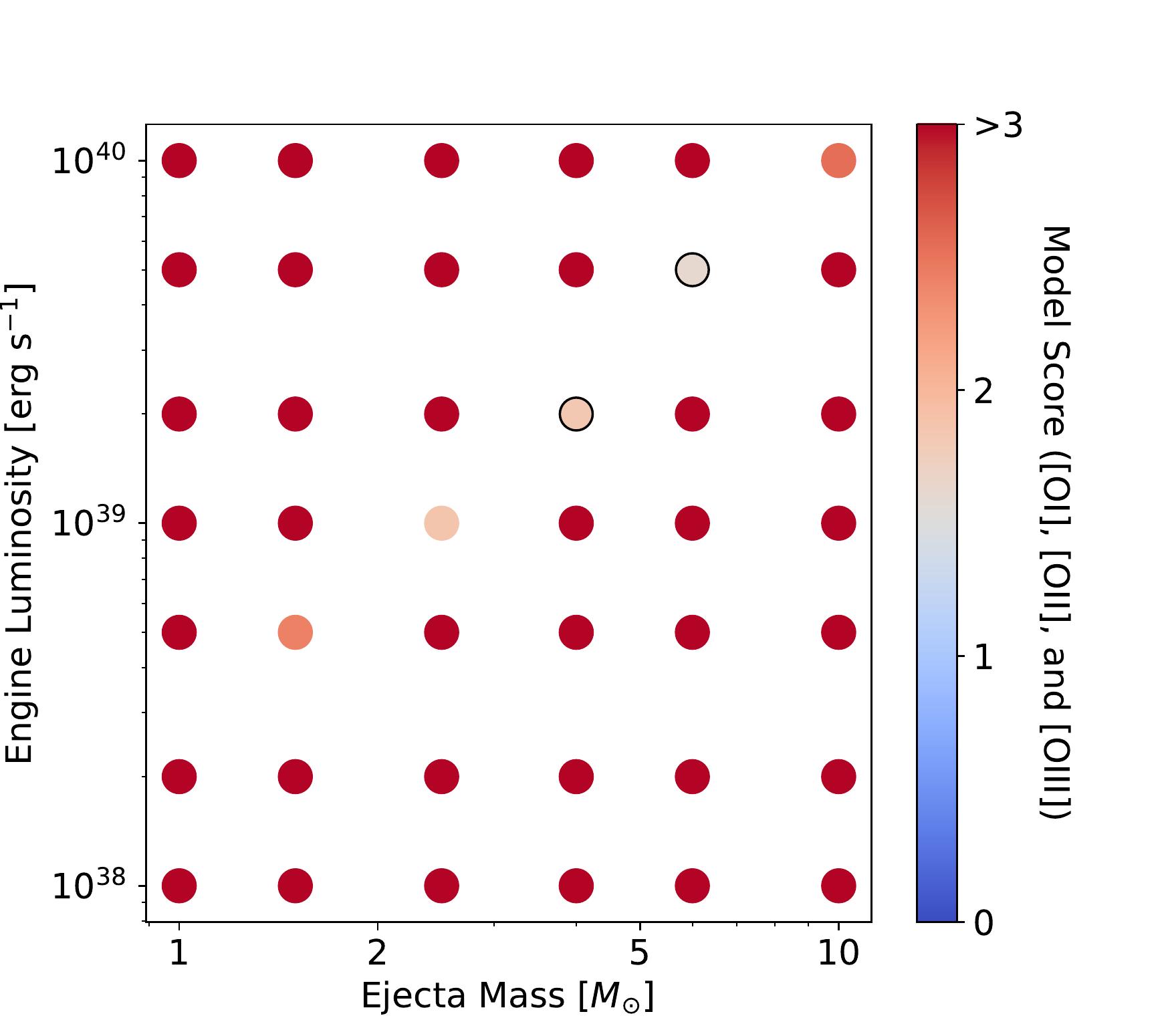}&
\includegraphics[width=1.1\linewidth]{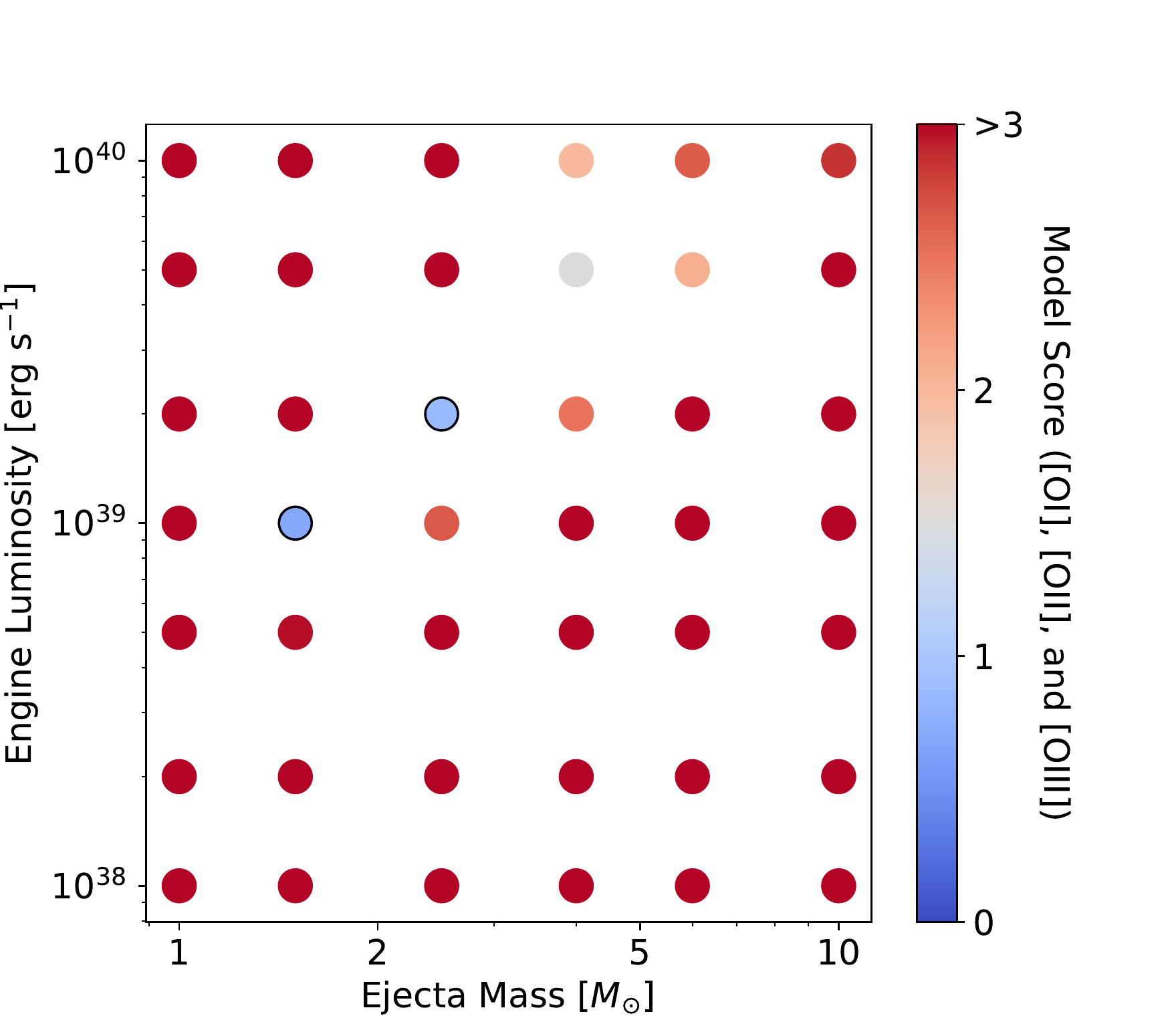}\\[-1.5ex]
\end{tabular}}
\caption{The goodness-of-fit score for each model for the pure oxygen composition at 6 years based on the [O I], [O II], and [O III] lines.  Lower scores indicate a better fit to the data (from Equation \ref{eqn:modscore}, a perfect fit has score 0, all three lines off by factor 2 has score 0.27, and all three lines off by factor 10 has score 3). The black circles indicate the two models with the lowest scores for each $T_{\rm PWN}$, which are plotted in Figure \ref{fig:o6y_spec}.}%
\label{fig:o6y_score}
\end{figure*}

The goodness of fit score for each model, generated from Equation \ref{eqn:modscore} with [O I], [O II], and [O III] contributing, is shown in Figure \ref{fig:o6y_score}.  None of the models with engine luminosity $< 5 \times 10^{38}$ erg s$^{-1}$ provide a very good fit to the data, highlighting the need for sufficient energy injection to reproduce the spectrum.  The best-fit models generally have higher ejecta masses and higher engine luminosities, but increasing $T_{\rm PWN}$ causes the best fit models to decrease in ejecta mass and engine luminosity.  However, within this simple model, we find that engine luminosities $\sim \left(1-5\right) \times 10^{39}$ erg s$^{-1}$ and ejecta masses $\sim 1.5-6$ $M_{\odot}$ reproduce the spectrum best.

\begin{figure*}
\newcolumntype{D}{>{\centering\arraybackslash} m{6cm}}
\makebox[\textwidth]{
\begin{tabular}{DDD}
\includegraphics[width=1.1\linewidth]{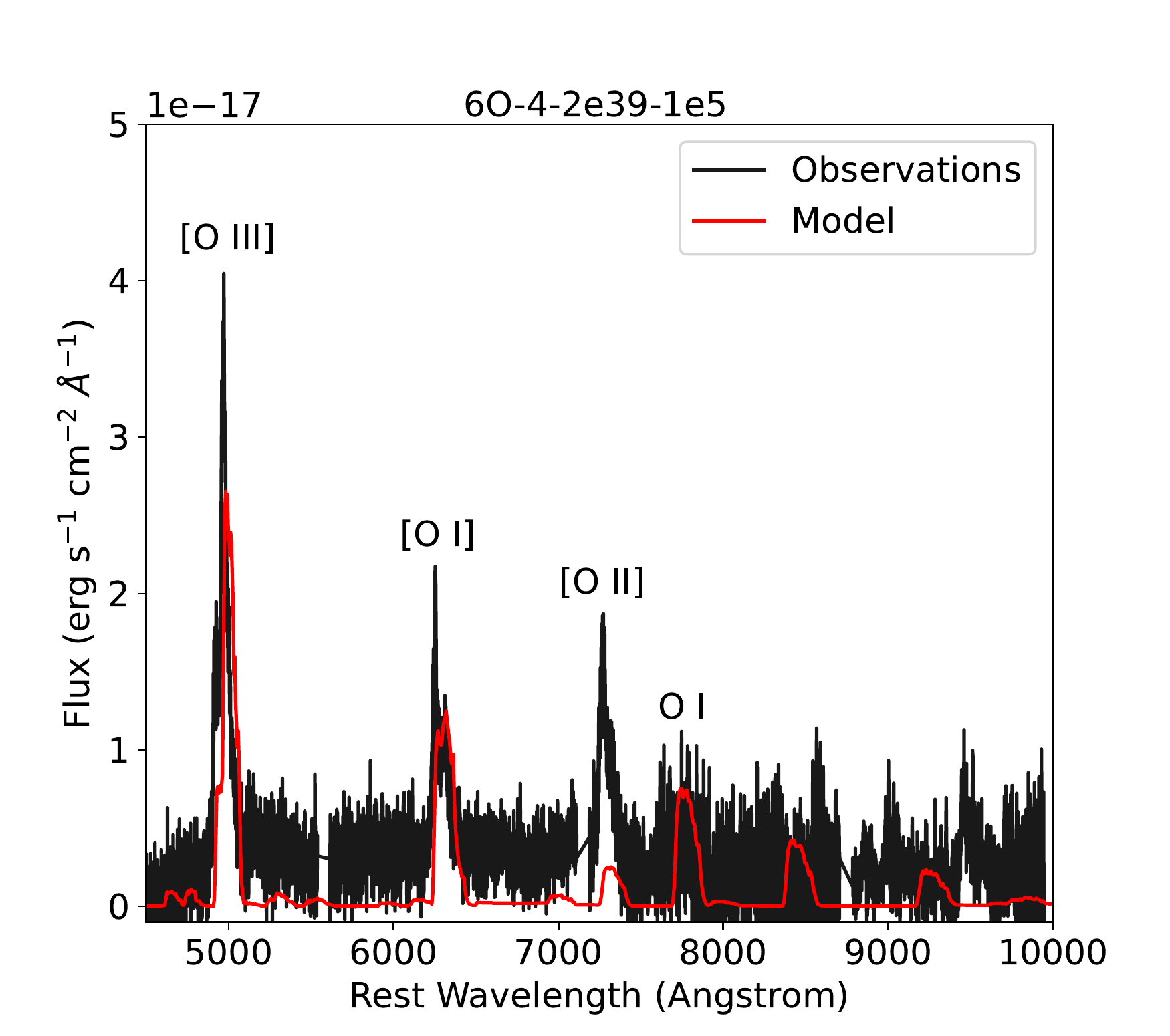}&
\includegraphics[width=1.1\linewidth]{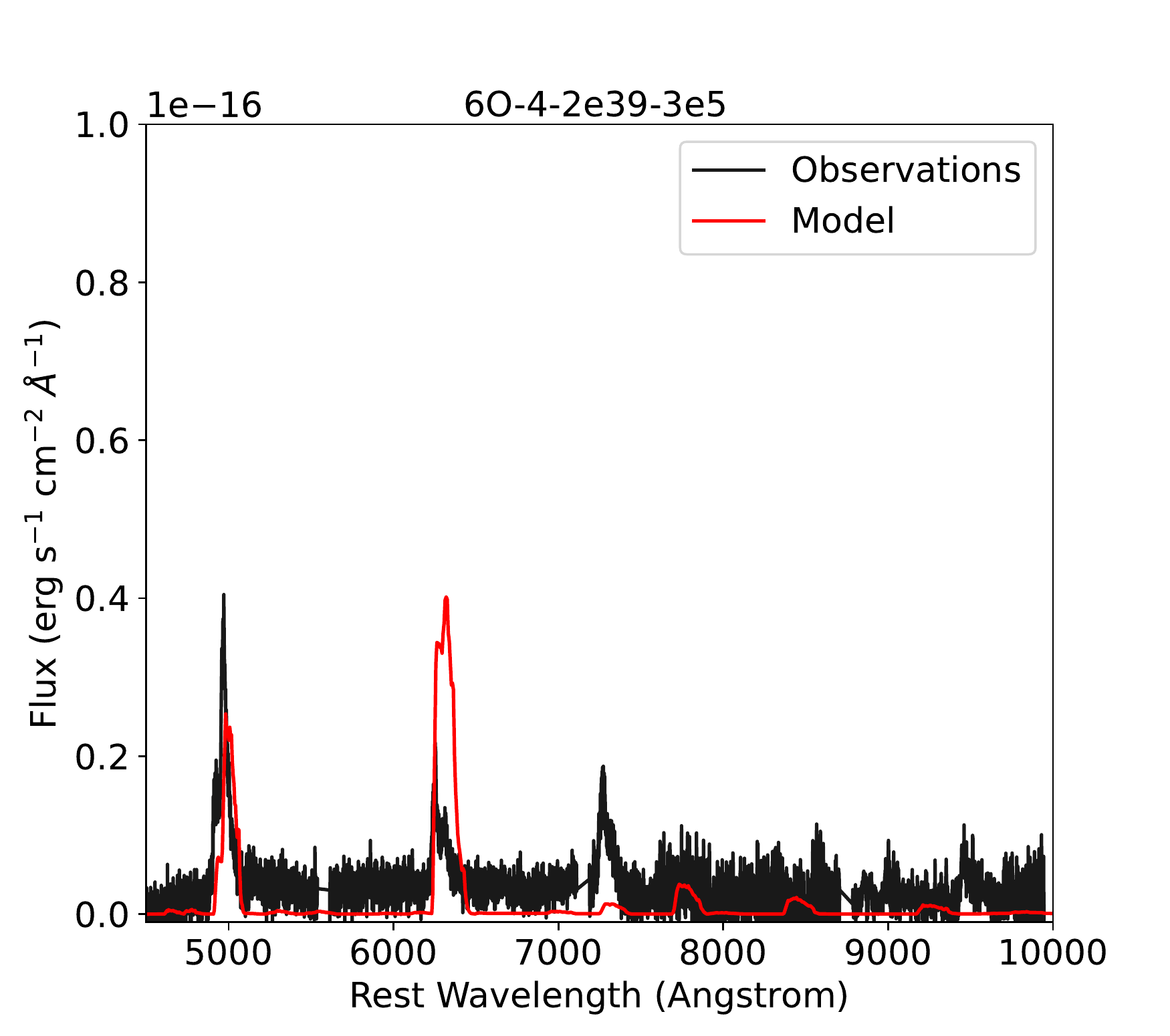}&
\includegraphics[width=1.1\linewidth]{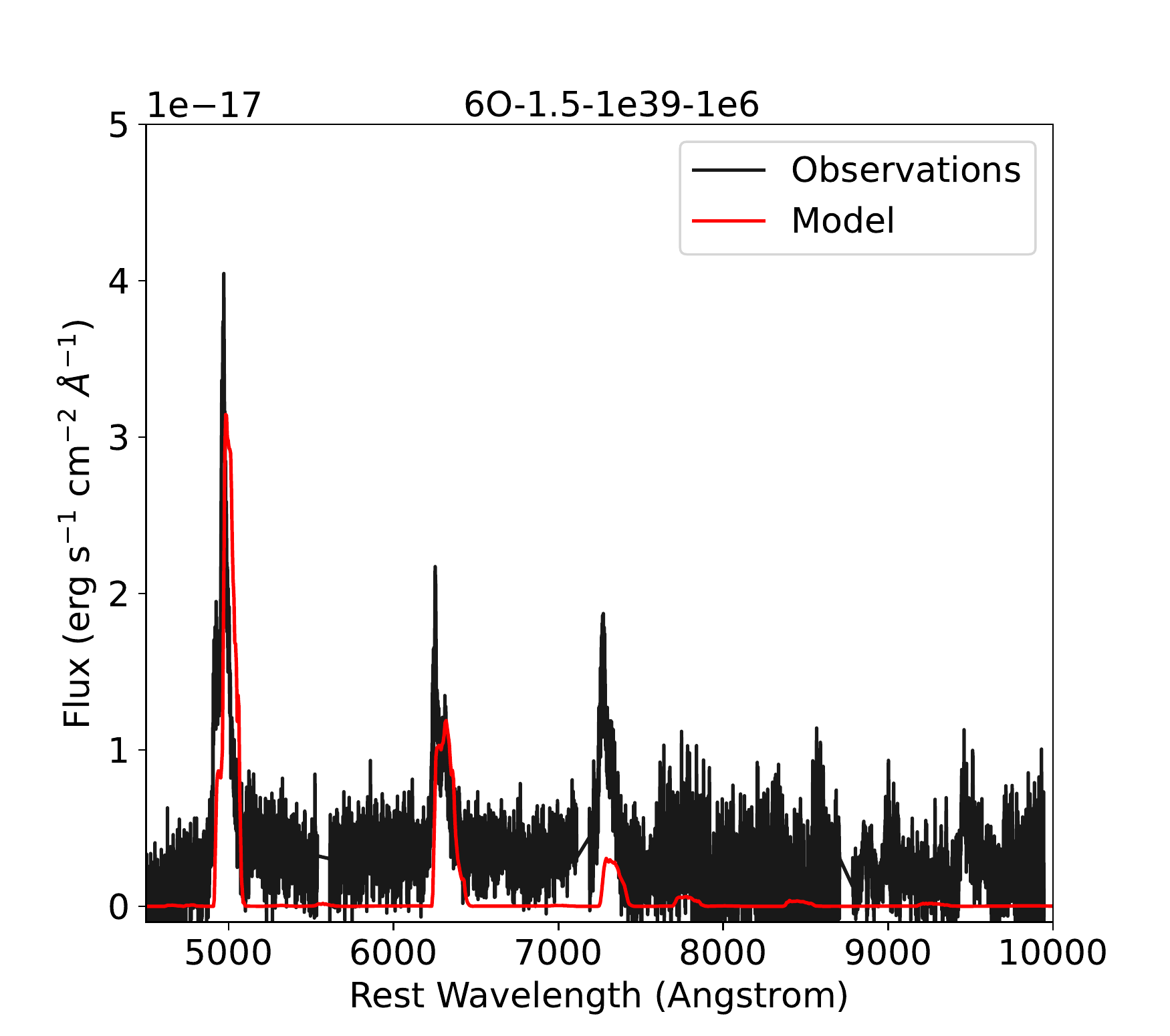}\\
\includegraphics[width=1.1\linewidth]{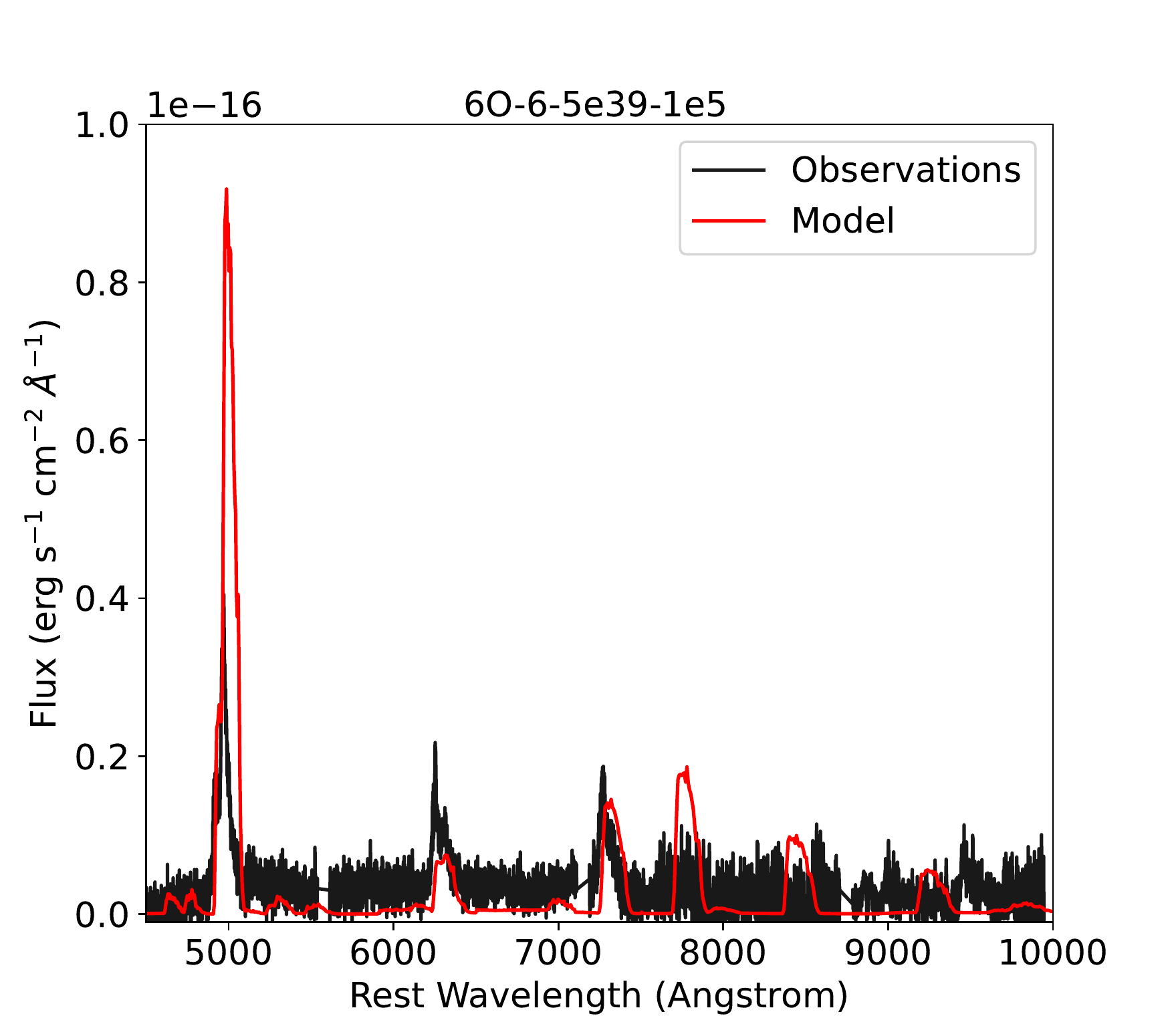}&
\includegraphics[width=1.1\linewidth]{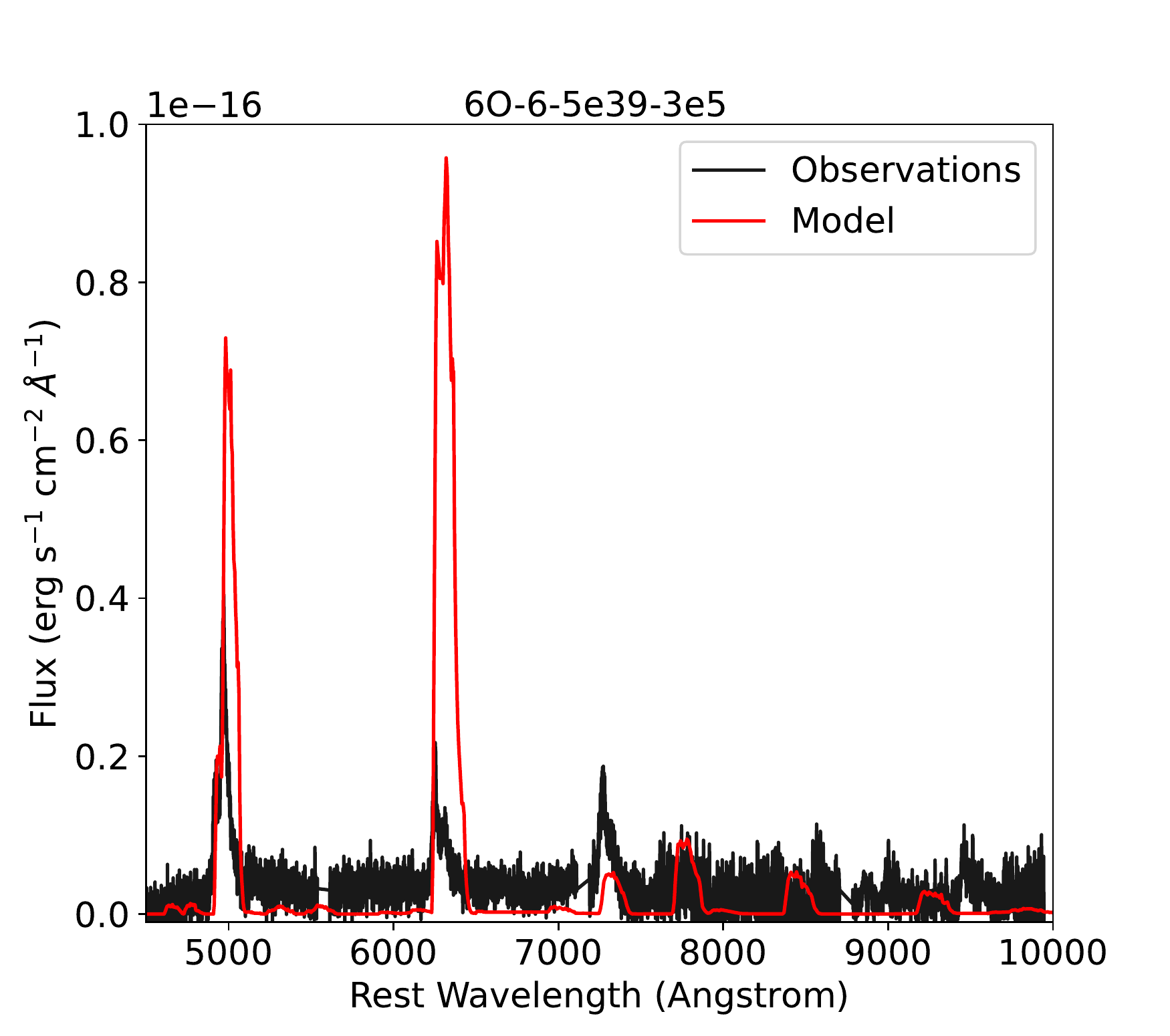}&
\includegraphics[width=1.1\linewidth]{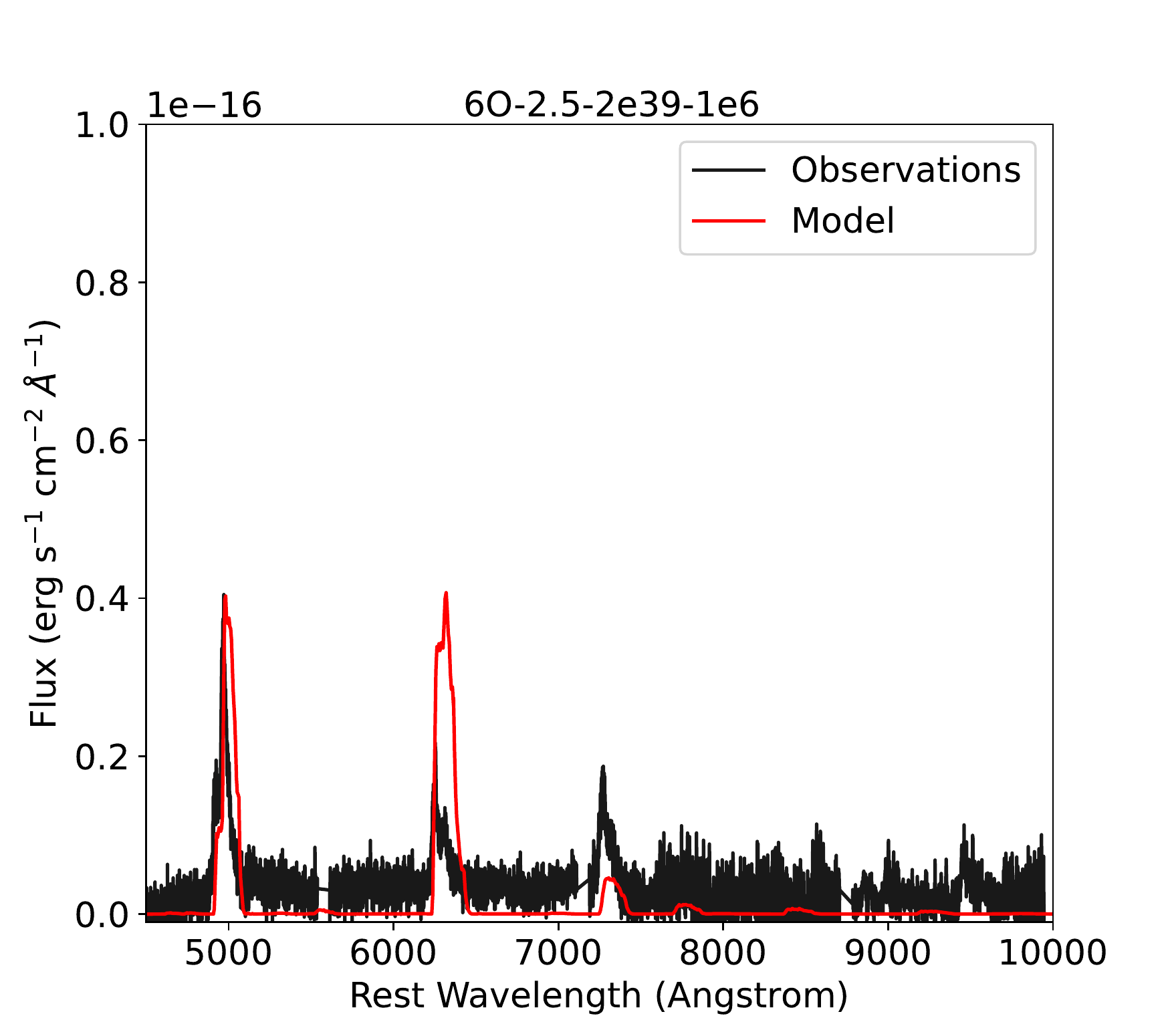}
\end{tabular}}
\caption{The two best-fitting dust-corrected model spectra to SN 2012au for each value of $T_{\rm PWN}$ at 6 years for a pure oxygen composition compared to the observed spectrum from \cite{Milisavljevic2018}.  Strong lines and features are labelled in the upper left plot.}%
\label{fig:o6y_spec}
\end{figure*}

The two best-fitting spectra for each value of $T_{\rm PWN}$ are shown in Figure \ref{fig:o6y_spec}.  Four of these six models are qualitatively unsatisfactory; 6O-4-2e39-3e5, 6O-6-5e39-3e5, and 6O-2.5-2e39-1e6 have an [O I] luminosity greater than their [O III] luminosity (contrary to observations), and 6O-6-5e39-1e5 has O I 7774 $\AA$ and 8446 $\AA$ emission significantly higher than the observed background.  The other two models, 6O-4-2e39-1e5 and 6O-1.5-1e39-1e6, are able to reproduce the [O III] and [O I] luminosities by within a factor of $\sim$ 2, although they both underestimate the [O II] luminosity - this may be due to possible contamination from the [Ca II] $\lambda \lambda$ 7291, 7323 doublet in the observed spectrum.  
We tentatively refer to the 6O-4-2e39-1e5 and 6O-1.5-1e39-1e6 the best-fit models for the pure oxygen composition at 6 years.

\subsubsection{1 Year}

Based on the best fit models at 6 years, we examine a smaller parameter space at 1 year to check if these models can also be consistent with observations at this phase.  The engine luminosity is scaled higher, assuming vacuum dipole spin-down, but the injection SED temperature range is kept the same.  The 1y grid is $L_{\rm PWN} = 3 \times 10^{40}$ - $3 \times 10^{42}$ erg s$^{-1}$, $T_{\rm PWN} = 10^5$ -  $10^6$ K, and $M_{\rm ej} = 1.5$ - $4 M_{\odot}$.

\begin{figure*}
\newcolumntype{D}{>{\centering\arraybackslash} m{6cm}}
\noindent
\makebox[\textwidth]{
\begin{tabular}{m{0.8cm} DDD}
& \boldsymbol{$T_{\rm PWN} = 10^5$} \textbf{ K} & \boldsymbol{$T_{\rm PWN} = 3 \times 10^5$} \textbf{ K} & \boldsymbol{$T_{\rm PWN} = 10^6$} \textbf{ K}\\
\textbf{O I}&
\includegraphics[width=1.1\linewidth]{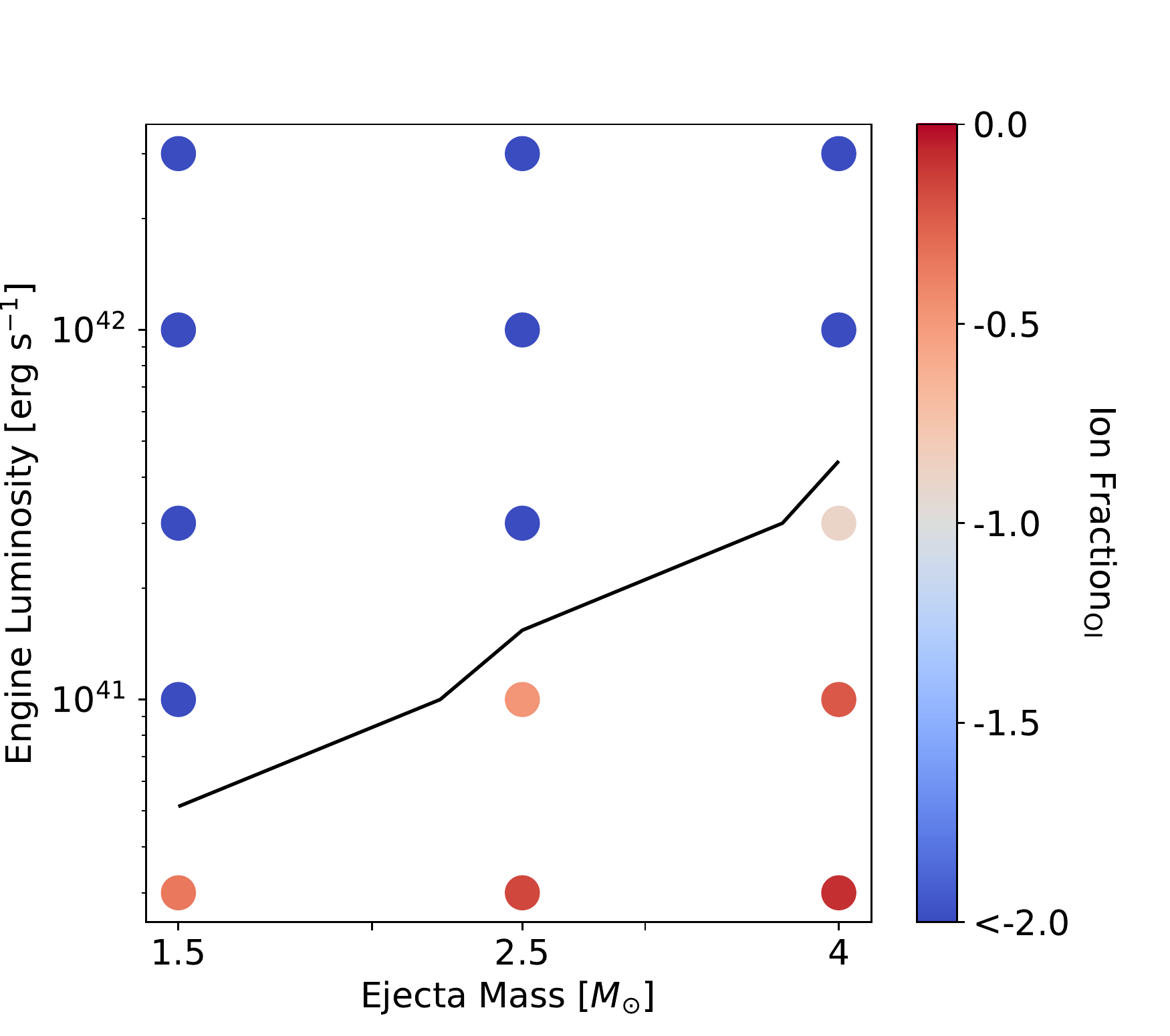}&
\includegraphics[width=1.1\linewidth]{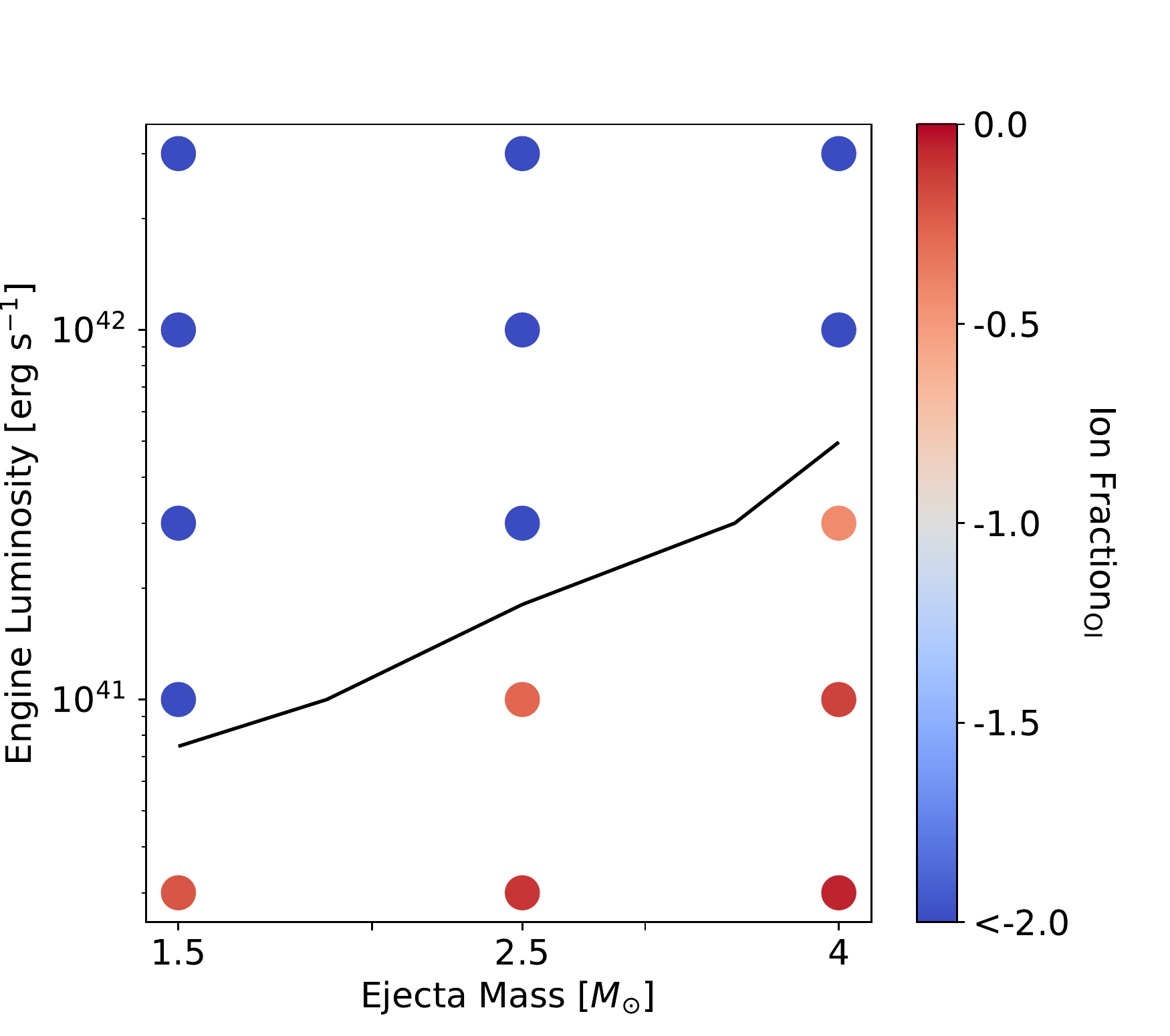}&
\includegraphics[width=1.1\linewidth]{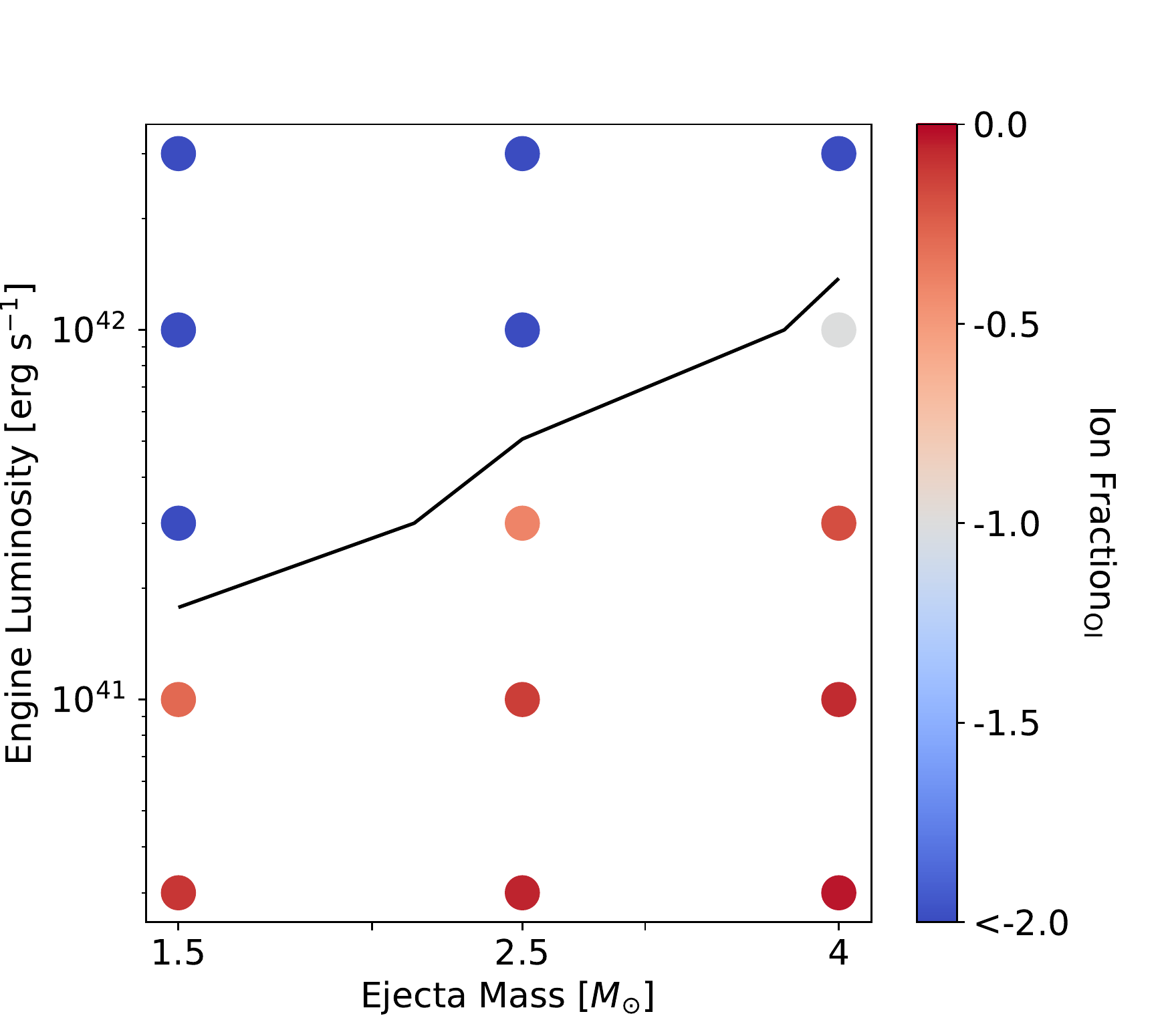}\\[-1.5ex]
\textbf{O II}&
\includegraphics[width=1.1\linewidth]{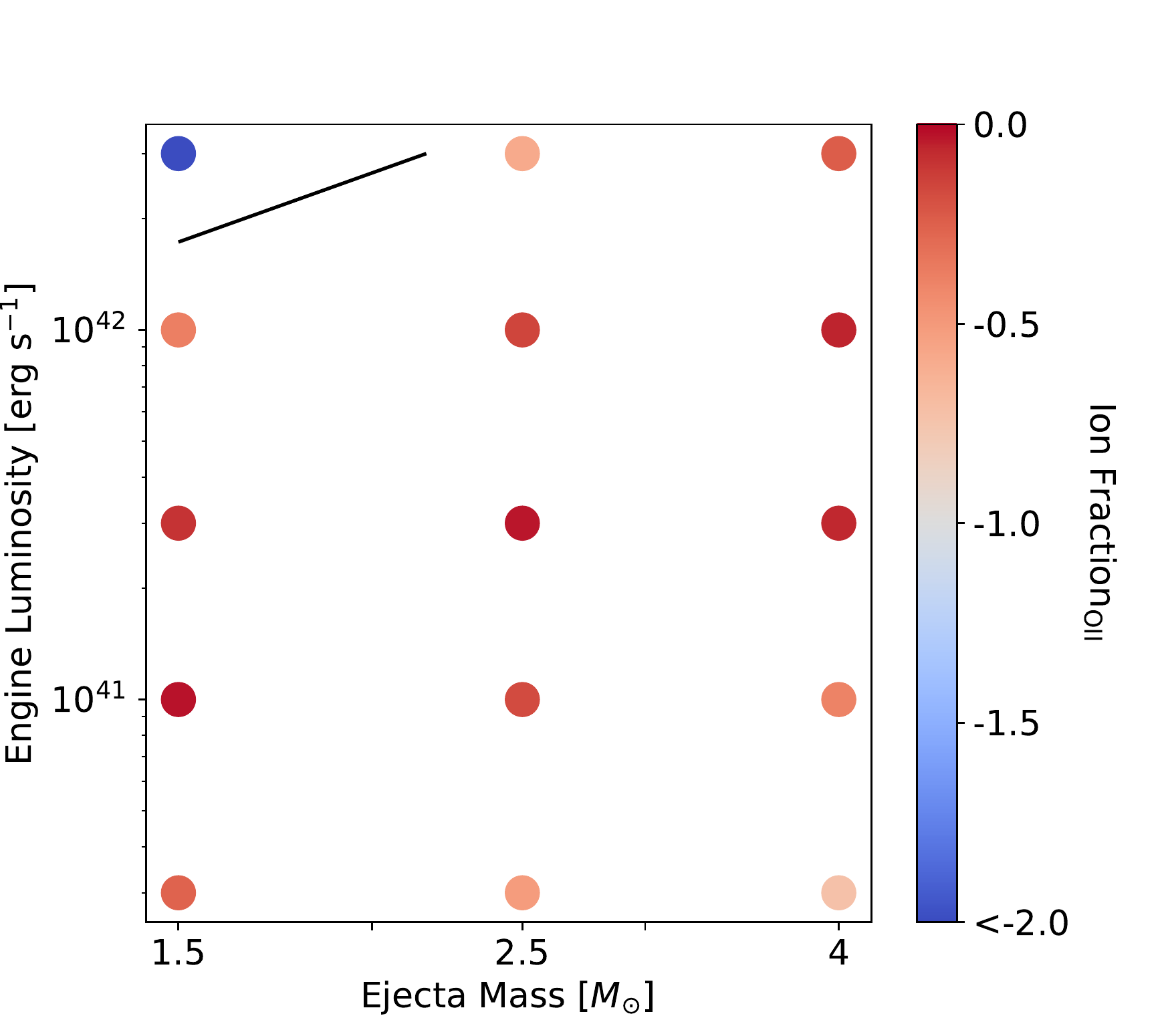}&
\includegraphics[width=1.1\linewidth]{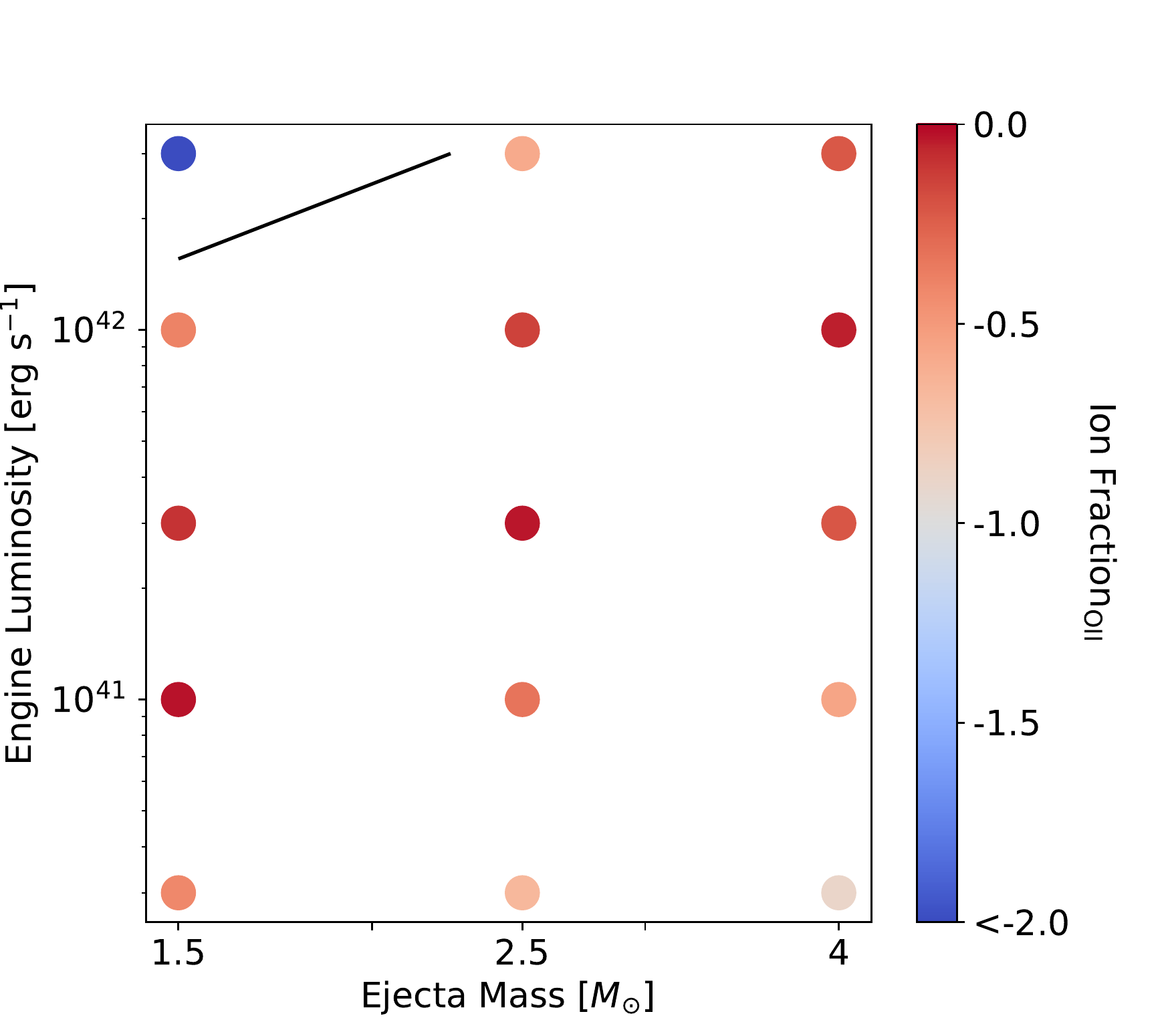}&
\includegraphics[width=1.1\linewidth]{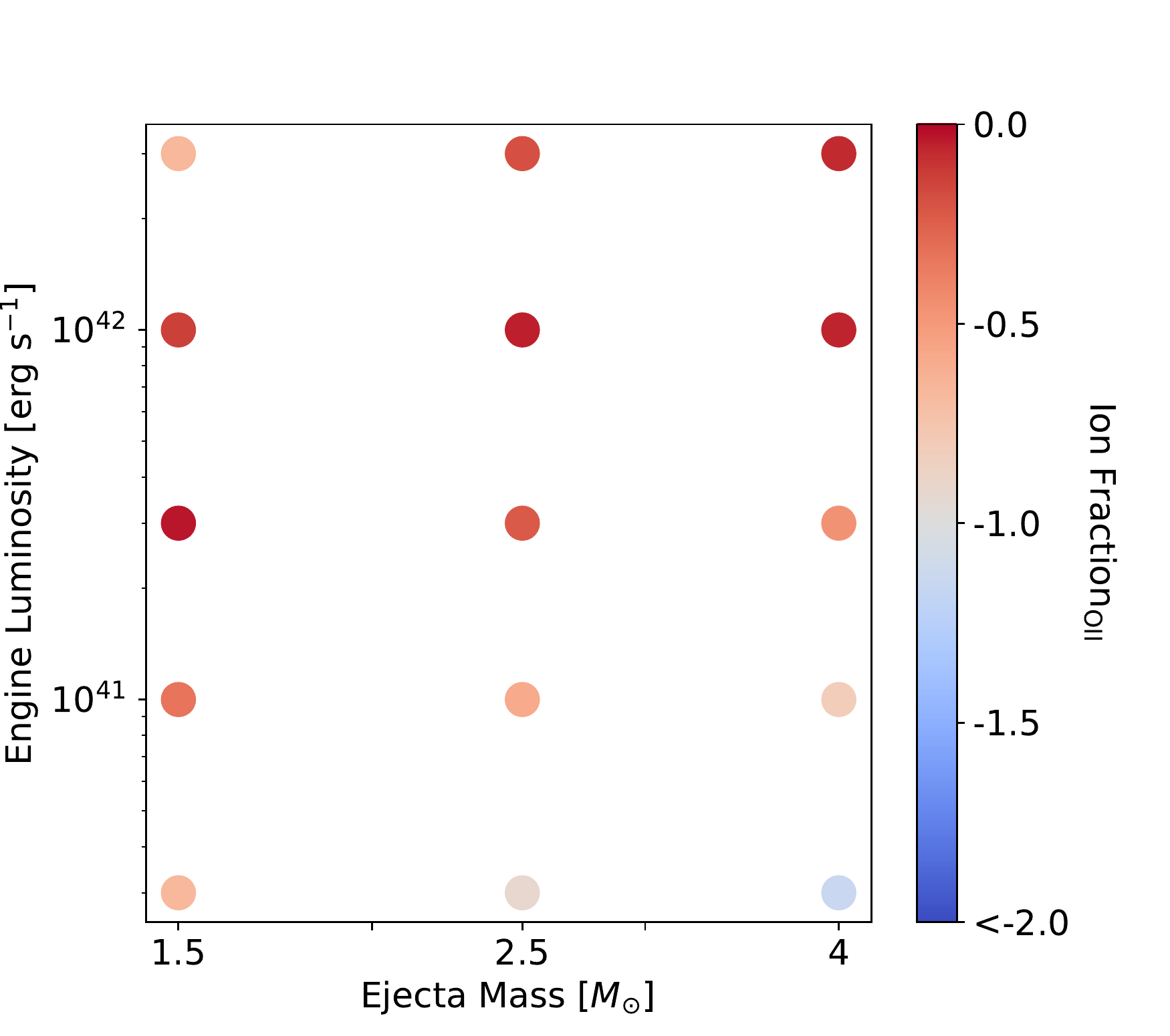}\\[-1.5ex]
\textbf{O III}&
\includegraphics[width=1.1\linewidth]{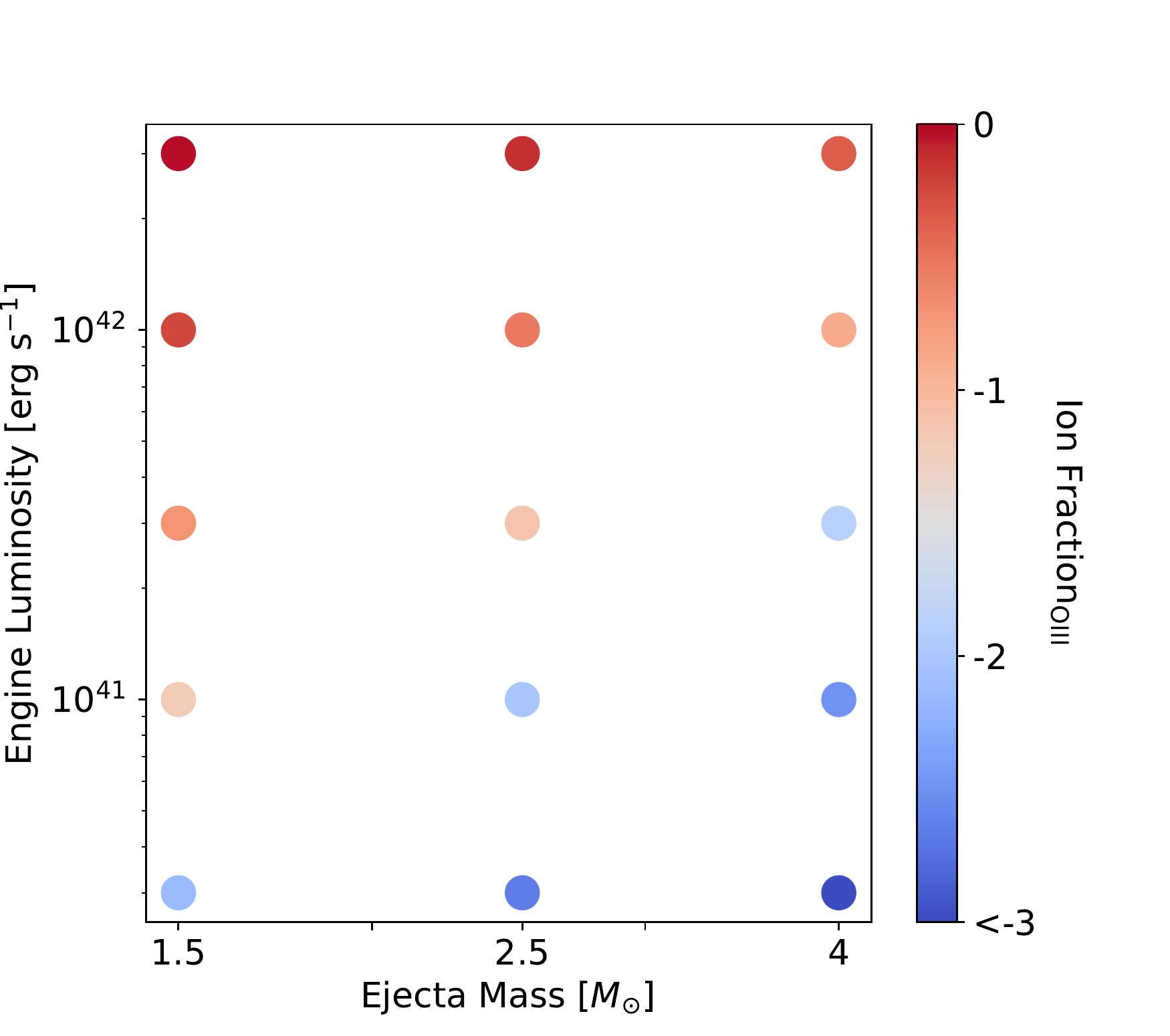}&
\includegraphics[width=1.1\linewidth]{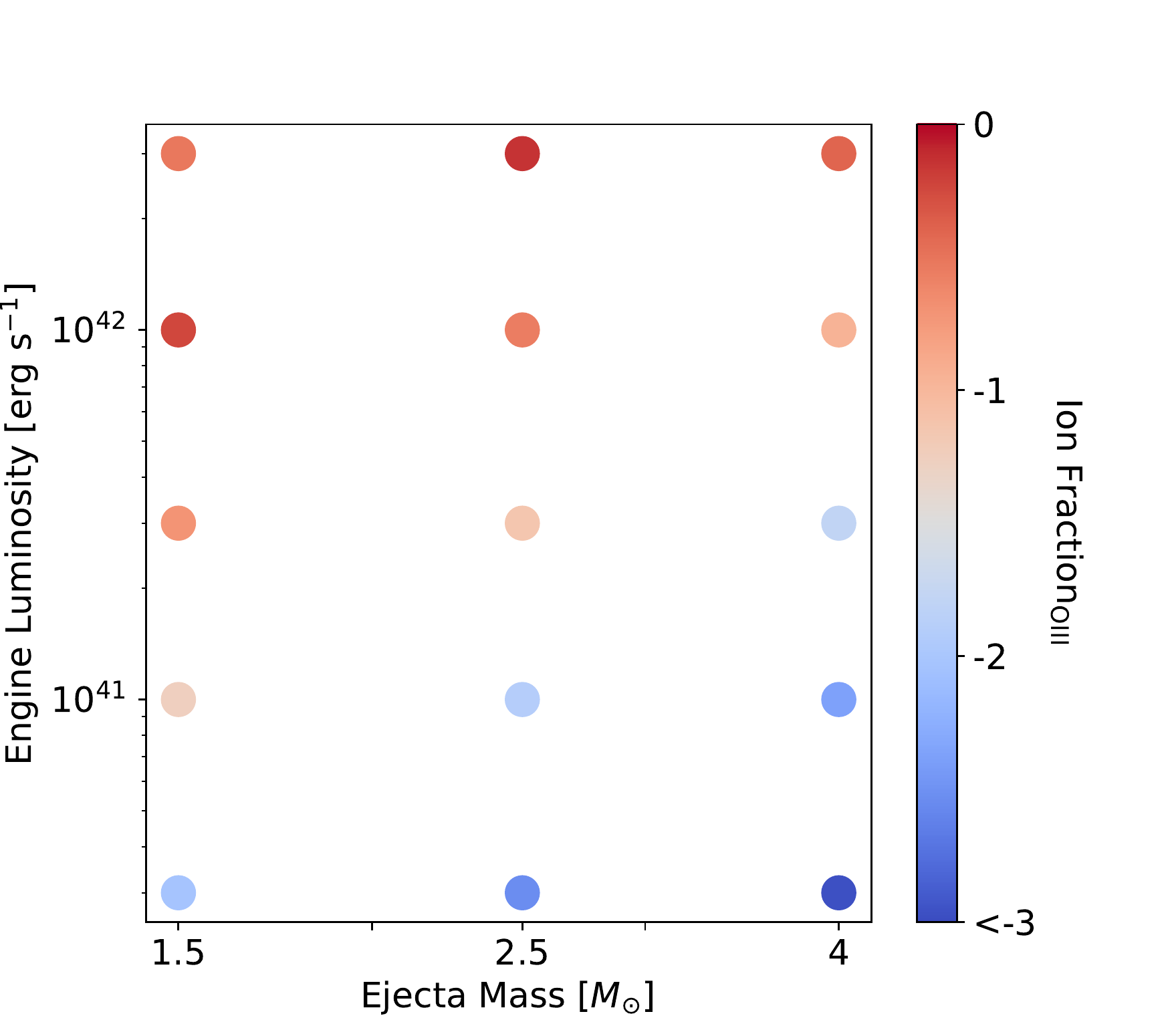}&
\includegraphics[width=1.1\linewidth]{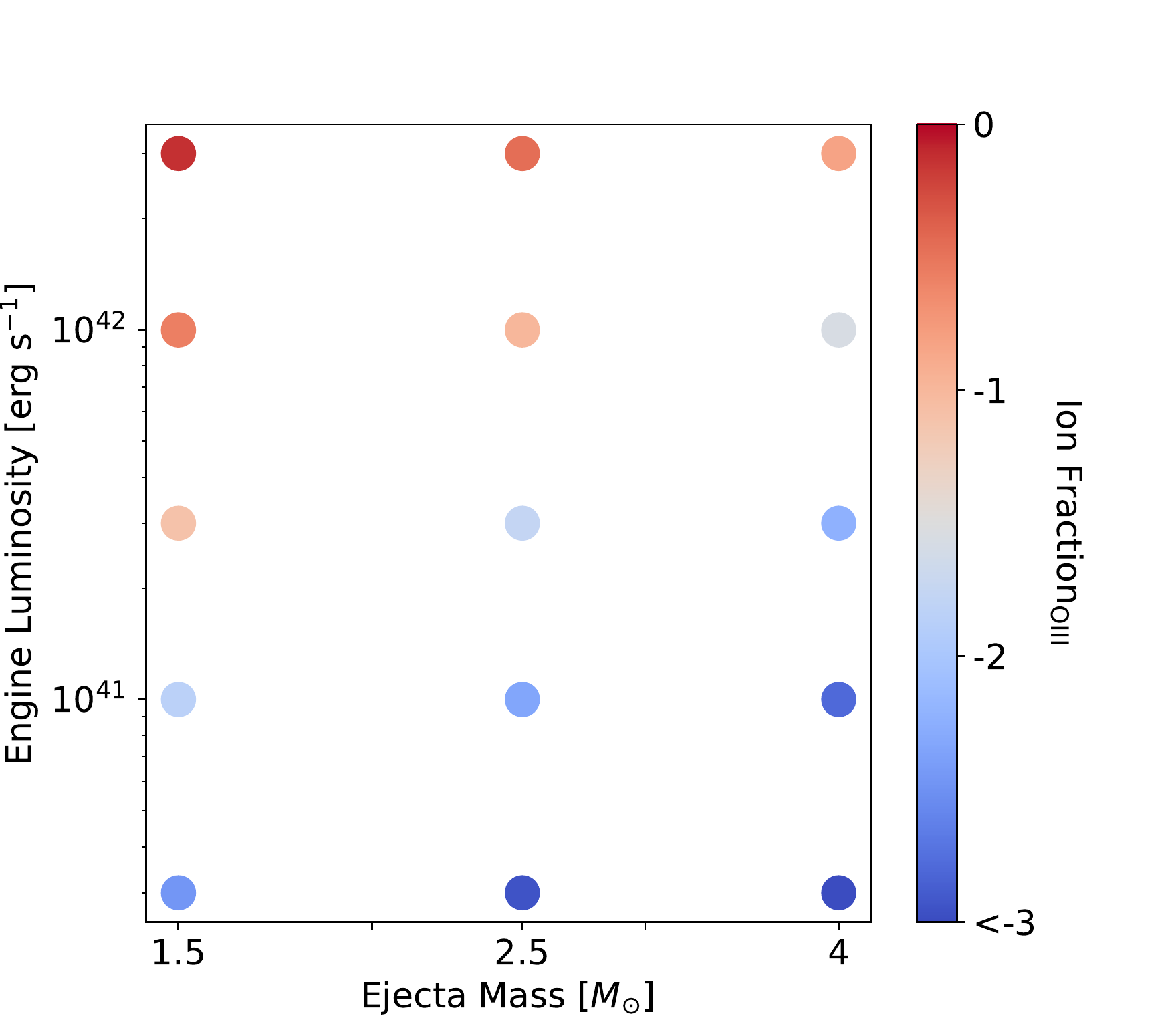}\\[-1.5ex]
\boldsymbol{$T_{\rm ej}$}&
\includegraphics[width=1.1\linewidth]{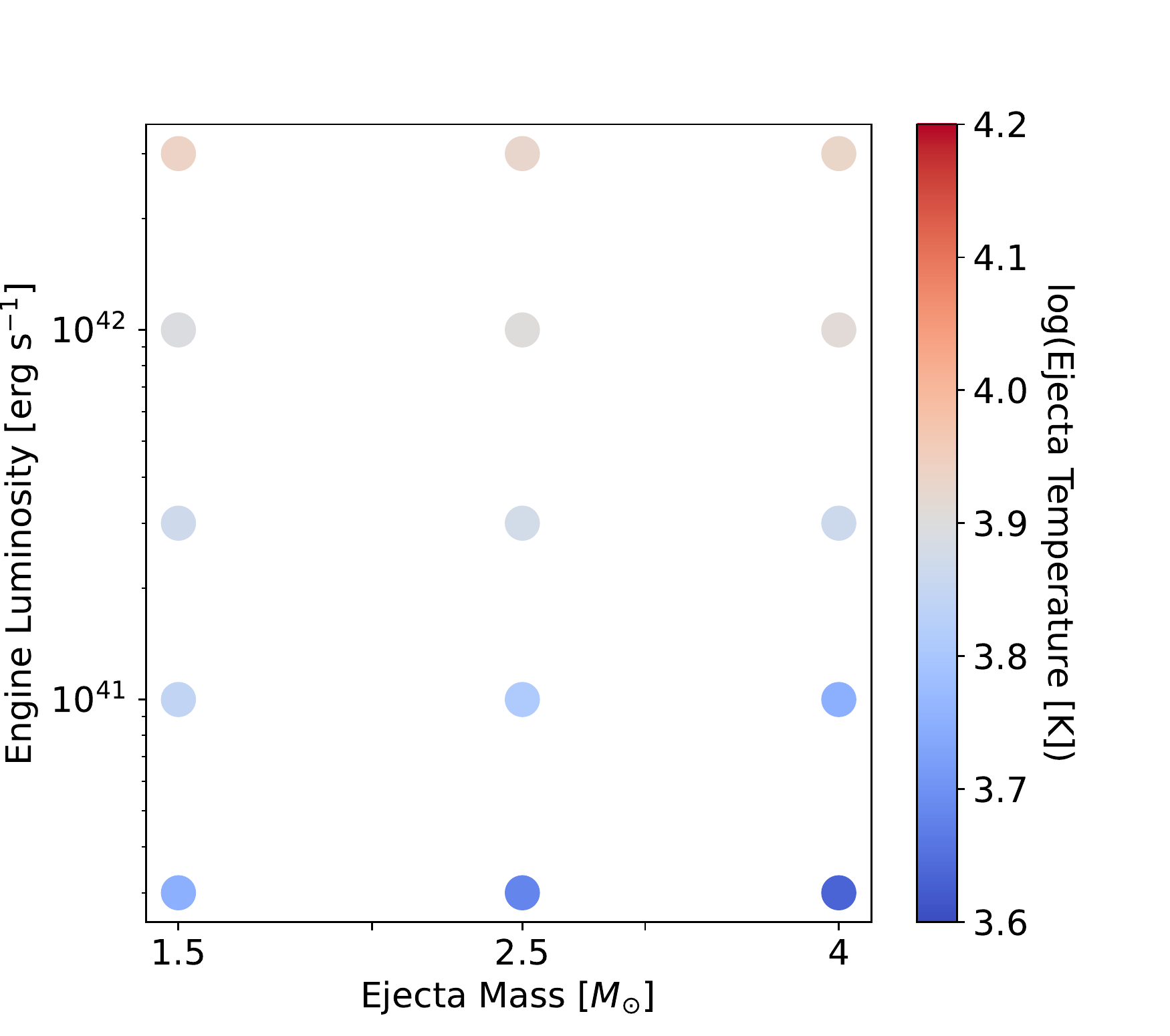}&
\includegraphics[width=1.1\linewidth]{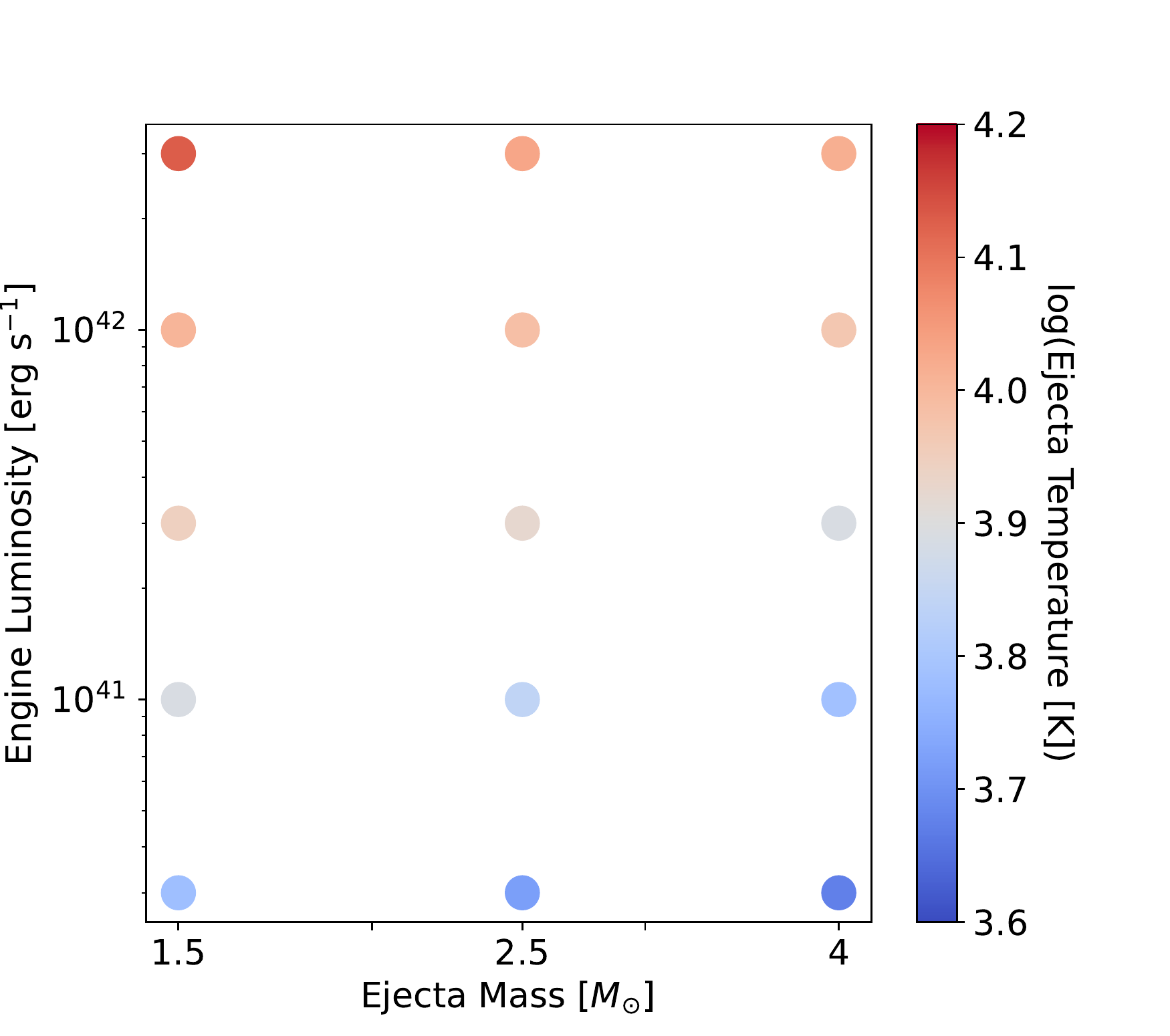}&
\includegraphics[width=1.1\linewidth]{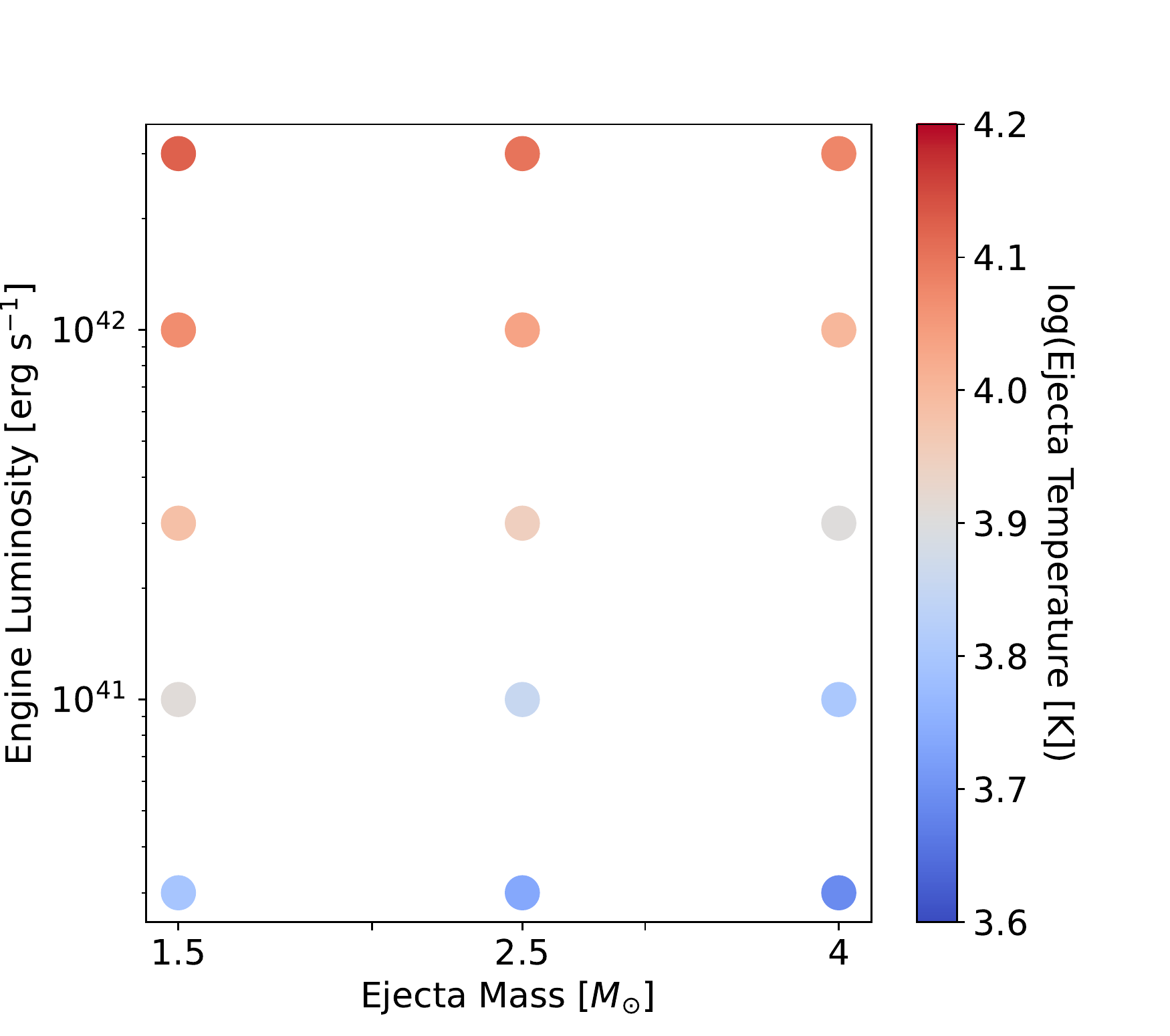}\\[-1.5ex]
\end{tabular}}
\caption{The ion fractions of O I (top), O II (second row), and O III (third row), and the ejecta temperature $T_{\rm ej}$ (bottom) in the simulations at 1 year for a pure oxygen composition at three different values of $T_{\rm PWN}$.  The black contour denotes the low ejecta mass, high engine luminosity regime where runaway ionization can occur for both O I and O II.}%
\label{fig:o1y_ionfrac}
\end{figure*}

\begin{figure*}
\newcolumntype{D}{>{\centering\arraybackslash} m{6cm}}
\noindent
\makebox[\textwidth]{
\begin{tabular}{m{1cm} DDD}
& \boldsymbol{$T_{\rm PWN} = 10^5$} \textbf{ K} & \boldsymbol{$T_{\rm PWN} = 3 \times 10^5$} \textbf{ K} & \boldsymbol{$T_{\rm PWN} = 10^6$} \textbf{ K}\\
\textbf{[O I]}&
\includegraphics[width=1.1\linewidth]{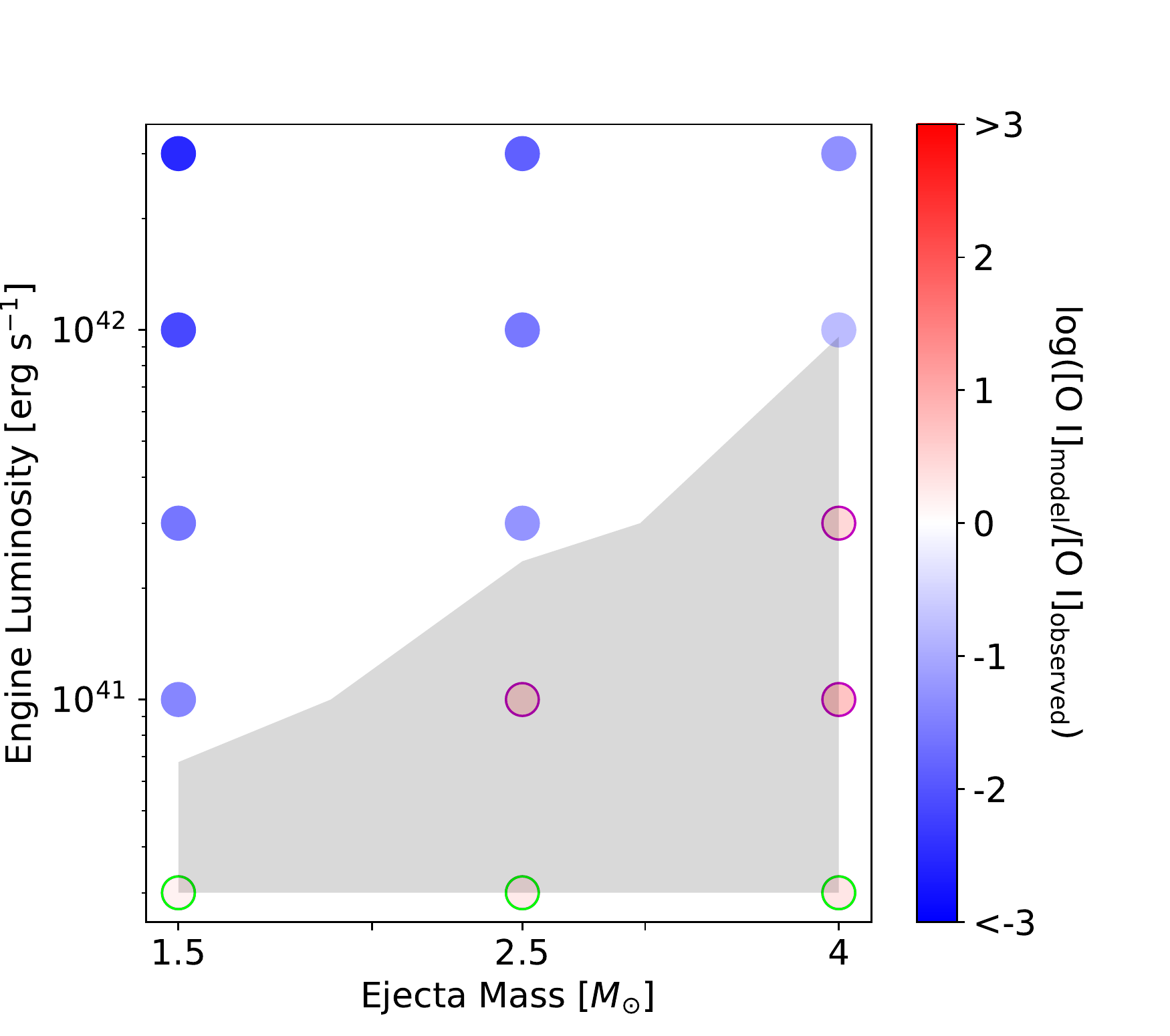}&
\includegraphics[width=1.1\linewidth]{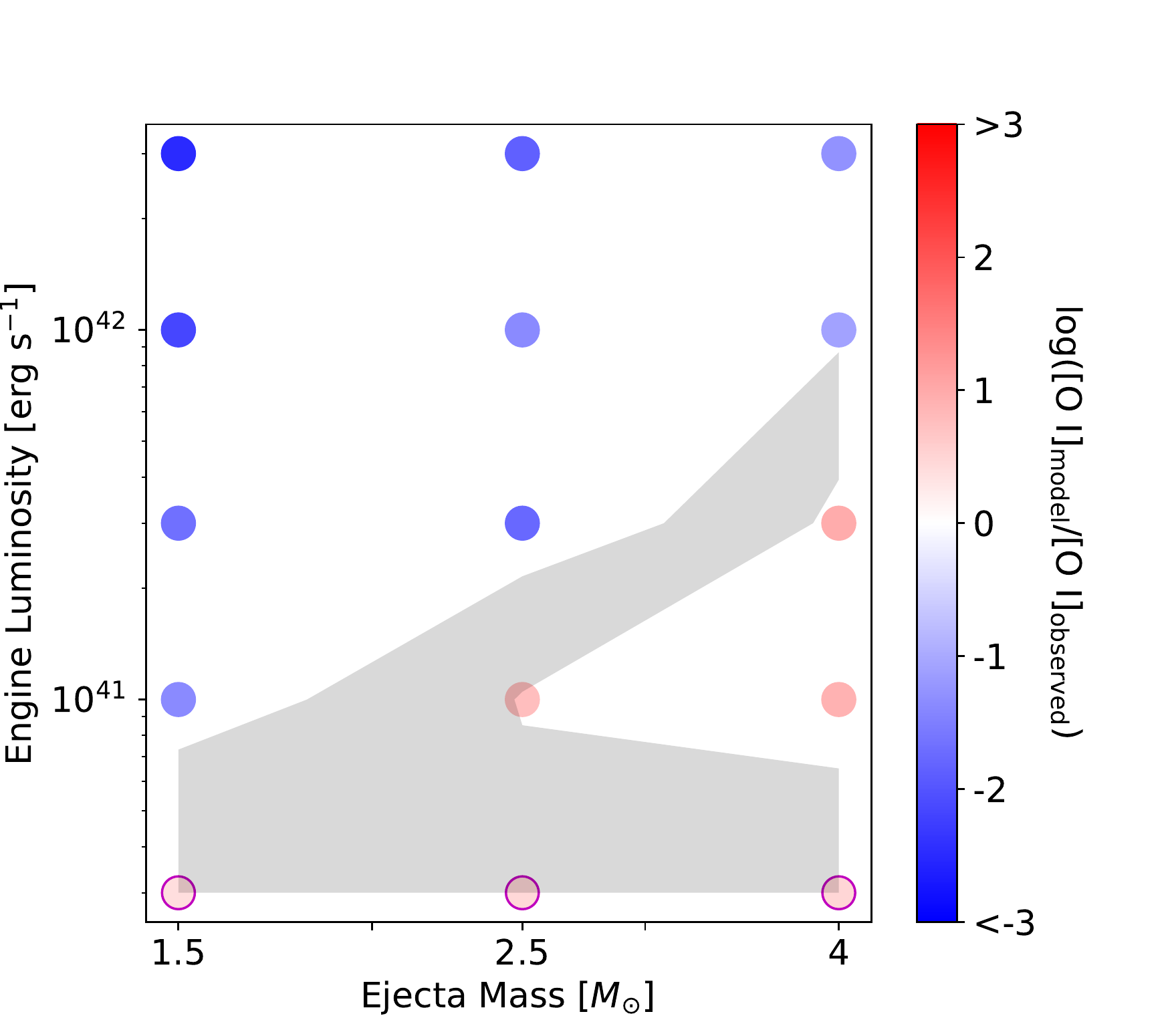}&
\includegraphics[width=1.1\linewidth]{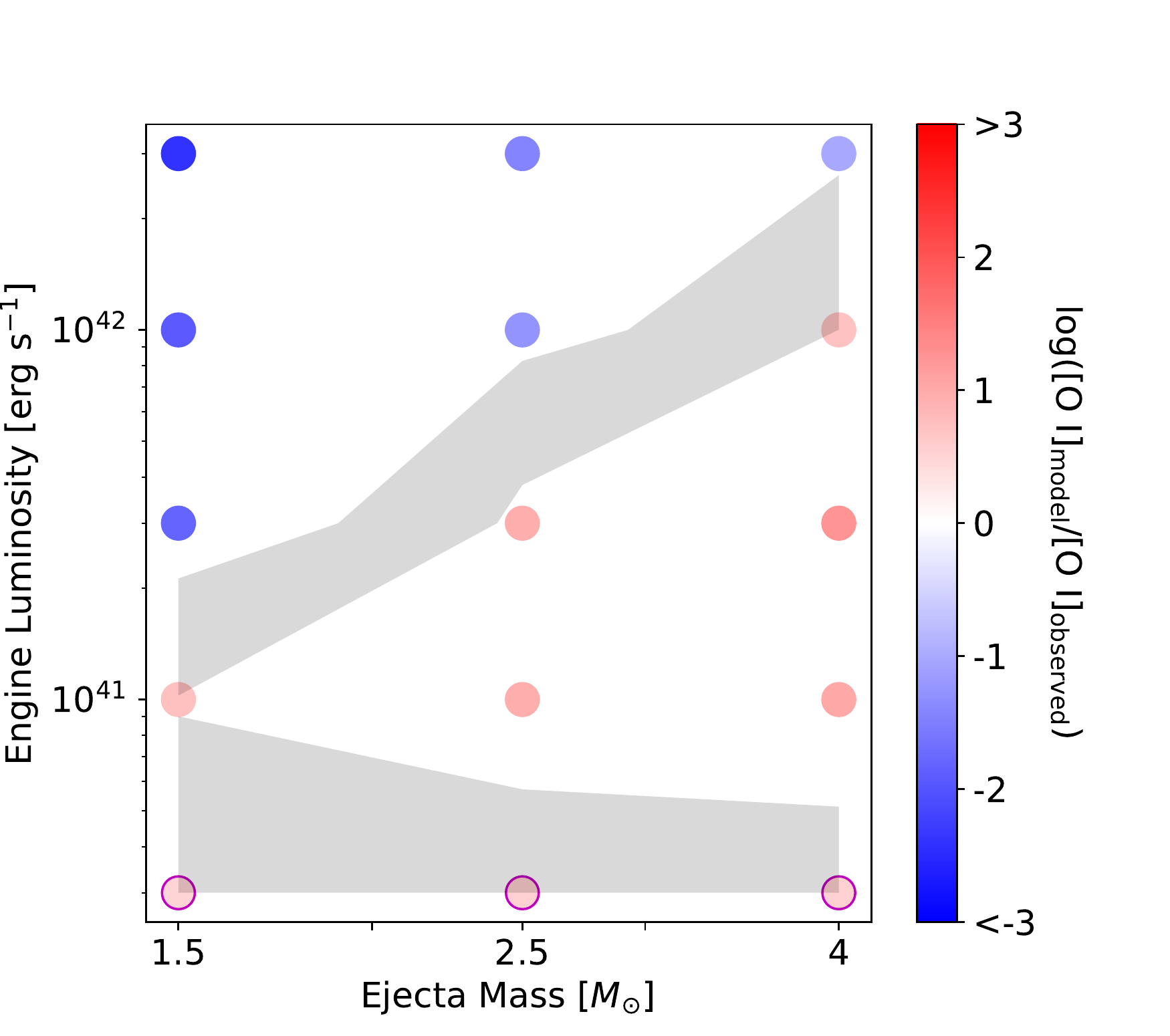}\\[-1.5ex]
\textbf{[O II]}&
\includegraphics[width=1.1\linewidth]{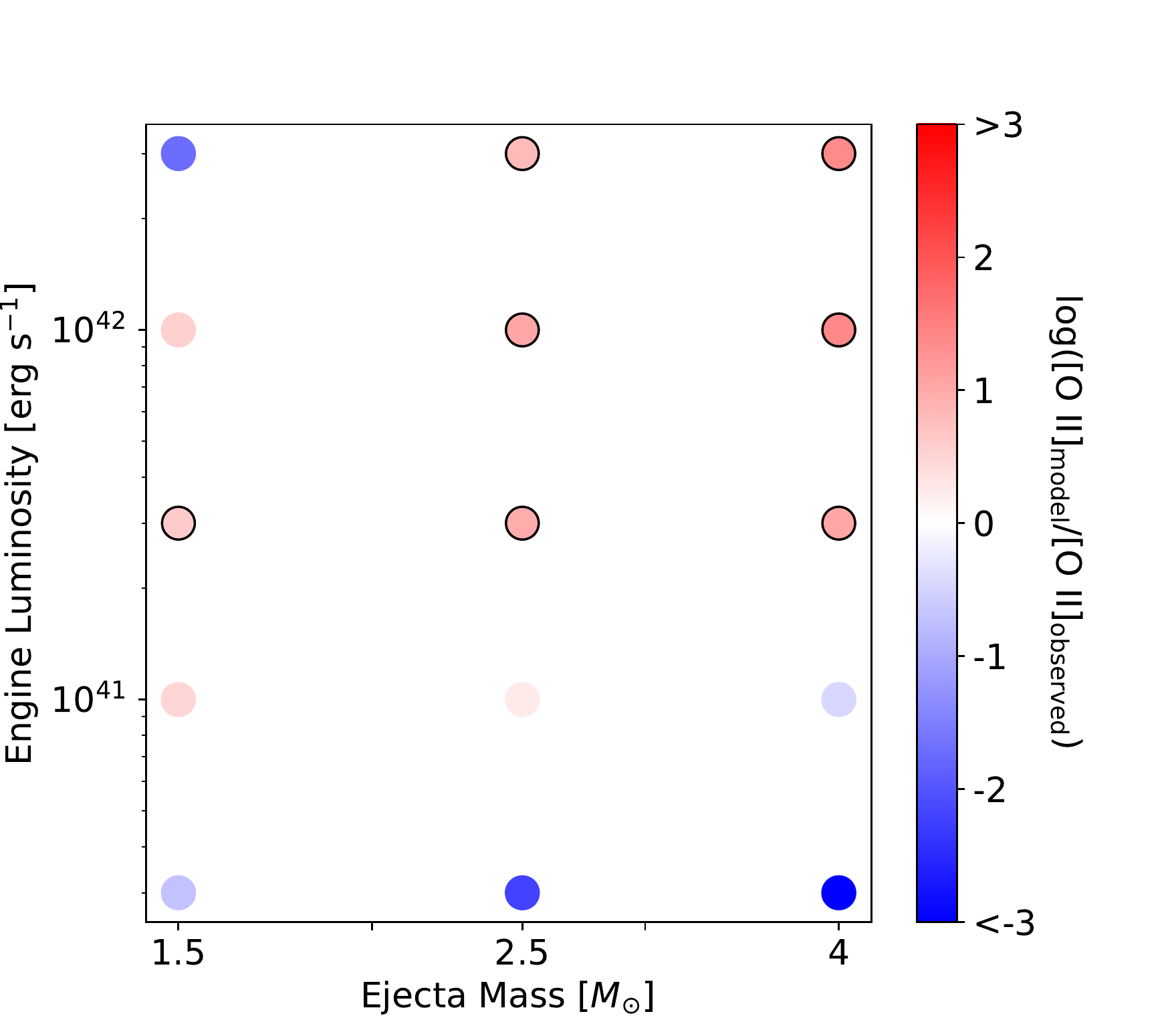}&
\includegraphics[width=1.1\linewidth]{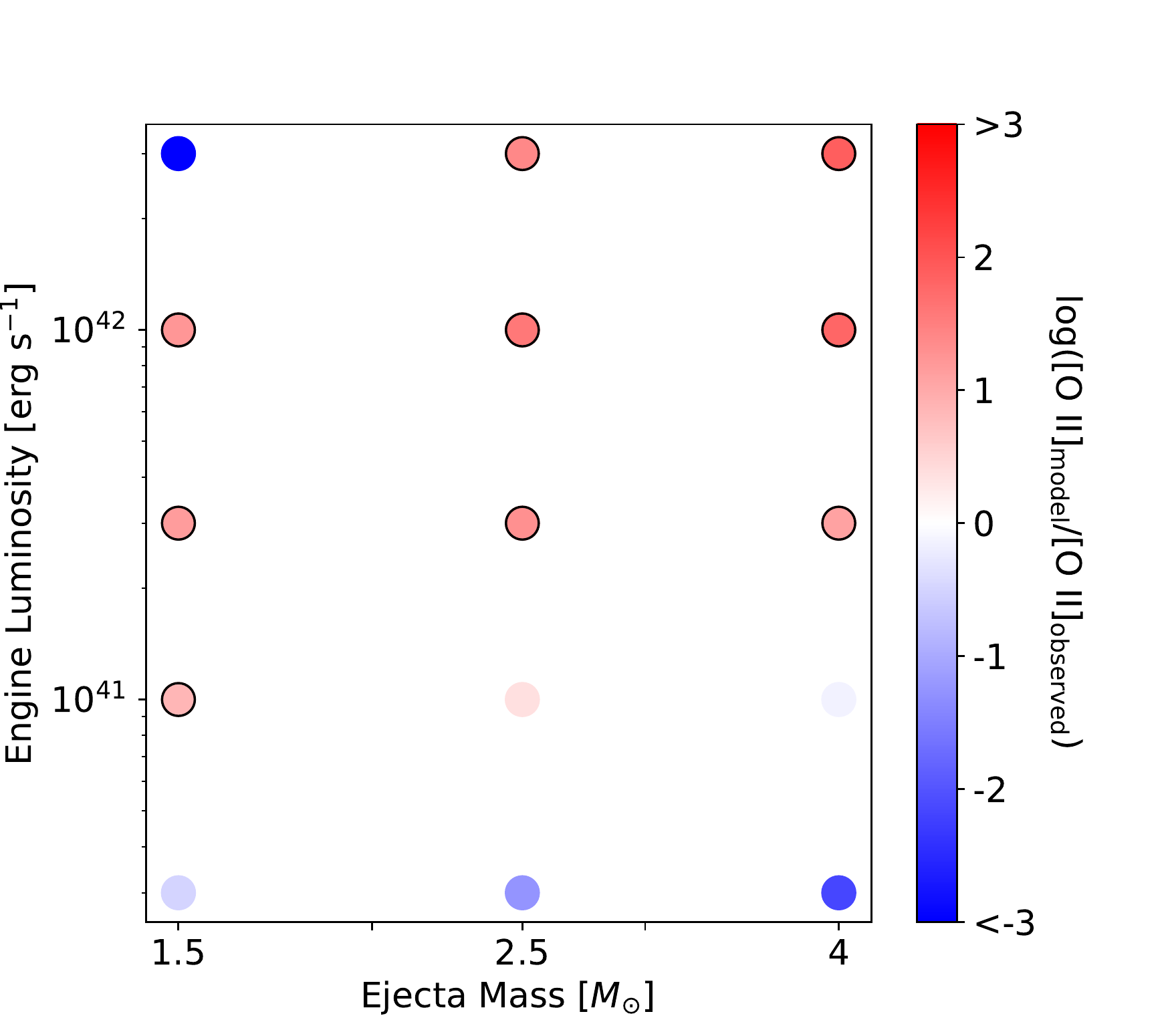}&
\includegraphics[width=1.1\linewidth]{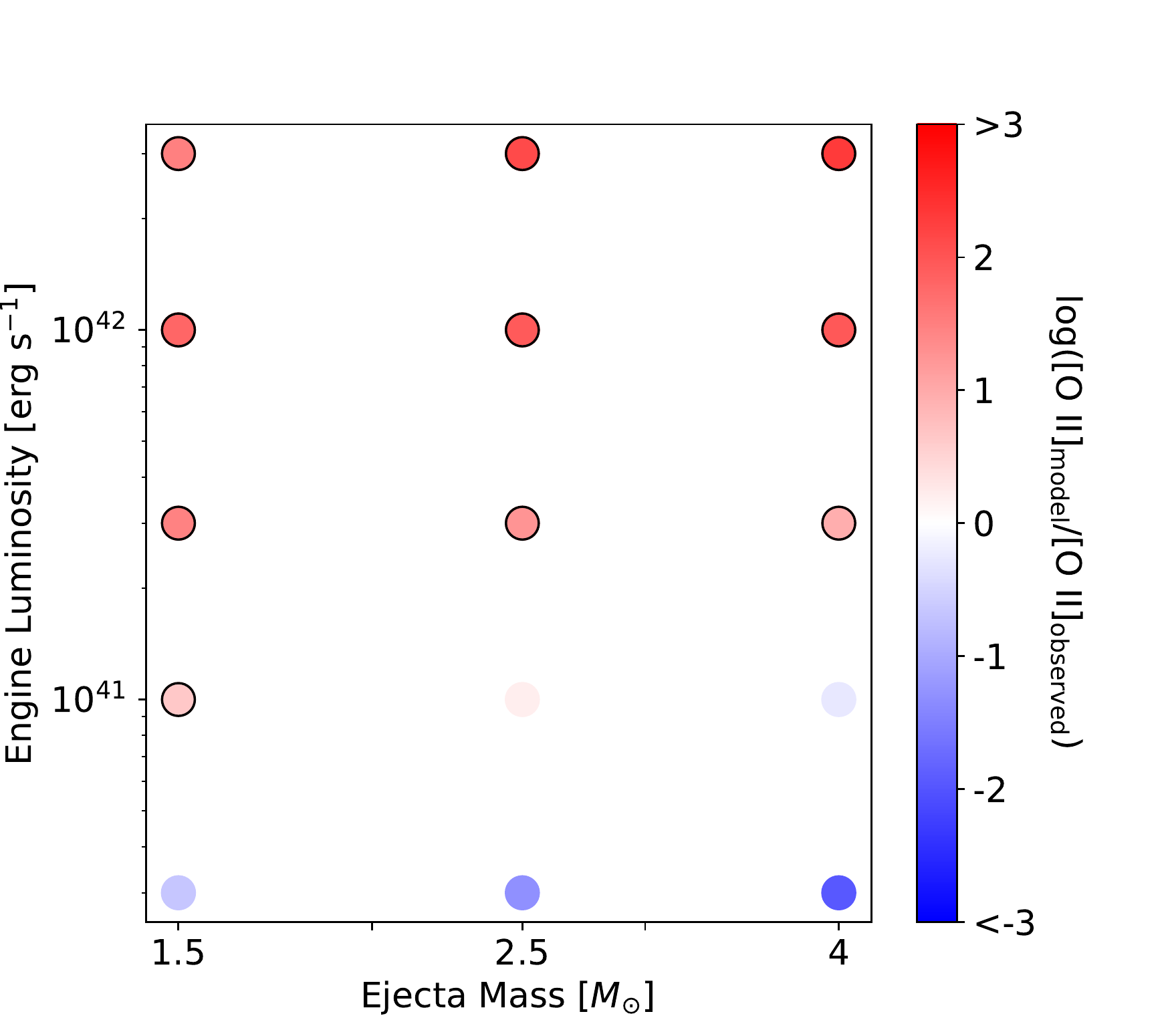}\\[-1.5ex]
\textbf{[O III]}&
\includegraphics[width=1.1\linewidth]{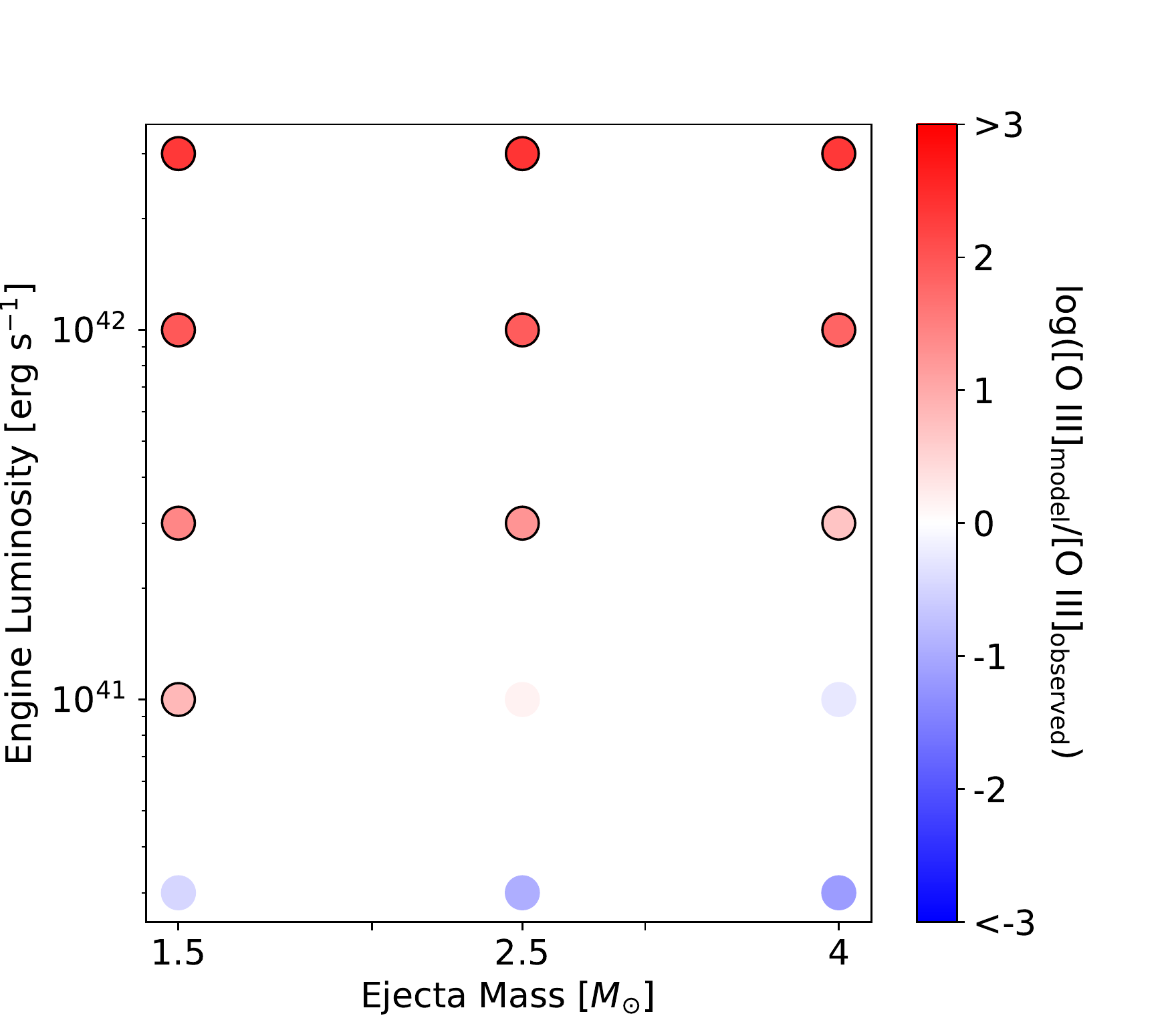}&
\includegraphics[width=1.1\linewidth]{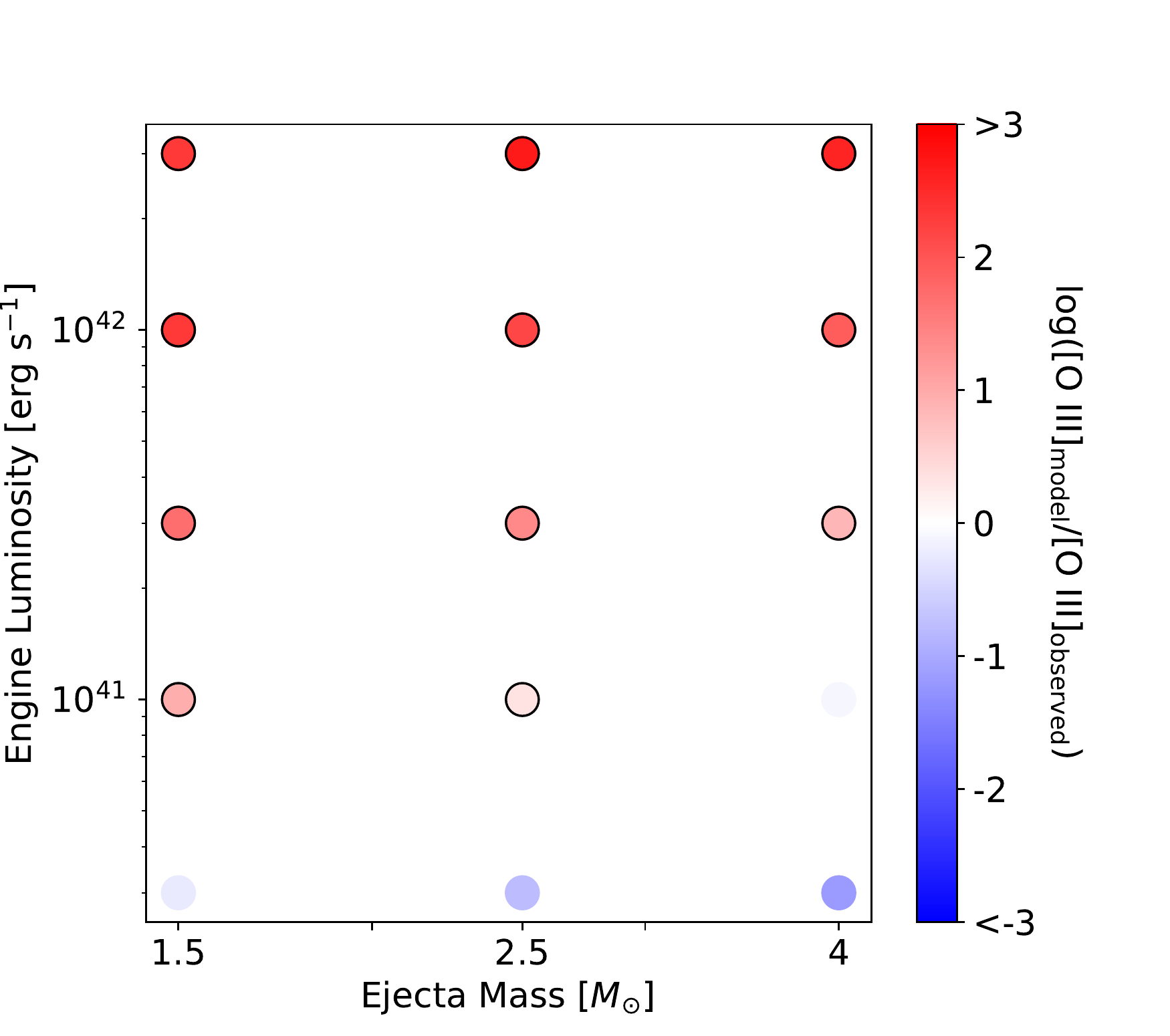}&
\includegraphics[width=1.1\linewidth]{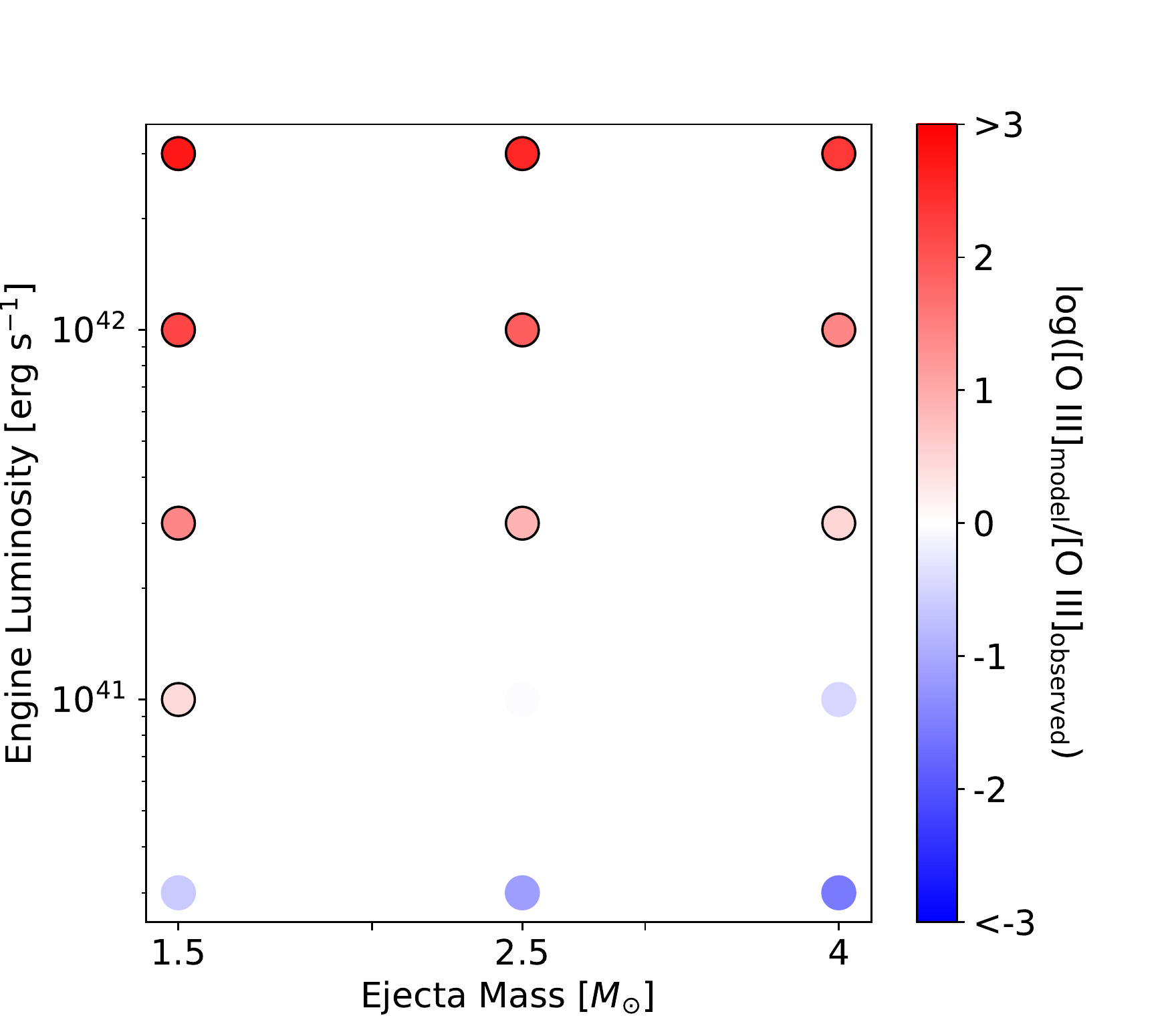}\\[-1.5ex]
\textbf{O I}&
\includegraphics[width=1.1\linewidth]{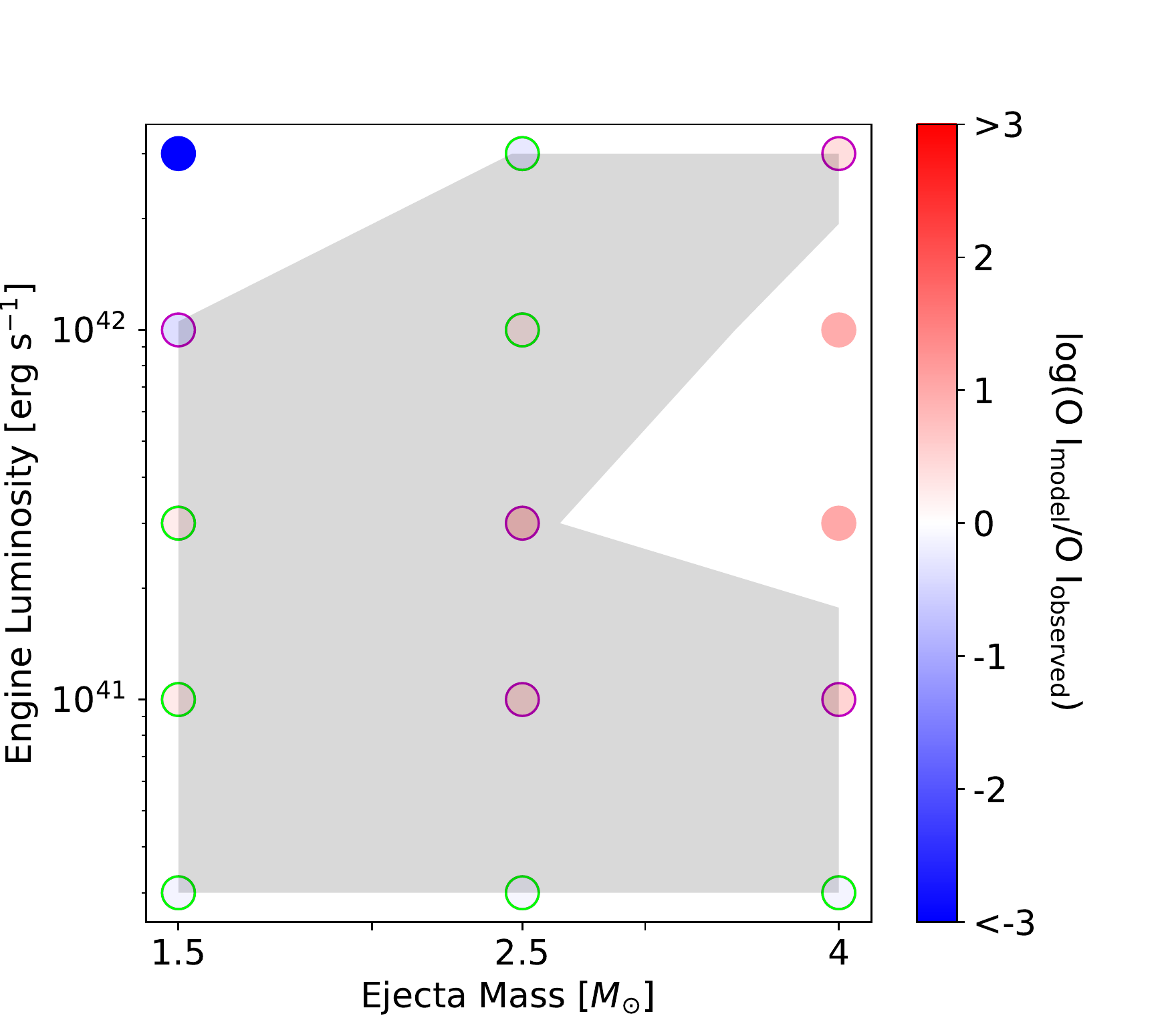}&
\includegraphics[width=1.1\linewidth]{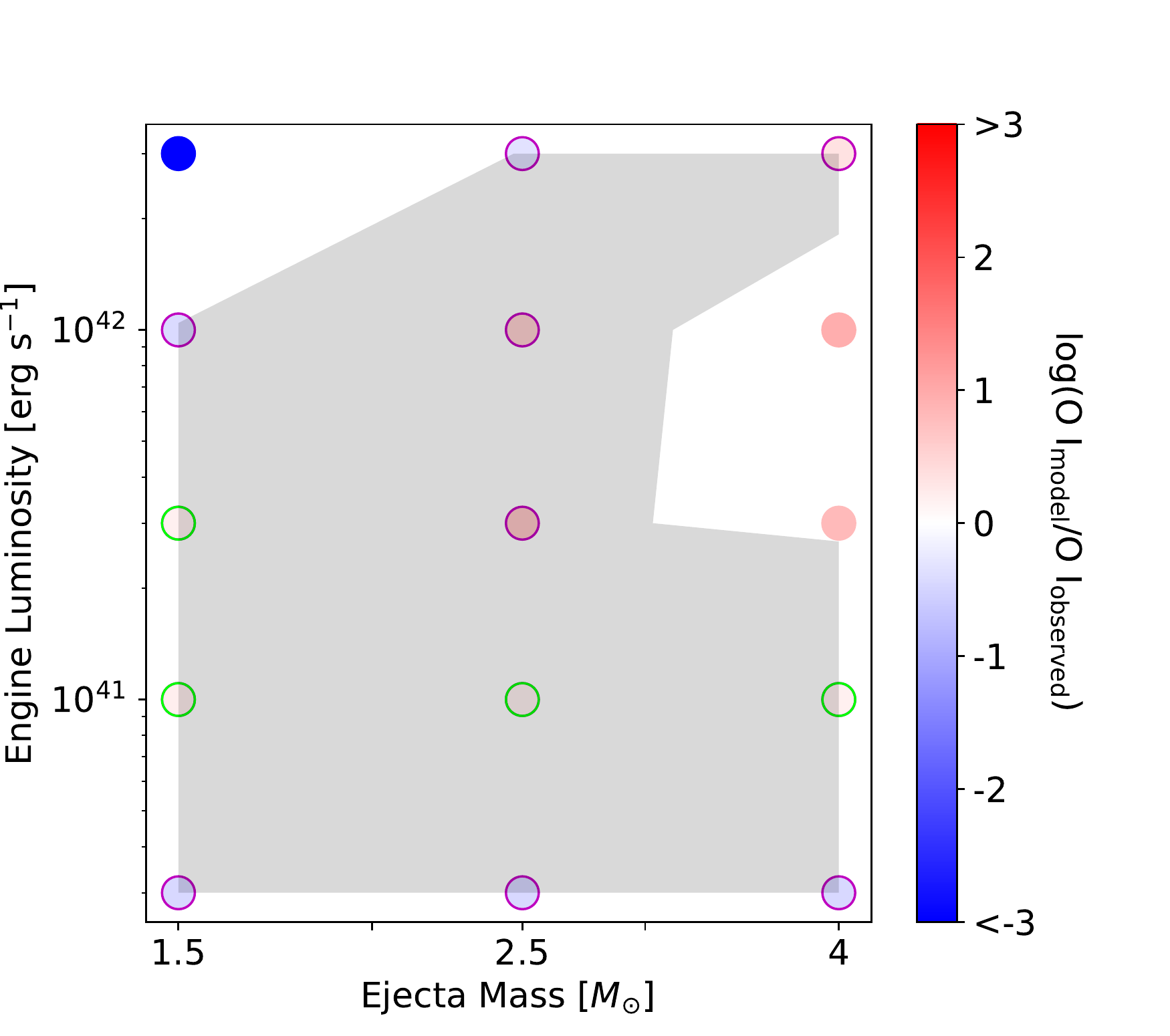}&
\includegraphics[width=1.1\linewidth]{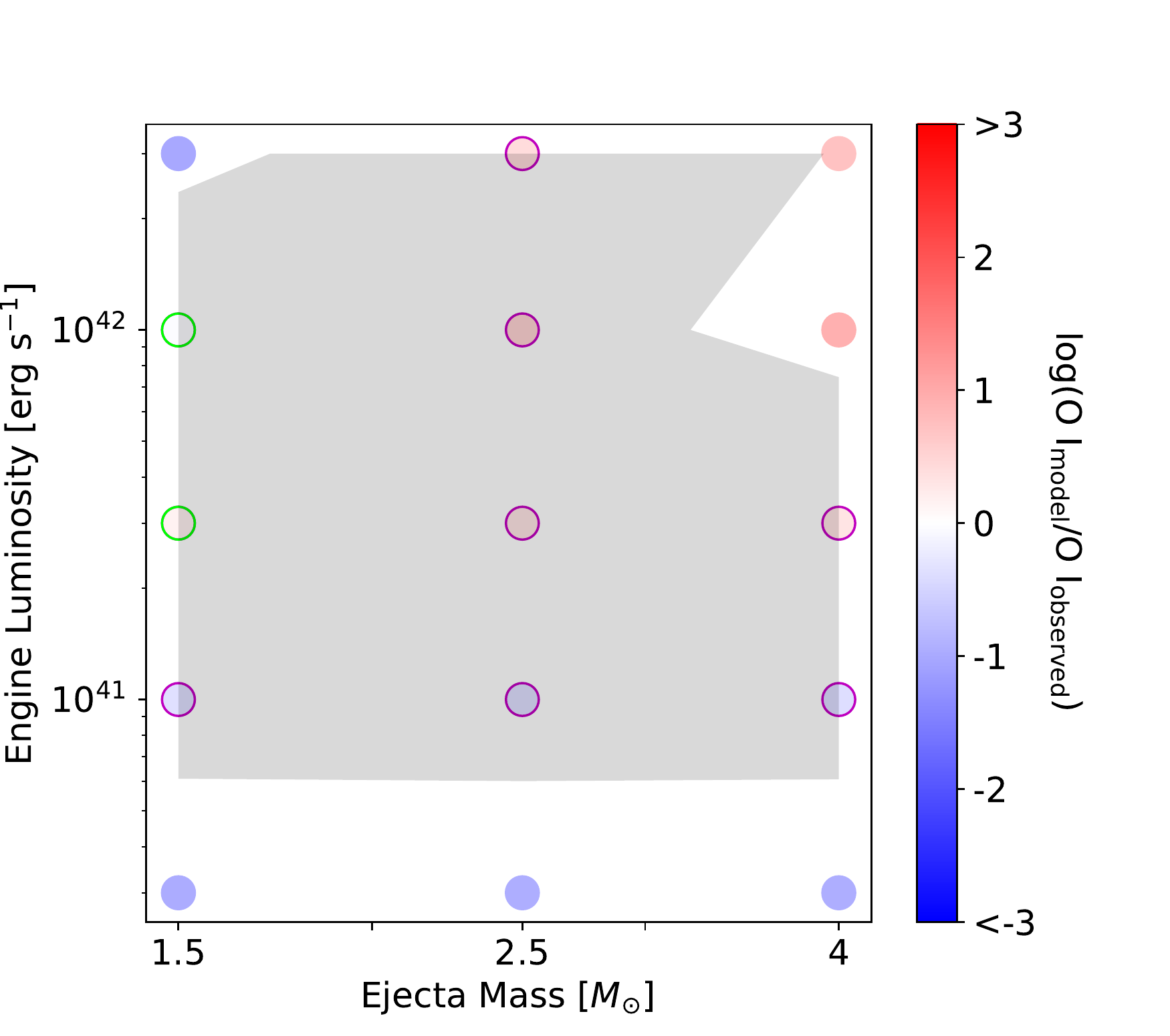}\\[-1.5ex]
\end{tabular}}
\caption{The luminosity of the model [O I] (top), [O II] (second row), [O III] (third row), and O I (bottom) lines in units of the observed line luminosities and limits for SN 2012au at 1 year for a pure oxygen composition at three different values of $T_{\rm PWN}$.  The green circled points represent where the model and observed values are within a factor of 2, the purple circles within a factor of 5, and the grey shaded region also within a factor of 5.  The black circled points (for [O II] and [O III]) represent where the model luminosity is more than a factor 2 larger than the observational limit.}%
\label{fig:o1y_linecomp}
\end{figure*}

The ejecta temperature and ion fractions for O I, O II, and O III at 1 year 
are shown in Figure \ref{fig:o1y_ionfrac} and the dust-corrected model line luminosities of [O I] and O I normalized to the observed line luminosities of 7.1 $\times$ 10$^{39}$ and 1.3 $\times$ 10$^{39}$ erg s$^{-1}$ respectively, and the [O II] and [O III] line luminosities normalized to the observational limits of 3.9 $\times$ 10$^{39}$ and 8.6 $\times$ 10$^{38}$ erg s$^{-1}$ respectively \citep[at 320 days post peak,][]{Milisavljevic2013}, are shown in Figure \ref{fig:o1y_linecomp}. 

The ionization state shows the same qualitative behaviour as at 6 years; with runaway ionization occurring for O I in the high engine luminosity, low ejecta mass regime, with this regime shrinking as $T_{\rm PWN}$ increases; and runaway ionization occurring for O II at the very highest engine luminosity and lowest ejecta mass in the model grid.  

The ejecta temperature at 1 year shows the same qualitative behaviour as at 6 years, increasing when the ejecta mass decreases, the engine luminosity increases, or the injection SED temperature increases. The temperature covers the range $4,000-15,000$ K. Relating back to Section \ref{sec:olines}, this means that we stay in the regime where a stronger [O II] than [O I] can only occur if O II is more abundant than O I. Furthermore, the [O III] to [O I] ratio roughly reflects their abundances.

The luminosity of [O I] 
most closely matches observations at low $L_{\rm PWN}$ and along a narrow strip in the parameter space near the edge of where runaway ionization occurs, which is not well resolved in our model grid.  The [O II] and [O III] luminosities trace the ionization factions fairly well and are high enough to exclude most of the high $L_{\rm PWN}$ portion of the parameter space, although the [O II] emission is much stronger for higher $T_{\rm PWN}$, which results in high $T_{\rm ej}$. The line luminosity for O I is fairly consistent with observations over most of the model grid, and is only weak when O II is experiencing runaway ionization.

\begin{figure*}
\newcolumntype{D}{>{\centering\arraybackslash} m{6cm}}
\noindent
\makebox[\textwidth]{
\begin{tabular}{DDD}
\boldsymbol{$T_{\rm PWN} = 10^5$} \textbf{ K} & \boldsymbol{$T_{\rm PWN} = 3 \times 10^5$} \textbf{ K} & \boldsymbol{$T_{\rm PWN} = 10^6$} \textbf{ K}\\
\includegraphics[width=1.1\linewidth]{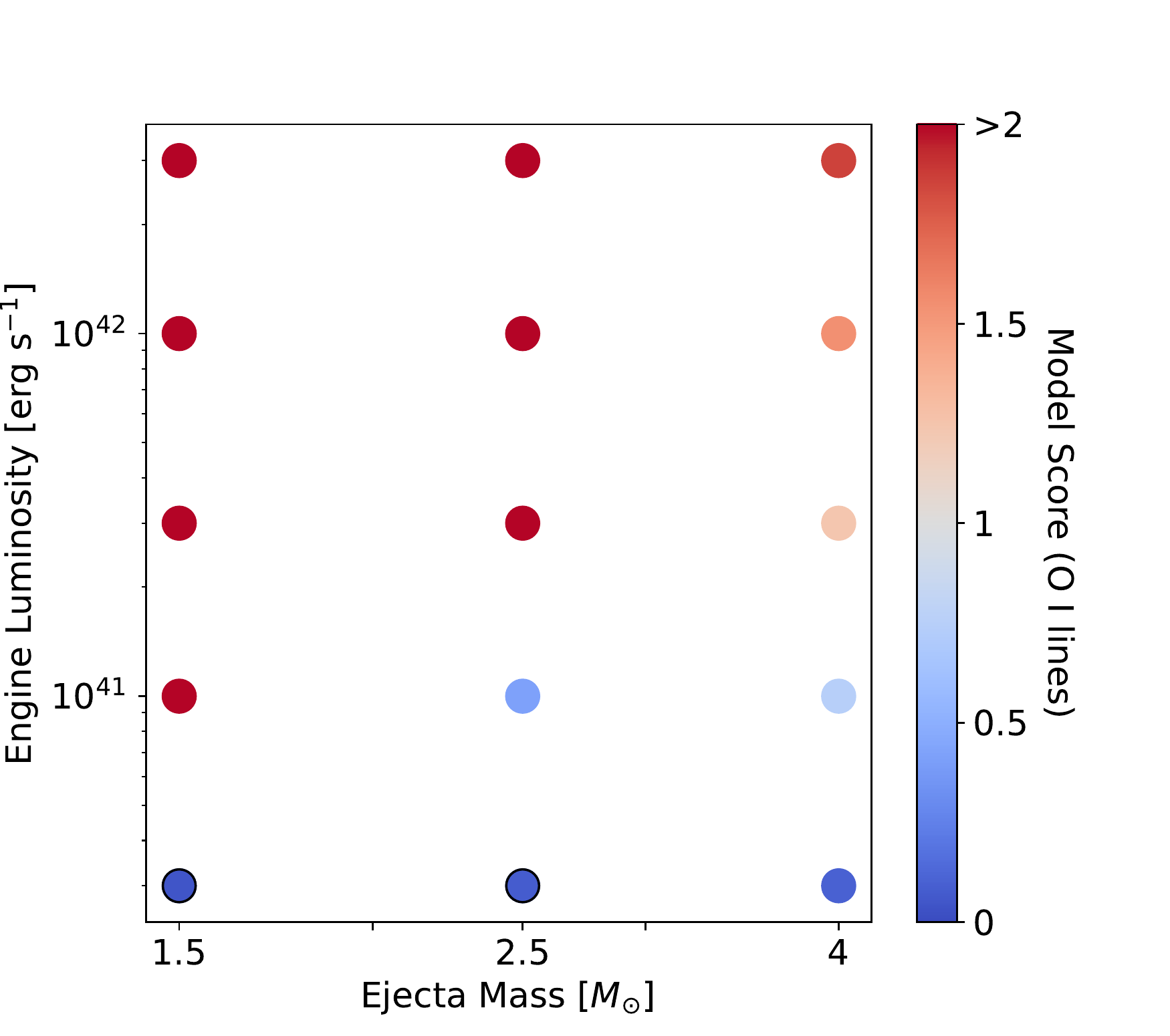}&
\includegraphics[width=1.1\linewidth]{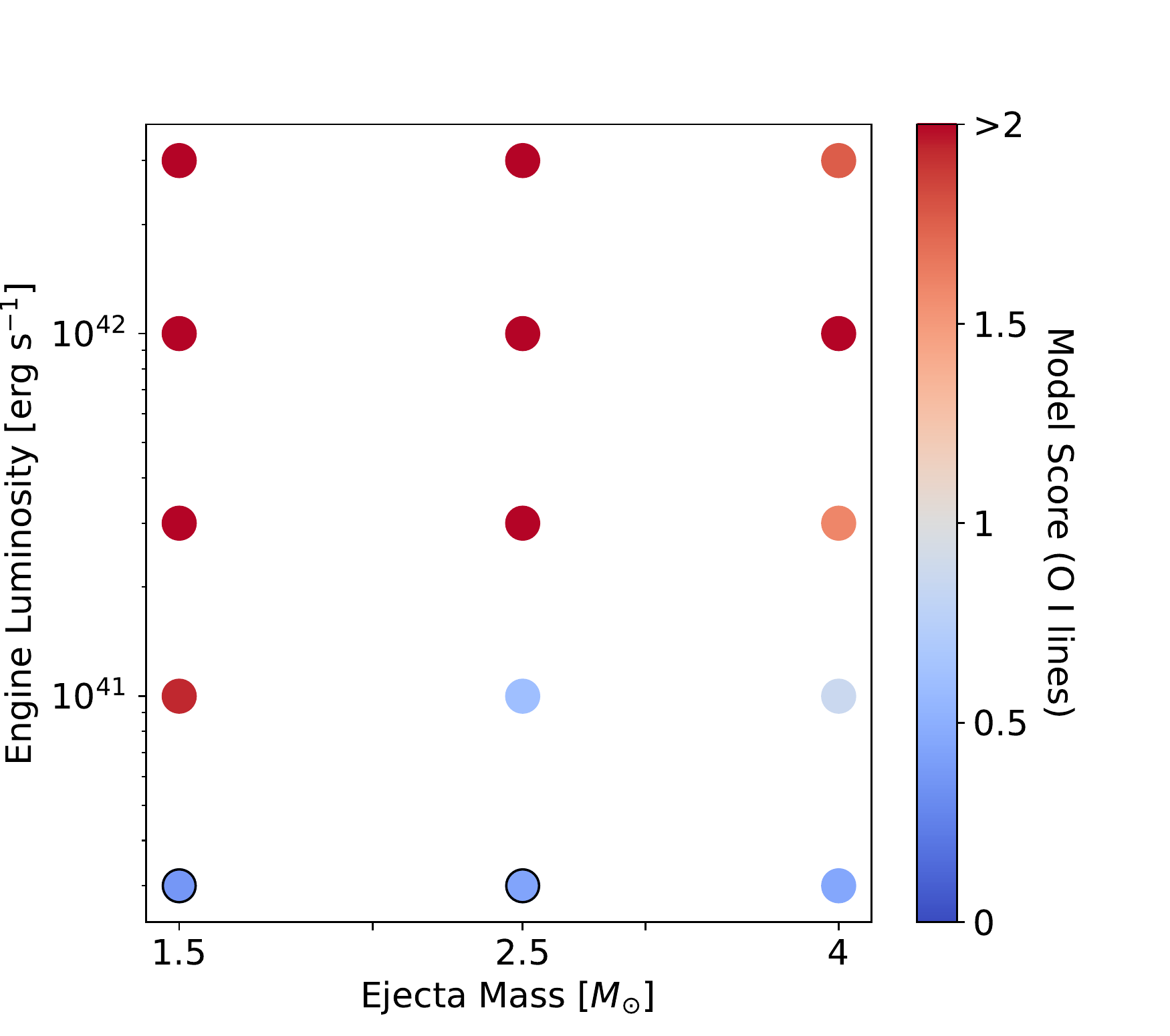}&
\includegraphics[width=1.1\linewidth]{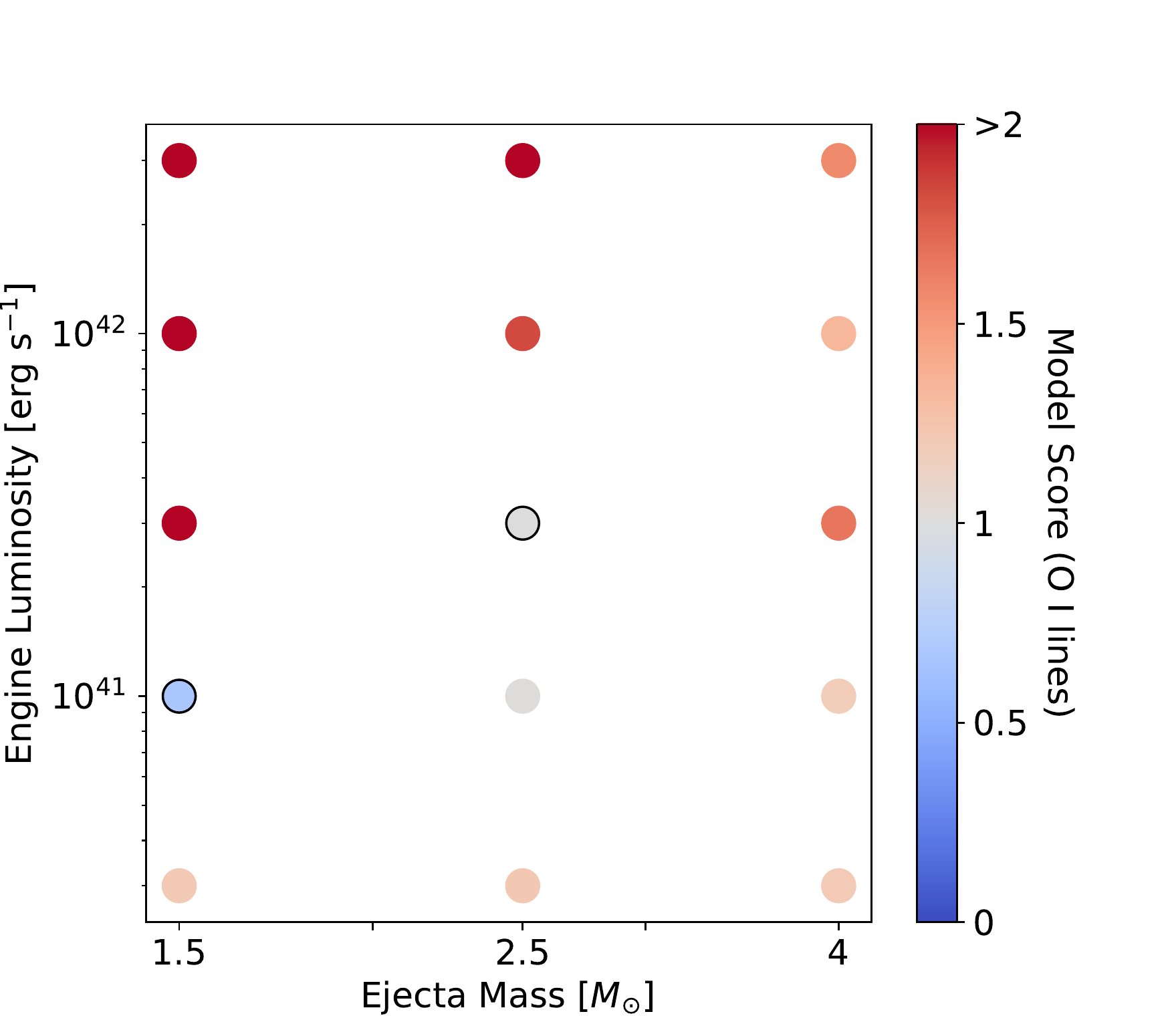}\\[-1.5ex]
\end{tabular}}
\caption{The goodness-of-fit score for each model for the pure oxygen composition at 1 year based on the [O I] and O I lines.  Lower scores indicate a better fit to the data (from Equation \ref{eqn:modscore}, a perfect fit has score 0, both lines off by factor 2 has score 0.18, and both lines off by factor 10 has score 2). The black circles indicate the two models with the lowest scores for each $T_{\rm PWN}$, which are plotted in Figure \ref{fig:o1y_spec}.}%
\label{fig:o1y_score}
\end{figure*}

\begin{figure*}
\newcolumntype{D}{>{\centering\arraybackslash} m{6cm}}
\noindent
\makebox[\textwidth]{
\begin{tabular}{DDD}
\includegraphics[width=1.1\linewidth]{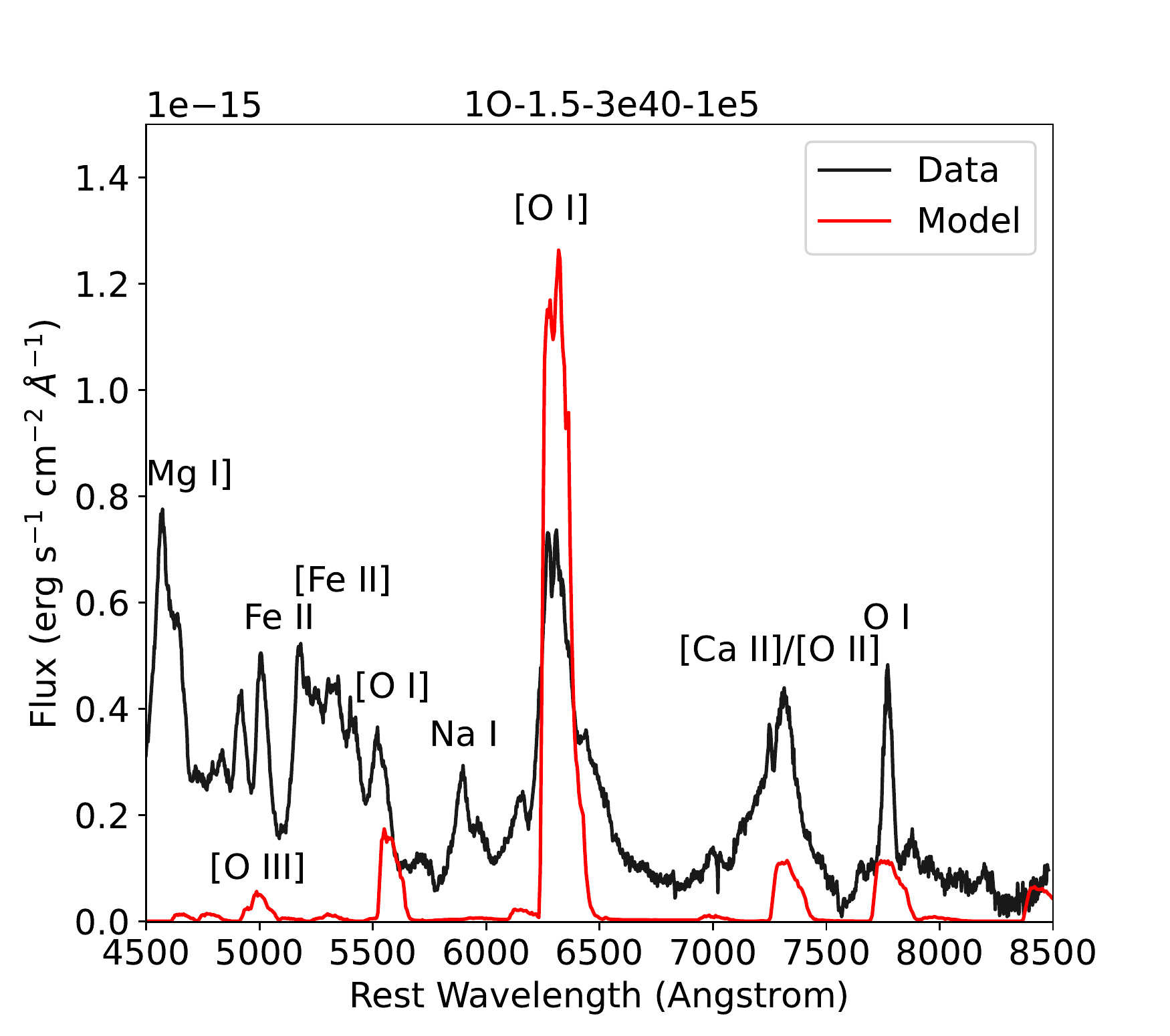}&
\includegraphics[width=1.1\linewidth]{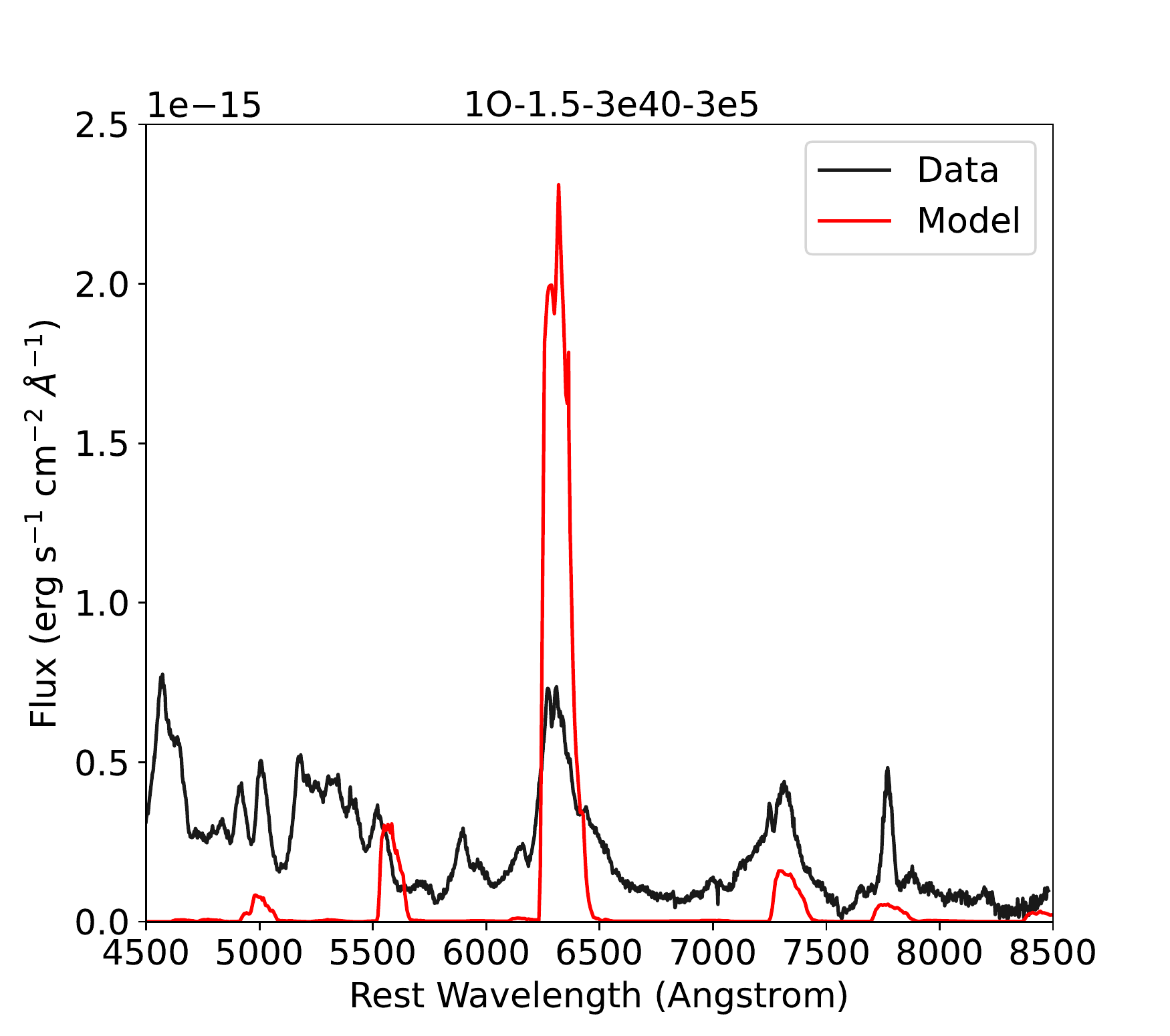}&
\includegraphics[width=1.1\linewidth]{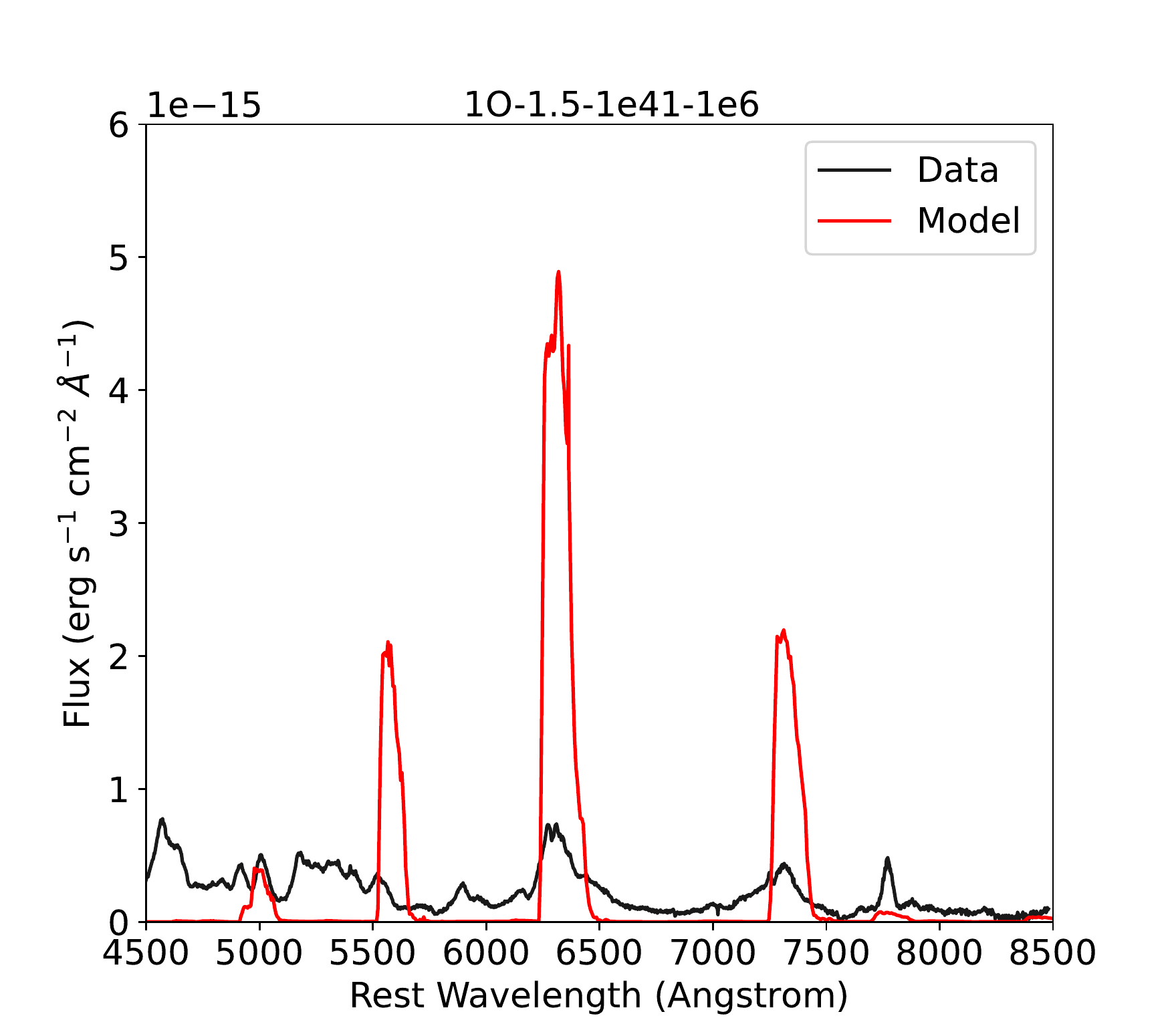}\\
\includegraphics[width=1.1\linewidth]{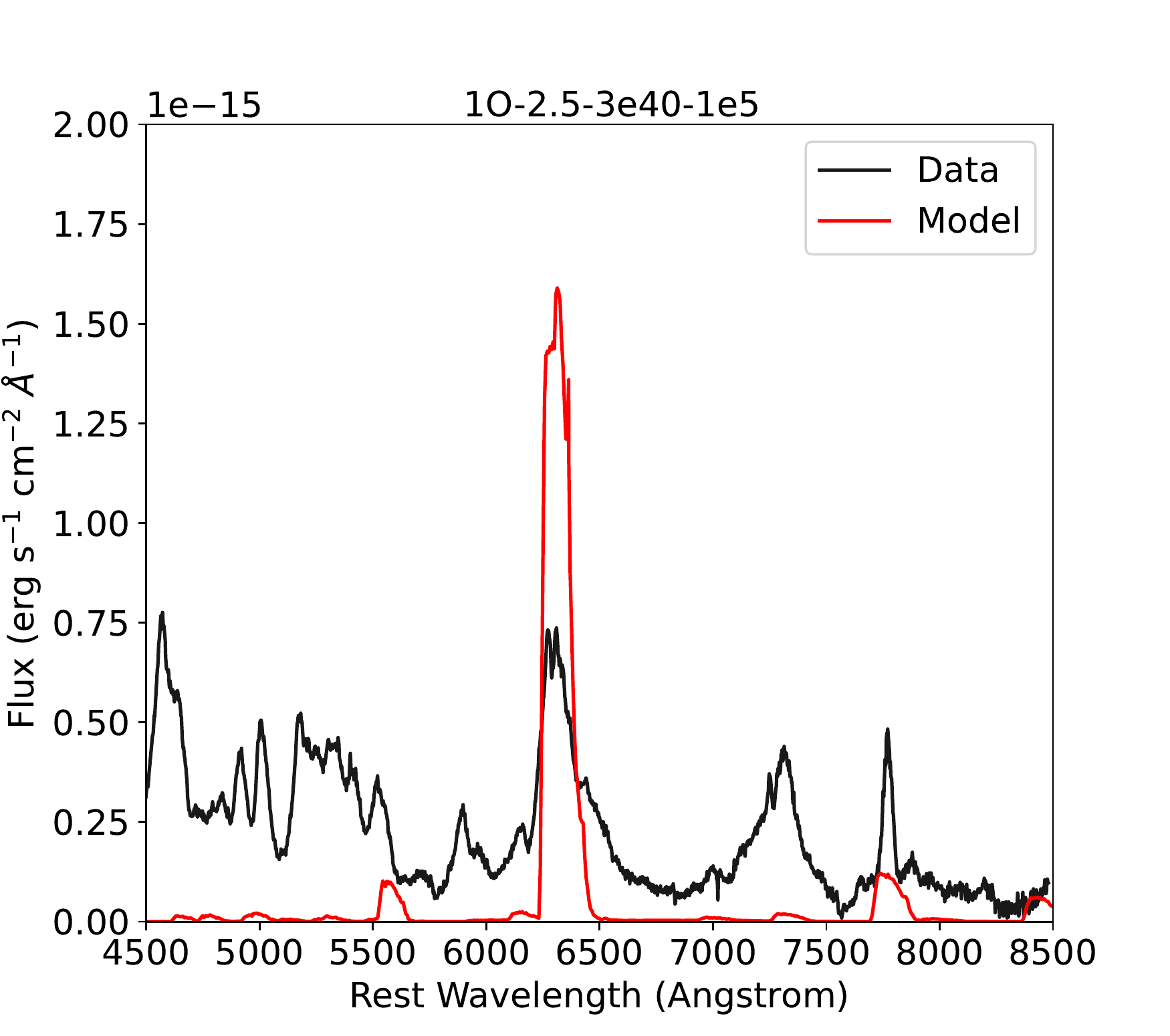}&
\includegraphics[width=1.1\linewidth]{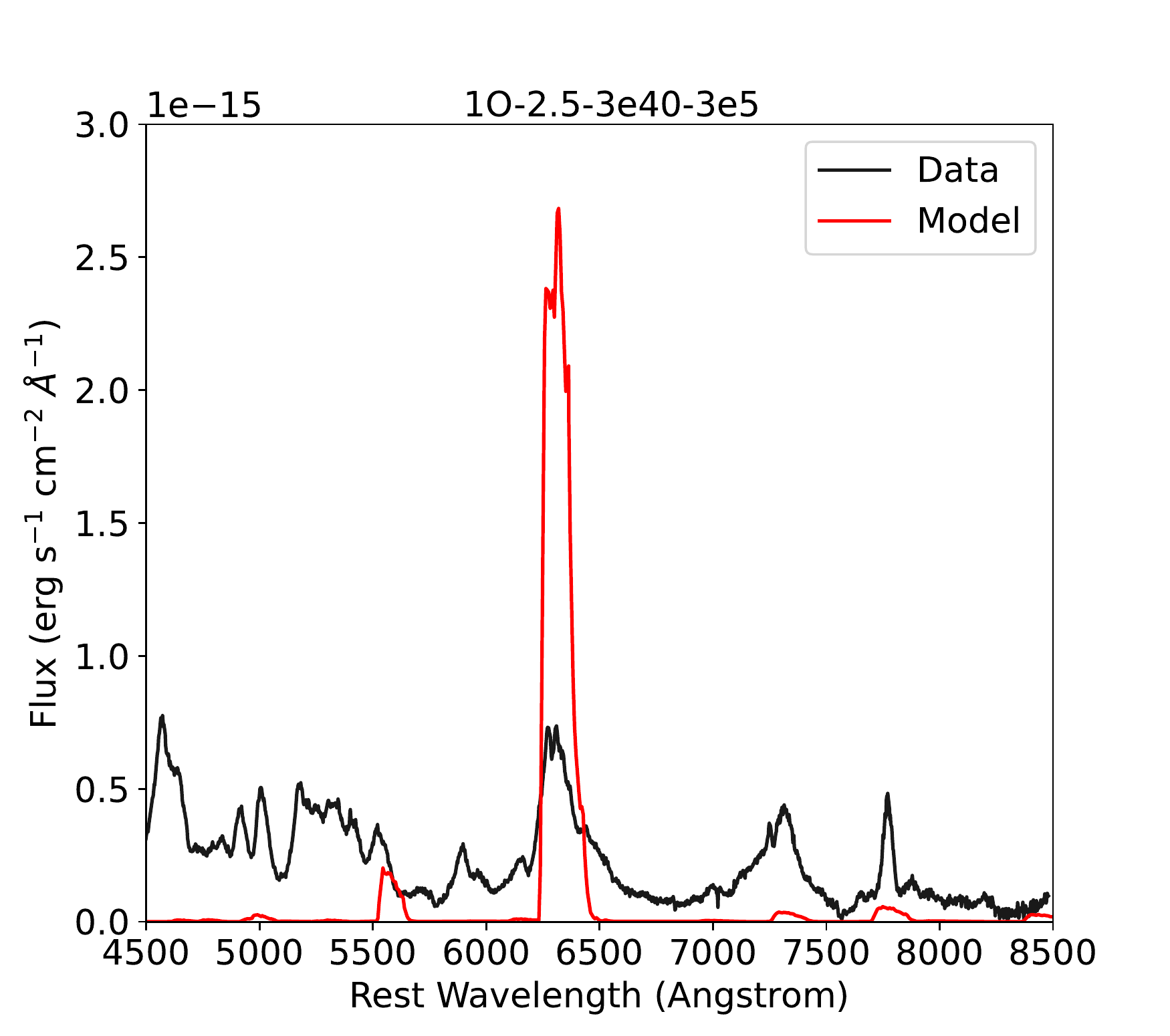}&
\includegraphics[width=1.1\linewidth]{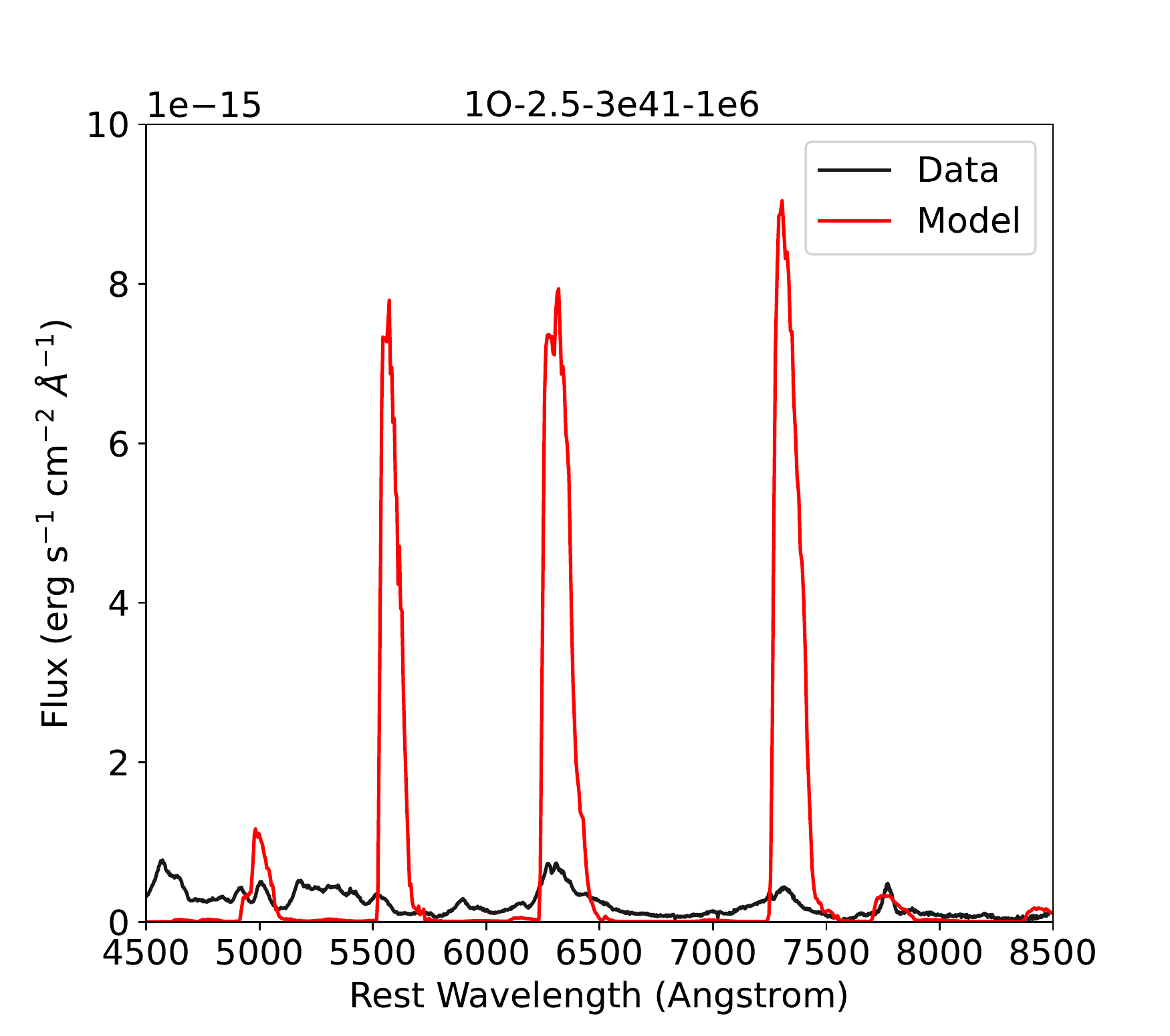}
\end{tabular}}
\caption{The two best-fitting dust-corrected model spectra to SN 2012au for each value of $T_{\rm PWN}$ at 1 years for a pure oxygen composition compared to the observed spectrum from \cite{Milisavljevic2013}.   Strong lines and features are labelled in the upper left plot.  
}%
\label{fig:o1y_spec}
\end{figure*}

The model scores based on the O I and [O I] lines (there are only upper limits for [O II] and [O III]) are shown in Figure \ref{fig:o1y_score}, and the two best-fitting spectra for each value of $T_{\rm PWN}$ are shown in Figure \ref{fig:o1y_spec}.  Due to our one zone approach, the model [O I] line is significantly narrower than the observed line, so the peak intensity must be significantly higher to compensate.  Spectra at $T_{\rm PWN} = 10^6$ K have strong [O II], [O III], and [O I] $\lambda$ 5577 $\AA$ emission, as well as [O I] emission much stronger than observed.  The models at $T_{\rm PWN} = 3 \times 10^5$ K show no obvious inadequacies, and fit the [O I] and O I lines within a factor of 5, but the models at $T_{\rm PWN} = 10^5$ K fit both lines within a factor of 2, making these the best fitting models at 1 year.


 While this is at the edge of the explored parameter space at 1 year, there are physical motivations for not extending the model grid to lower ejecta masses, engine luminosities, and injection SED temperatures.  An ejecta mass of 1.5 $M_{\odot}$ is not atypical of a SN Ic-BL \citep{Taddia2019}, but is lower than previous studies have inferred for SN 2012au \citep{Milisavljevic2013, Takaki2013, Pandey2021} as well as the lower limit that could reproduce the spectrum at 6 years.  A lower engine luminosity at 1 year would imply that the spin-down rate would have to be slower than vacuum dipole and have a breaking index $n < 2.5$, which is lower than that inferred for most pulsars \citep{Parthasarathy2020} and for potential millisecond magnetars that have driven GRBs \citep{Lasky2017, SasmazMus2019}.  Finally, a lower injection SED temperature is inconsistent with the results found at 6 years regarding the O I line emission; a PWN is also expected to emit lower energy photons at late times as the energy injection decreases and the nebula itself expands (see Equation \ref{eqn:nub}), so a lower $T_{\rm PWN}$ at early times seems unphysical.  It is likely that the PWN emits mainly in hard X-rays at 1 year, with a corresponding $T_{\rm PWN}$ $\approx 10^7$ K, but we can not yet probe that regime due to the lack of inner shell processes in \texttt{SUMO} (see Section \ref{sec:modsetup}).

\subsection{Realistic Composition}

\subsubsection{6 Years}

\begin{figure*}
\newcolumntype{D}{>{\centering\arraybackslash} m{6cm}}
\noindent
\makebox[\textwidth]{
\begin{tabular}{m{0.8cm} DDD}
& \boldsymbol{$T_{\rm PWN} = 10^5$} \textbf{ K} & \boldsymbol{$T_{\rm PWN} = 3 \times 10^5$} \textbf{ K} & \boldsymbol{$T_{\rm PWN} = 10^6$} \textbf{ K}\\
\textbf{O I}&
\includegraphics[width=1.1\linewidth]{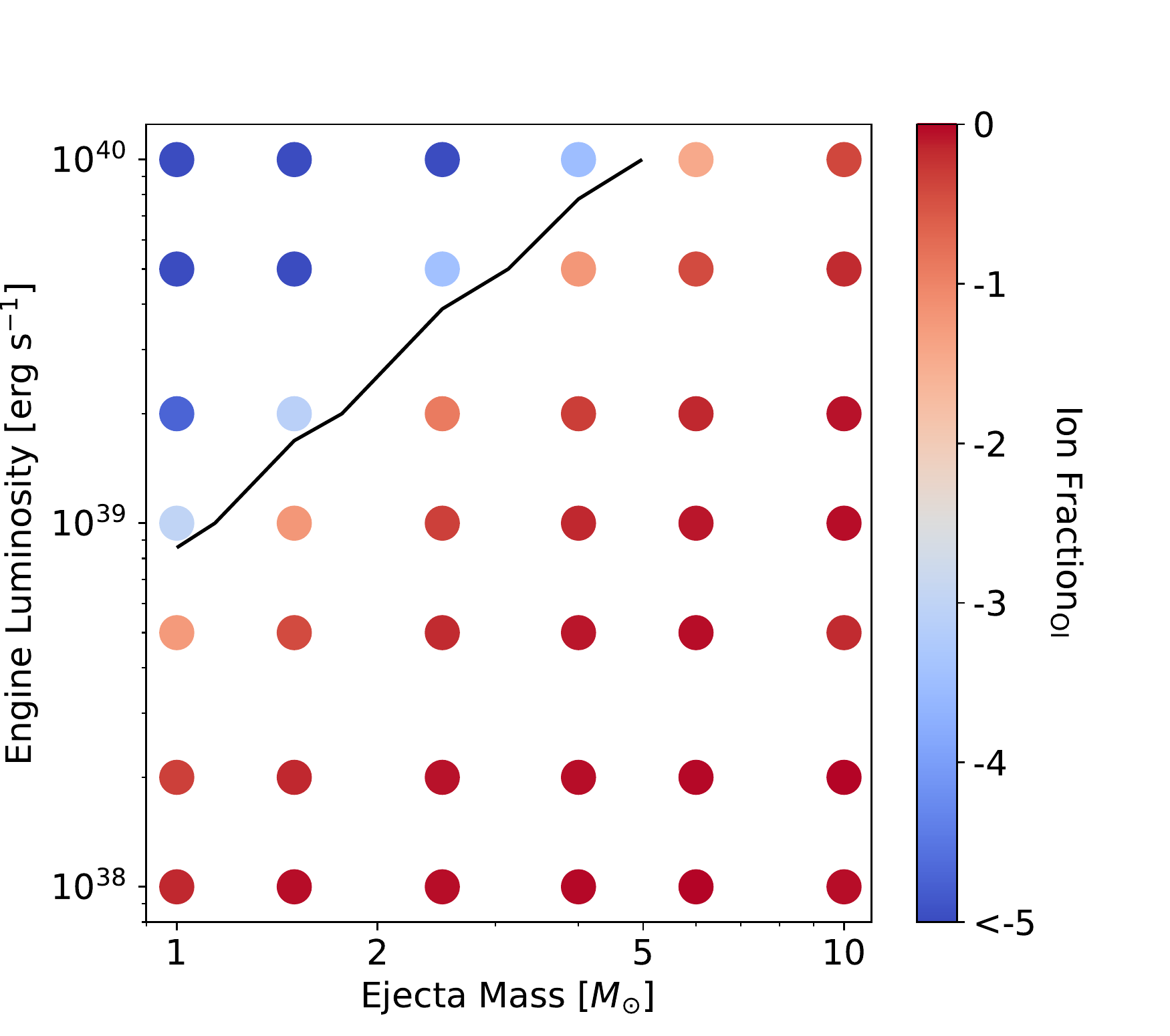}&
\includegraphics[width=1.1\linewidth]{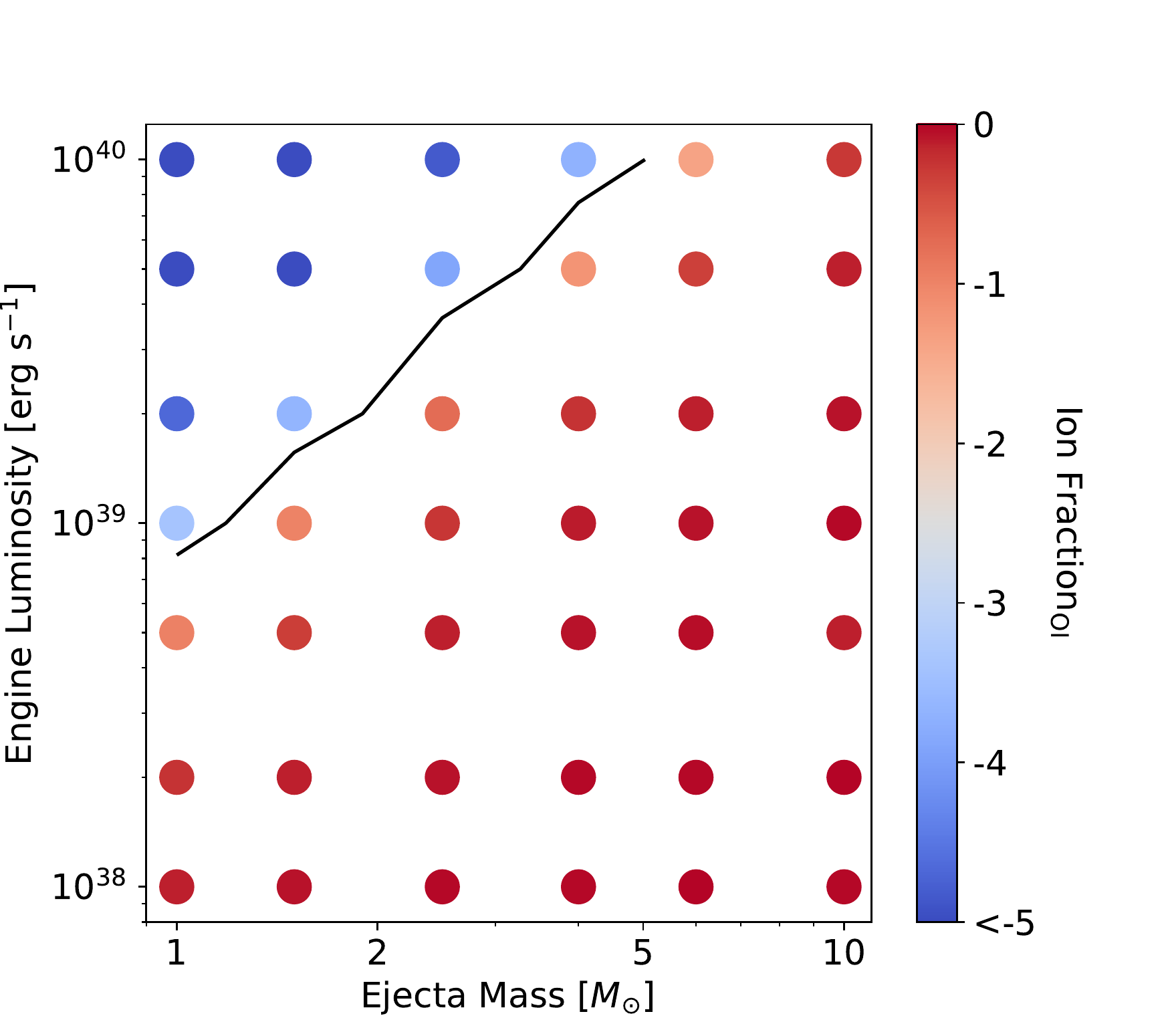}&
\includegraphics[width=1.1\linewidth]{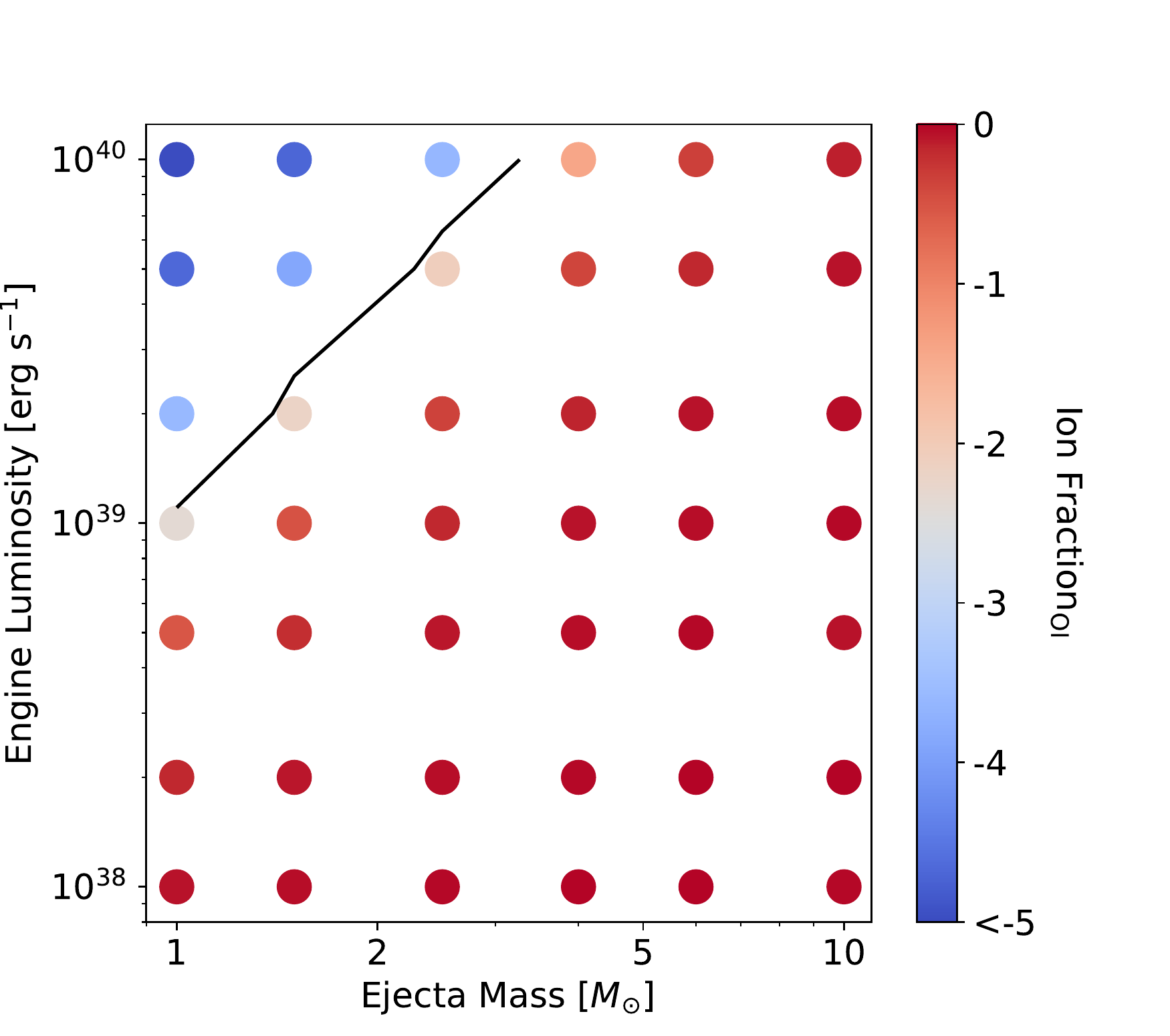}\\[-1.5ex]
\textbf{O II}&
\includegraphics[width=1.1\linewidth]{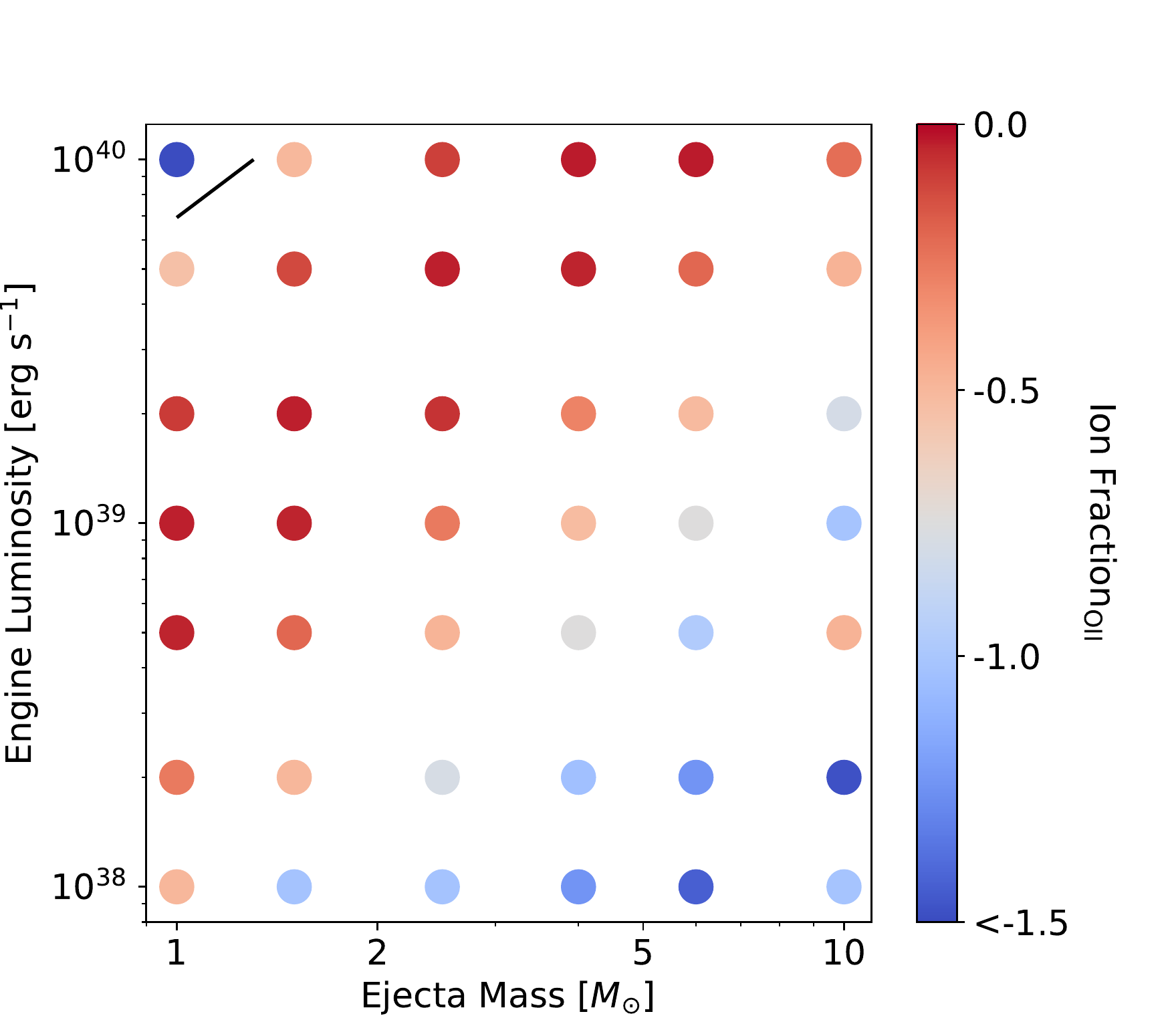}&
\includegraphics[width=1.1\linewidth]{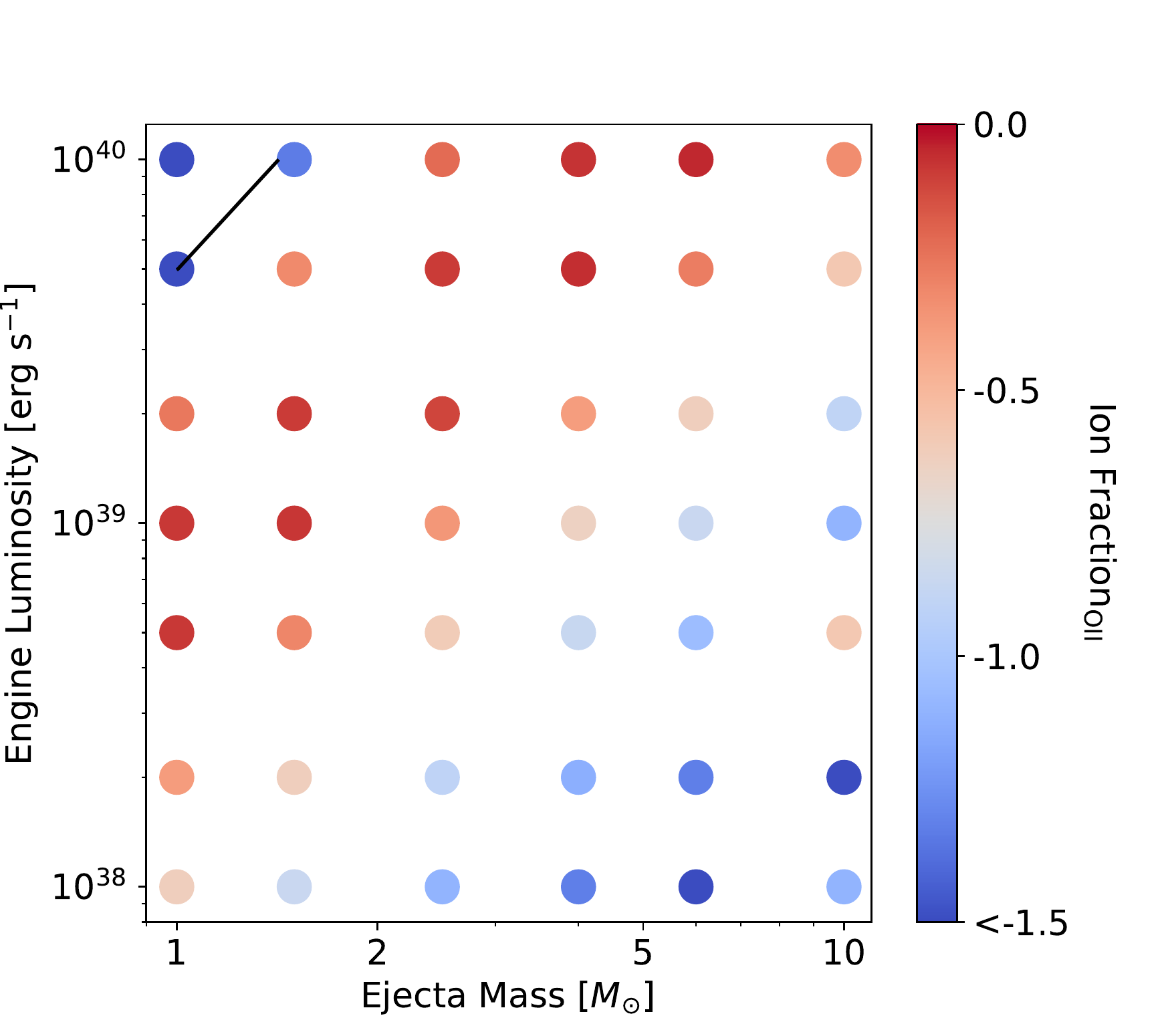}&
\includegraphics[width=1.1\linewidth]{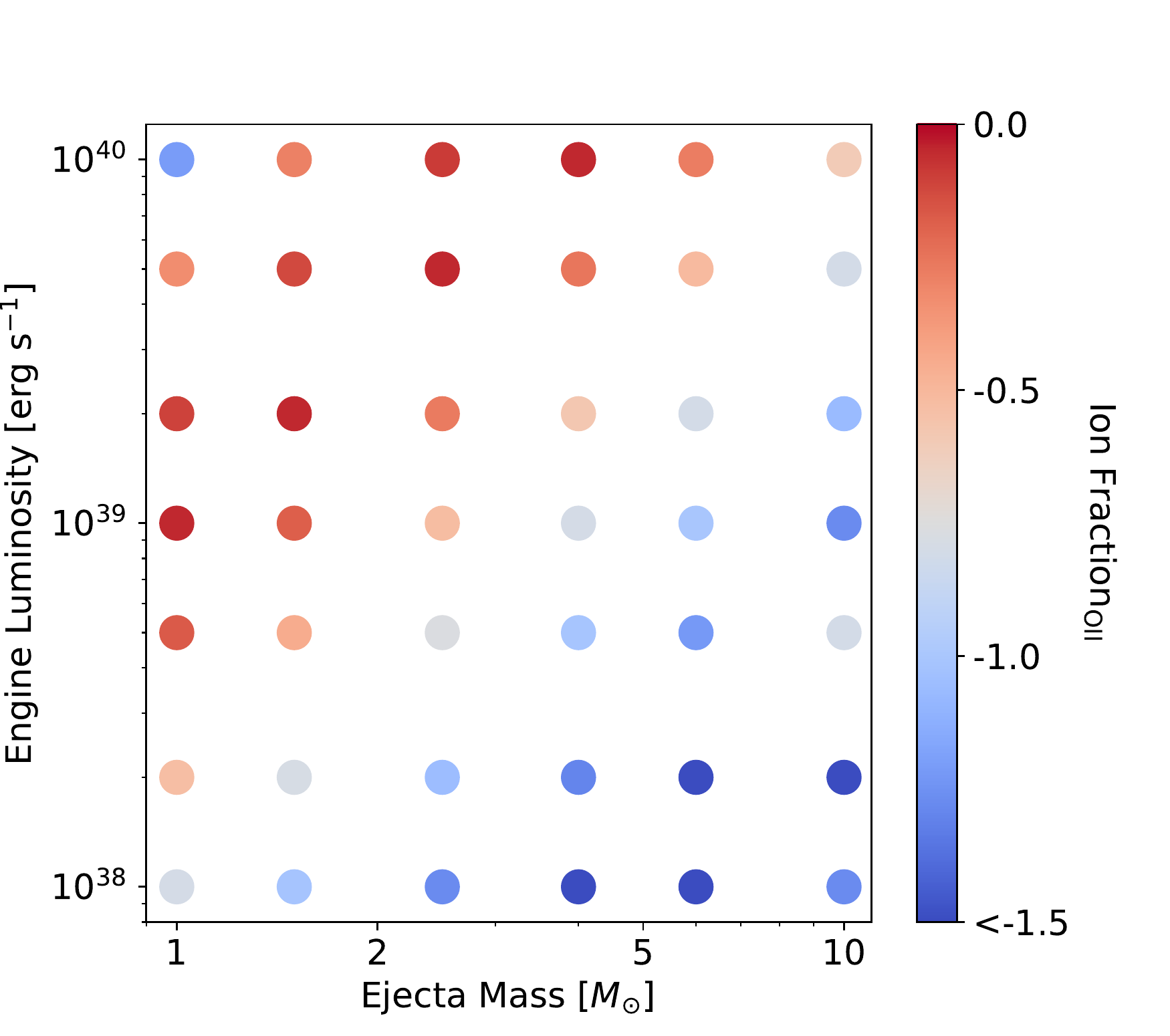}\\[-1.5ex] 
\textbf{O III}&
\includegraphics[width=1.1\linewidth]{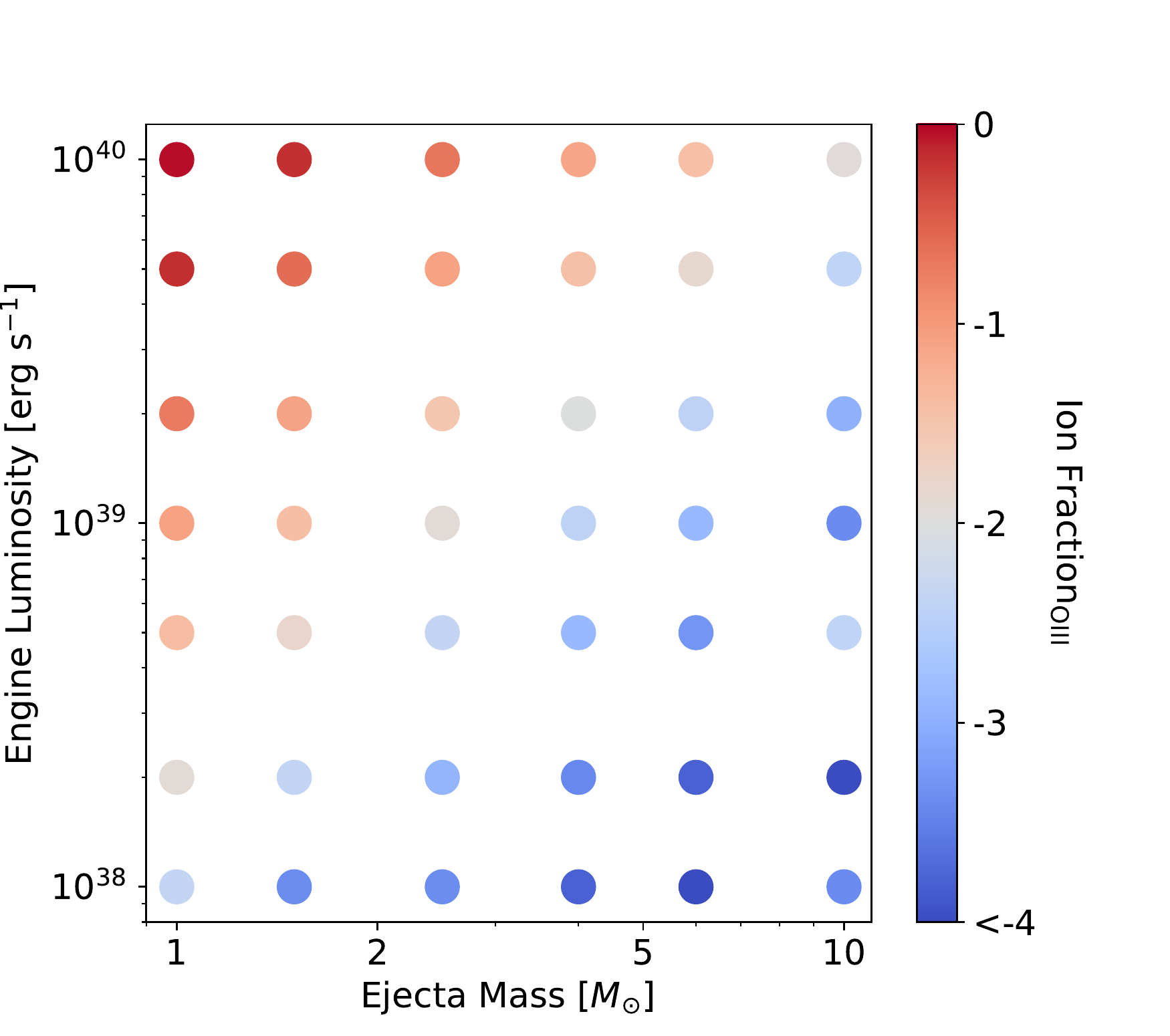}&
\includegraphics[width=1.1\linewidth]{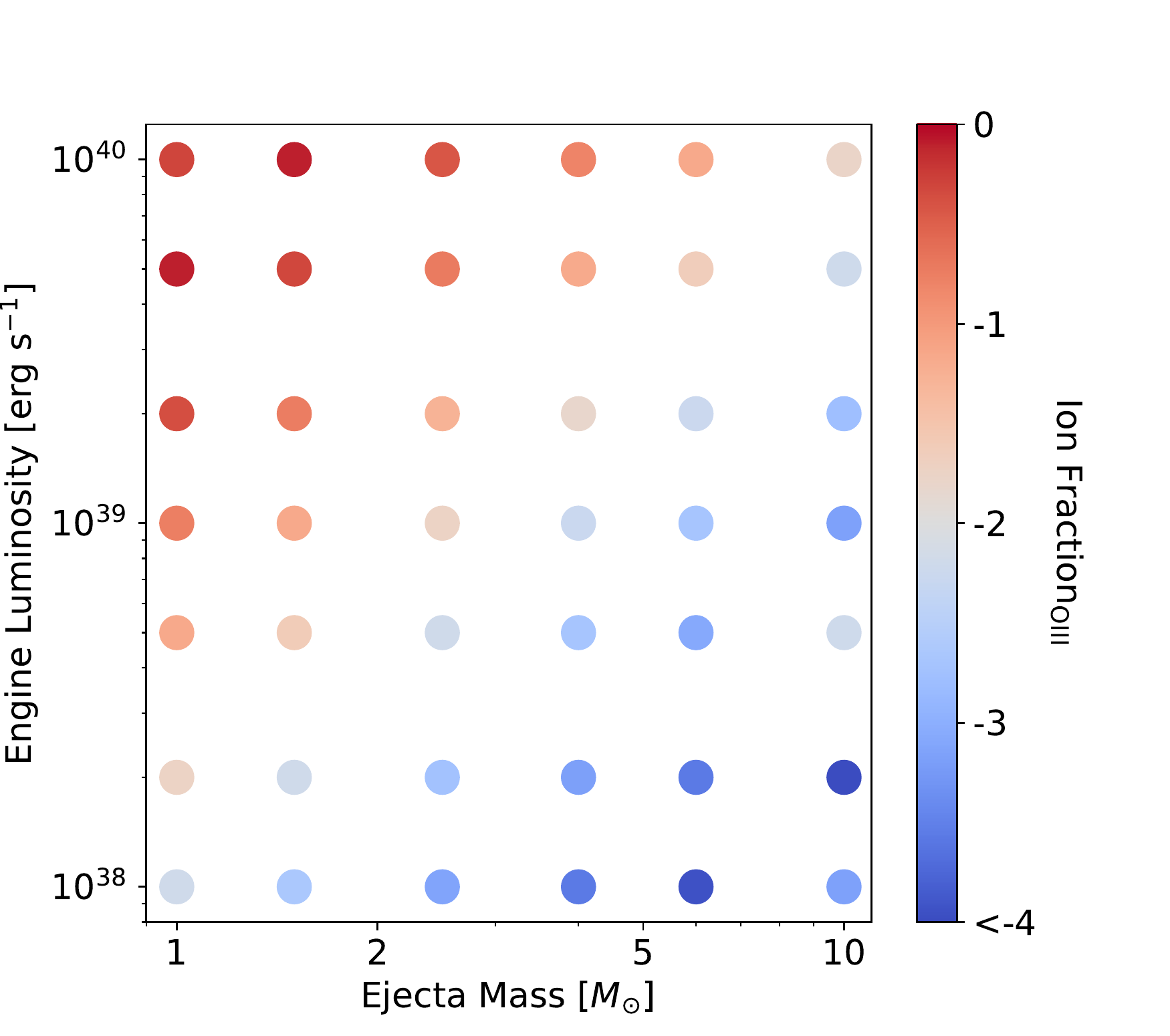}&
\includegraphics[width=1.1\linewidth]{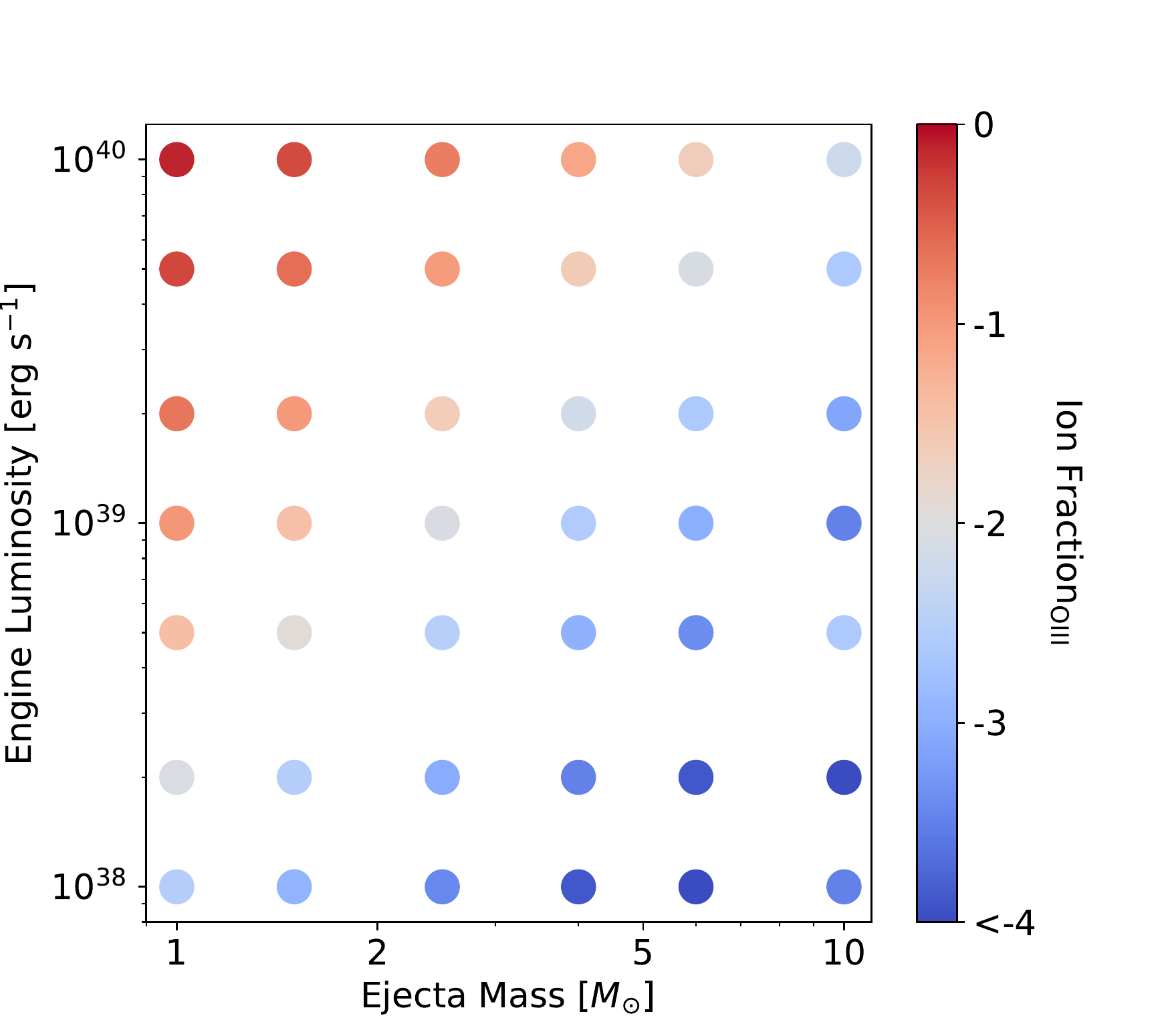}\\[-1.5ex] 
\boldsymbol{$T_{\rm ej}$}&
\includegraphics[width=1.1\linewidth]{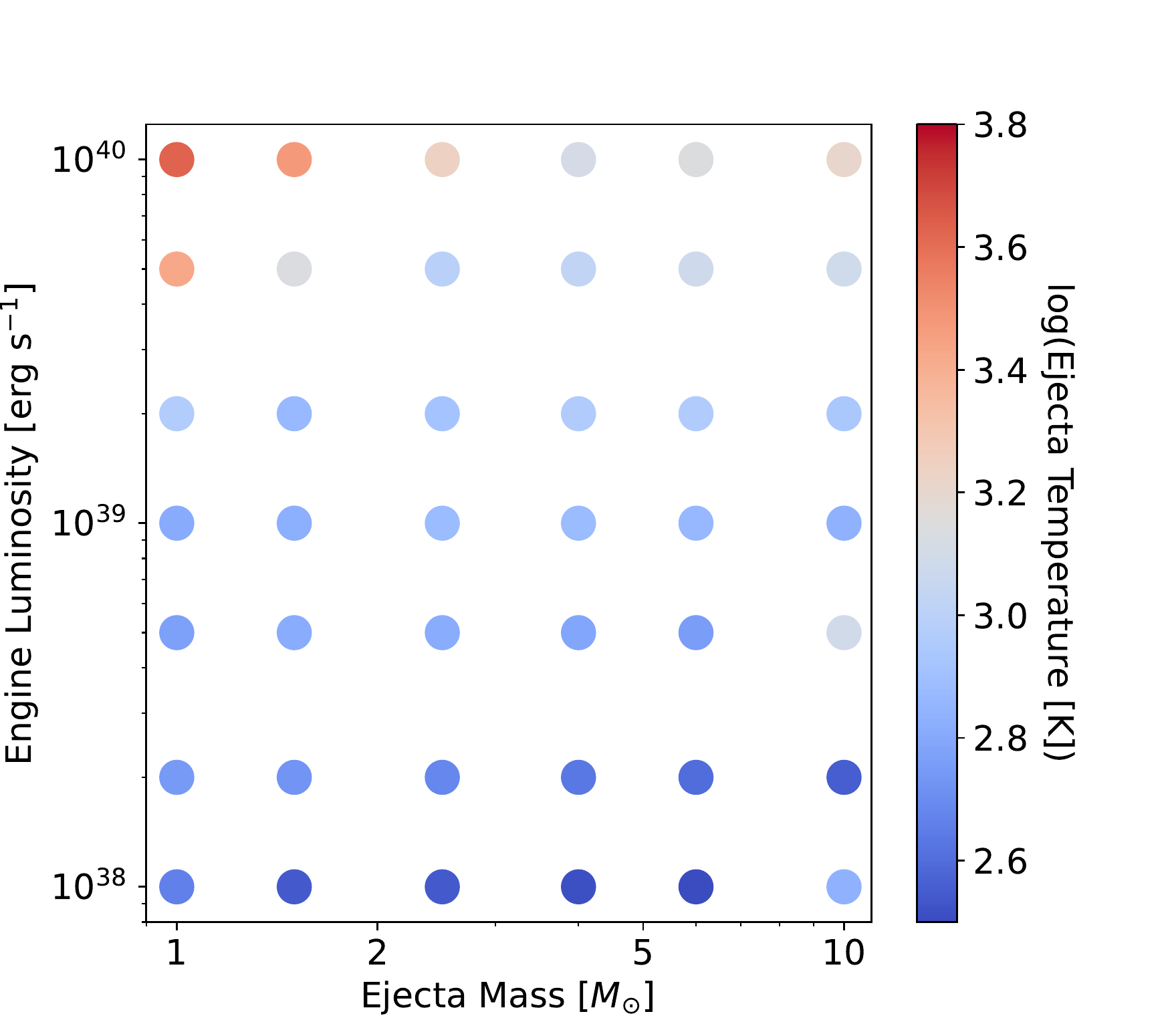}&
\includegraphics[width=1.1\linewidth]{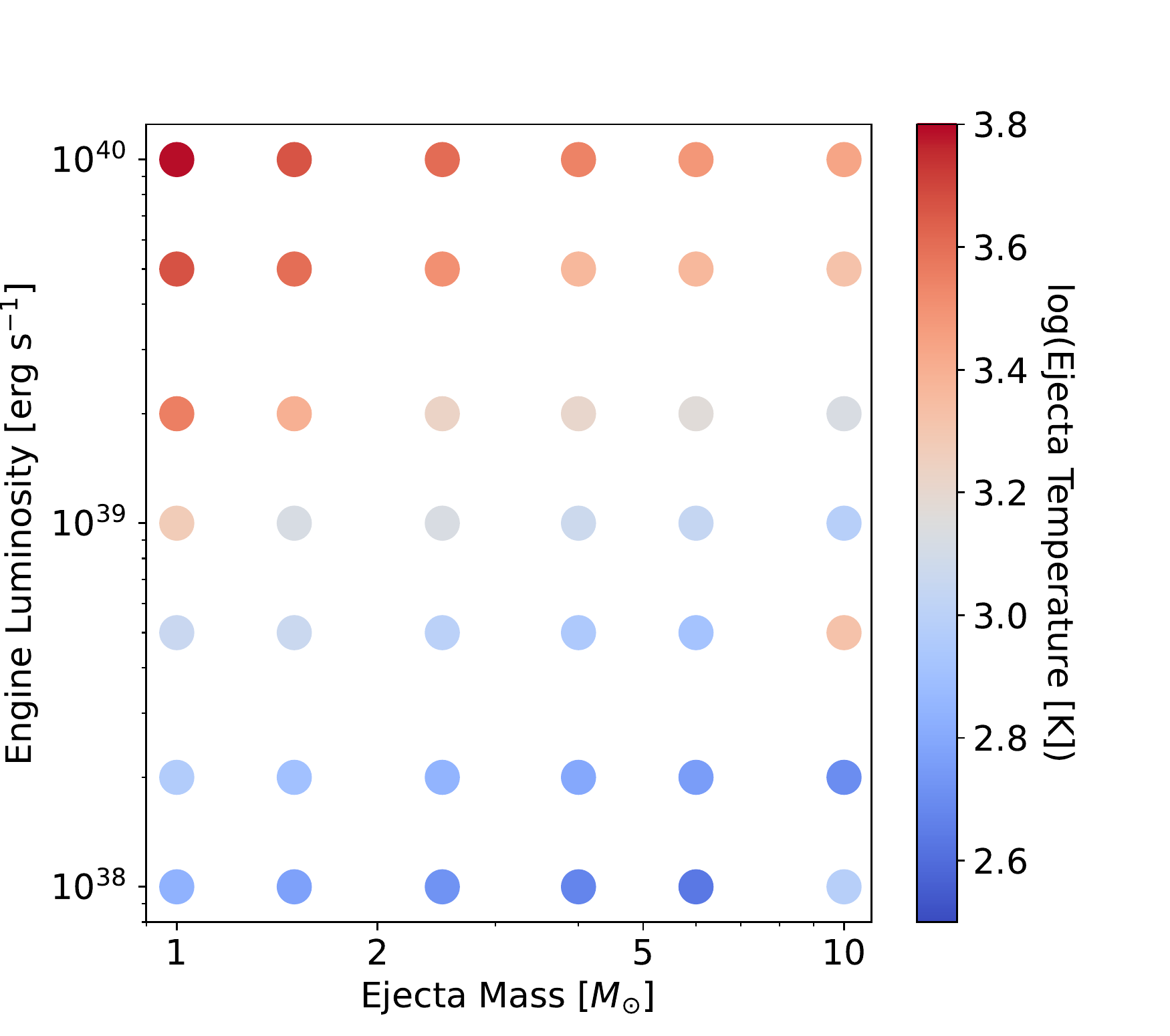}&
\includegraphics[width=1.1\linewidth]{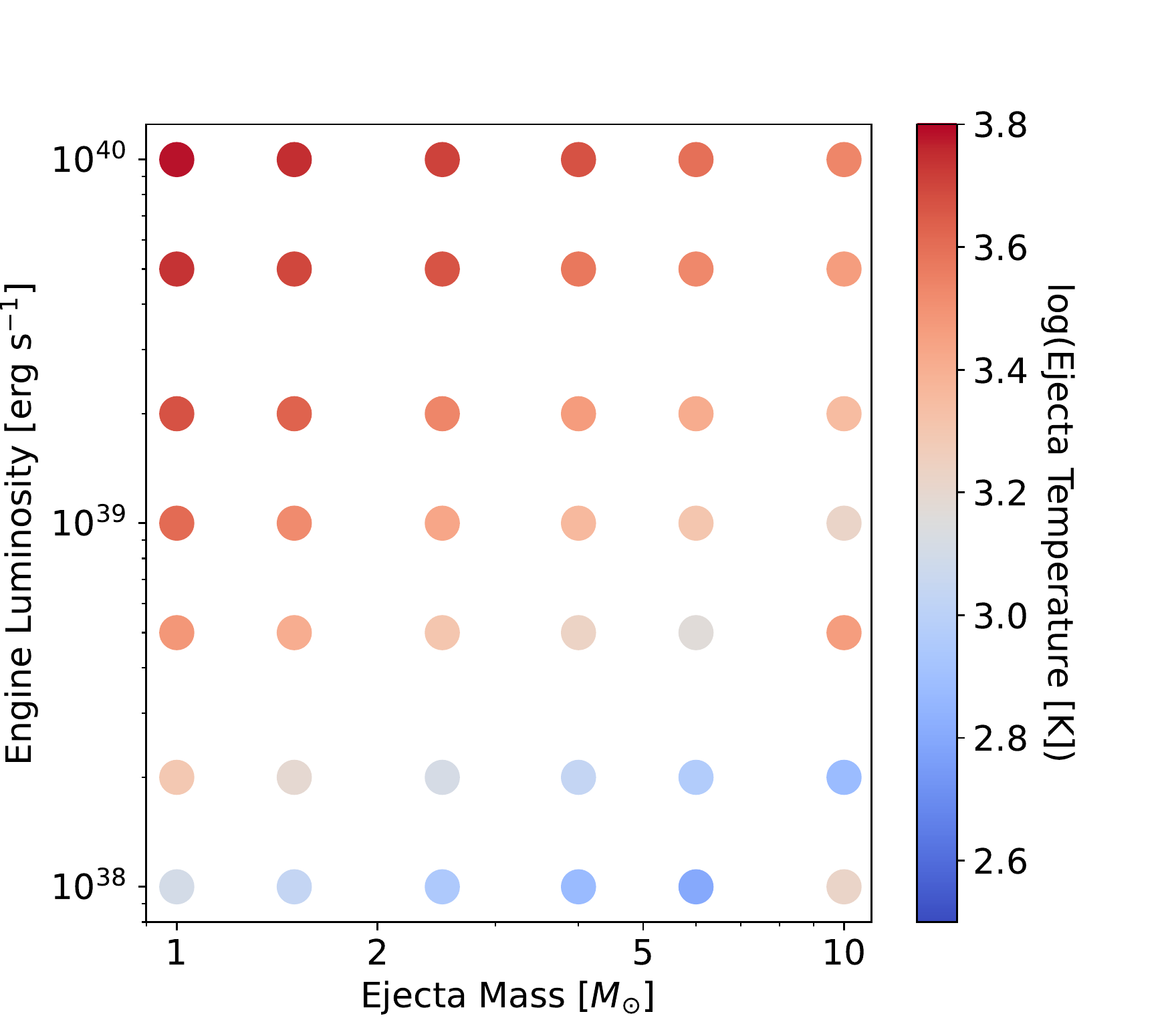}\\[-1.5ex]
\end{tabular}}
\caption{The ion fractions of O I (top), O II (second row), and O III (third row), and the ejecta temperature $T_{\rm ej}$ (bottom) in the simulations at 6 years for the realistic composition at three different values of $T_{\rm PWN}$.  The black contour denotes the low ejecta mass, high engine luminosity regime where runaway ionization can occur for both O I and O II.}%
\label{fig:rc6y_ionfrac}
\end{figure*}

We now examine the results from simulations using the realistic stripped-envelope supernova composition from \cite{Woosley2007}.  The oxygen ion fractions and ejecta temperature at 6 years are shown in Figure \ref{fig:rc6y_ionfrac}. They display similar behaviour as in the pure oxygen model (Figure \ref{fig:o6y_ionfrac}), with the most notable differences being that O I runaway ionization happens at higher engine luminosities and lower ejecta masses, and that the ejecta temperature is much cooler (the range is here $400-6300$ K compared to $1600-6300$ K for pure oxygen) and shows a stronger dependence on $T_{\rm PWN}$.  

\begin{figure*}
\newcolumntype{D}{>{\centering\arraybackslash} m{6cm}}
\noindent
\makebox[\textwidth]{
\begin{tabular}{m{1cm} DDD}
& \boldsymbol{$T_{\rm PWN} = 10^5$} \textbf{ K} & \boldsymbol{$T_{\rm PWN} = 3 \times 10^5$} \textbf{ K} & \boldsymbol{$T_{\rm PWN} = 10^6$} \textbf{ K}\\
\textbf{[O I]}&
\includegraphics[width=1.1\linewidth]{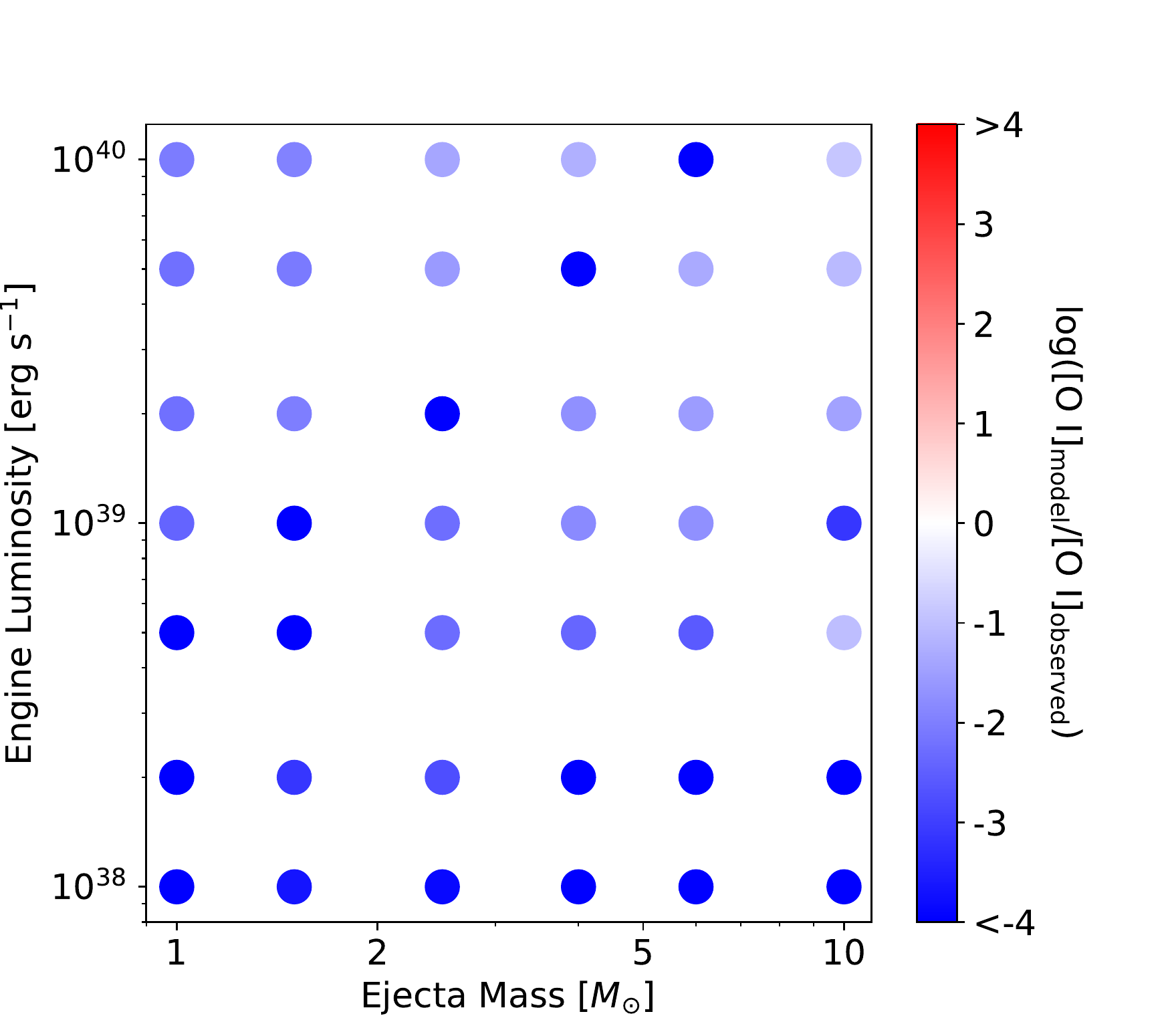}&
\includegraphics[width=1.1\linewidth]{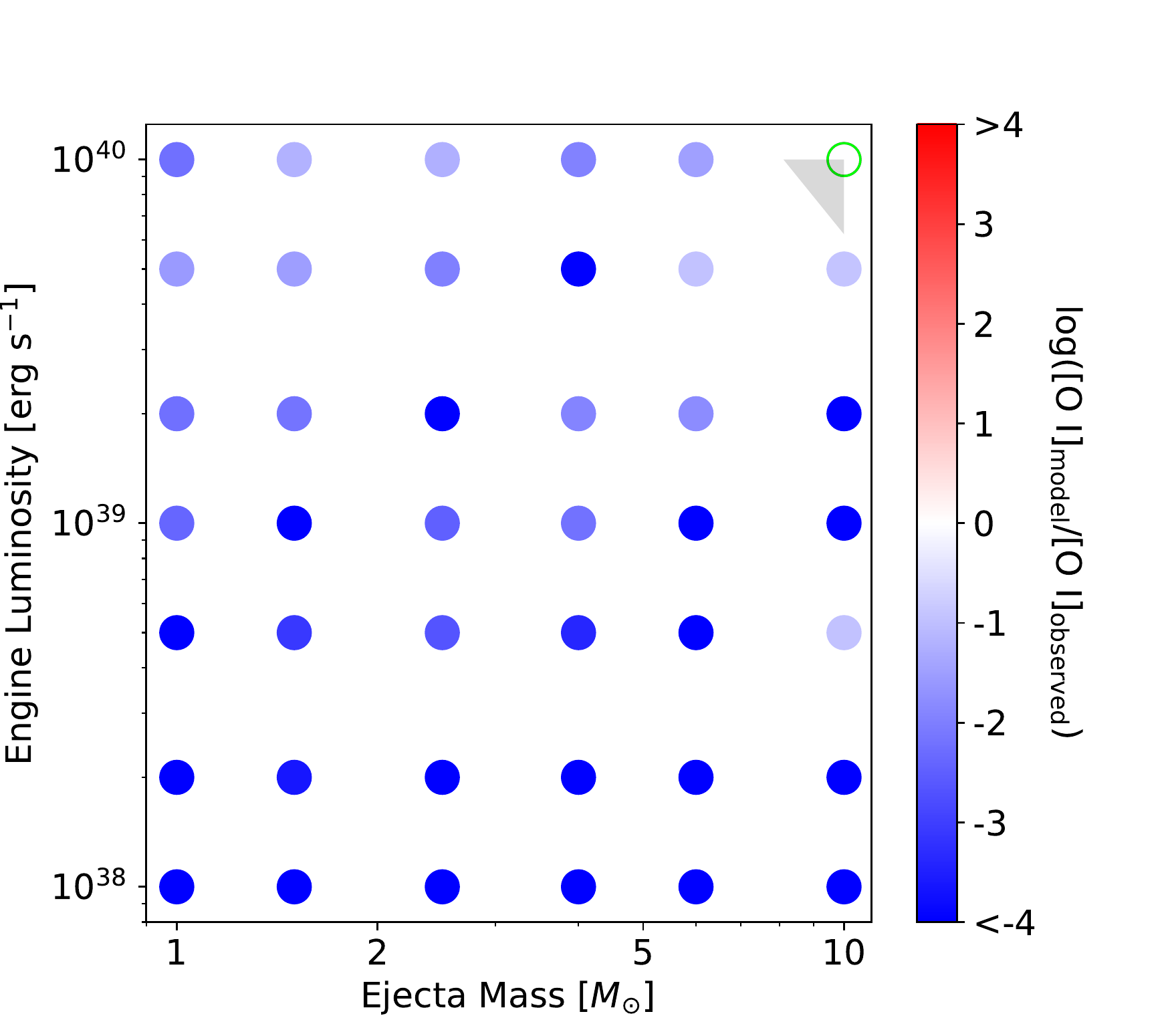}&
\includegraphics[width=1.1\linewidth]{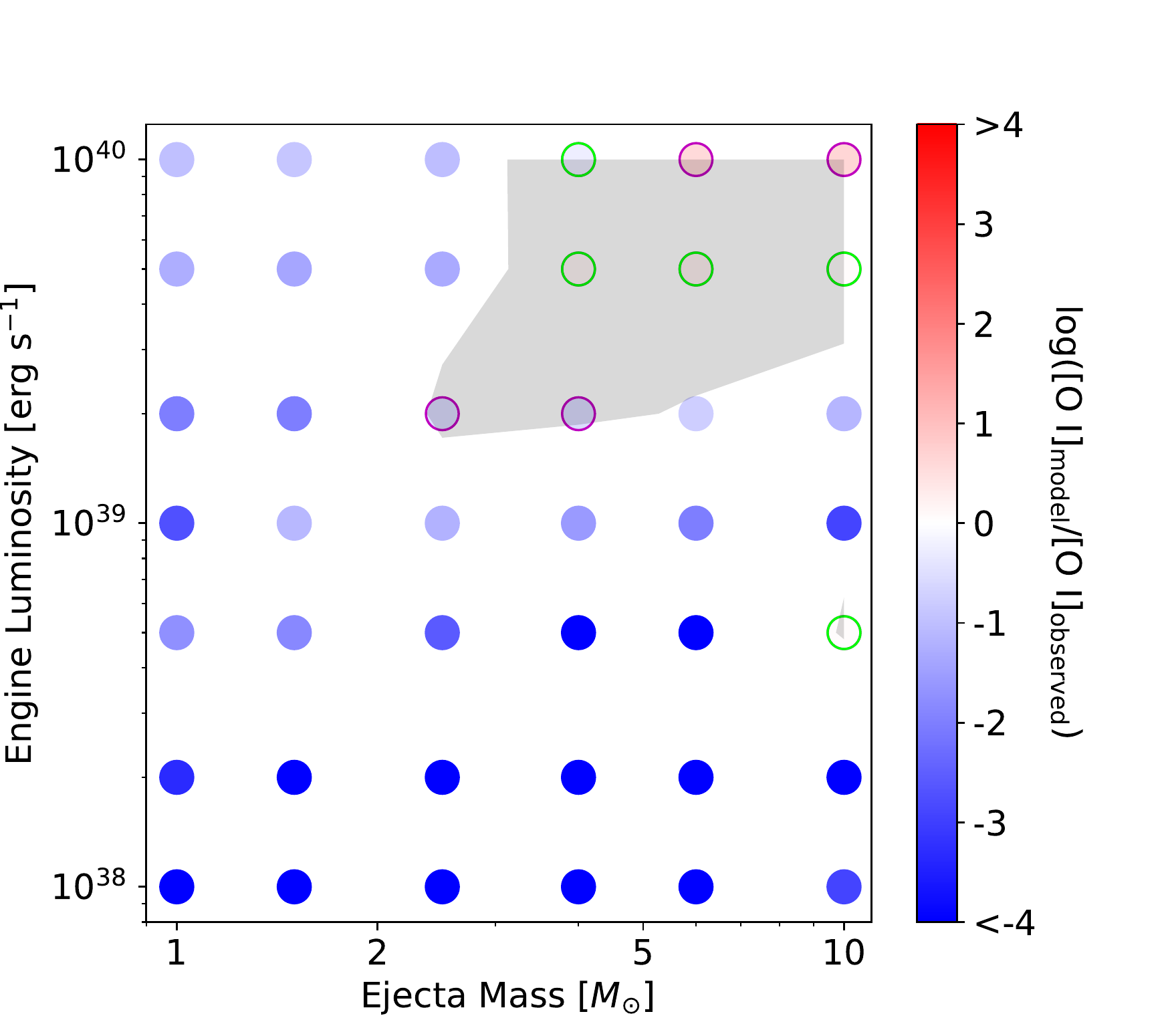}\\[-1.5ex]
\textbf{[O II]}&
\includegraphics[width=1.1\linewidth]{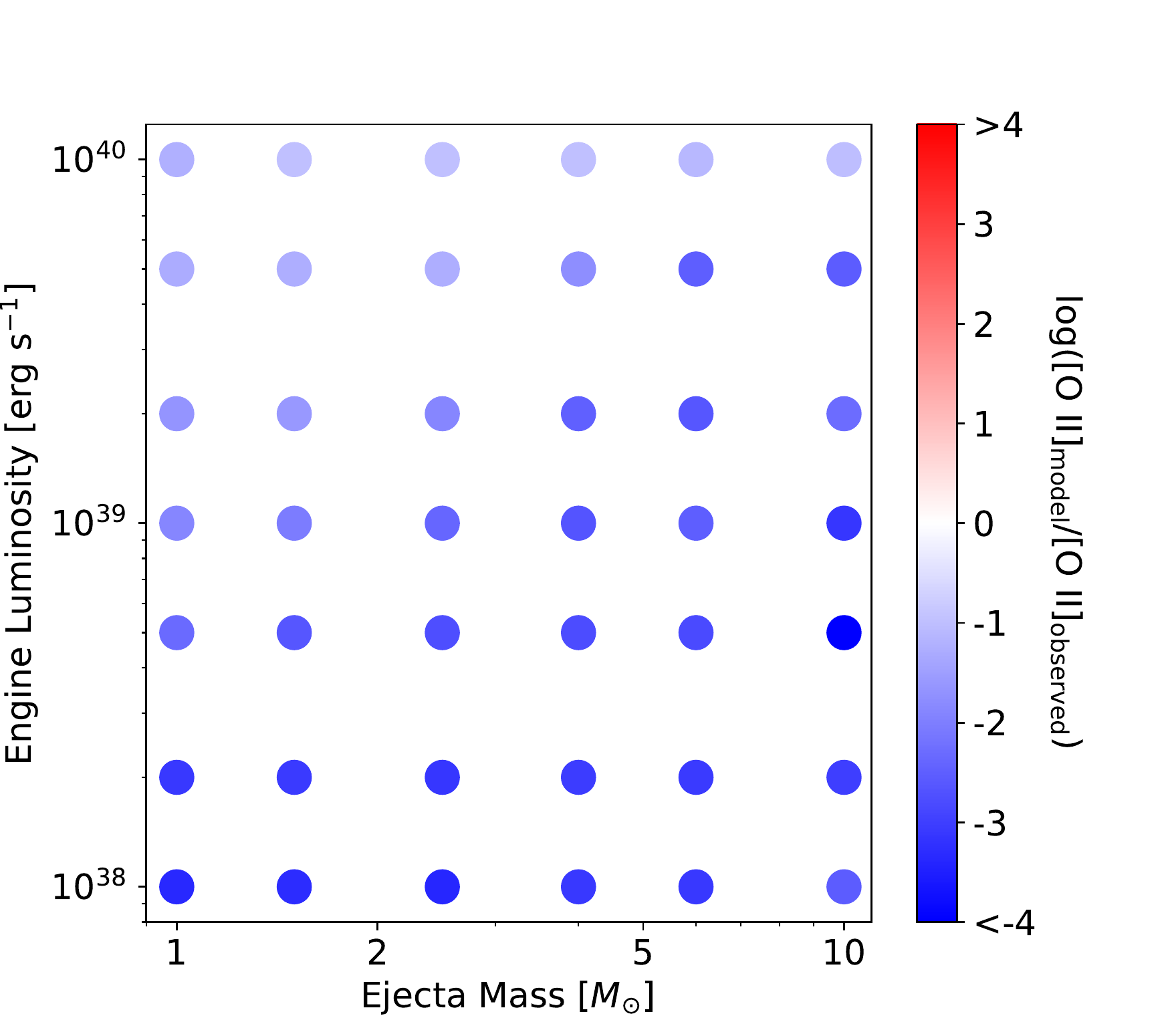}&
\includegraphics[width=1.1\linewidth]{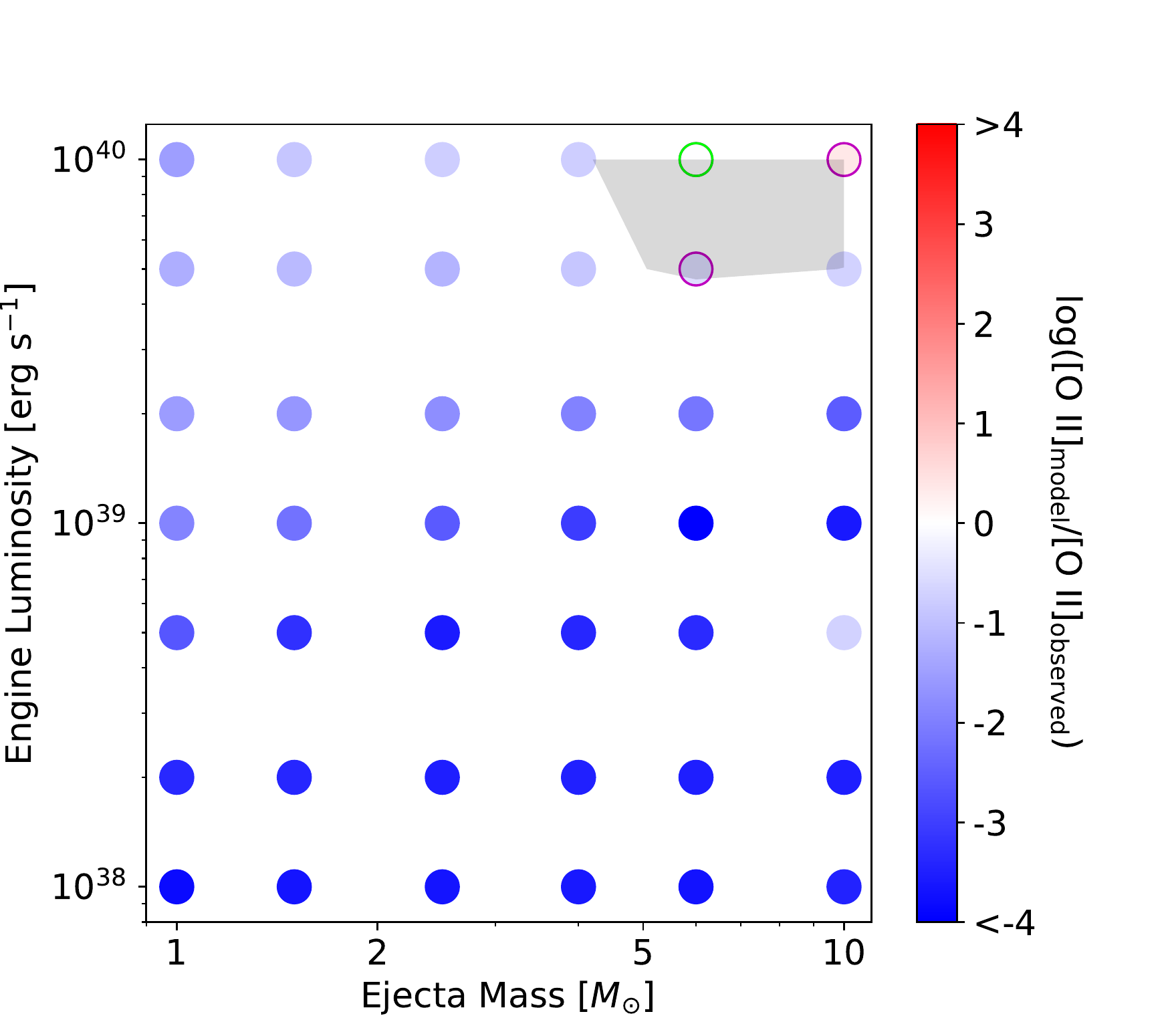}&
\includegraphics[width=1.1\linewidth]{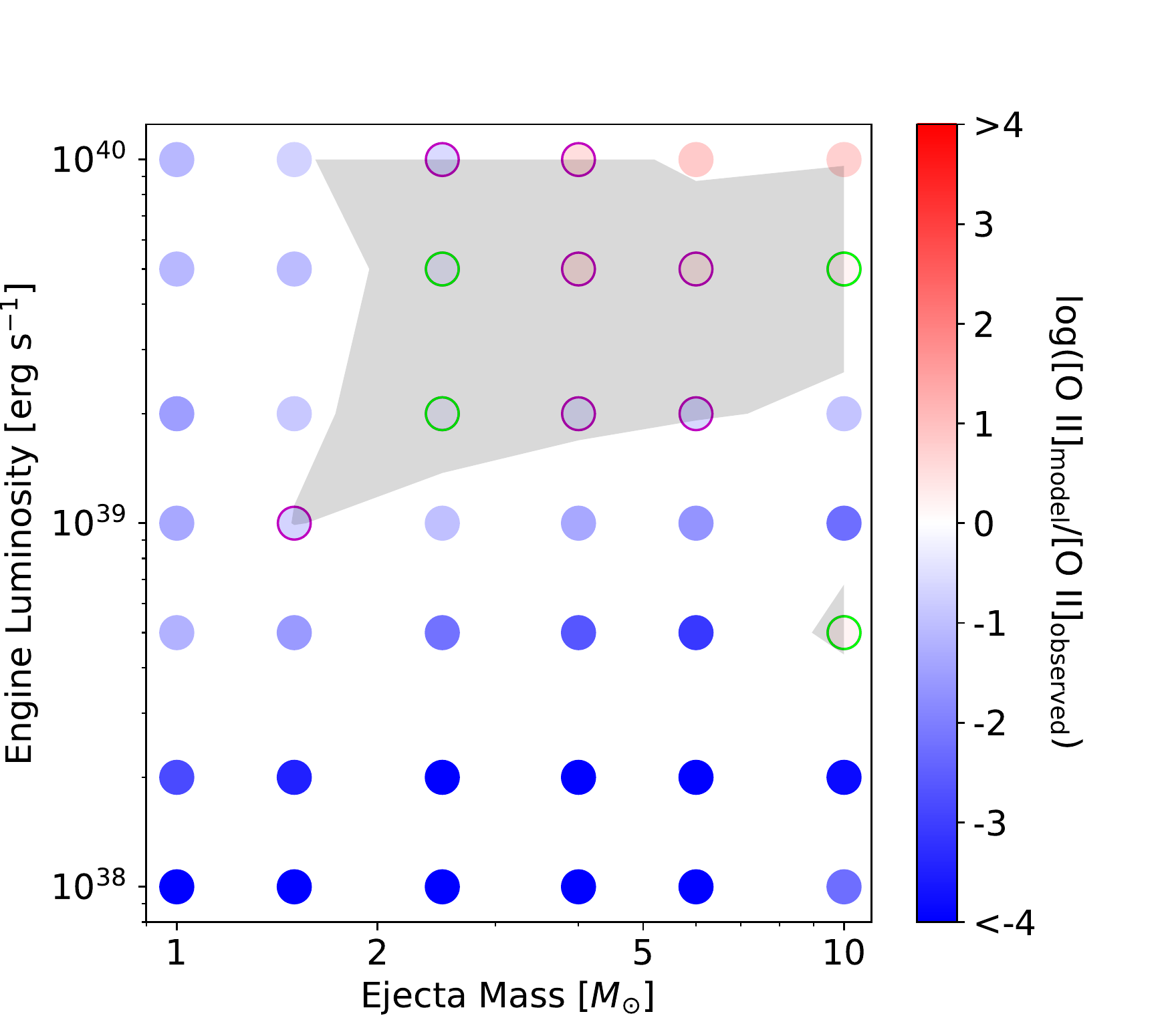}\\[-1.5ex]
\textbf{[O III]}&
\includegraphics[width=1.1\linewidth]{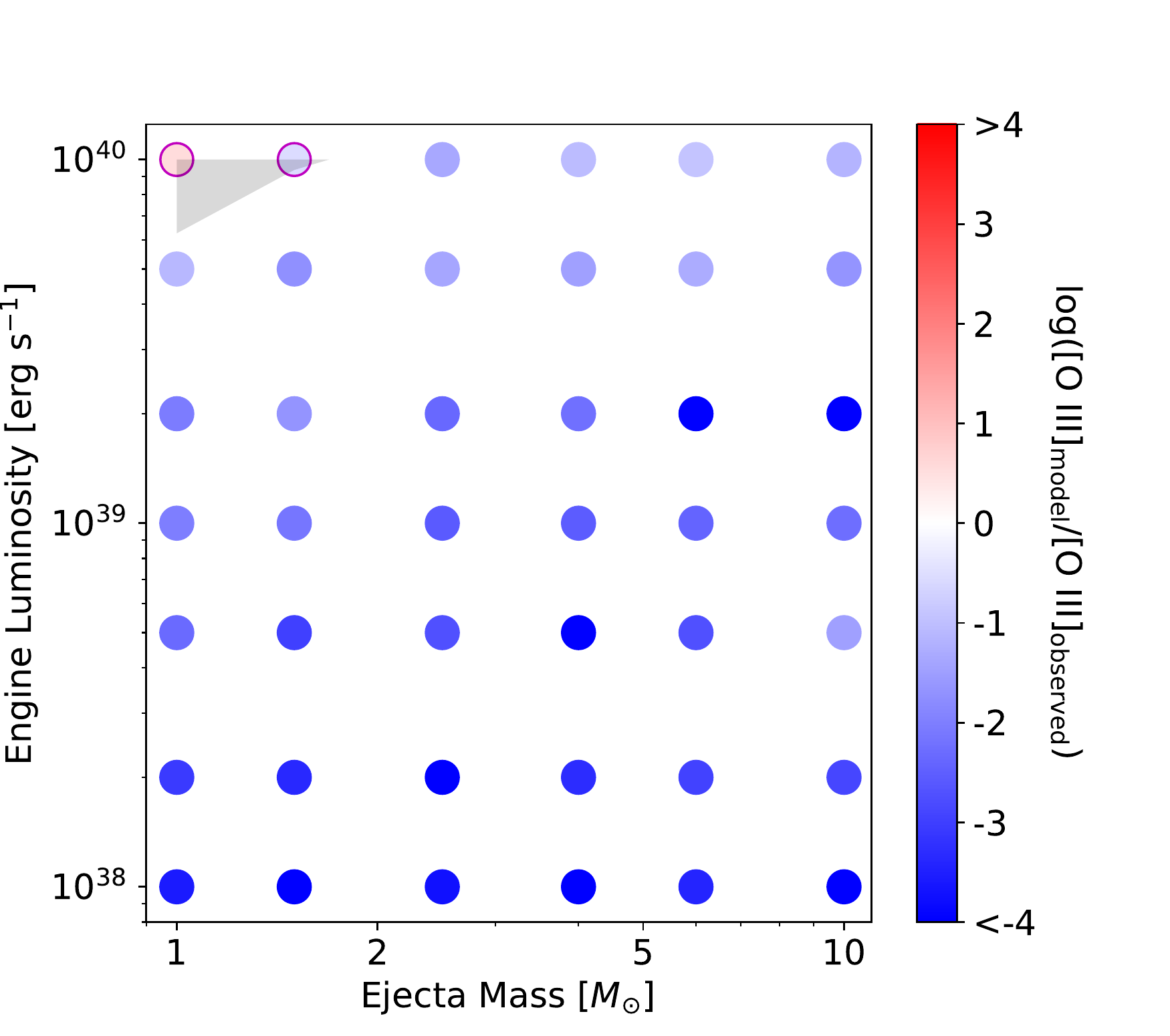}&
\includegraphics[width=1.1\linewidth]{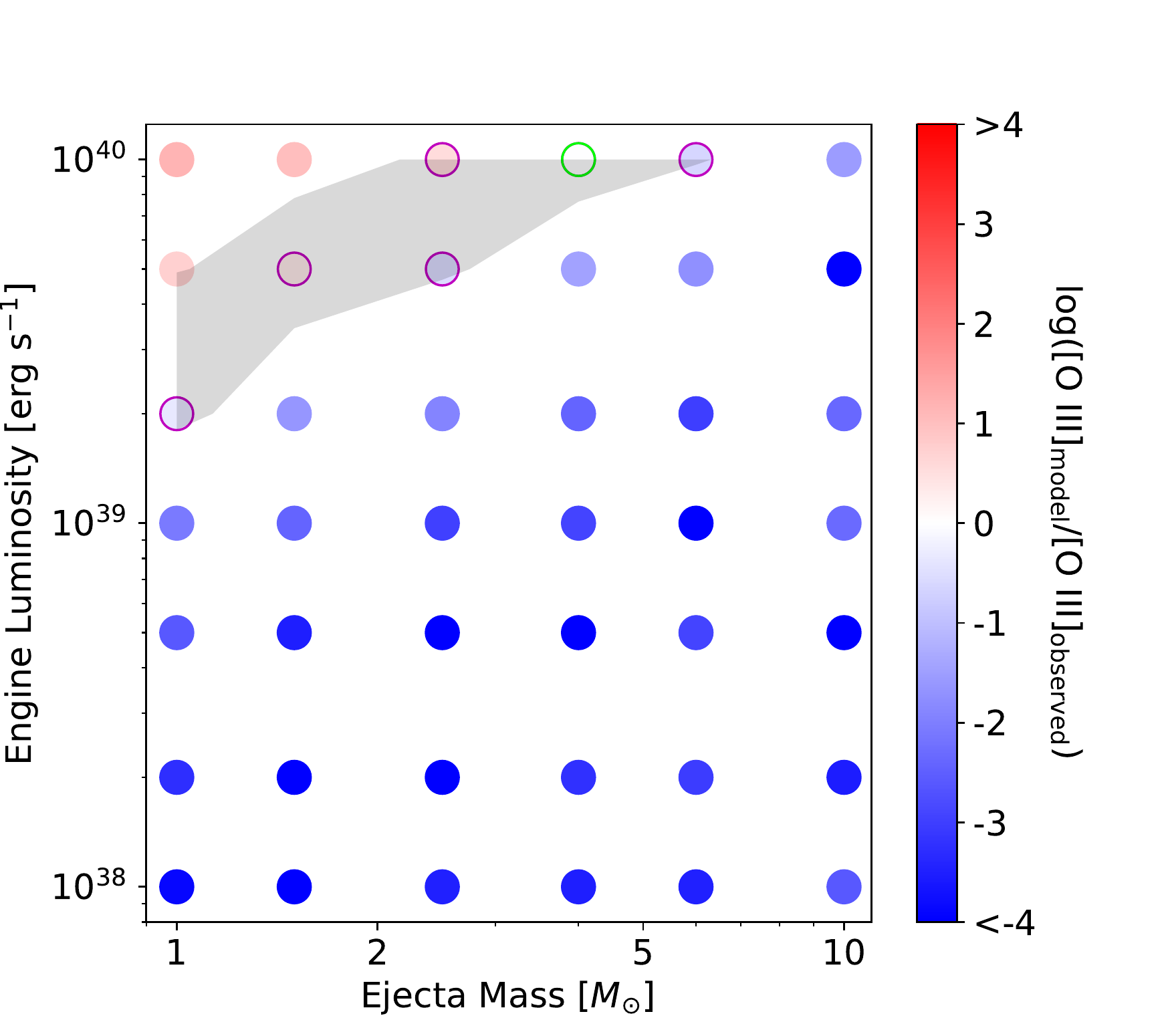}&
\includegraphics[width=1.1\linewidth]{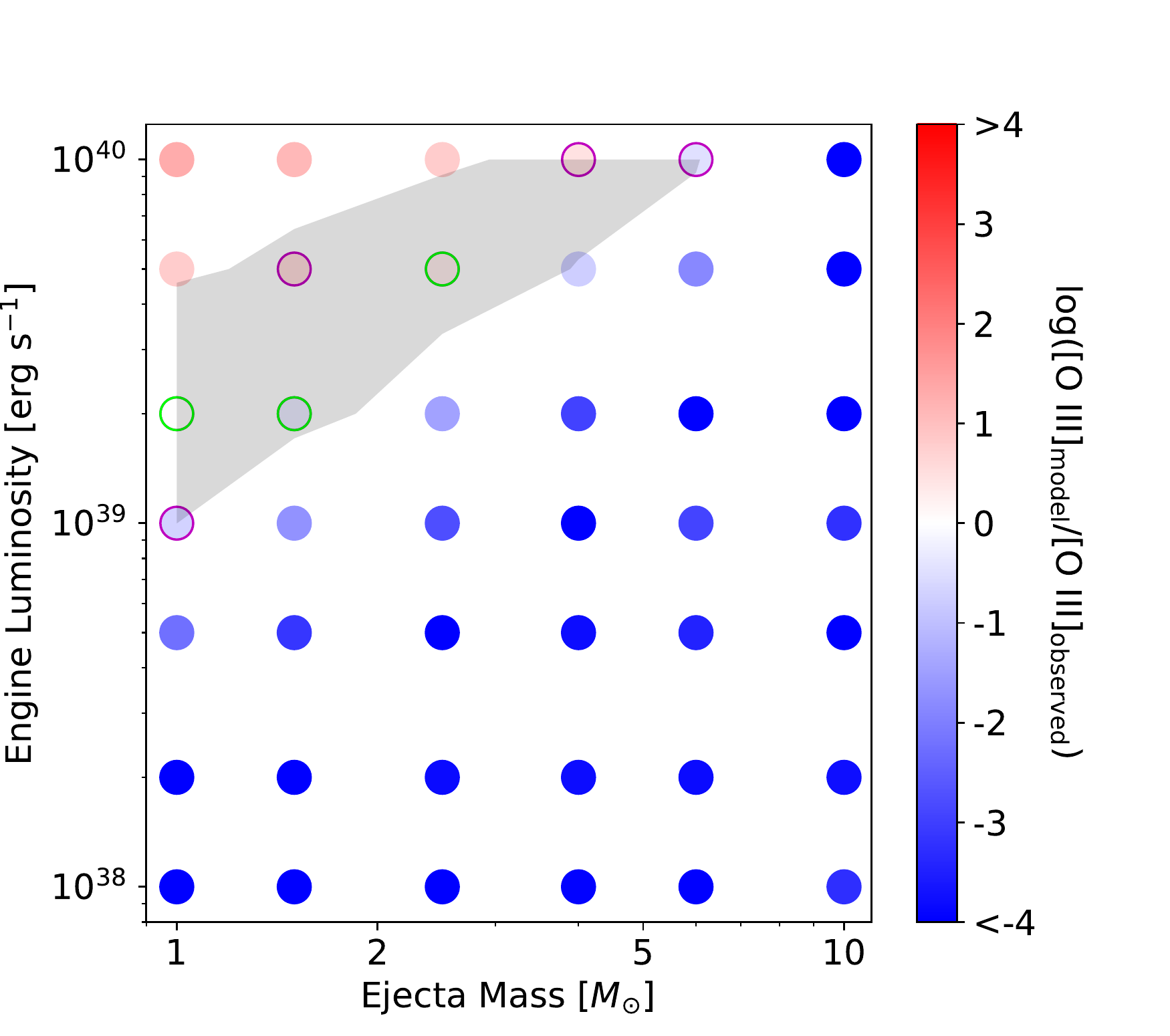}\\[-1.5ex]
\textbf{O I}&
\includegraphics[width=1.1\linewidth]{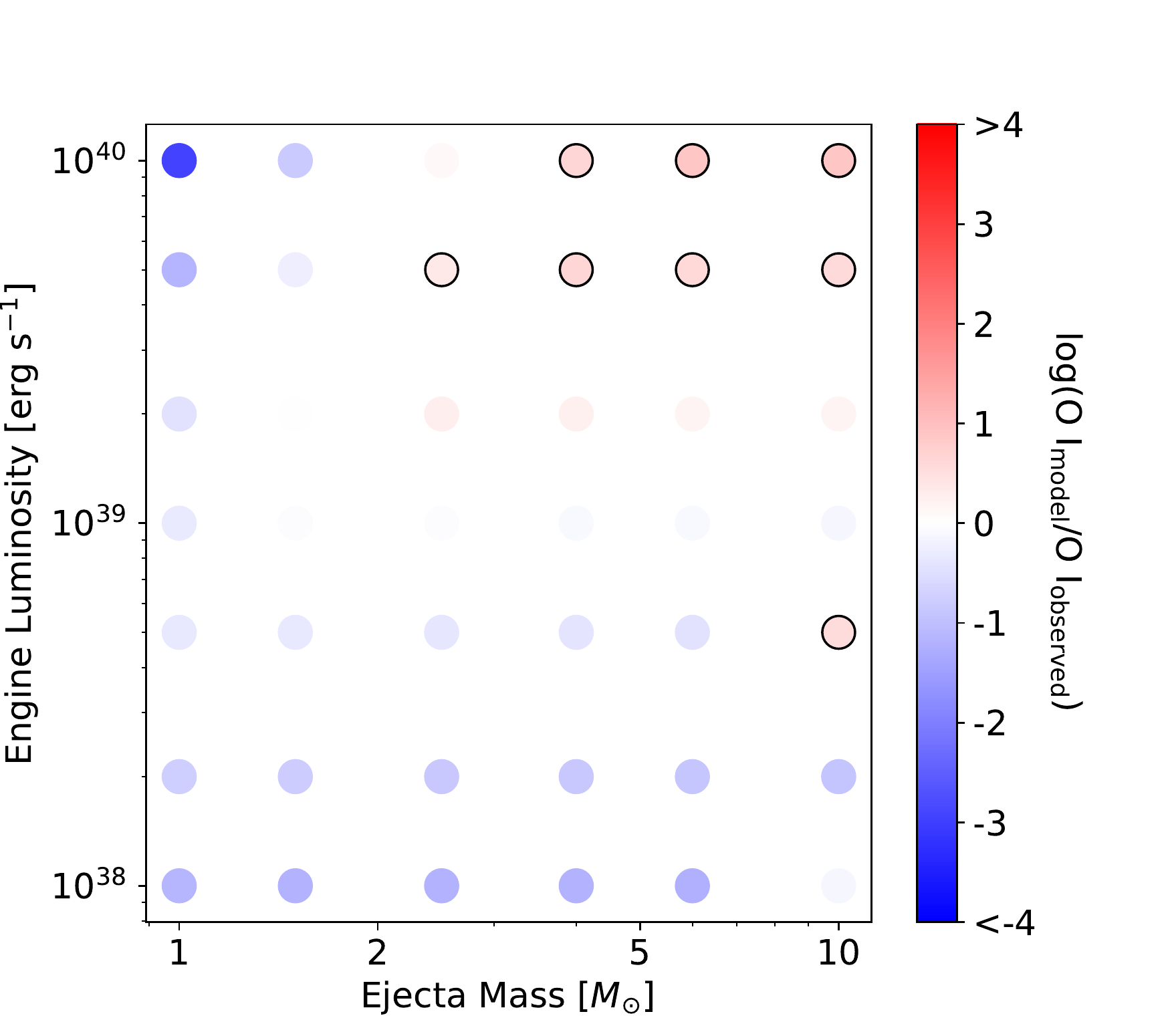}&
\includegraphics[width=1.1\linewidth]{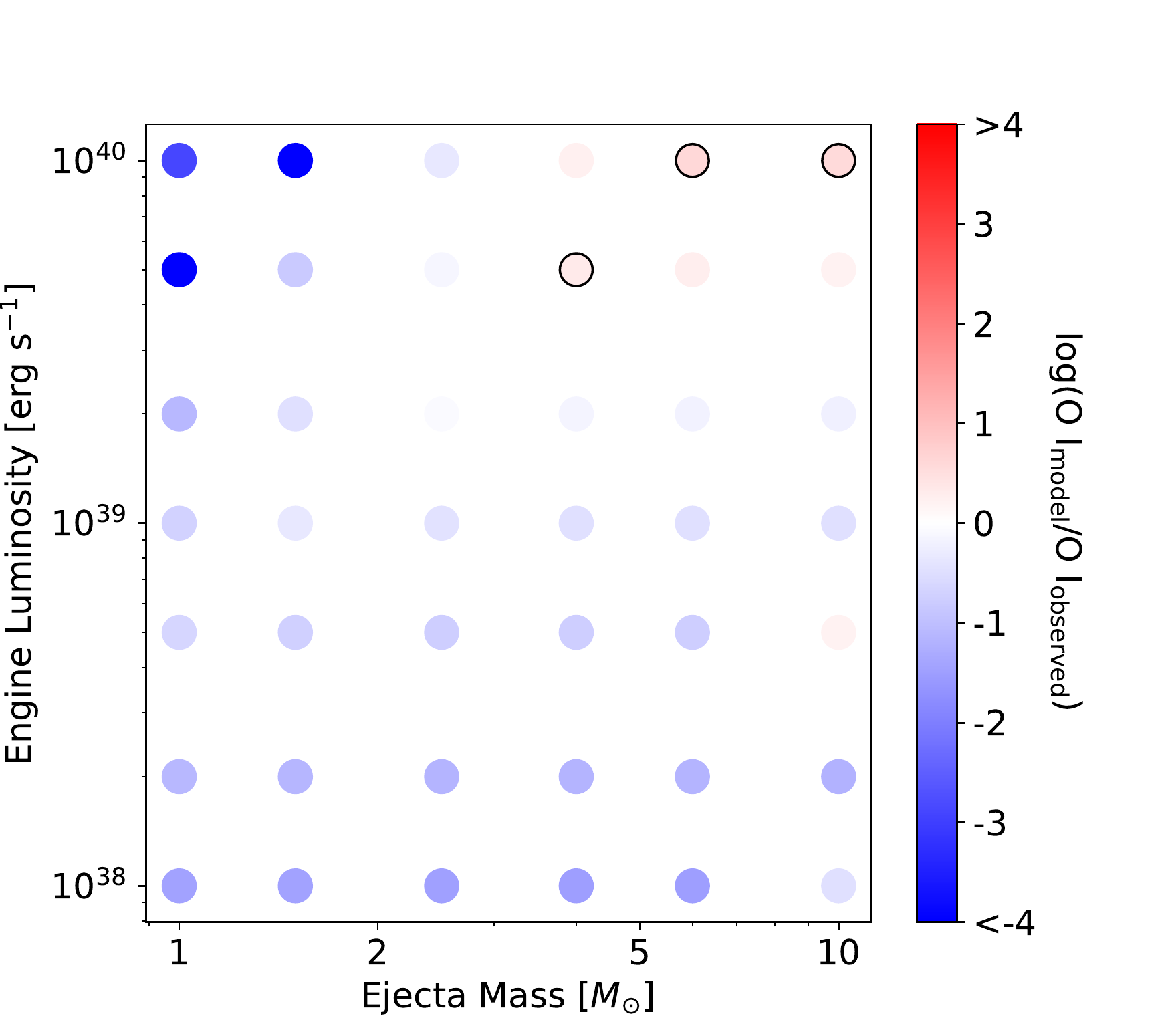}&
\includegraphics[width=1.1\linewidth]{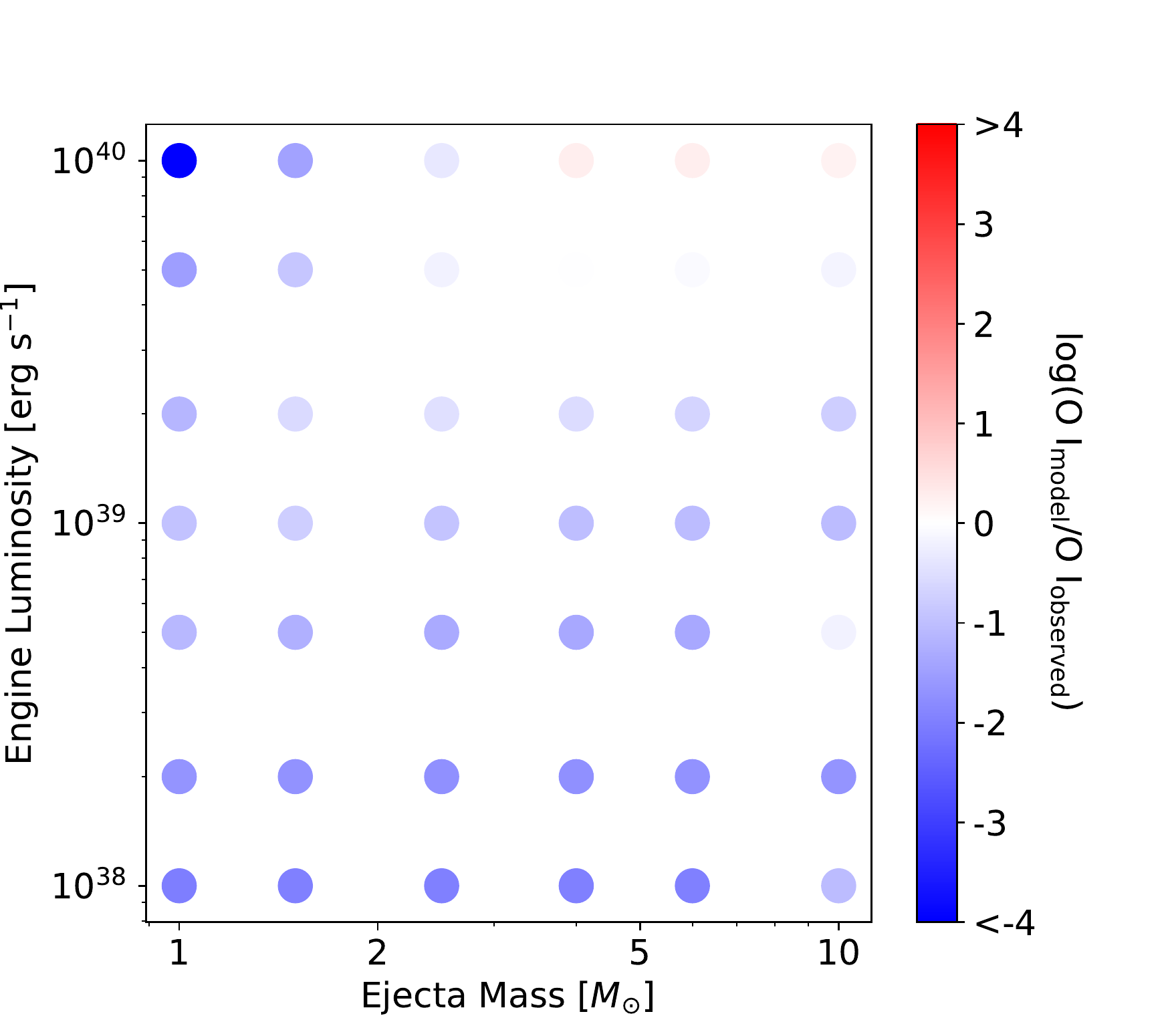}\\[-1.5ex]
\end{tabular}}
\caption{The luminosity of the model [O I] (top), [O II] (second row), [O III] (third row), and O I (bottom) lines compared to the observed line luminosities and luminosity limits from SN 2012au at 6 years for the realistic composition at three different values of $T_{\rm PWN}$.  The green circled points represent where the model and observed values are within a factor of 2, the purple circles within a factor of 5, and the grey shaded region also within a factor of 5.  The black circled points represent where the model luminosity is more than a factor 2 larger than the observational limit.  Note that the grid includes luminosity from all elements in the wavelength regions where these lines are emitted.}%
\label{fig:rc6y_linecomp}
\end{figure*}

The continuum-subtracted luminosities for the wavelength regions of each line, shown in Figure \ref{fig:rc6y_linecomp}, are much lower than in the pure oxygen case. This is driven by lower temperatures, as other elements now compete for the cooling and emission. 
Engine luminosity values close to 10$^{39}$ erg s$^{-1}$, that were sufficient to power the [O I] and [O III] lines in the oxygen composition, are now insufficient, and values closer to 10$^{40}$ erg s$^{-1}$ are now preferred for all three lines.  The [O I] line now fits best in the high mass, high luminosity regime, and only at high injection SED temperature, although it is possible that there could be a larger region of parameter space with higher ejecta masses and engine luminosities that may fit the [O I] line at lower values of $T_{\rm PWN}$.  The [O II] line needs similar luminosities as the pure oxygen model to fit observations, but do not require as large ejecta masses, and there are no good fits at $T_{\rm PWN} = 10^5$ K, just like [O I].  The [O III] line, similar to the [O I] line, now fits at high luminosities, and the best-fit region is at increasing ejecta mass and decreasing engine luminosity as injection SED temperature increases.  The O I line luminosity is again too high in the high mass, high luminosity regime, although the line luminosity decreases as injection SED temperature increases, leaving the entire parameter grid being feasible at $T_{\rm PWN} = 10^6$ K.  Overall, the realistic composition significantly favours higher values of $T_{\rm PWN}$, as at $T_{\rm PWN} = 10^5$ K none of the line luminosities are fit within a factor of 2. This demonstrates the sensitivity of the emission to the microscopic composition.

\begin{figure*}
\newcolumntype{D}{>{\centering\arraybackslash} m{6cm}}
\noindent
\makebox[\textwidth]{
\begin{tabular}{DDD}
\boldsymbol{$T_{\rm PWN} = 10^5$} \textbf{ K} & \boldsymbol{$T_{\rm PWN} = 3 \times 10^5$} \textbf{ K} & \boldsymbol{$T_{\rm PWN} = 10^6$} \textbf{ K}\\
\includegraphics[width=1.1\linewidth]{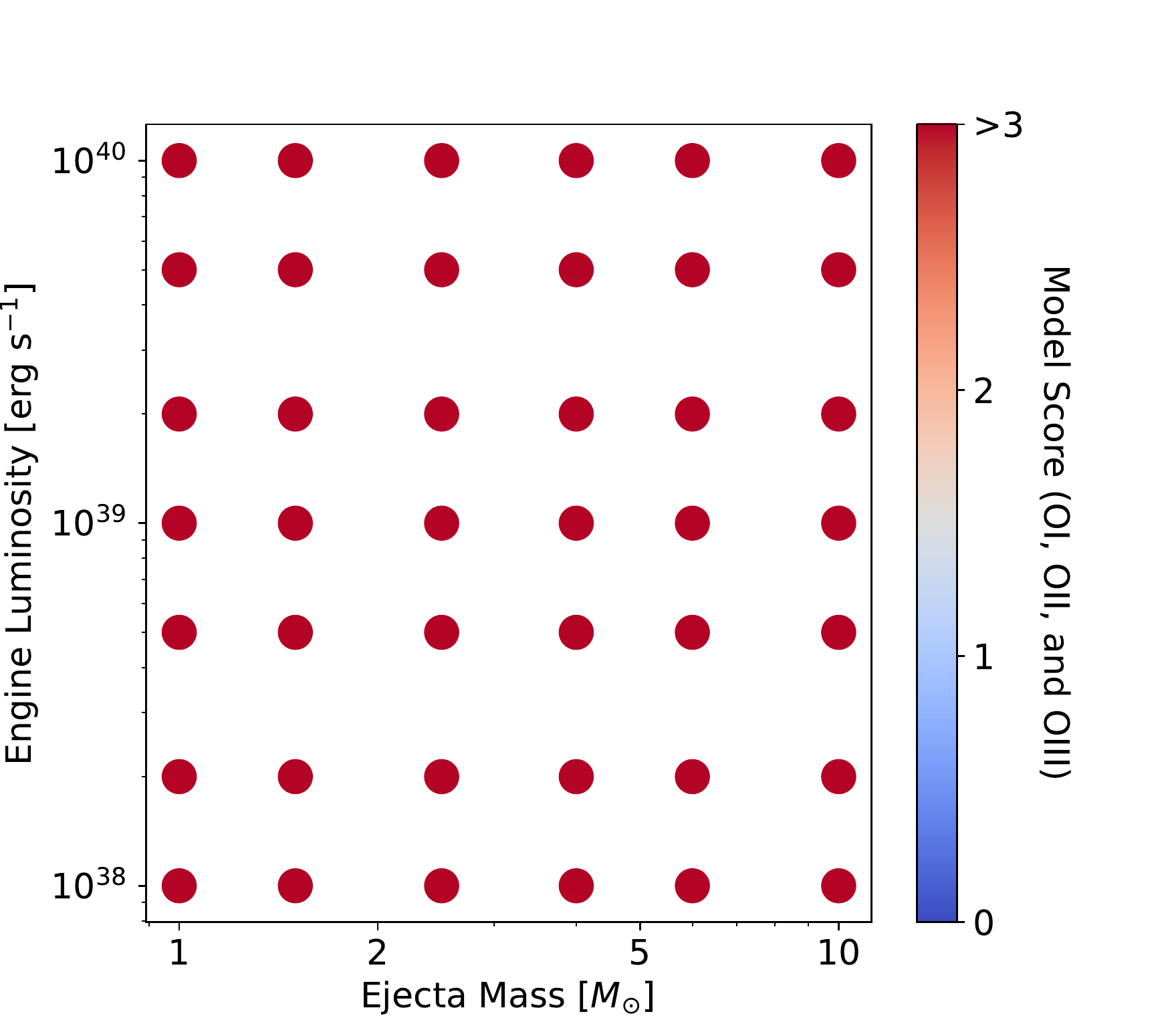}&
\includegraphics[width=1.1\linewidth]{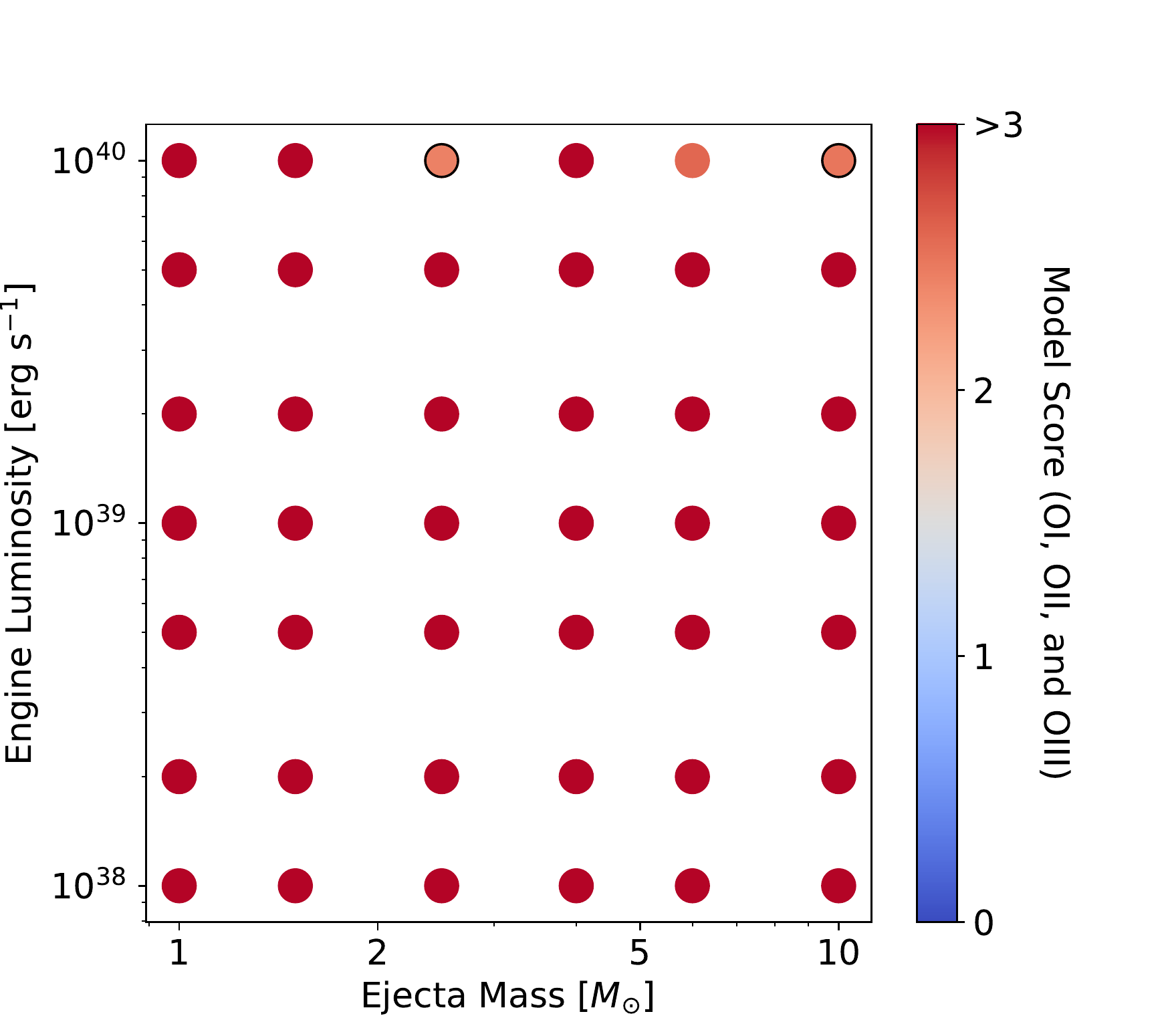}&
\includegraphics[width=1.1\linewidth]{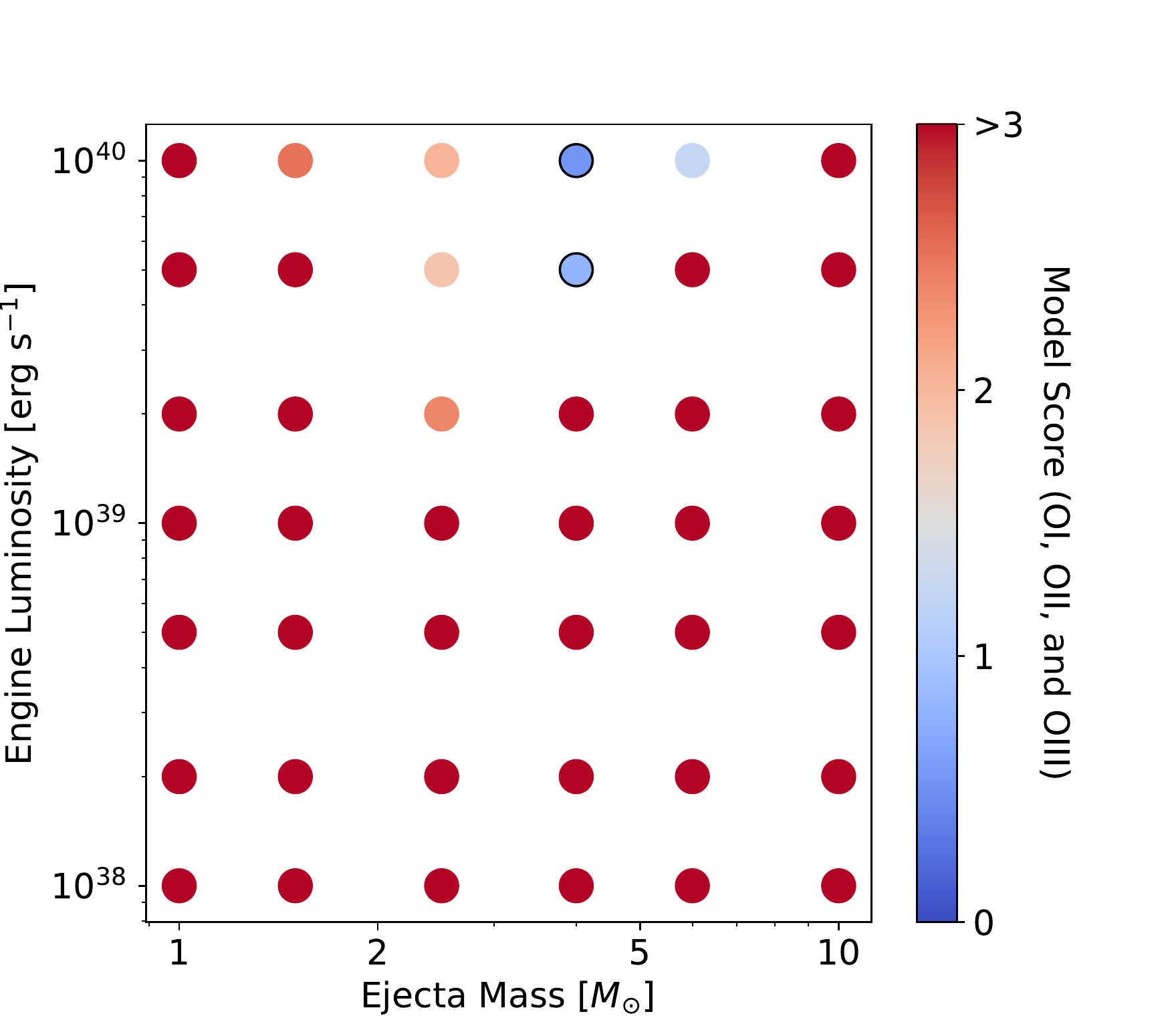}\\[-1.5ex]
\end{tabular}}
\caption{The goodness-of-fit score for each model in the realistic composition at 6 years based on the [O I], [O II], and [O III] lines.  Lower scores indicate a better fit to the data (from Equation \ref{eqn:modscore}, a perfect fit has score 0, all three lines off by factor 2 has score 0.27, and all three lines off by factor 10 has score 3). The black circles indicate the two models with the lowest scores for each $T_{\rm PWN}$, although $T_{\rm PWN} = 10^5$ K has no models with scores $<$ 3, and $T_{\rm PWN} = 3 \times 10^5$ K has no models with scores $<$ 2.  The best fitting models for $T_{\rm PWN} = 10^6$ K are plotted in Figure \ref{fig:rc6y_spec}.}%
\label{fig:rc6y_score}
\end{figure*}

\begin{figure*}
\newcolumntype{D}{>{\centering\arraybackslash} m{8cm}}
\noindent
\makebox[\textwidth]{
\begin{tabular}{DDD}
\includegraphics[width=1.1\linewidth]{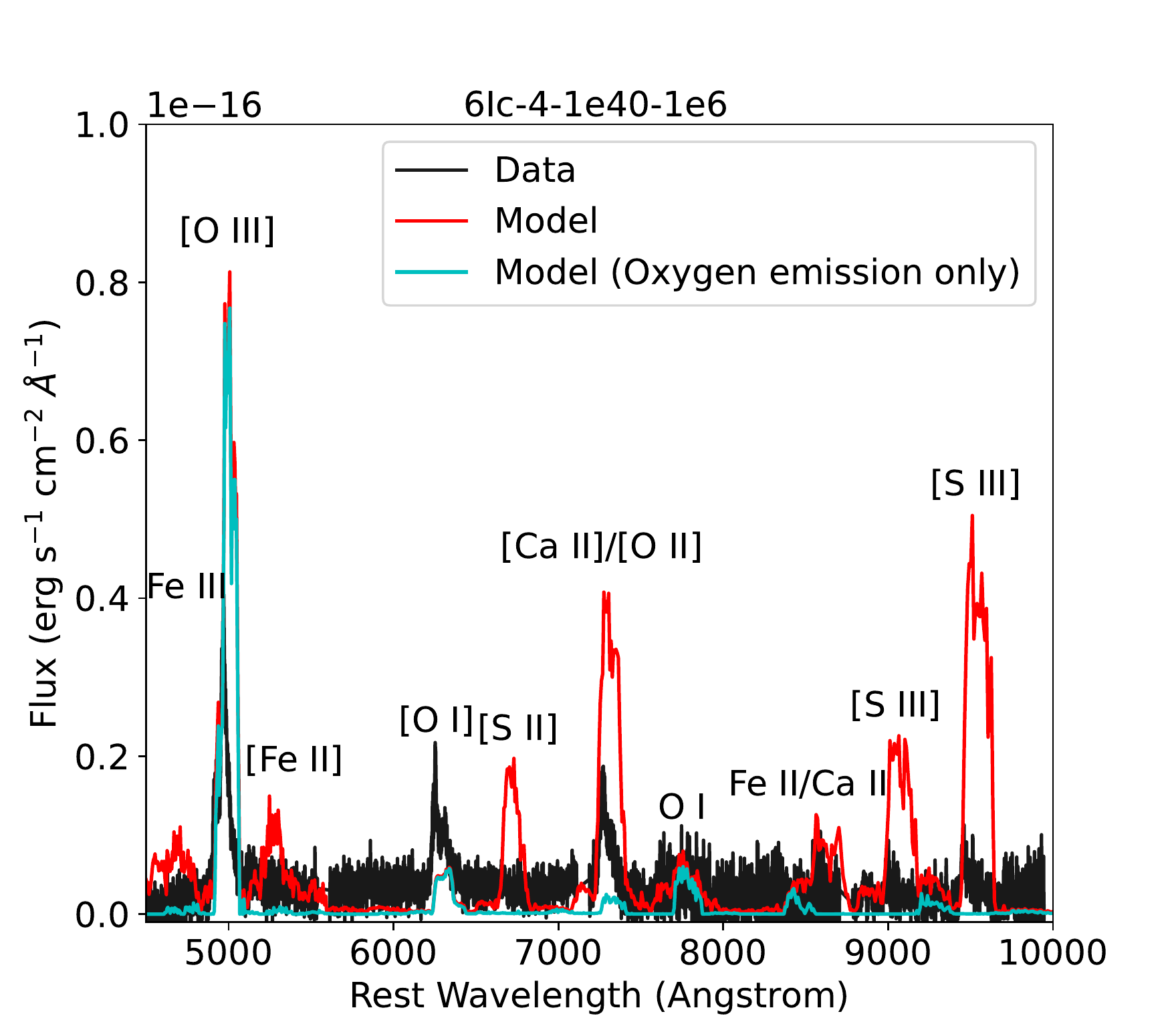}&
\includegraphics[width=1.1\linewidth]{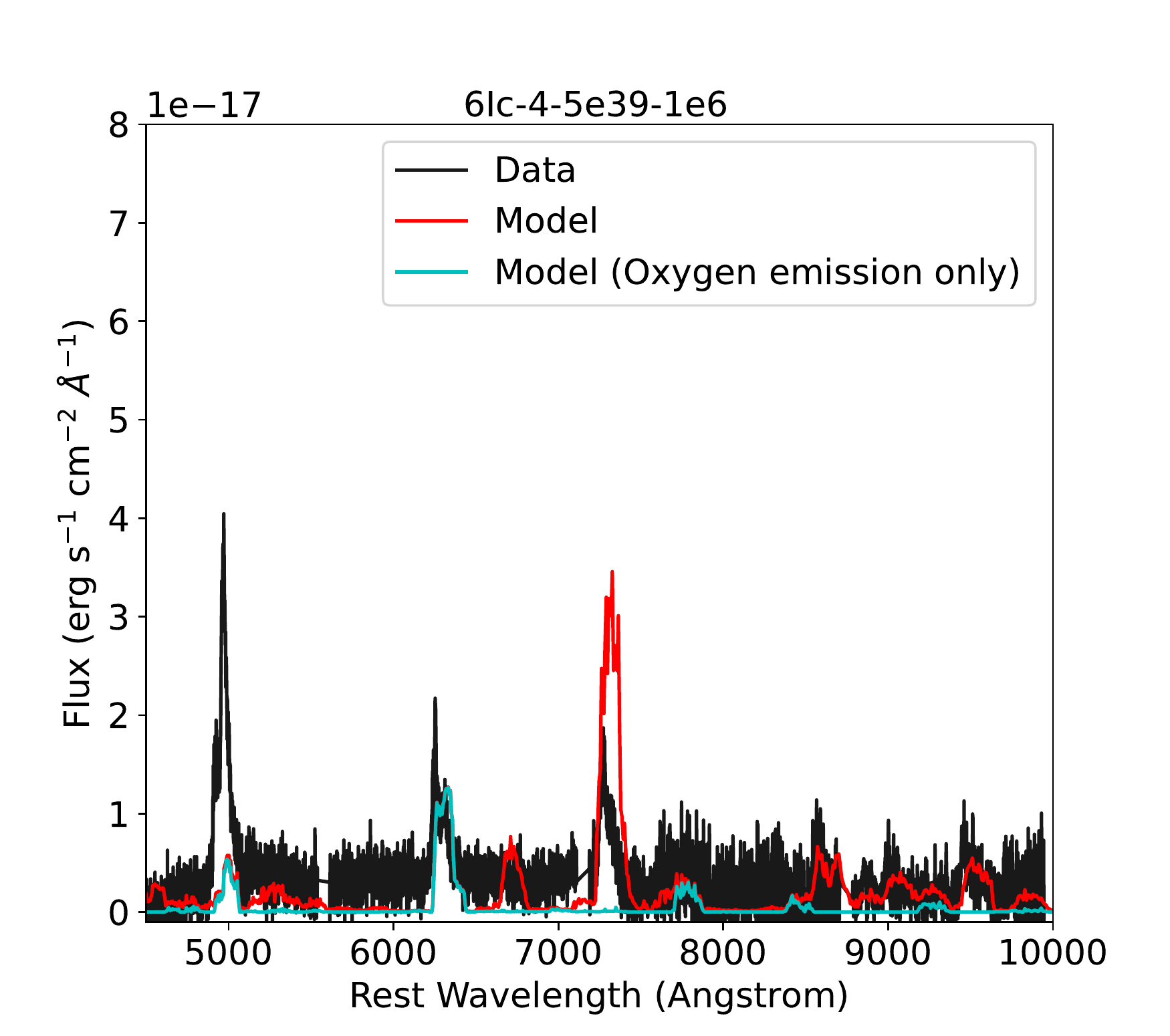}\\
\end{tabular}}
\caption{The two best-fitting dust-corrected model spectra for $T_{\rm PWN}=10^6$ K at 6 years for the realistic composition compared to the observed spectrum from \cite{Milisavljevic2018}.  The total model emission is shown in red, while the emission from only oxygen is shown in cyan.   Strong lines and features are labelled on the left plot.}%
\label{fig:rc6y_spec}
\end{figure*}

The model scores based on the [O I], [O II], and [O III] lines are shown in Figure \ref{fig:rc6y_score}. $L_{\rm PWN} < 5 \times 10^{39}$ erg s$^{-1}$ are almost completely ruled out, and there are no low model scores at $T_{\rm PWN} \leq 3 \times 10^5$ K.  The two best fitting spectra for $T_{\rm PWN}$ = $10^6$ K are shown in Figure \ref{fig:rc6y_spec}.  Both models predict high luminosity around 7300 $\AA$, but this is coming from [Ca II] instead of [O II].  6Ic-4-5e39-1e6 underestimates the [O III] luminosity, while 6Ic-4-1e40-1e6 overestimates it by about the same factor, and both are within a factor of 2 for the [O I] luminosity. These models also predict other detectable spectral features, particularly in the 6Ic-4-1e40-1e6 model, including the [S III] $\lambda \lambda$ 9069, 9531 doublet, of which the 9531 $\AA$ peak was observed; the S II $\lambda \lambda$ 6717, 6730 doublet; mixed Fe II and Fe III features around 4700 and 5300 $\AA$; and a mixed Fe II and Ca II feature around 8600 $\AA$.  The prevalence of these features could either be due to differences in composition or the extent of mixing between our models and SN 2012au.  When examining only the oxygen emission, the models are roughly comparable in how well they reproduce the spectrum, but due to the prevalence of other features in 6Ic-4-1e40-1e6, we tentatively consider 6Ic-4-5e39-1e6 the best-fit model in this composition; however, this model is much less capable of reproducing the observed line ratios than 6O-4-2e39-1e5 and 6O-1.5-1e39-1e6, the best fitting models in the pure oxygen composition.

The high $L_{\rm PWN}$ needed to match the observed oxygen lines in the realistic composition compared to the pure oxygen composition is due to high levels of cooling outside of the observed wavelength range, which results is a much lower ejecta temperature (Figure \ref{fig:rc6y_ionfrac}).  For example, in the 6Ic-4-5e39-1e6 model, the emission from 4500-10000 $\AA$ accounts for $\sim$ 50$\%$ of the total cooling in the ejecta, with the rest coming mostly from Ne II, Ar II, Ar III, Fe II, Fe III, Ni II, and Ni III in the near- and mid-IR, as well as Mg II, S II and Ca II in the UV.  In lower $L_{\rm PWN}$ models, like the Ic composition-equivalent of the pure oxygen composition best-fit models, 6Ic-4-2e39-1e5 and 6Ic-1.5-1e39-1e6, $<$ 5$\%$ of cooling is in the optical band, with most of the emission coming from Ca II, and most of the rest of the emission is in the IR. Ne II is a particularly strong coolant in these models, with $>$ 50$\%$ of energy loss in the ejecta coming from the [Ne II] 12.8 $\mu$m line.  This suggests that pulsar-driven supernovae with strong mixing may be extremely bright IR sources at very late times, particularly in mid-IR. These could be potentially interesting targets for the James Webb Space Telescope (JWST) as a test of both mixing in the supernova as well as the strength of the pulsar wind nebula, although emission from molecules \citep{Liljegren2022} or pulsar-heated dust \citep{Omand2019} could make the interpretation more complicated.  

\subsubsection{1 Year}

\begin{figure*}
\newcolumntype{D}{>{\centering\arraybackslash} m{6cm}}
\noindent
\makebox[\textwidth]{
\begin{tabular}{m{0.8cm} DDD}
& \boldsymbol{$T_{\rm PWN} = 10^5$} \textbf{ K} & \boldsymbol{$T_{\rm PWN} = 3 \times 10^5$} \textbf{ K} & \boldsymbol{$T_{\rm PWN} = 10^6$} \textbf{ K}\\
\textbf{O I}&
\includegraphics[width=1.1\linewidth]{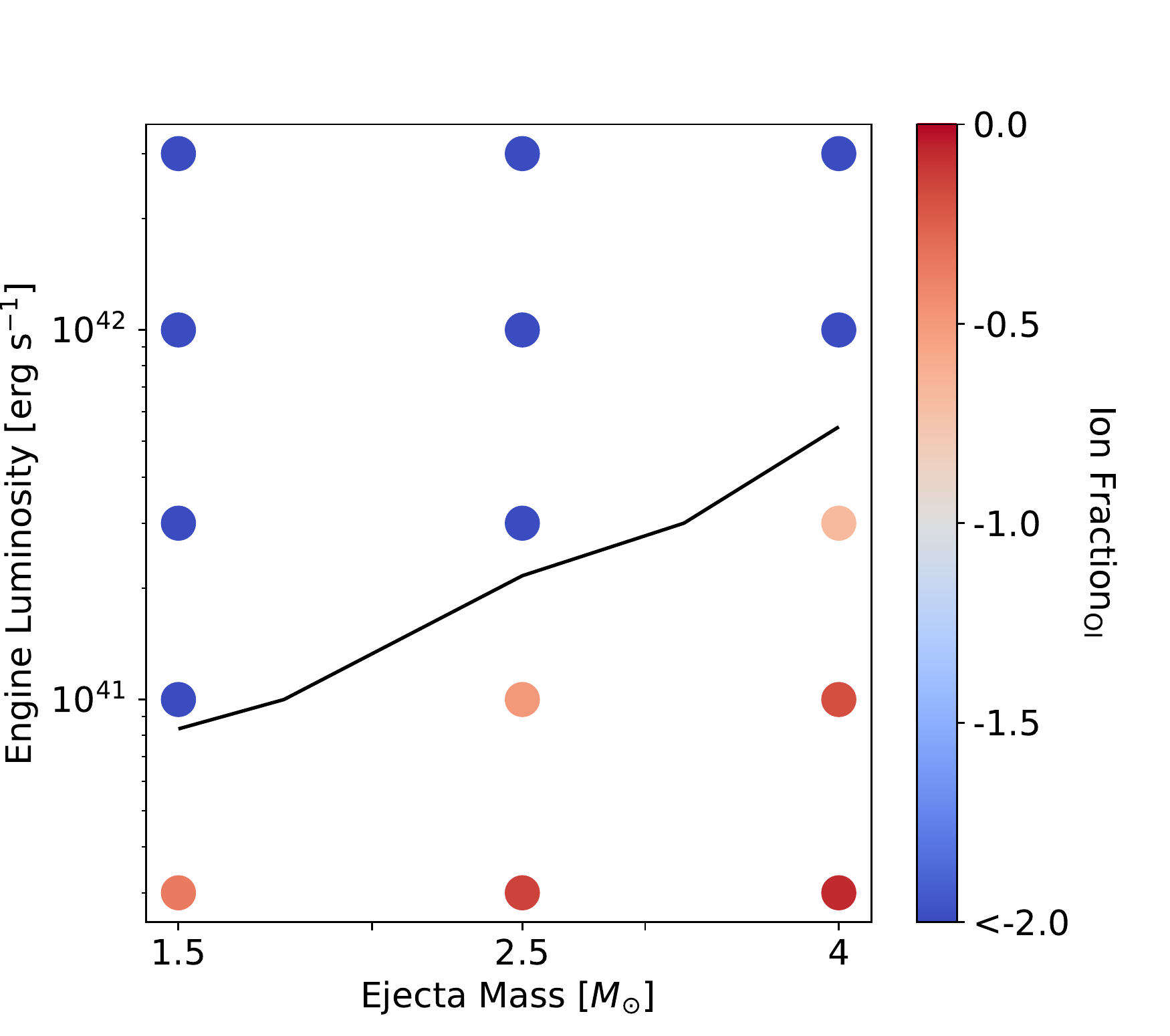}&
\includegraphics[width=1.1\linewidth]{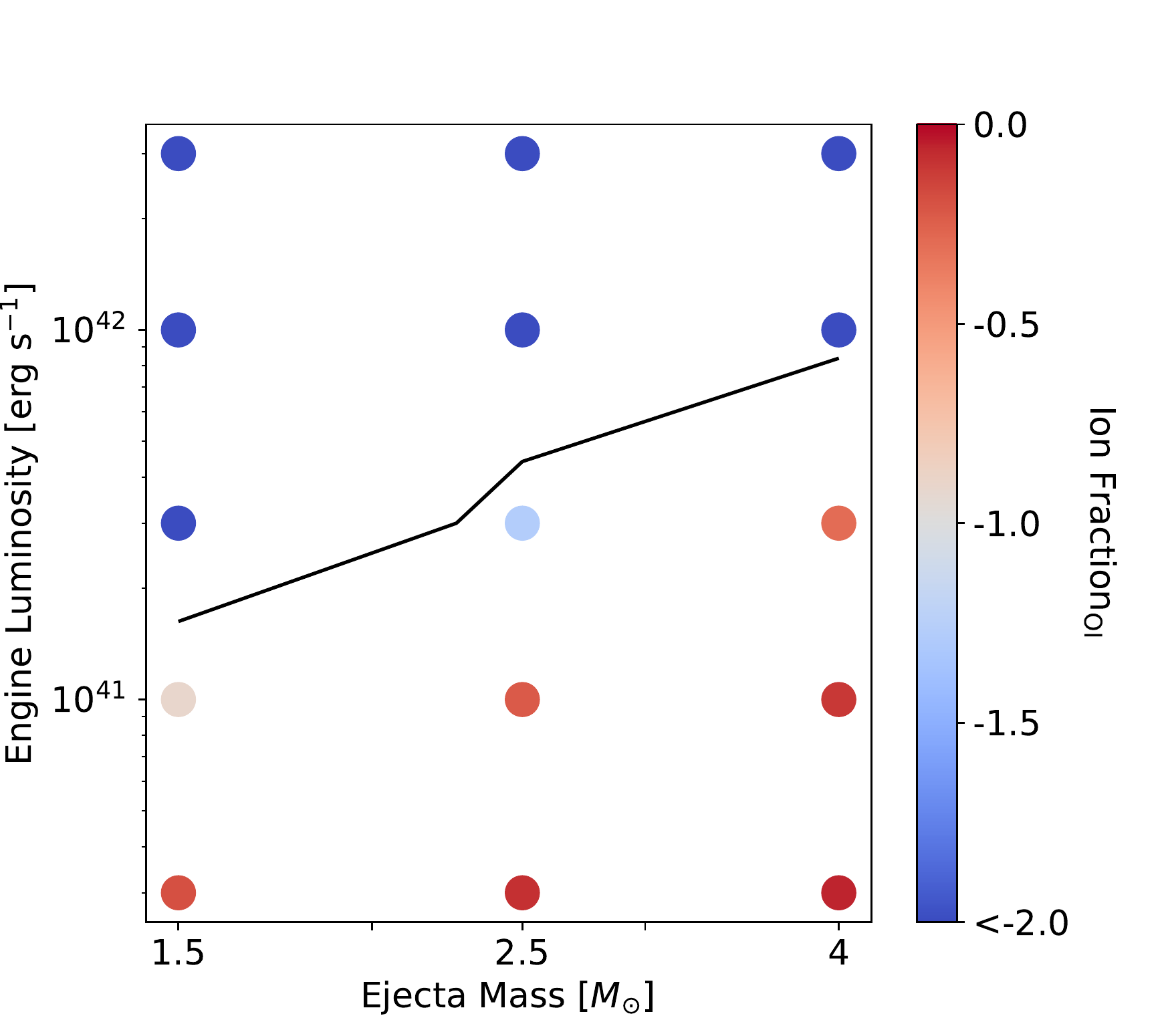}&
\includegraphics[width=1.1\linewidth]{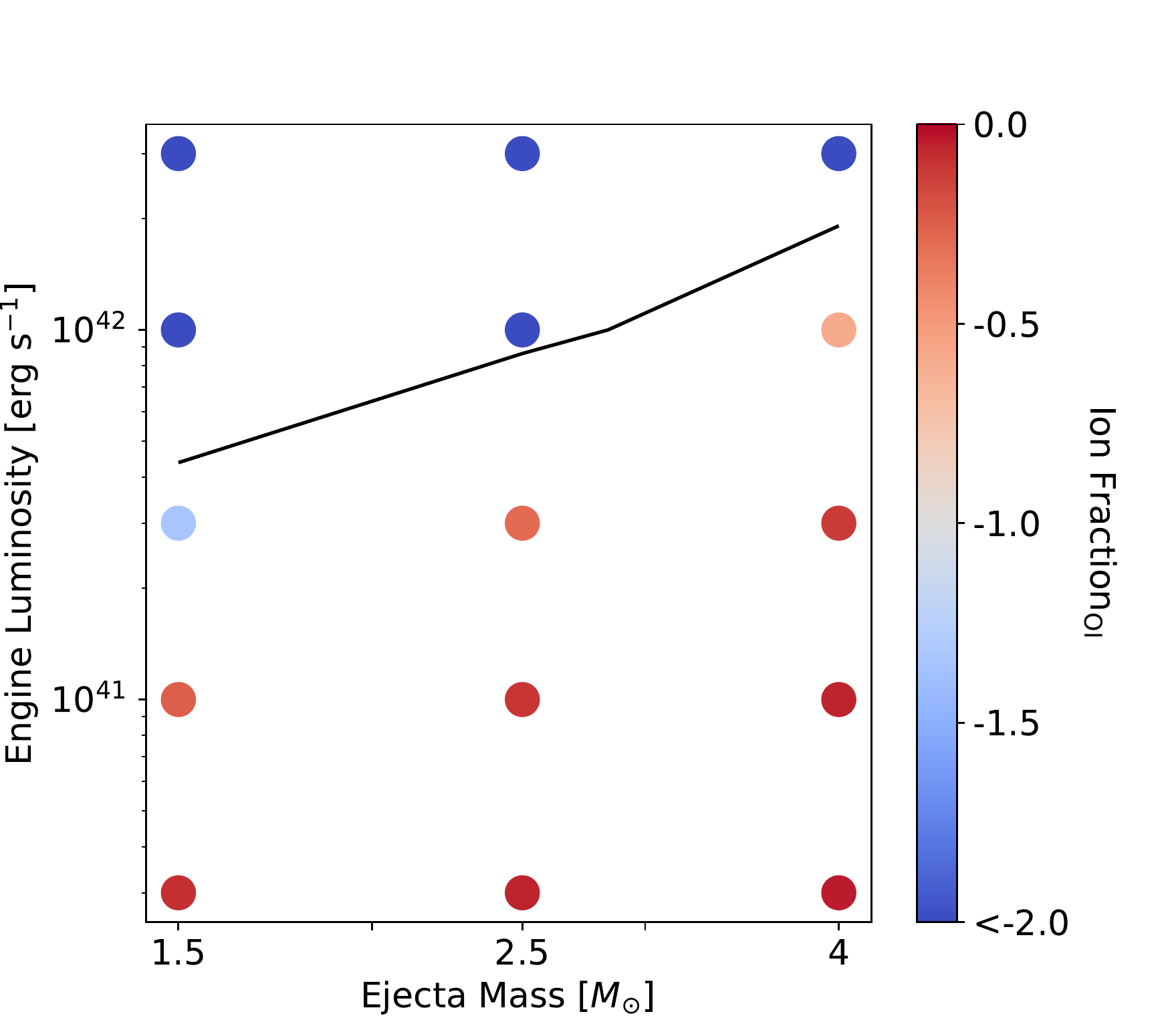}\\[-1.5ex]
\textbf{O II}&
\includegraphics[width=1.1\linewidth]{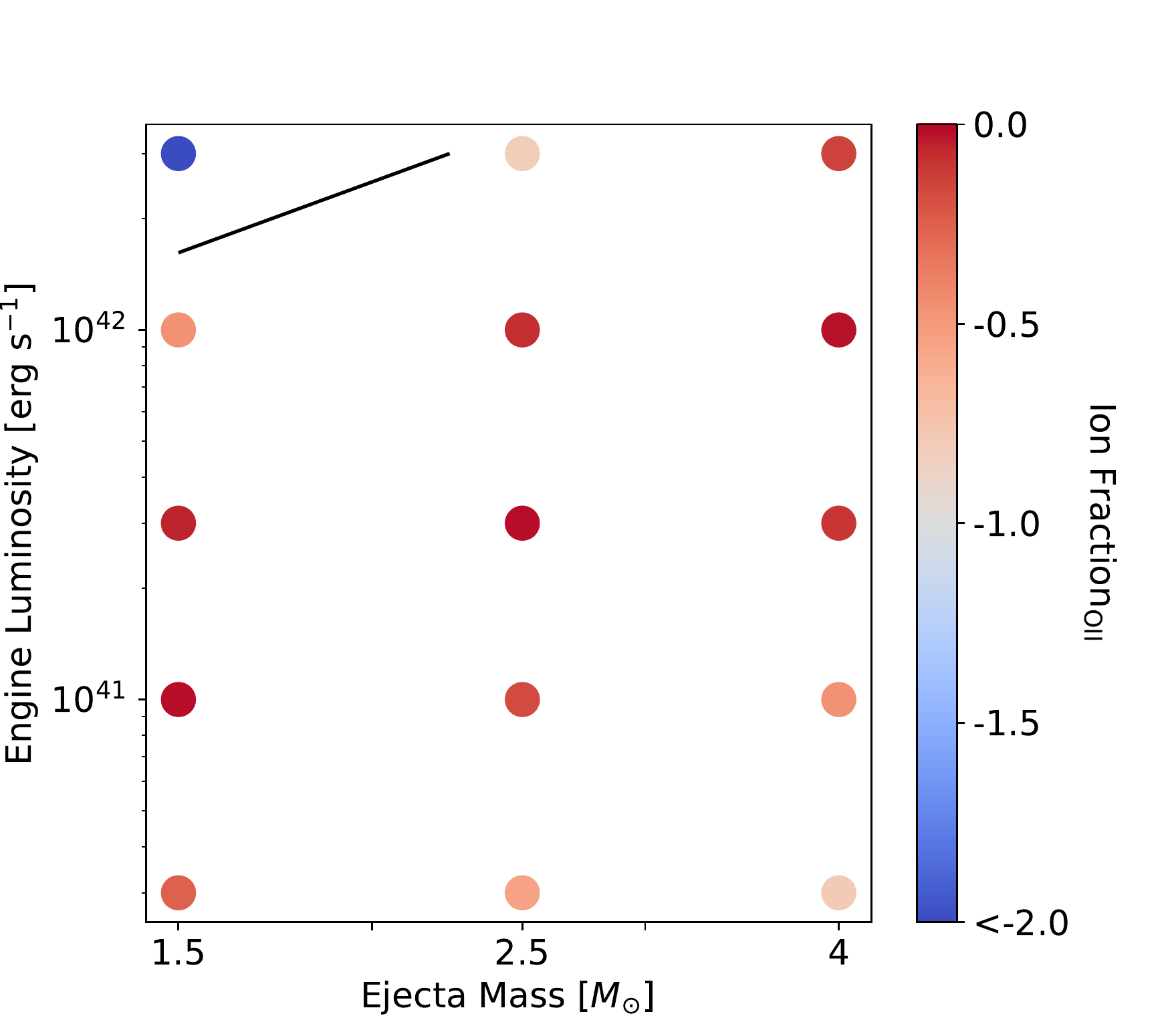}&
\includegraphics[width=1.1\linewidth]{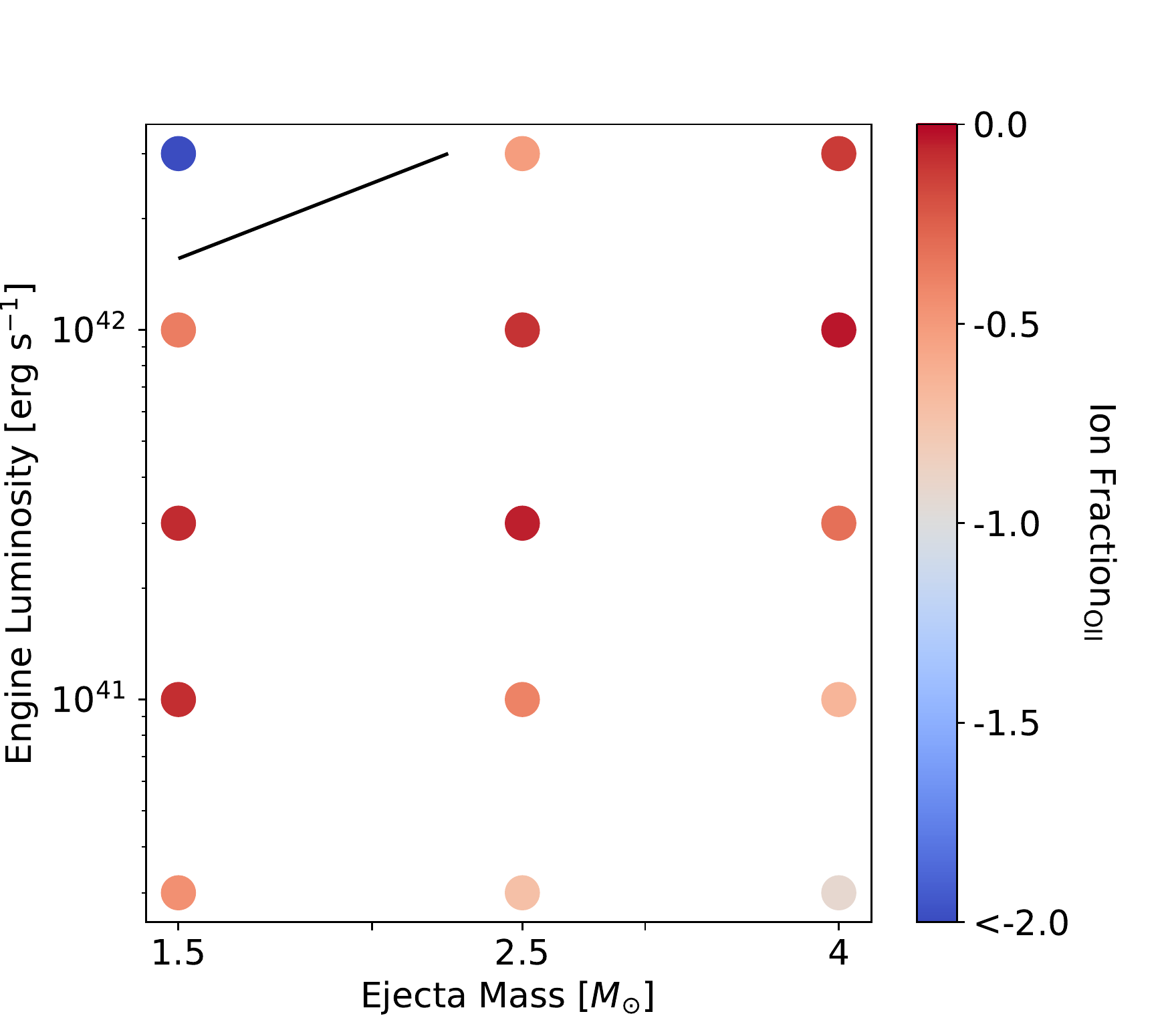}&
\includegraphics[width=1.1\linewidth]{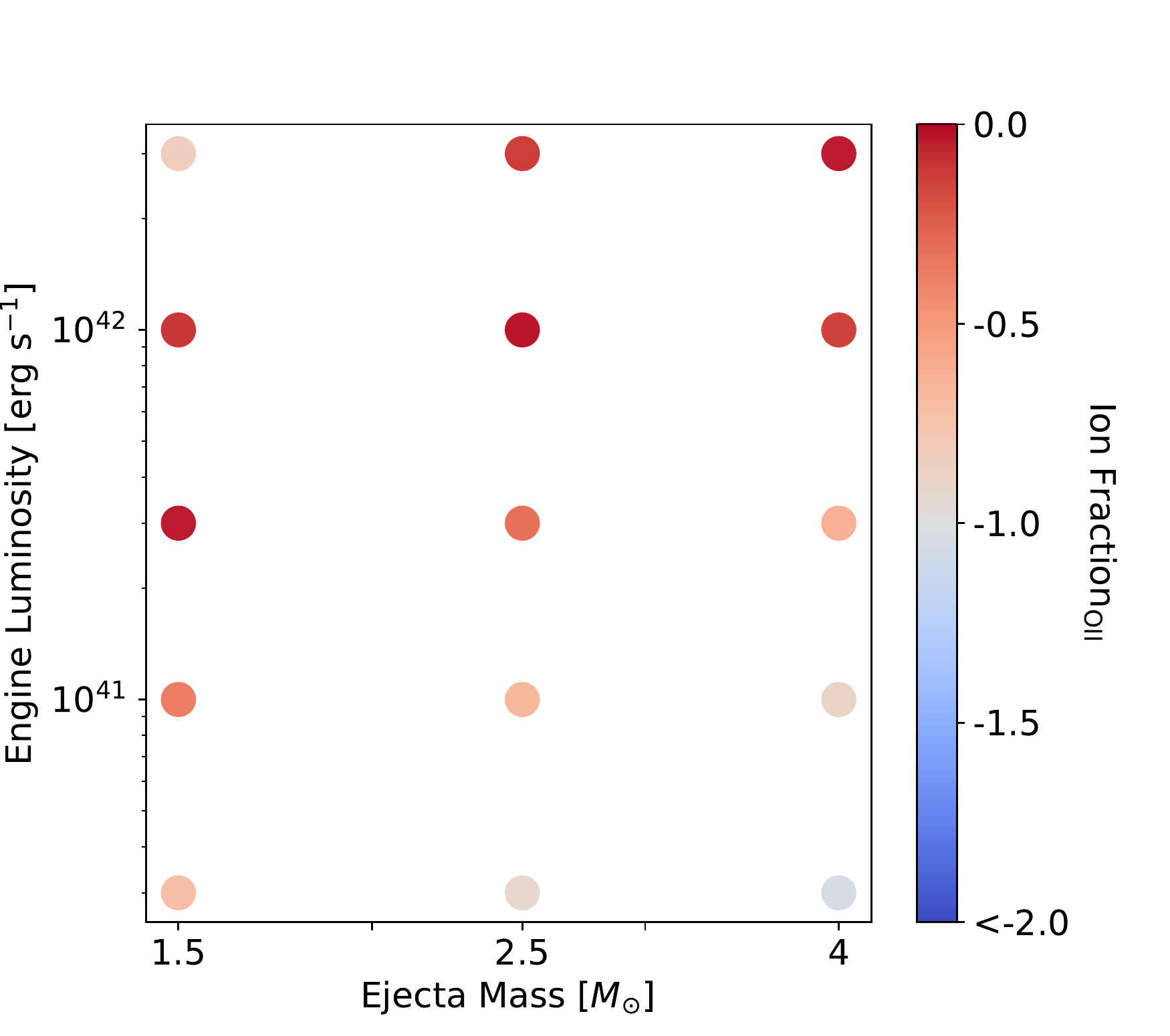}\\[-1.5ex]
\textbf{O III}&
\includegraphics[width=1.1\linewidth]{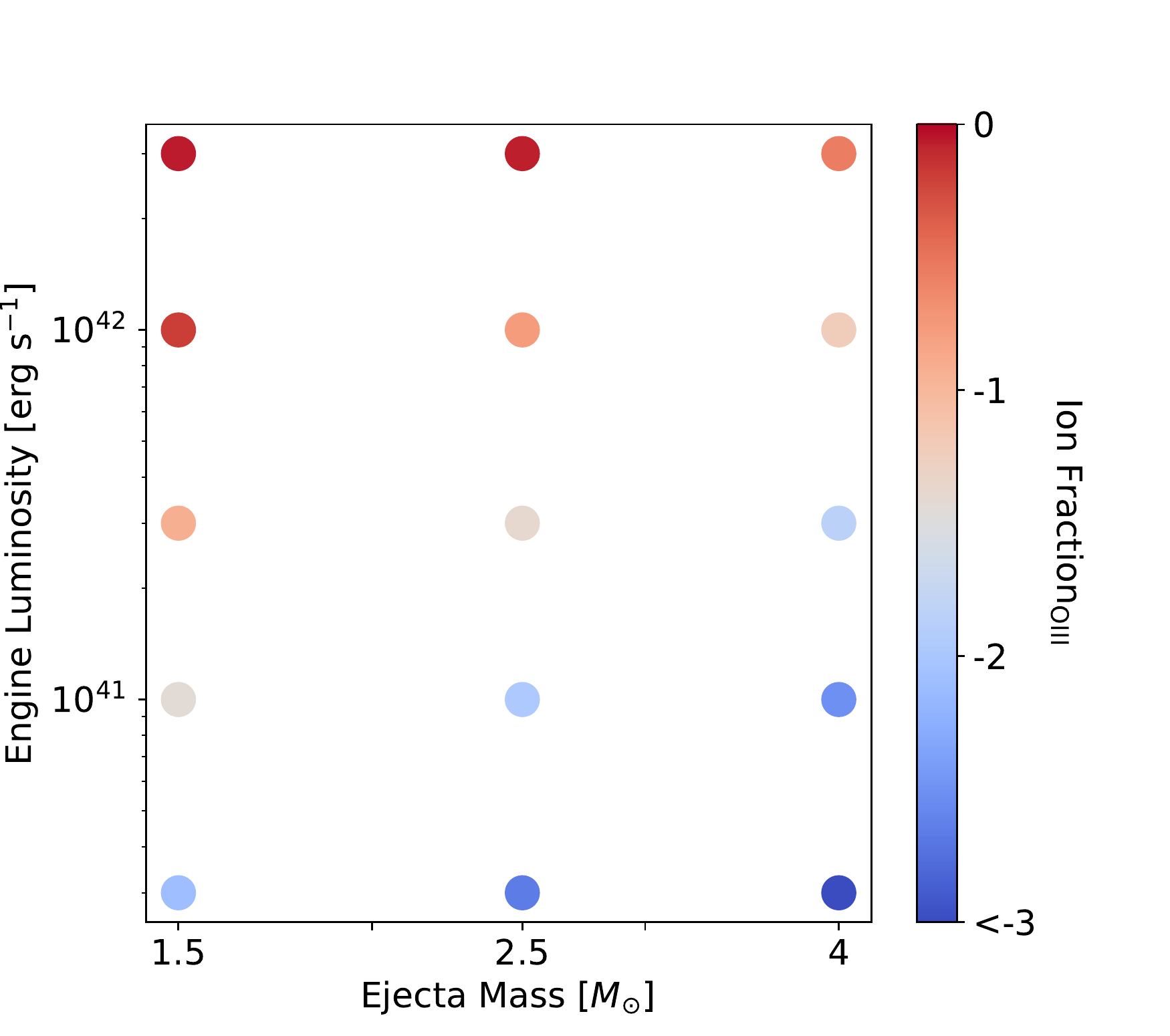}&
\includegraphics[width=1.1\linewidth]{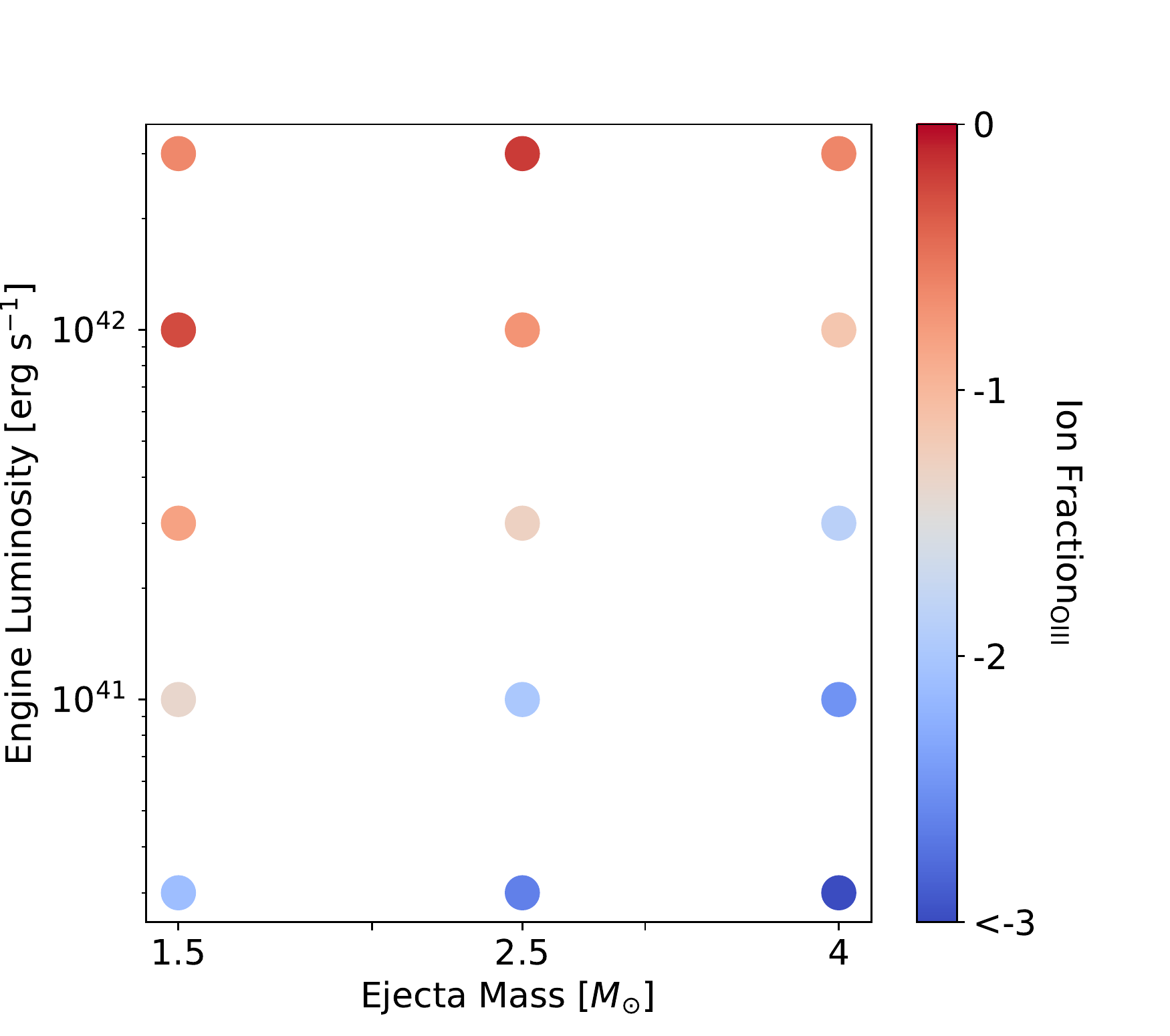}&
\includegraphics[width=1.1\linewidth]{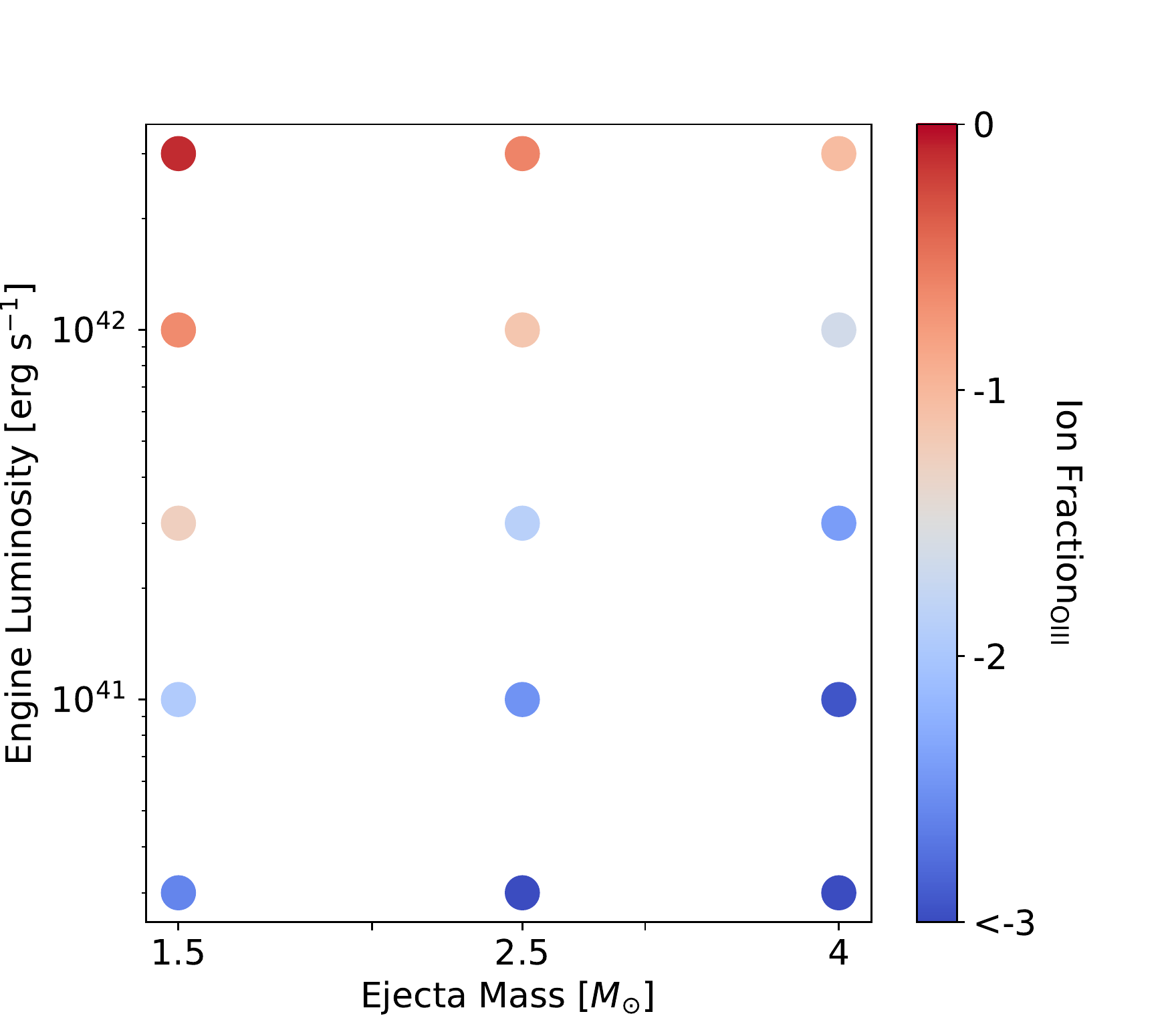}\\[-1.5ex]
\boldsymbol{$T_{\rm ej}$}&
\includegraphics[width=1.1\linewidth]{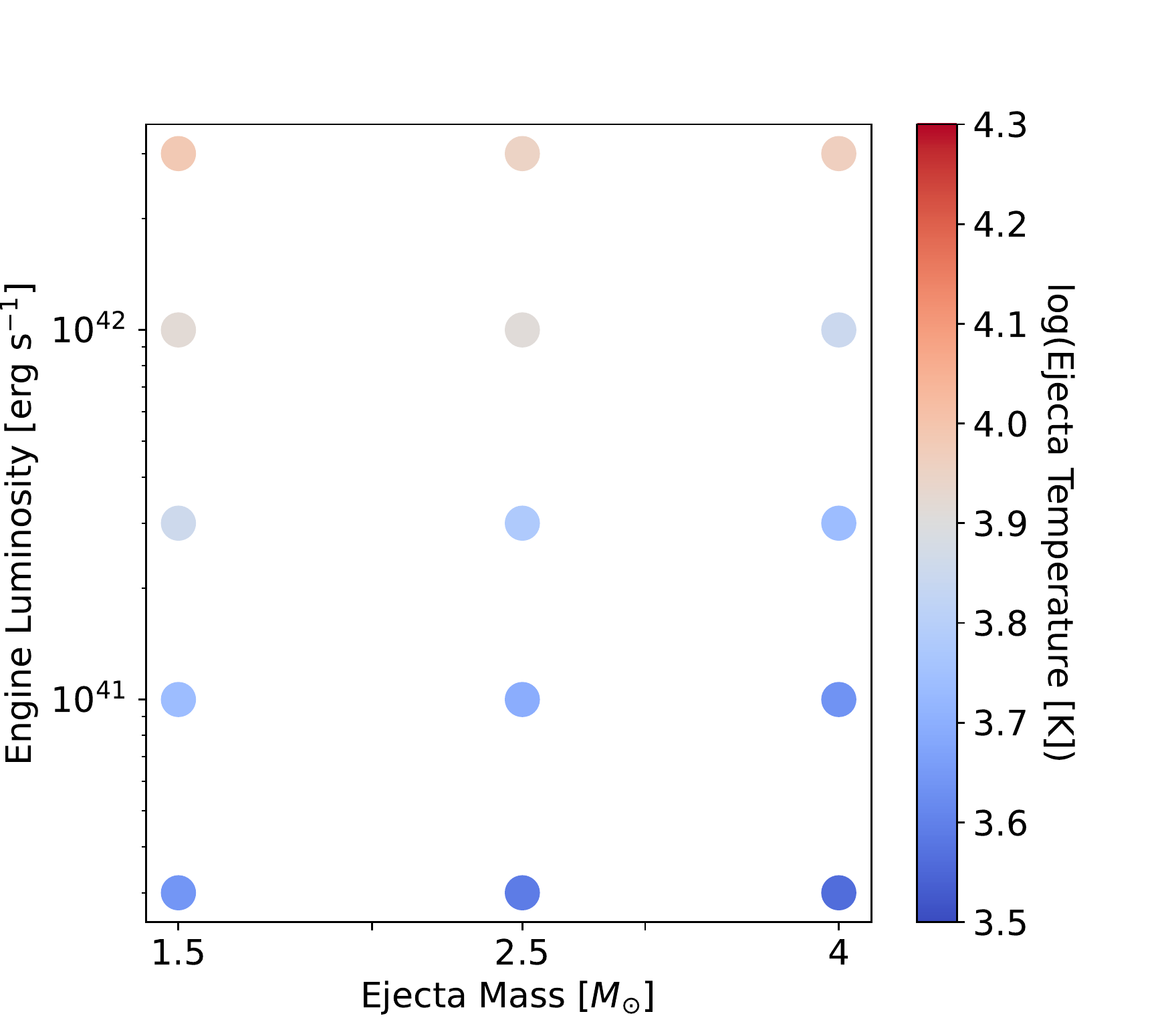}&
\includegraphics[width=1.1\linewidth]{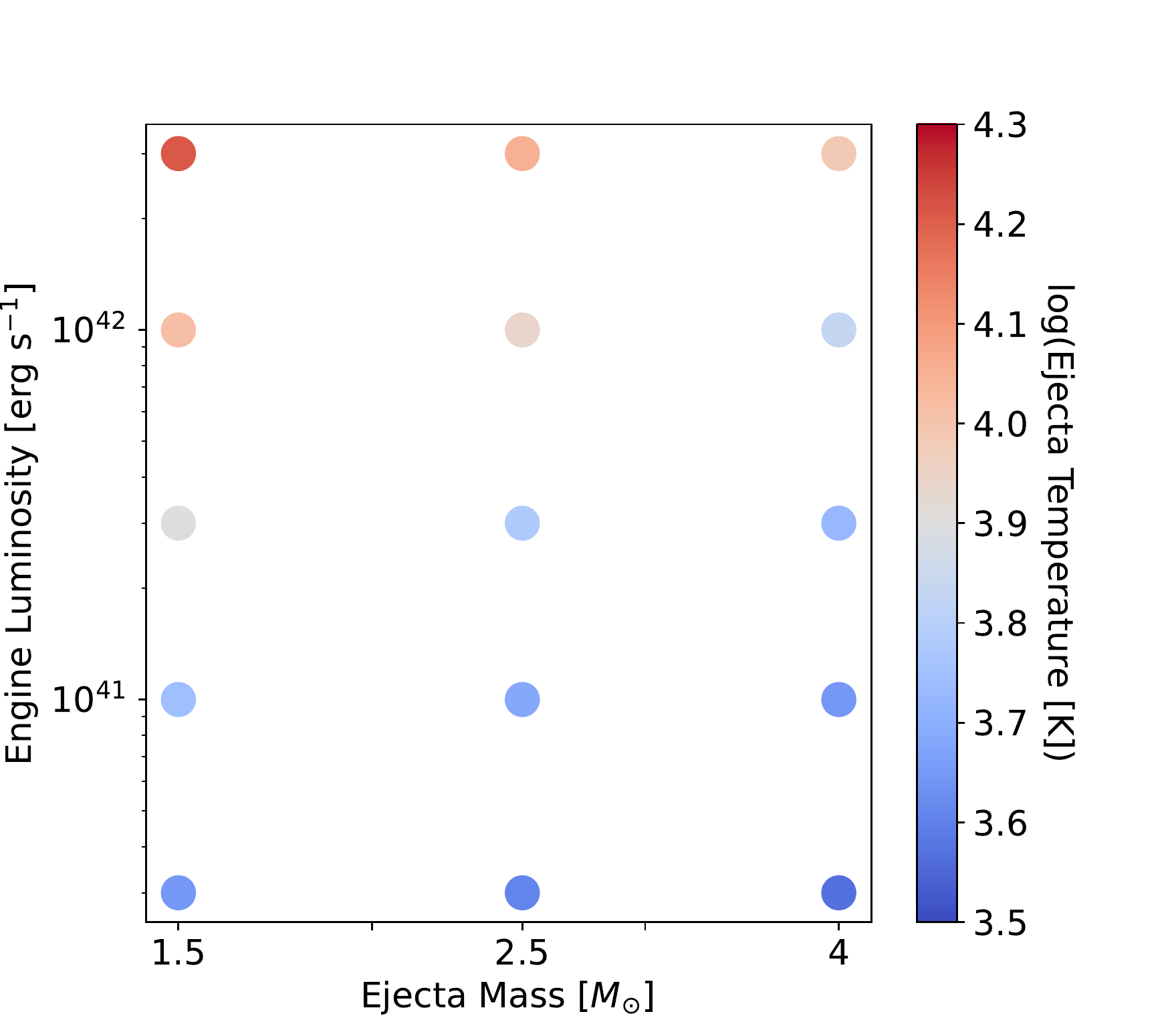}&
\includegraphics[width=1.1\linewidth]{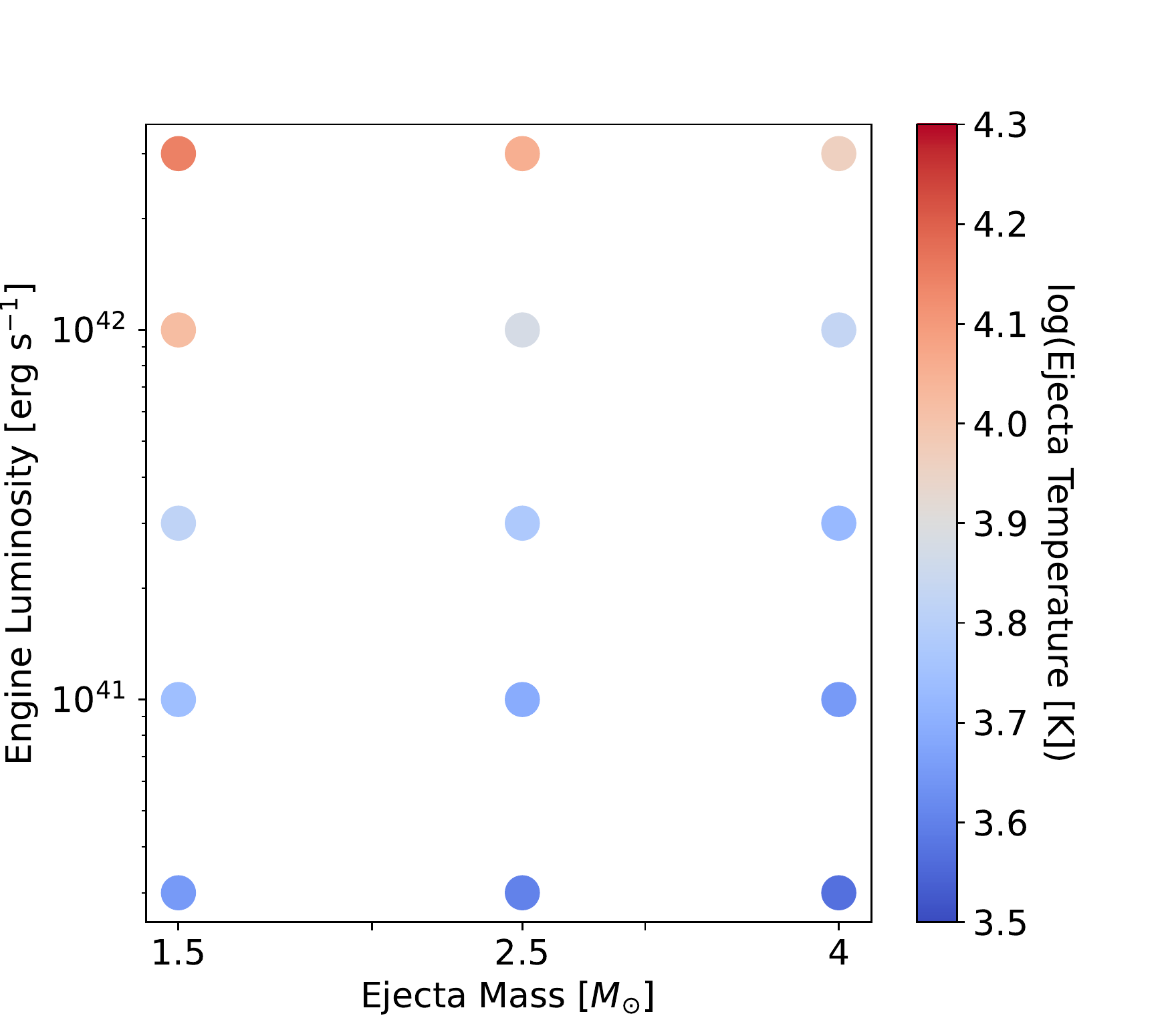}\\[-1.5ex]
\end{tabular}}
\caption{The ion fractions of O I (top), O II (second row), and O III (third row), and the ejecta temperature $T_{\rm ej}$ (bottom) in the simulations at 1 year for a realistic composition at three different values of $T_{\rm PWN}$.  The black contour denotes the low ejecta mass, high engine luminosity regime where runaway ionization can occur for both O I and O II.}%
\label{fig:rc1y_ionfrac}
\end{figure*}

\begin{figure*}
\newcolumntype{D}{>{\centering\arraybackslash} m{6cm}}
\noindent
\makebox[\textwidth]{
\begin{tabular}{m{1cm} DDD}
& \boldsymbol{$T_{\rm PWN} = 10^5$} \textbf{ K} & \boldsymbol{$T_{\rm PWN} = 3 \times 10^5$} \textbf{ K} & \boldsymbol{$T_{\rm PWN} = 10^6$} \textbf{ K}\\
\textbf{[O I]}&
\includegraphics[width=1.1\linewidth]{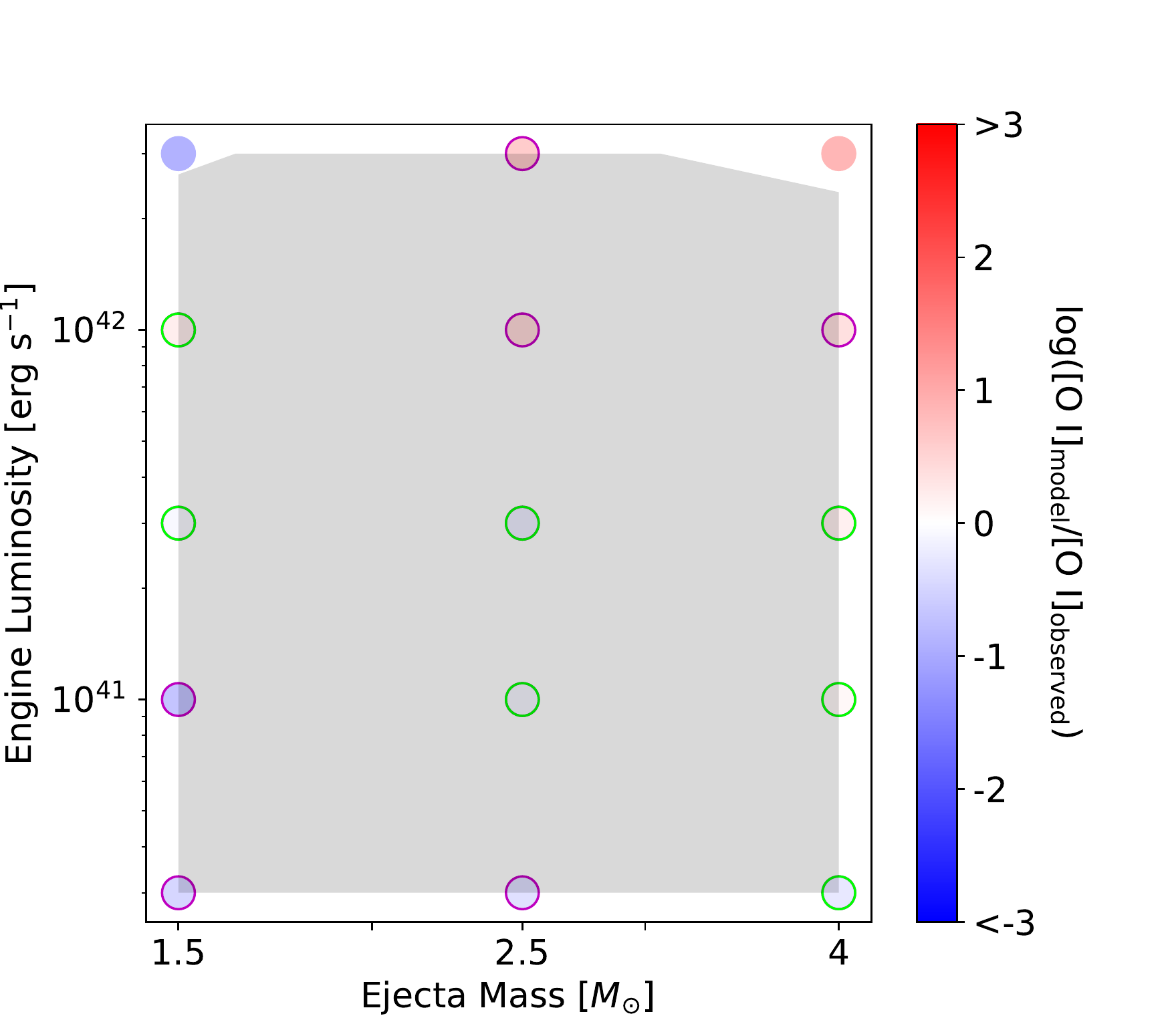}&
\includegraphics[width=1.1\linewidth]{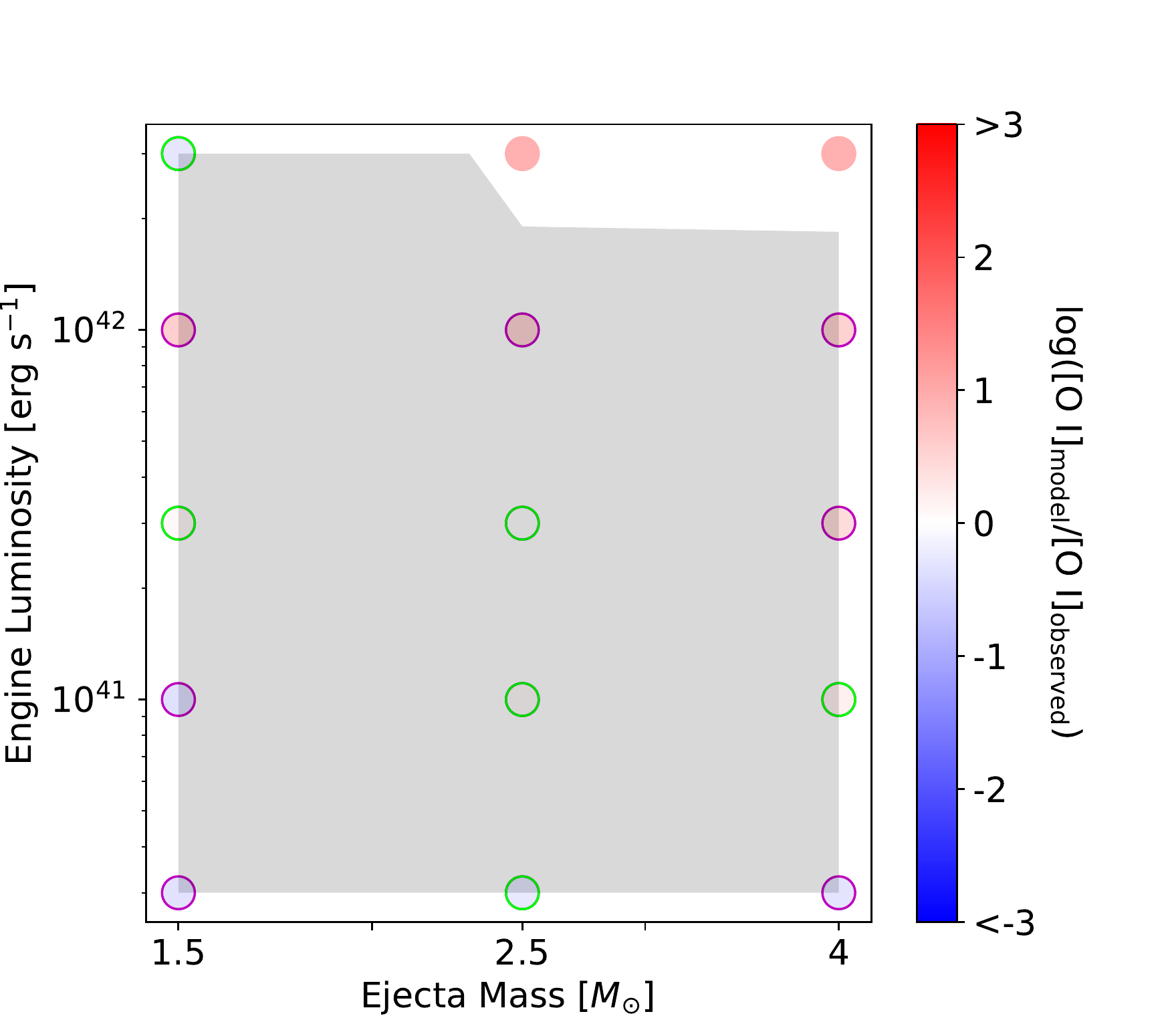}&
\includegraphics[width=1.1\linewidth]{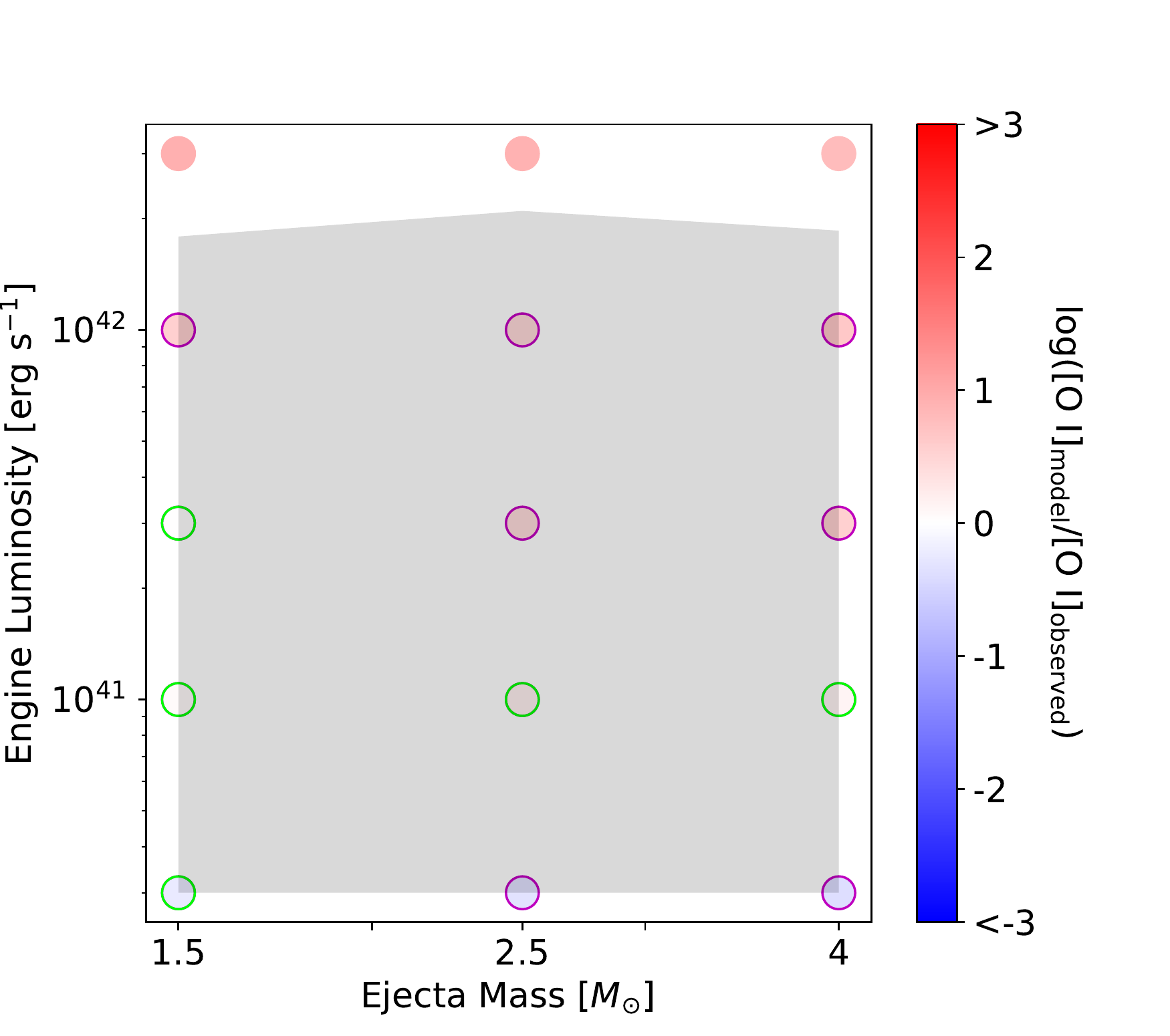}\\[-1.5ex]
\textbf{[O II]}&
\includegraphics[width=1.1\linewidth]{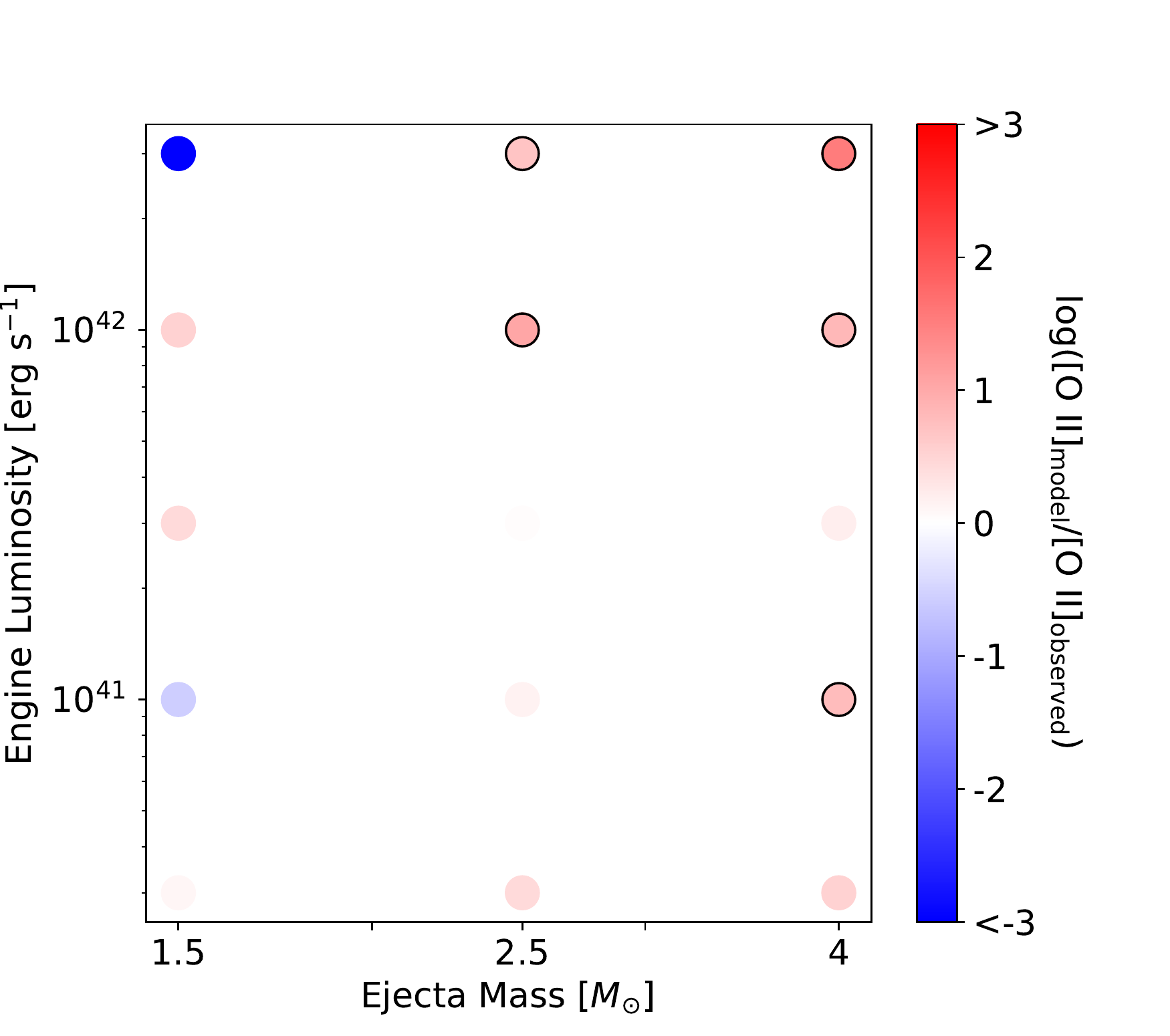}&
\includegraphics[width=1.1\linewidth]{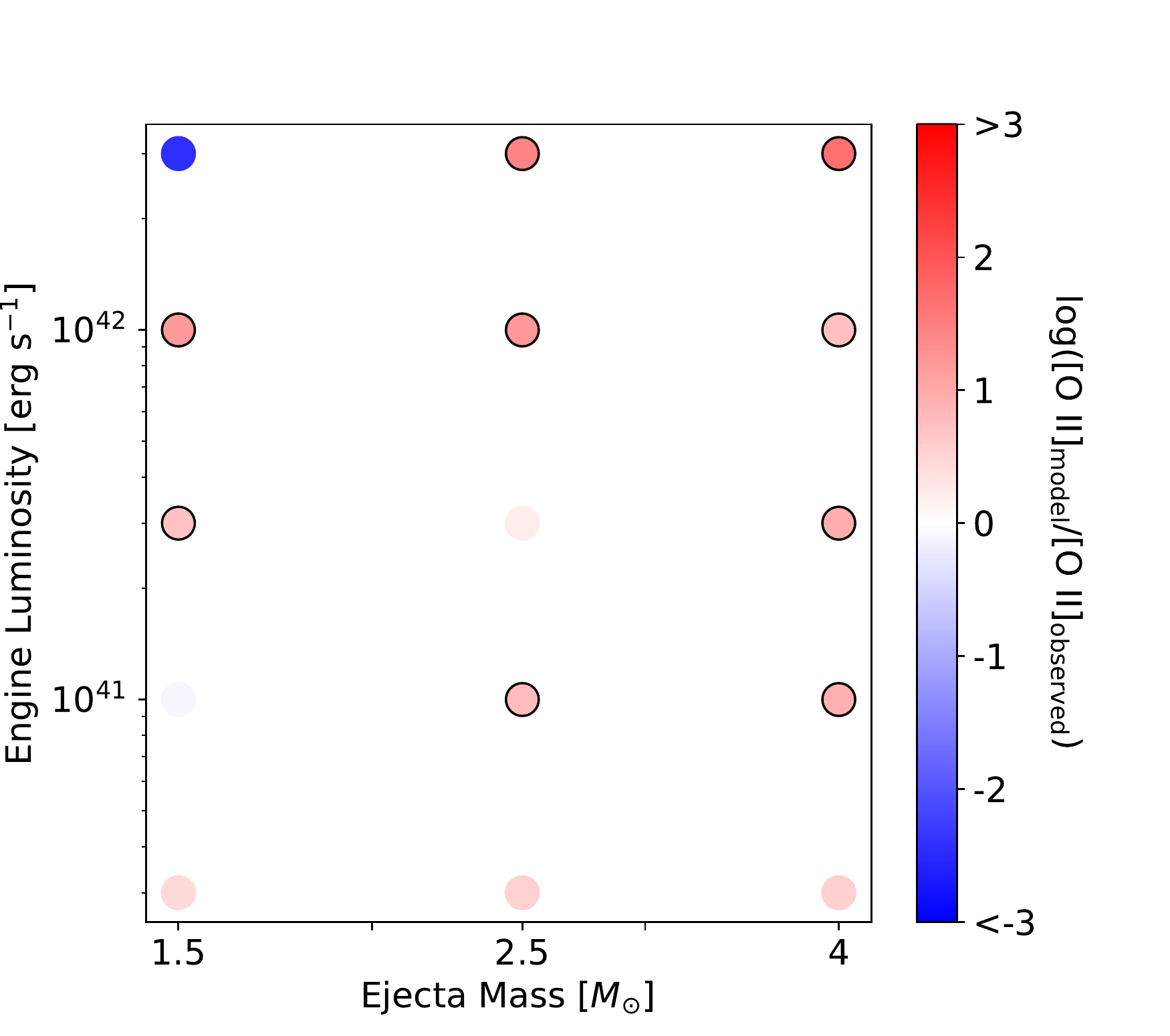}&
\includegraphics[width=1.1\linewidth]{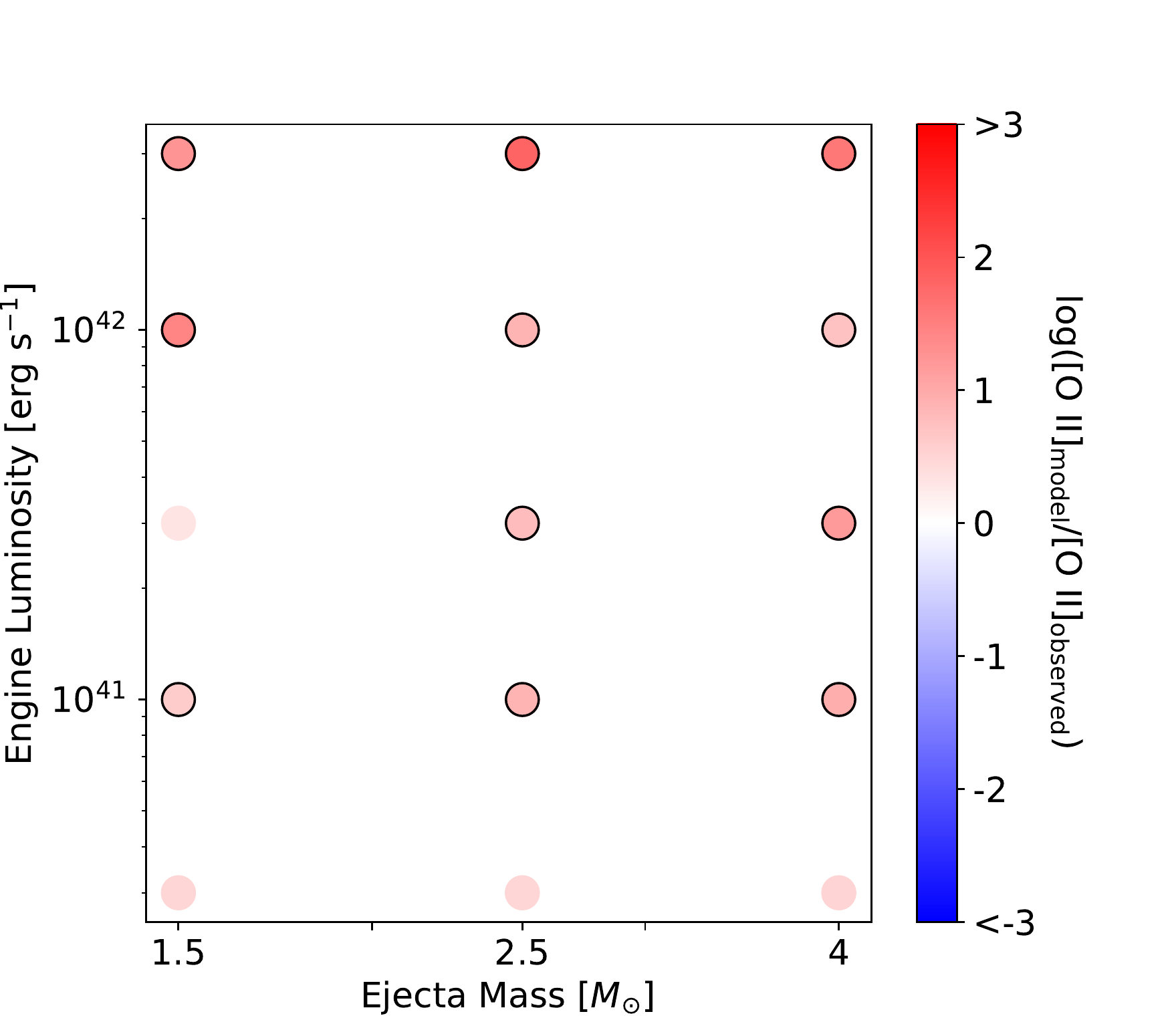}\\[-1.5ex]
\textbf{[O III]}&
\includegraphics[width=1.1\linewidth]{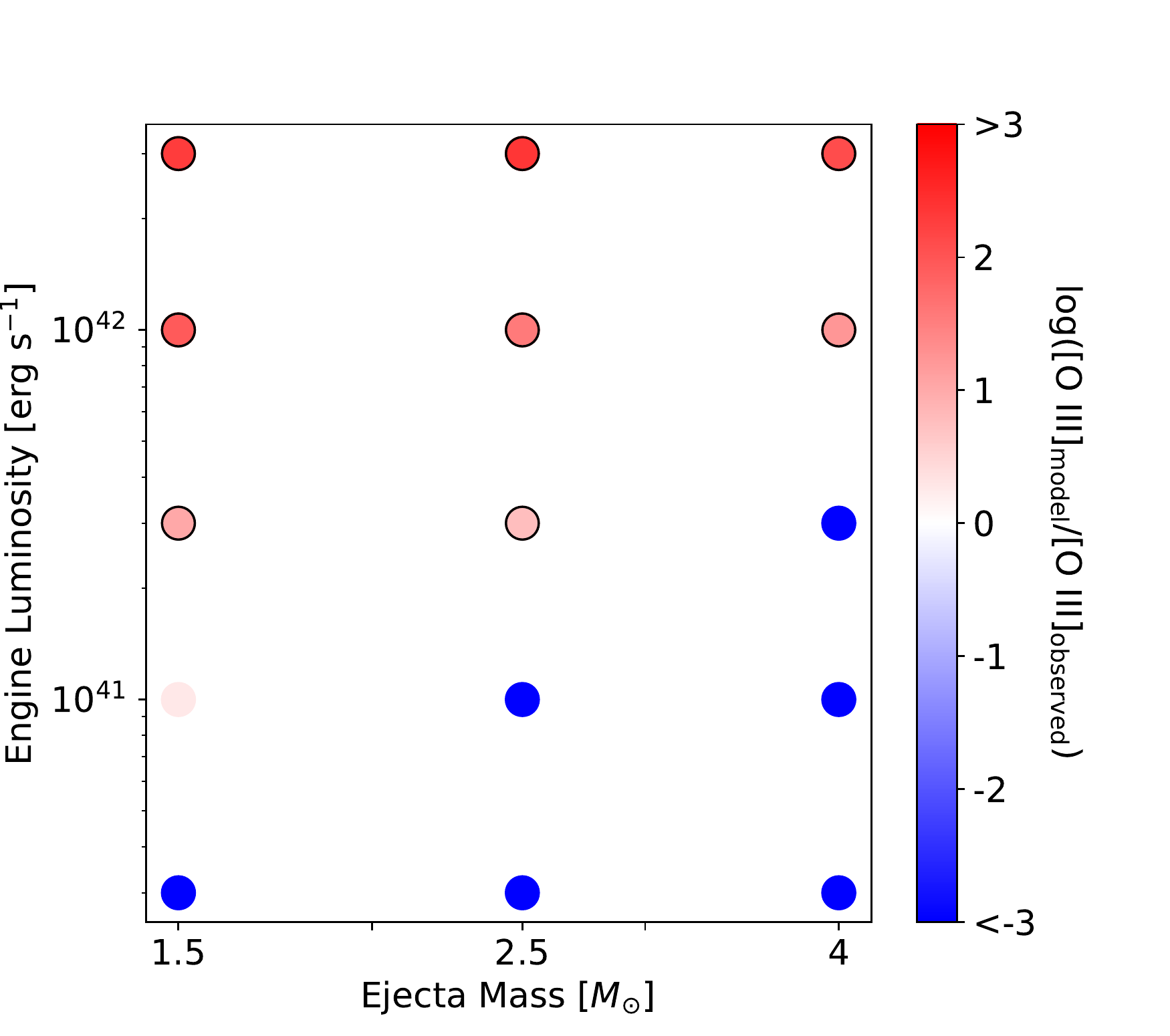}&
\includegraphics[width=1.1\linewidth]{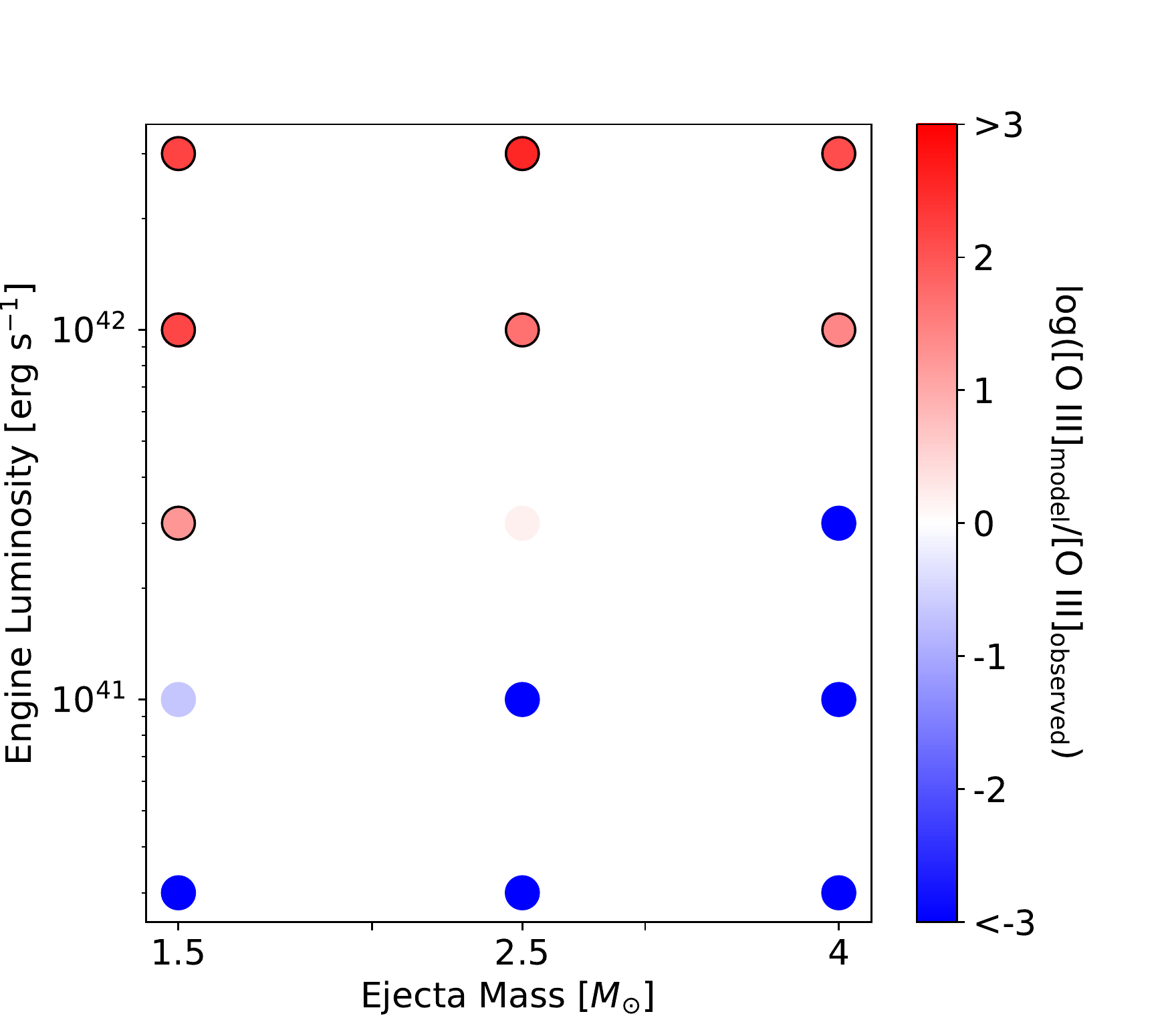}&
\includegraphics[width=1.1\linewidth]{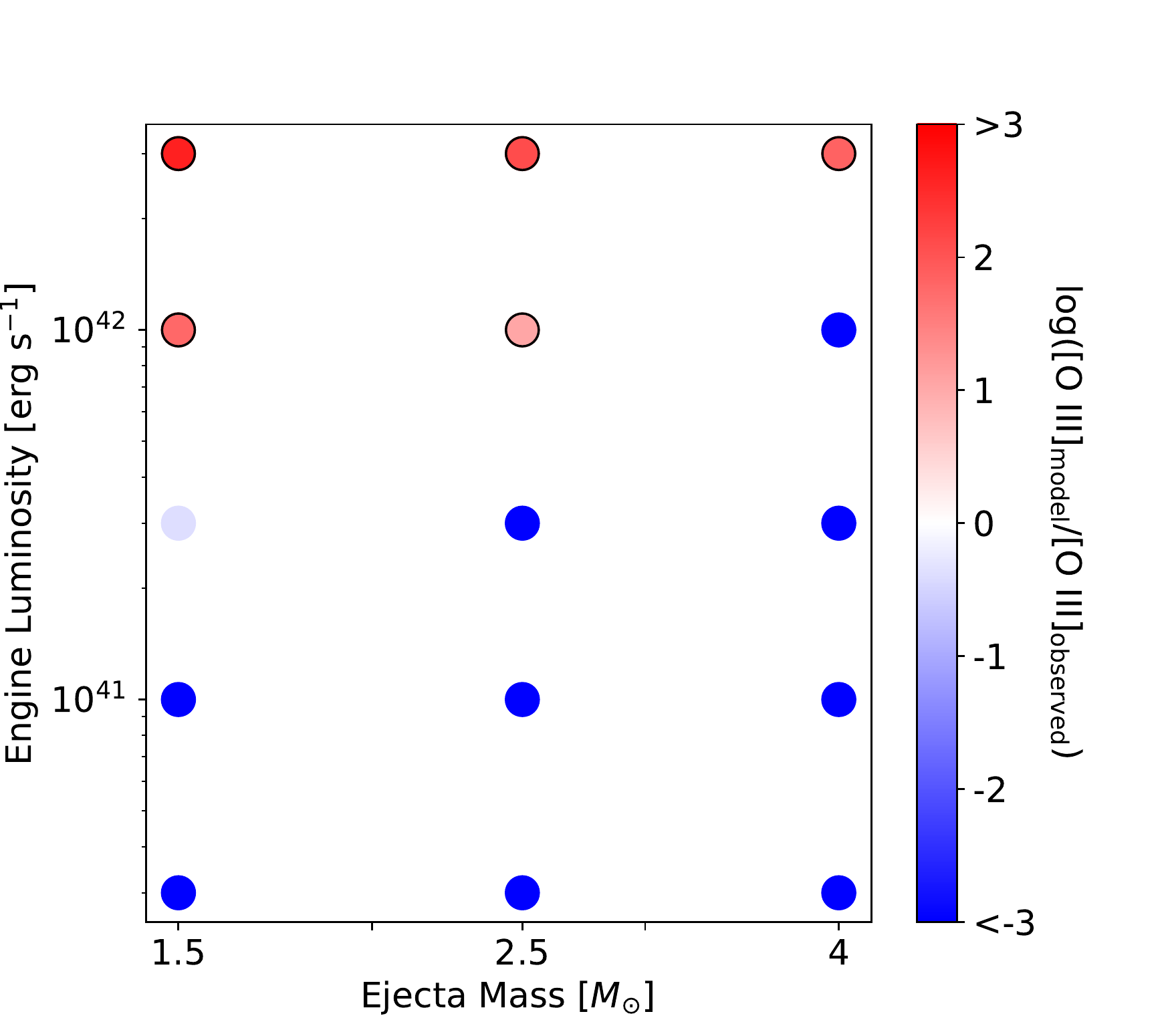}\\[-1.5ex]
\textbf{O I}&
\includegraphics[width=1.1\linewidth]{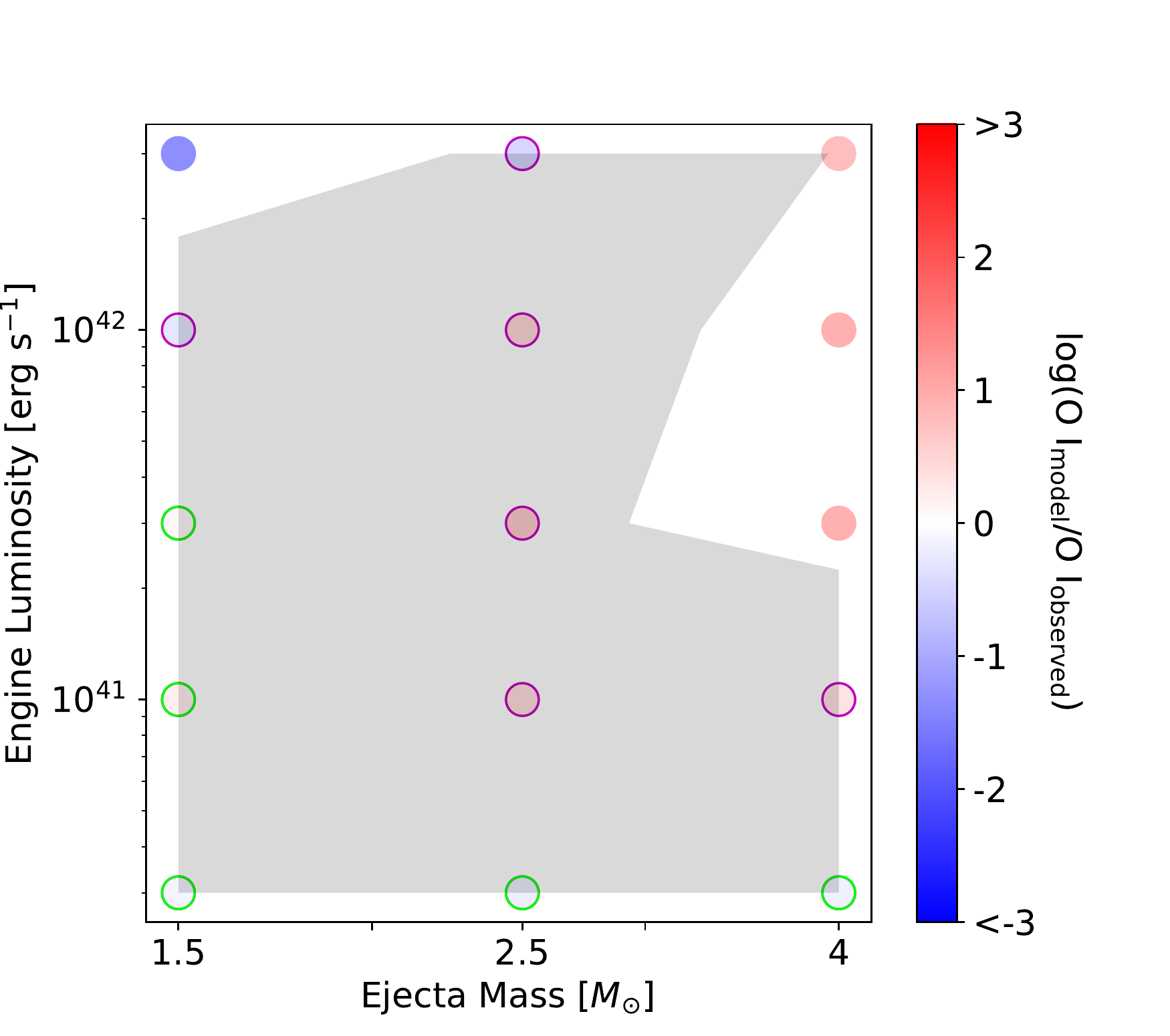}&
\includegraphics[width=1.1\linewidth]{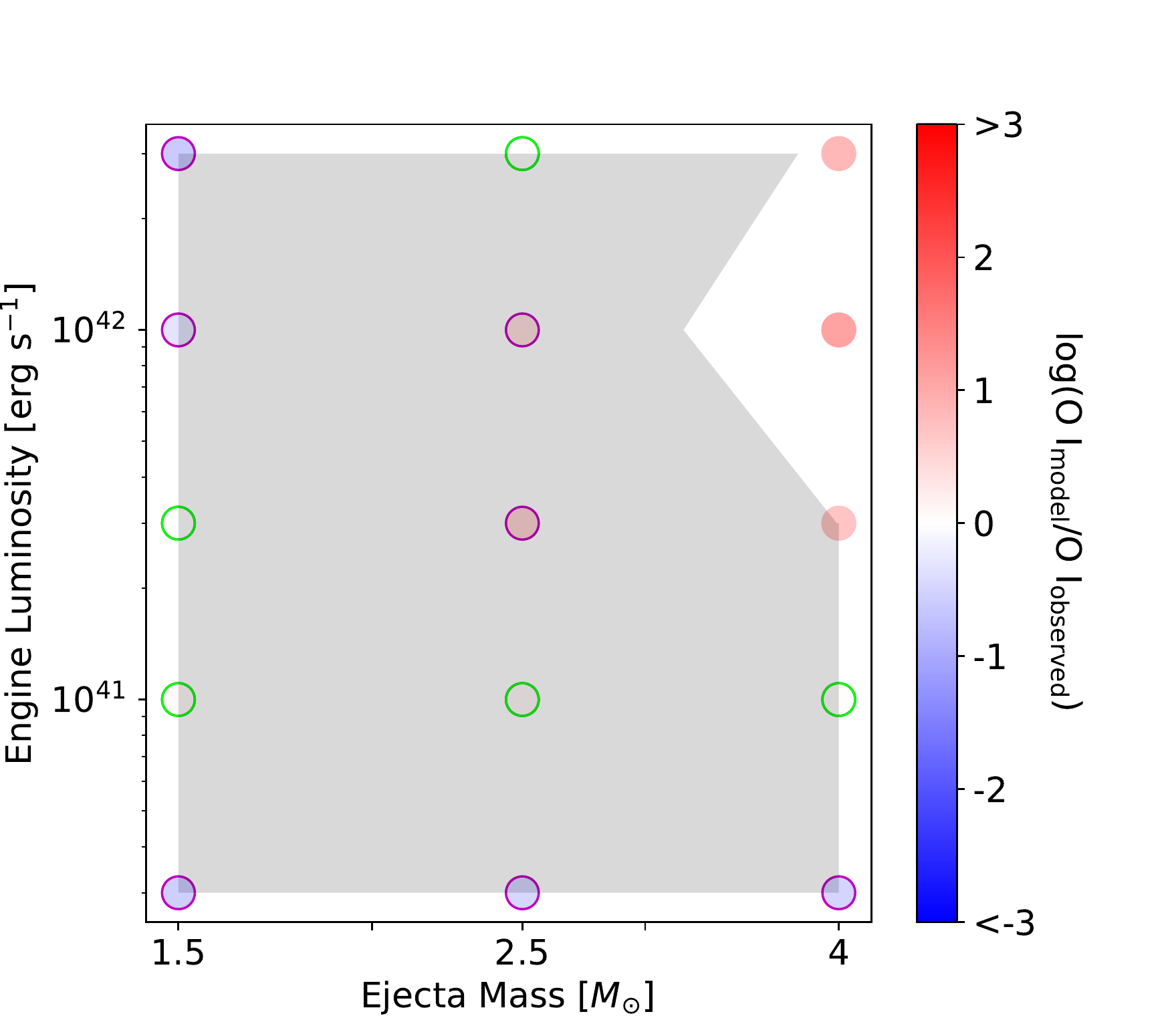}&
\includegraphics[width=1.1\linewidth]{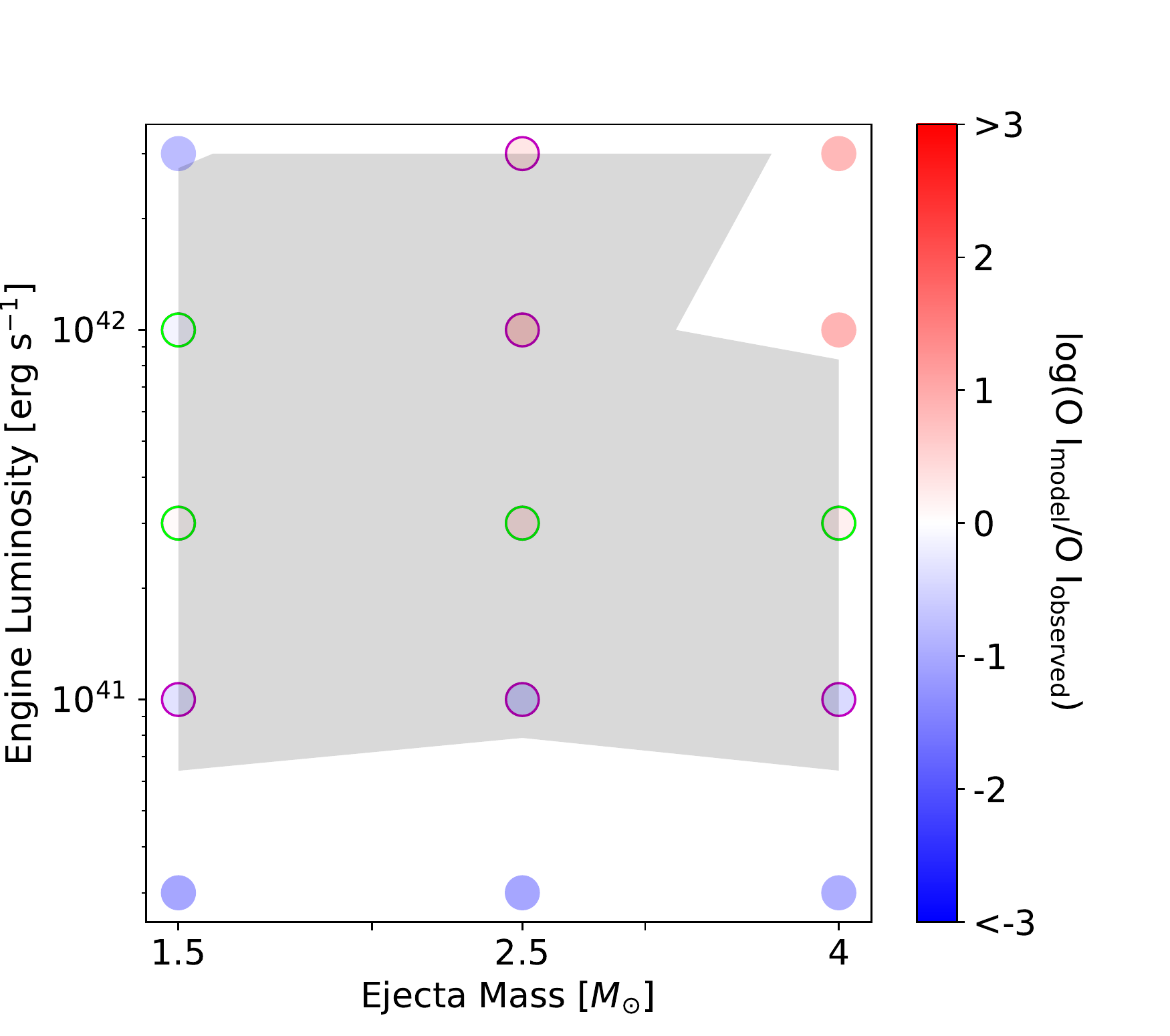}\\[-1.5ex]
\end{tabular}}
\caption{The luminosity of the model [O I] (top), [O II] (second row), [O III] (third row), and O I (bottom) lines in units of the observed line luminosities and limits of SN 2012au at 1 year for the realistic composition at three different values of $T_{\rm PWN}$.  The green circled points represent where the model and observed values are within a factor of 2, the purple circles within a factor of 5, and the grey shaded region also within a factor of 5.  The black circled points (for O I) represent where the model luminosity is more than a factor 2 larger than the observational limit.  Note that the grid includes luminosity from all elements in the wavelength regions where these lines are emitted.}%
\label{fig:rc1y_linecomp}
\end{figure*}

The oxygen ion fractions and ejecta temperature for the realistic composition at 1 year, using the same parameter grid as the pure oxygen composition, are shown in Figure \ref{fig:rc1y_ionfrac}, and the normalized line luminosities are shown in Figure \ref{fig:rc1y_linecomp}.  The ejecta temperatures are not significantly cooler than for the pure oxygen composition, in contrast to the differences seen at 6 years.  Examining the line luminosities for the [O II] and [O III] wavelength regions shows that the high $L_{\rm PWN}$ portion of  parameter space is excluded due to the high luminosity of these two lines, similarly to the pure oxygen composition, although the high luminosity at 7300 $\AA$ could also be due to [Ca II].  

\begin{figure*}
\newcolumntype{D}{>{\centering\arraybackslash} m{6cm}}
\noindent
\makebox[\textwidth]{
\begin{tabular}{DDD}
\boldsymbol{$T_{\rm PWN} = 10^5$} \textbf{ K} & \boldsymbol{$T_{\rm PWN} = 3 \times 10^5$} \textbf{ K} & \boldsymbol{$T_{\rm PWN} = 10^6$} \textbf{ K}\\
\includegraphics[width=1.1\linewidth]{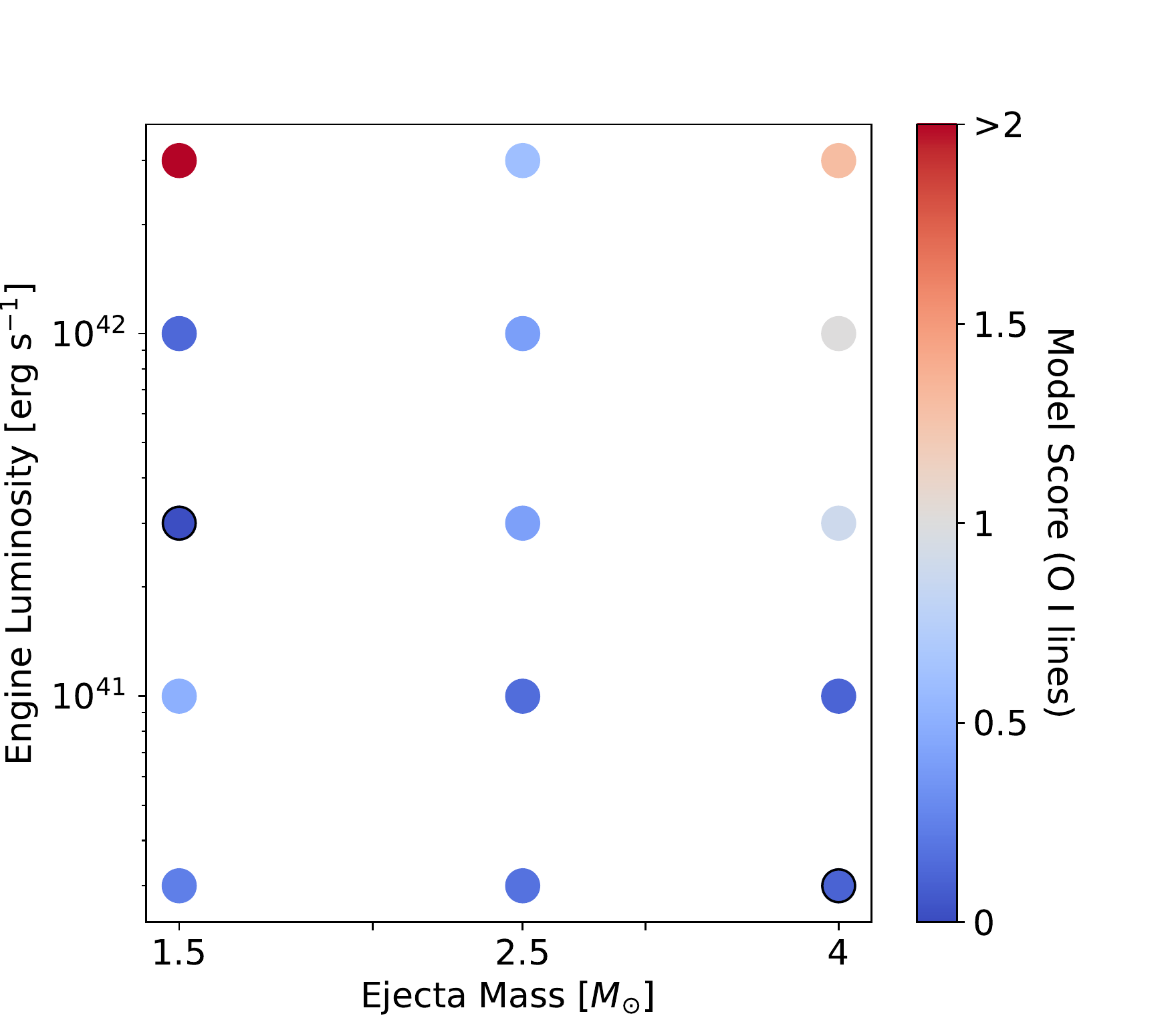}&
\includegraphics[width=1.1\linewidth]{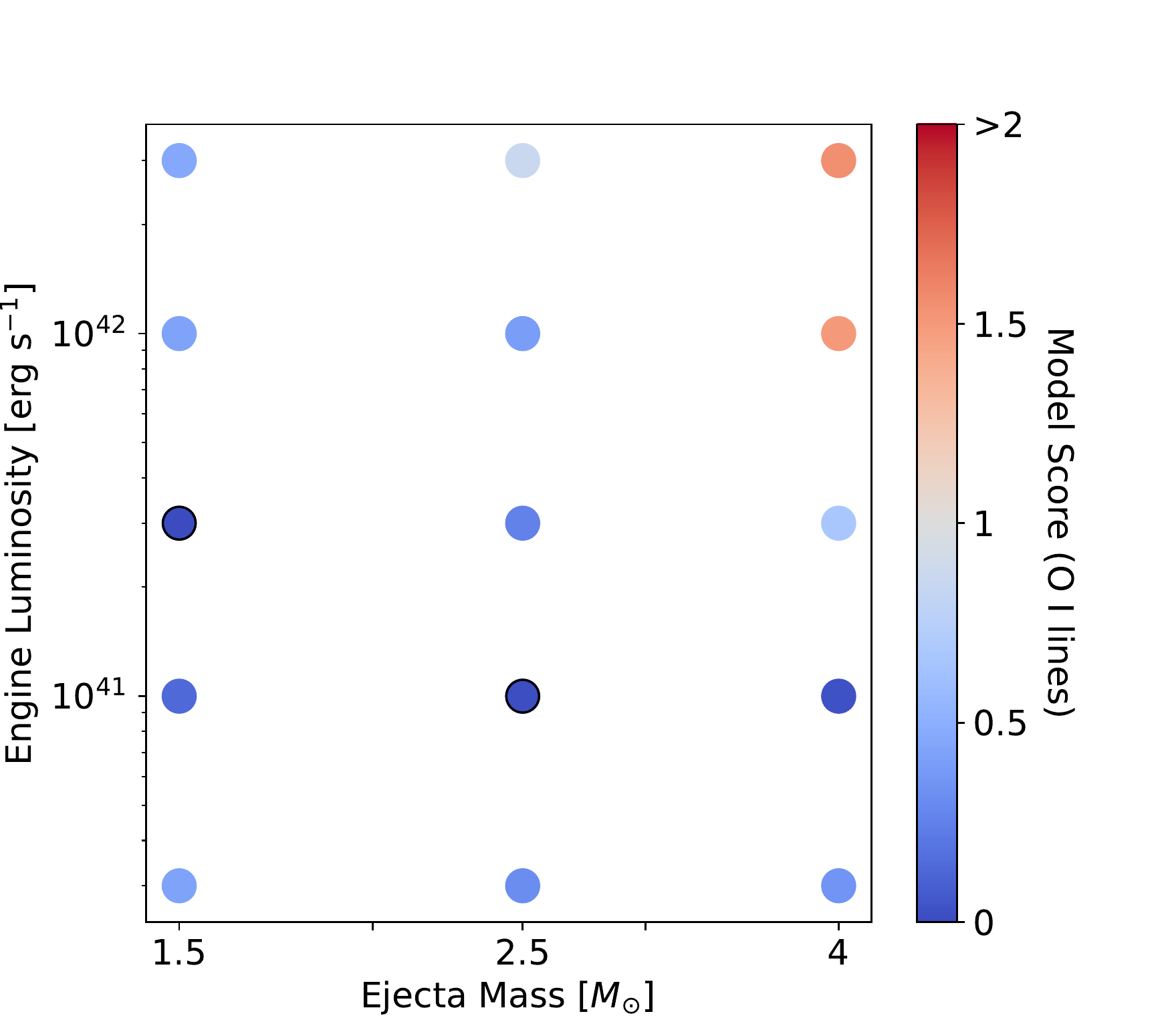}&
\includegraphics[width=1.1\linewidth]{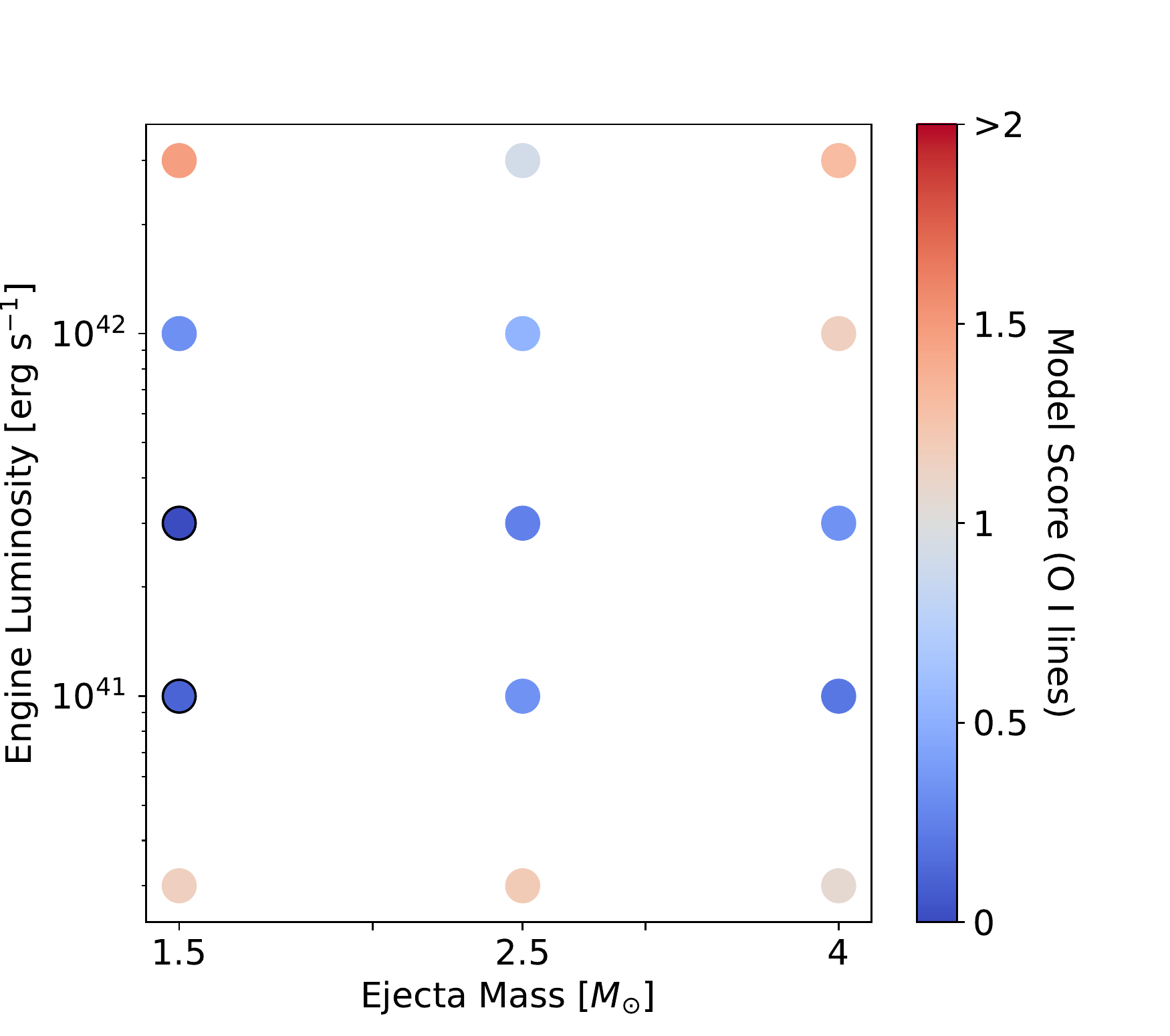}\\[-1.5ex]
\end{tabular}}
\caption{The goodness-of-fit score for each model in the realistic composition at 1 year based on the [O I] and O I lines.  Lower scores indicate a better fit to the data (from Equation \ref{eqn:modscore}, a perfect fit has score 0, both lines off by factor 2 has score 0.18, and both lines off by factor 10 has score 2). The black circles indicate the two models with the lowest scores for each $T_{\rm PWN}$, which are plotted in Figure \ref{fig:rc1y_spec}.}%
\label{fig:rc1y_score}
\end{figure*}

\begin{figure*}
\newcolumntype{D}{>{\centering\arraybackslash} m{6cm}}
\noindent
\makebox[\textwidth]{
\begin{tabular}{DDD}
\includegraphics[width=1.1\linewidth]{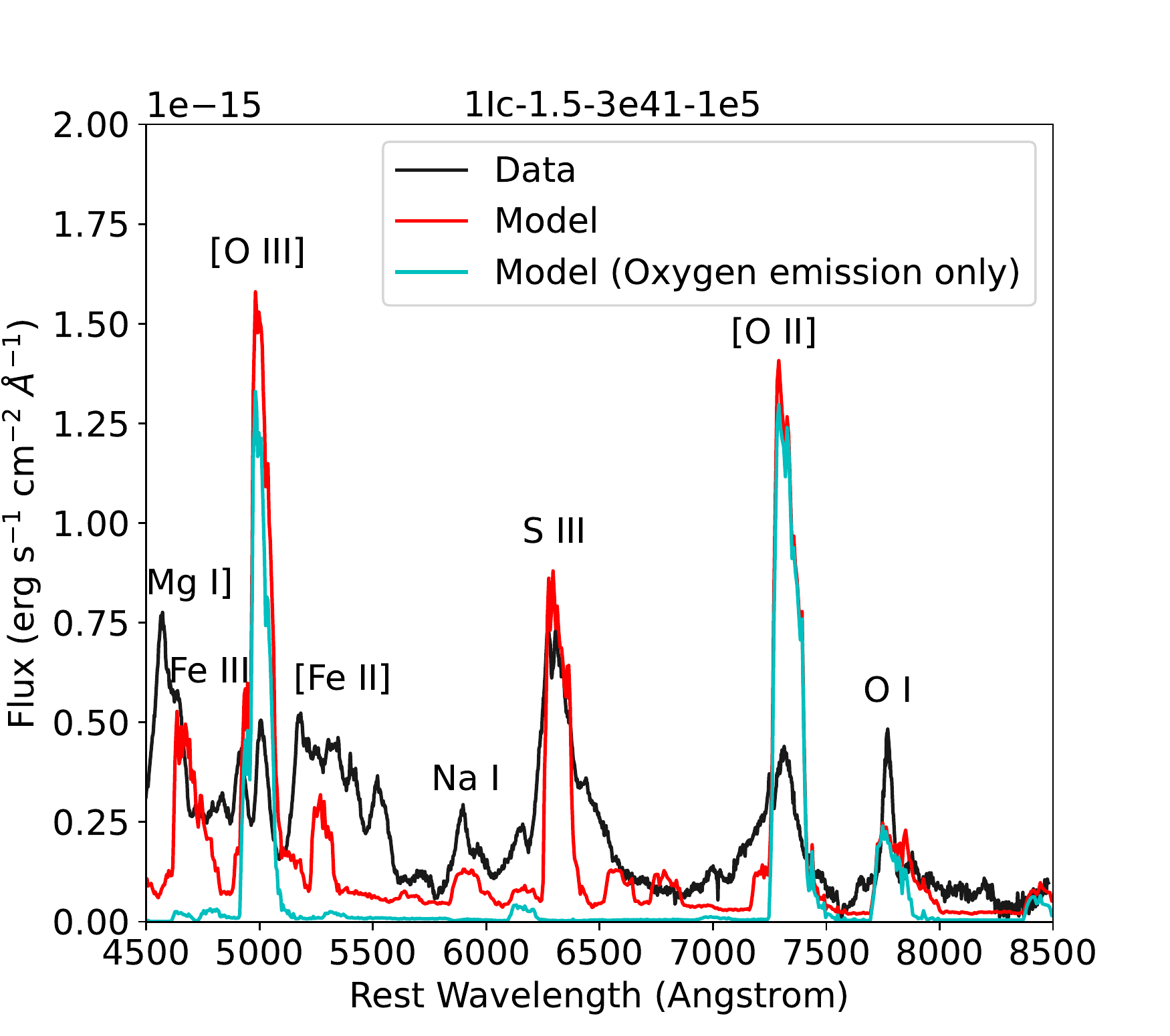}&
\includegraphics[width=1.1\linewidth]{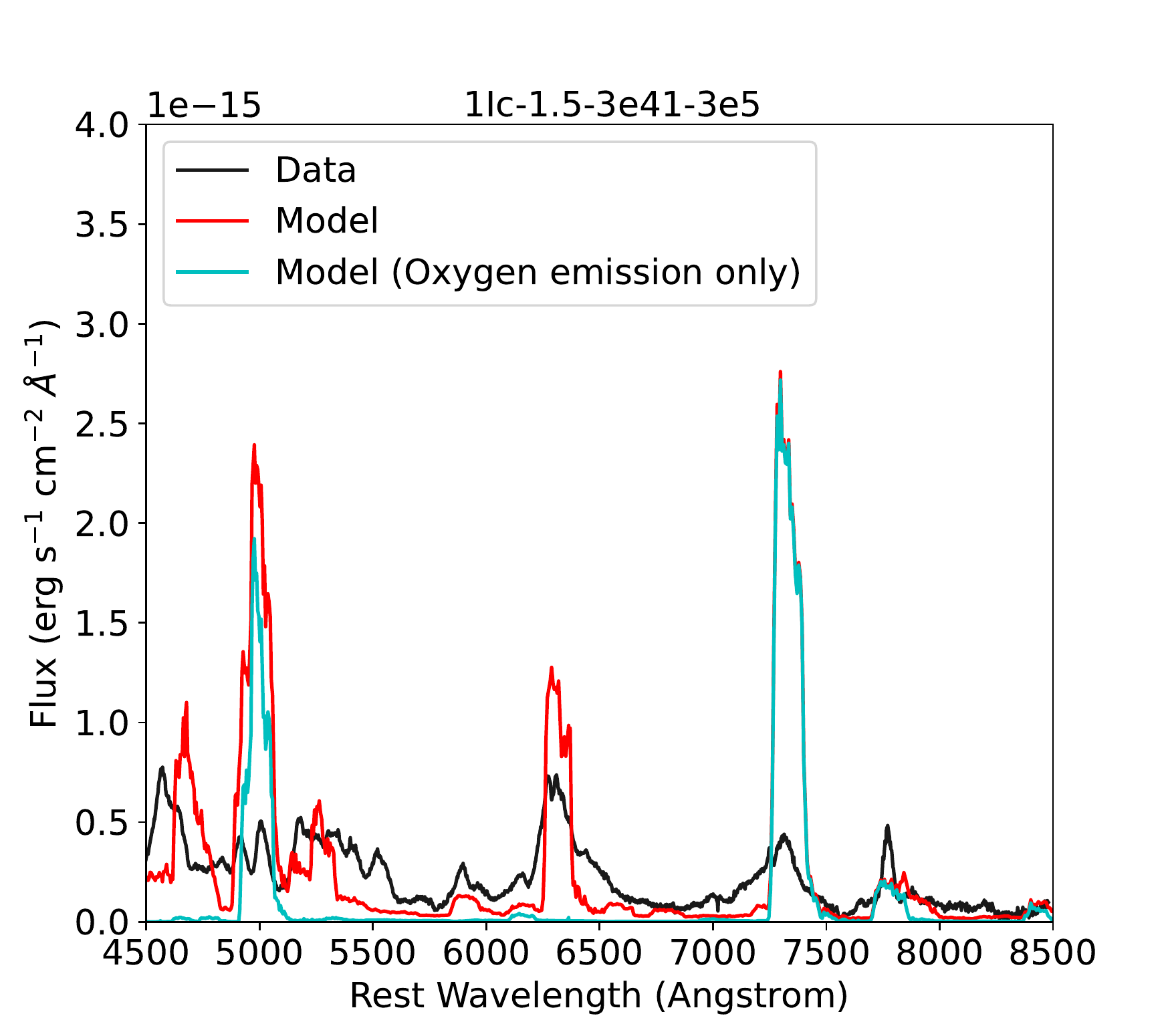}&
\includegraphics[width=1.1\linewidth]{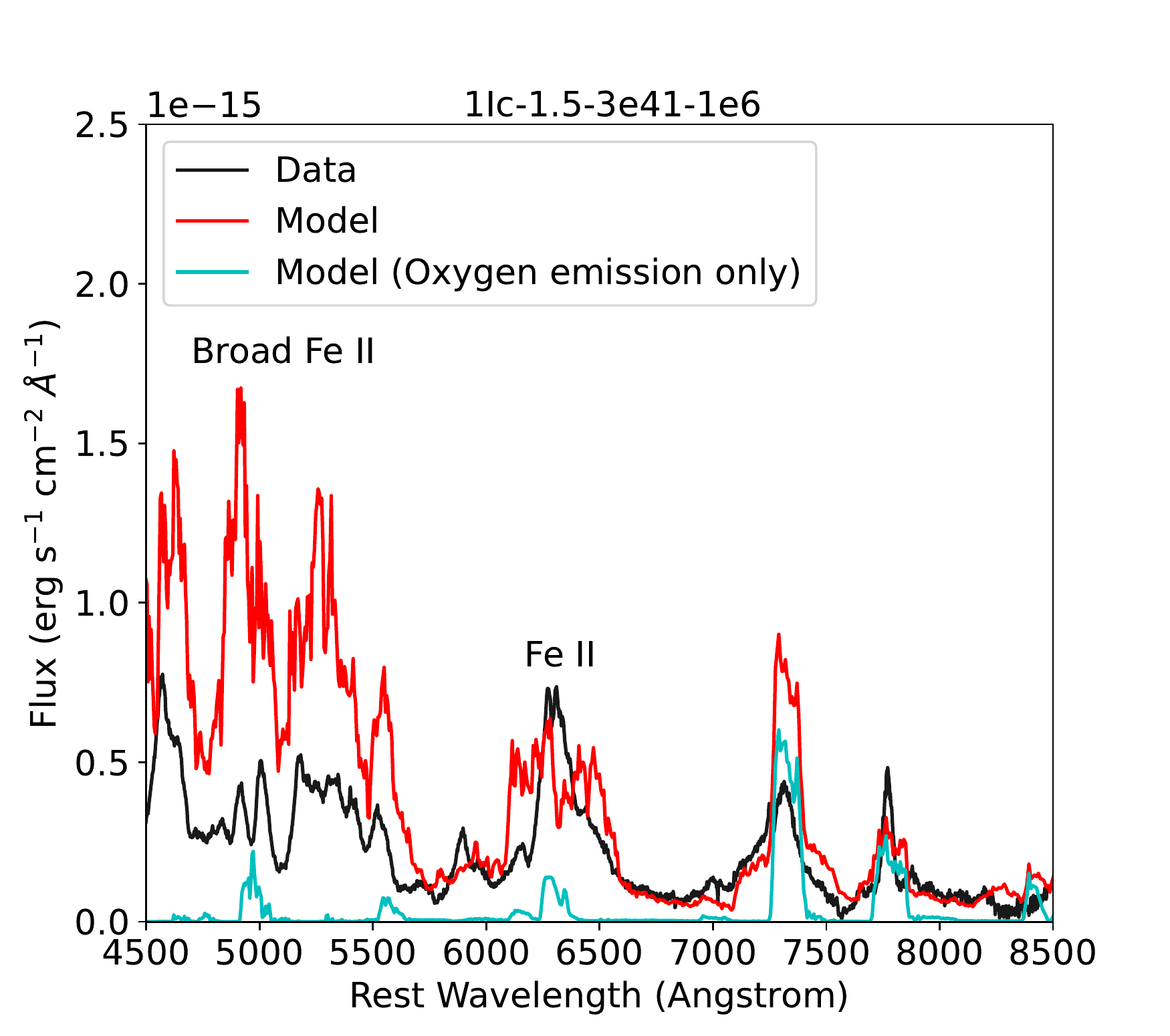}\\
\includegraphics[width=1.1\linewidth]{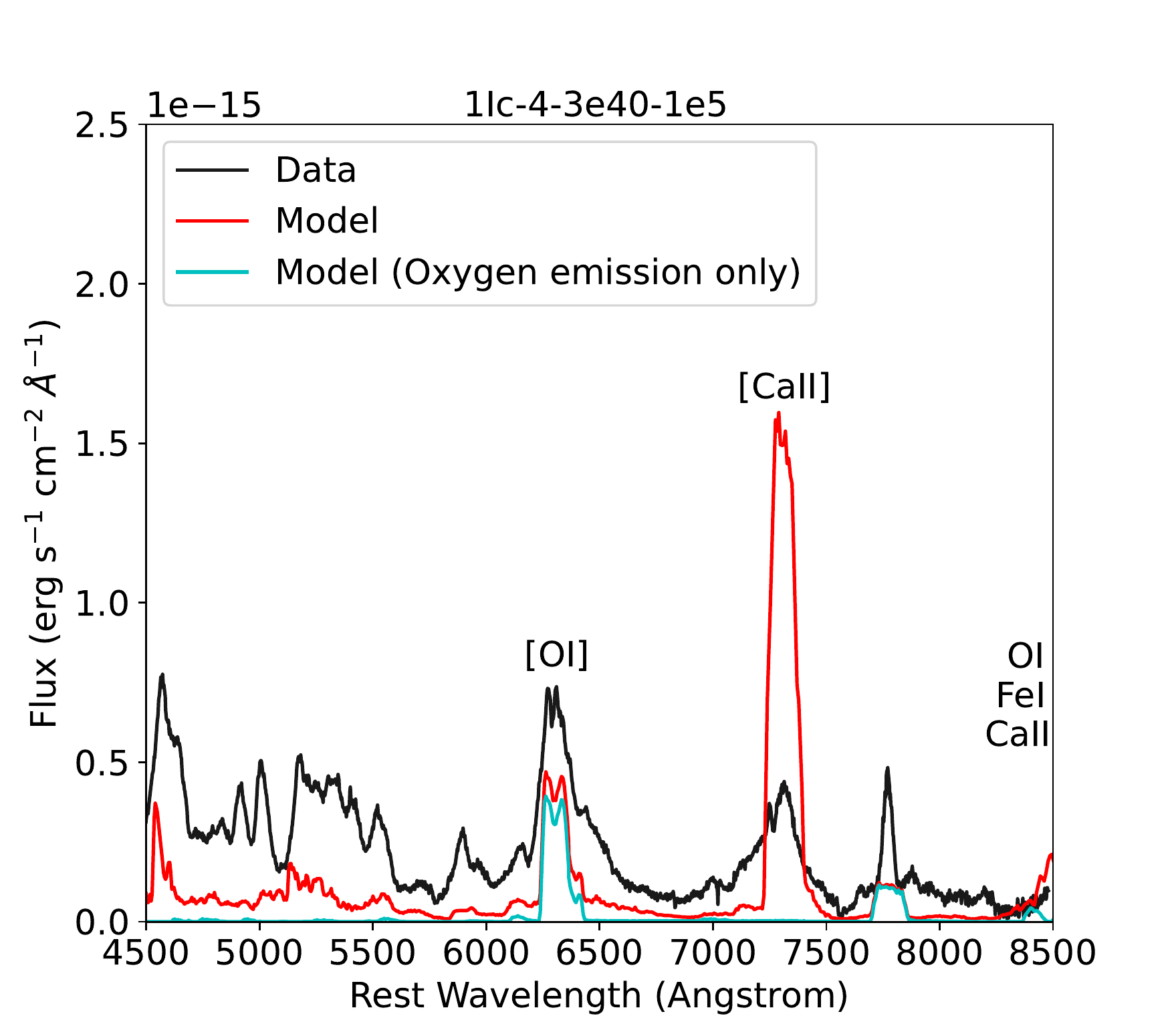}&
\includegraphics[width=1.1\linewidth]{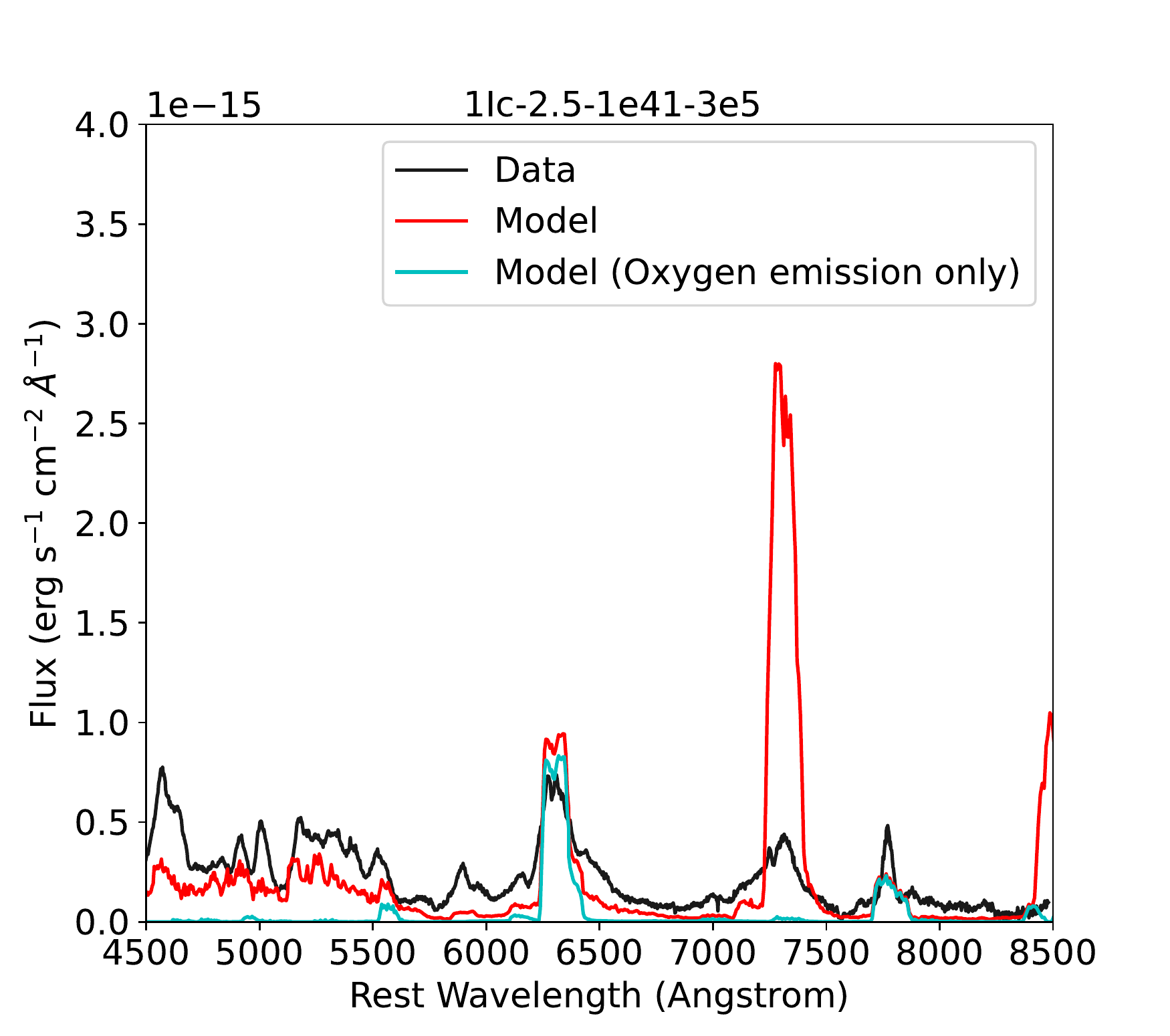}&
\includegraphics[width=1.1\linewidth]{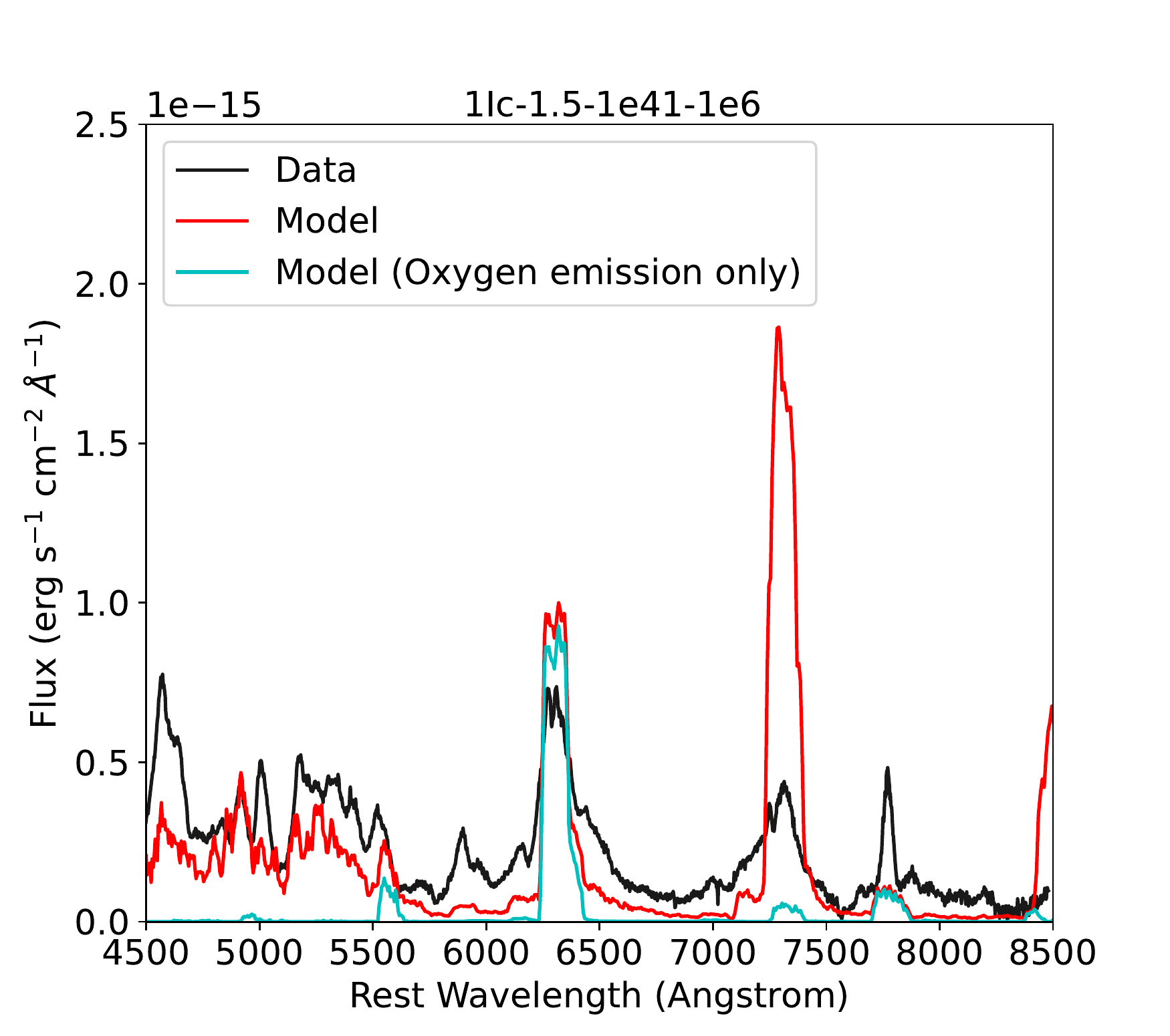}
\end{tabular}}
\caption{The two best-fitting dust-corrected model spectra to SN 2012au for each value of $T_{\rm PWN}$ at 1 years for the realistic composition compared to the observed spectrum from \cite{Milisavljevic2013}.  The total model emission is shown in red, while the emission from only oxygen is shown in cyan. Strong lines and features are labelled in the upper left plot.}%
\label{fig:rc1y_spec}
\end{figure*}

The model scores based on the O I and [O I] lines are shown in Figure \ref{fig:rc1y_score}, and two best-fitting spectra for each value of $T_{\rm PWN}$ are shown in Figure \ref{fig:rc1y_spec}. The emission at 7300 $\AA$ can either be from [O II] or [Ca II] for different parameters, with the models showing [O II] emission having higher values of $T_{\rm ej}$.  The emission at 6300 $\AA$ can also come from different sources, with different models having [O I], S III, or broad Fe II features dominating the emission; this differs from \cite{Dessart2019}, where [O I] lines were only present with significant clumping.  Most models predict the observed broad Fe I and Fe II features below 5500 $\AA$, although the level of agreement between the model and observation is difficult to quantify over a broad continuum, but some models also predict unobserved features at higher wavelengths, which is likely due to a mix of the Fe I $\lambda$ 8350 $\AA$ line, O I $\lambda$ 8446 $\AA$ line, and broad Ca II NIR triplet around 8500 $\AA$.  1Ic-1.5-3e41-1e5, 1Ic-1.5-3e41-3e5, and 1Ic-1.5-3e41-1e6 all predict strong emission below 5500 $\AA$, either from [O III], Fe II/Fe III, or both.  None of the models can reproduce all the features of the spectrum without predicting strong unobserved lines except for 1Ic-4-3e40-1e5, which underestimates the low wavelength emission and overproduces [Ca II], so we do not consider any model to be a best-fit model for the realistic composition at 1 year.  Despite this, it's worth noting that the model scores over the entire parameter space (Figure \ref{fig:rc1y_score}) are lower than in the pure oxygen case (Figure \ref{fig:o1y_score}), meaning that the realistic case does a better job at reproducing the two lines the score is evaluated for, even if neither composition can reproduce the observed spectrum very well.  Improved treatment of treatment of x-rays and inner-shell processes (see Section \ref{sec:modsetup}) and more complex, informed mixing and clumping (see Section \ref{sec:mixclmp}) will be important to accurately determine the temperature and ionization balance of the ejecta at this epoch, particularly for heavier elements.

\subsection{Multi-Zone Modelling} \label{sec:multiz}

To explore how emission changes in multi-zone vs one-zone models, 
 we constructed a multi-zone model using the parameters from one of the highest continuum optical depth 
pure oxygen models, 1O-4-1e42-1e5, 
 which showed strong [O II] and [O III] emission, as well as O I $\lambda$ 7774 and an [O III] $\lambda$ 4363.  
The ejecta was divided into 1, 5, 10, and 20 zones of equal mass, using a density distribution $\rho \propto v^{-6}$ \citep{Suzuki2017, Suzuki2019}; one other model was constructed with non-uniform zones to try and resolve how deep the PWN radiation was penetrating into the ejecta.  The inner ejecta velocity is 2000 km s$^{-1}$, as before, while the outer velocity is increased so the outer zones can still exist given the density profile.

\begin{figure}
    \centering
    \includegraphics[width=\linewidth]{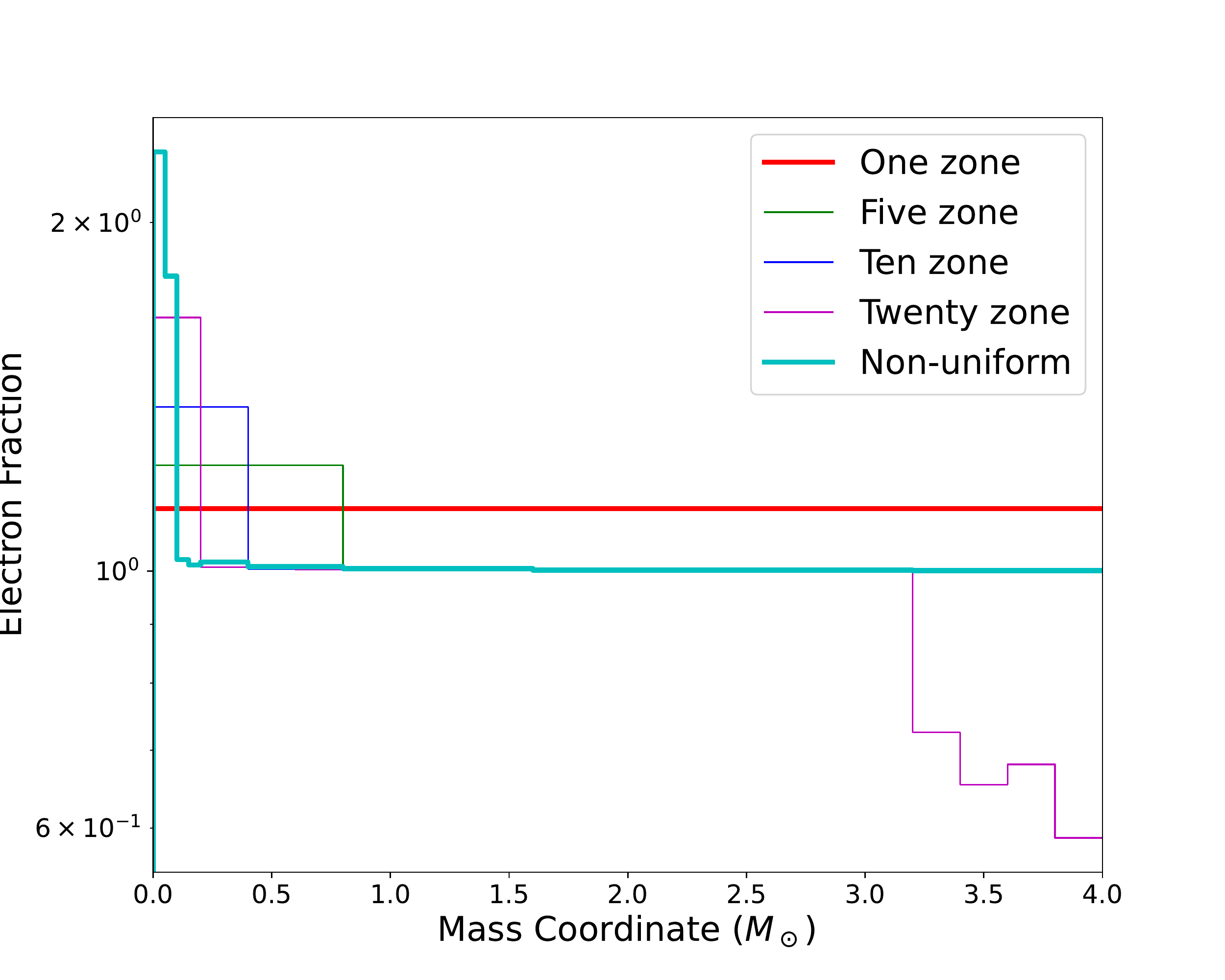}
    \\
    \includegraphics[width=\linewidth]{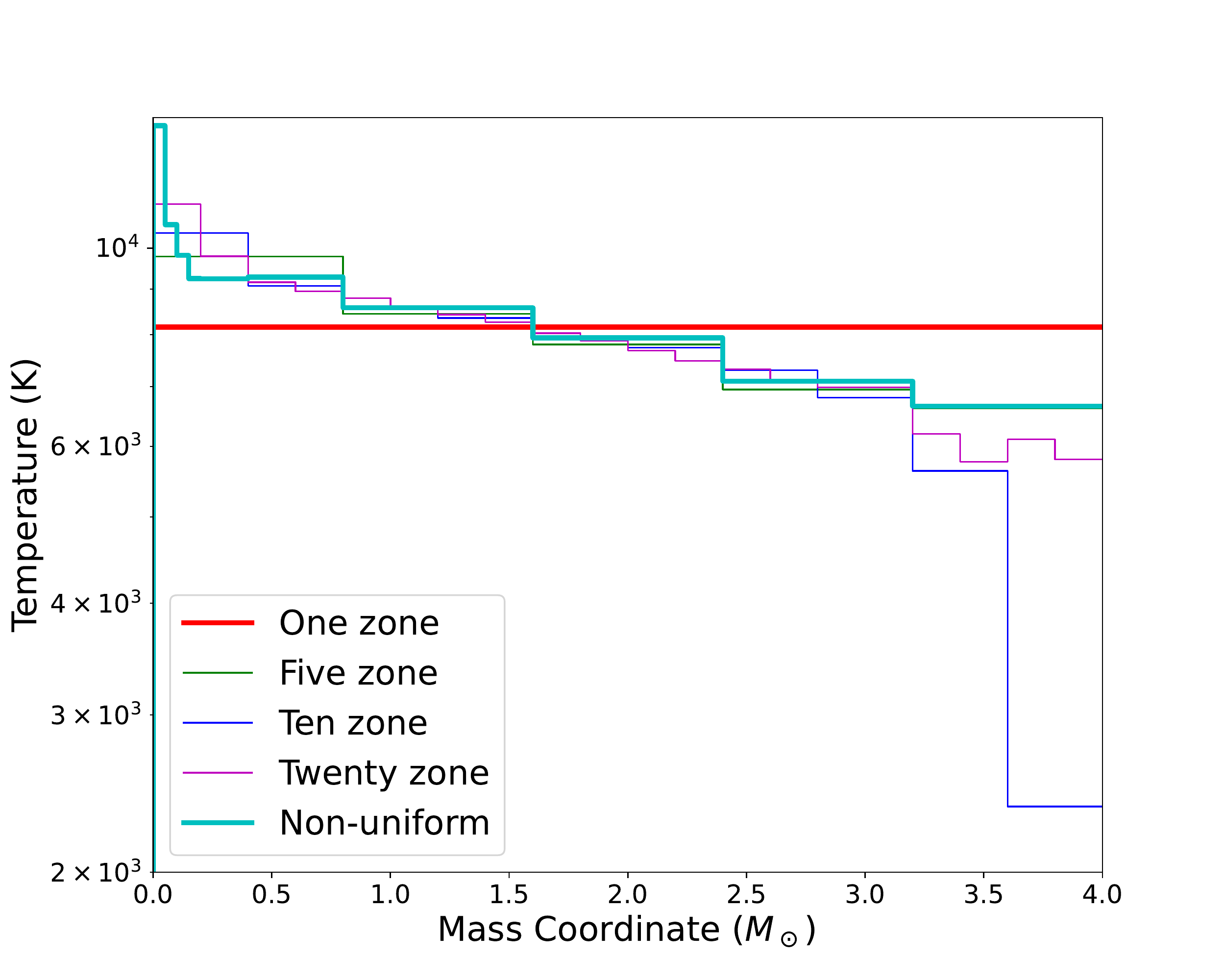}
    \caption{Electron fraction (top) and ejecta temperature (bottom) for the multi-zone 1O-4-1e42-1e5 models.  Only the non-uniform model can resolve the highly-ionized inner ejecta region that is close to the PWN.}
    \label{fig:multixet}
\end{figure}

The electron fraction and temperature for these models are shown in Figure \ref{fig:multixet}.  Only the non-uniform model can fully resolve the highly-ionized inner ejecta region that is close to the PWN; this shows that in this model, only the inner $\sim$ 0.1 $M_\odot$, with material travelling between 2000 and 2010 km s$^{-1}$, is highly ionized.  The innermost zone is heavily ionized, being mostly doubly ionized, but with a significant amount of triply ionized material as well, while the zones outside that are primarily dominated by O II.  The temperature of this innermost region is also higher than the surrounding region by about 30$\%$. The temperature distribution throughout the ejecta can be resolved fairly well by even a 5-zone model, ranging from $\sim$ 10 000 K in the inner regions to $\sim$ 7000 K in the outer regions.

This test shows that in a 1-D stratification, there are clear gradients both in ionization and temperature, as obtained also by \citet{Chevalier1992}. Because the density is higher at the inner edge, it is not obvious that material there should become more ionized and hotter (higher density typically favours more neutral gas and lower temperatures, all else the same). The model indicates that proximity to the ionization source wins out over density effects in setting conditions. Models in which this holds then predict more narrow line profiles for lines that become strong at high ionization and/or high temperature. That would mean [O II] and [O III] being more narrow than [O I].

\begin{figure}
    \centering
    \includegraphics[width=\linewidth]{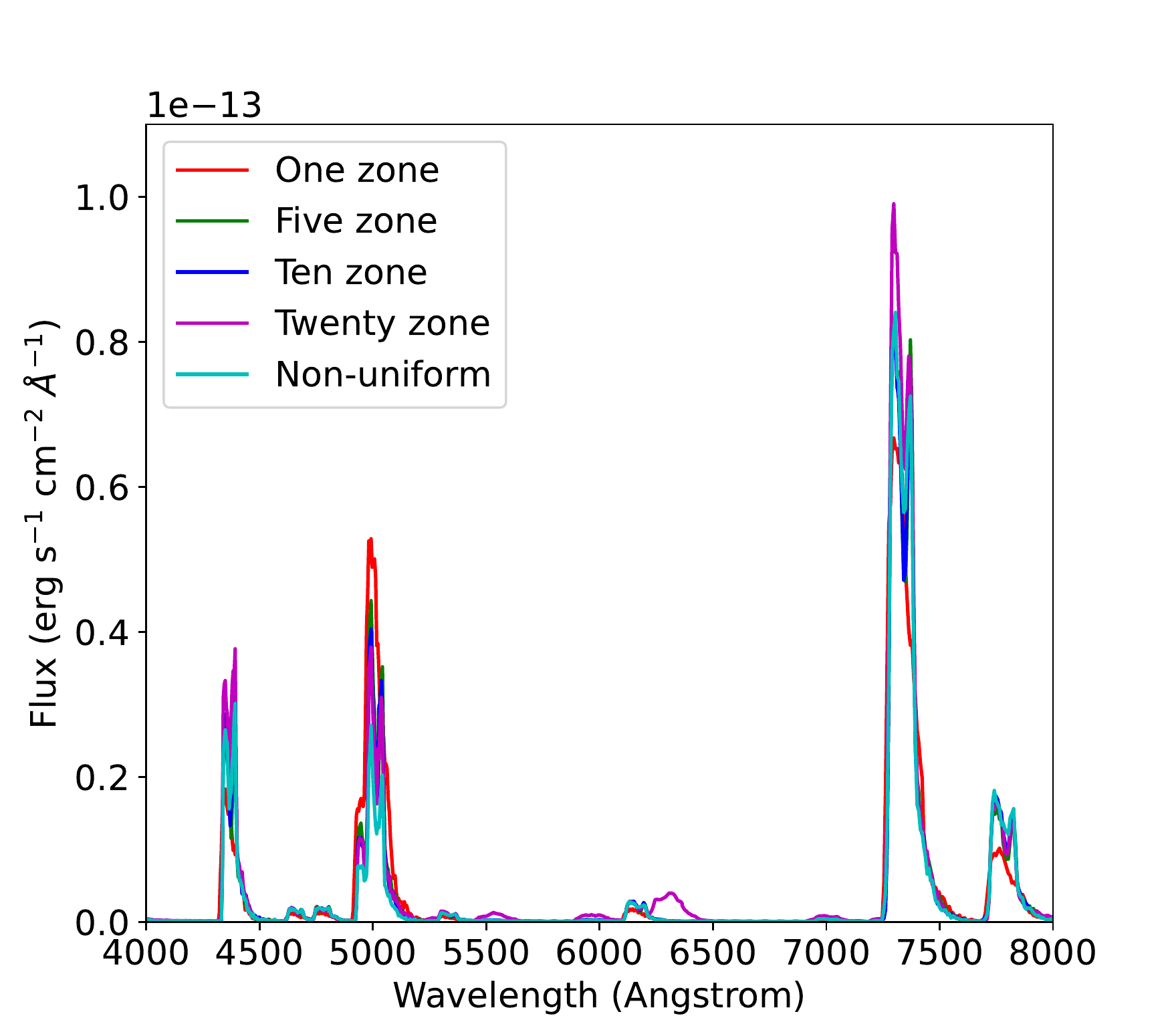}
    \caption{Spectra for the multi-zone 1O-4-1e42-1e5 models.  The resolution of the ejecta can change line luminosities by up to a factor $\sim$ 2 if the ionization fraction of the emitting ion is strongly influenced by the resolution of the ejecta.}
    \label{fig:multispec}
\end{figure}

The spectra of these multi-zone models are shown in Figure \ref{fig:multispec}.  The line luminosities for [O I] and [O II] are not heavily affected by the number of zones, while the luminosity of the [O III] lines decreases by a factor $\sim$ 2 as the resolution of the inner zone increases.  Considering that this model has one of the highest optical depths of any model studied, and thus should have some of the highest discrepancies between one-zone and multi-zone models, this indicates that the accuracy of one-zone modelling is a factor of a few. As discussed in Section \ref{sec:modsetup}, it is also not fully clear whether stratified 1D models are in fact more accurate representations of real PWNe than 1-zone models.

\section{Discussion} \label{sec:dis}

\subsection{Implications for Initial Conditions, the Pulsar Wind Nebula Spectrum, and the Radio Counterpart}  \label{sec:imp}

One interesting possibility for nebular spectral modelling of magnetar-driven supernovae is to combine with and cross-check other models to verify inferred parameters or predict multi-wavelength signals.  The most common way to infer magnetar and ejecta parameters is with Bayesian inference on photometric data using light-curve models \citep[e.g][]{Nicholl2017, Villar2018, Chen2023}. This can be done quickly on large samples and is sensitive to things like the magnetar spin period, which nebular spectra are not (Equation \ref{eqn:lpwntnum}), as well as magnetic field, and ejecta mass, among others; however, it is not sensitive to the PWN SED, which we have shown is important to understanding nebular spectra.

\begin{figure}
    \centering
    \includegraphics[width=\linewidth]{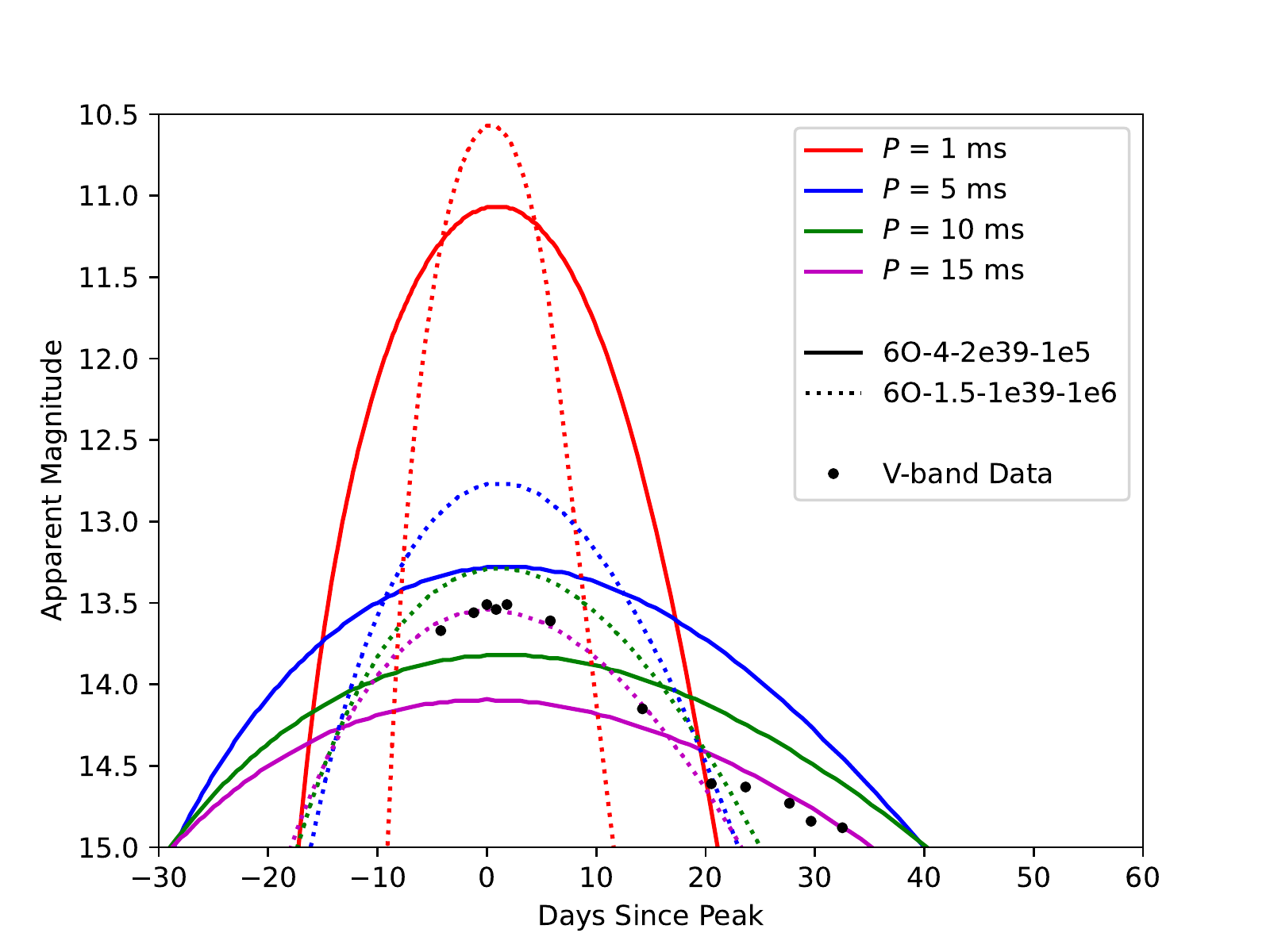}
    \caption{Light curve models for different spin periods with ejecta masses and magnetic fields derived from the 6O-4-2e39-1e5 and 6O-1.5-1e39-1e6 models, compared with V-band photometric data of SN 2012au from \cite{Milisavljevic2013}.  The most consistent model has a spin period $P_0$ of 15 ms and has parameters derived from the 6O-1.5-1e39-1e6 model.}
    \label{fig:lccomp}
\end{figure}

We use the two best-fit models from the pure oxygen composition at 6 years, 6O-4-2e39-1e5 and 6O-1.5-1e39-1e6, to show how these results could be combined with other methods.  Figure \ref{fig:lccomp} shows light curves generated using the model from \cite{Kashiyama2016} with spin period $P_0$ = 1, 5, 10, and 15 ms and $B$ and $M_{\rm ej}$ given by the models and Equation \ref{eqn:lpwntnum} compared to V-band photometric data from SN 2012au \citep{Milisavljevic2013}. The initial dipolar magnetic fields for these two models are $3 \times 10^{14}$ G for the 6O-4-2e39-1e5 model and $4 \times 10^{14}$ G for the 6O-1.5-1e39-1e6 model.  Around the light curve peak, the best fit light curve has $P_0 = 15$ ms, $B = 4 \times 10^{14}$ G, and $M_{\rm ej} = 1.5 M_\odot$.  This model has a spin-down time of 4.3 days, less than the rise time of the model, and total rotational energy of $1.1 \times 10^{50}$ erg, less than a typical supernova explosion energy of $10^{51}$ erg. 
These parameters are similar to those found by \cite{Pandey2021}, although with a lower ejecta mass.  

The injection SED temperature of $10^6$ K from the 6O-1.5-1e39-1e6 model can be used to infer the electron injection Lorentz factor $\gamma_b$ by assuming that the peak of the model blackbody distribution corresponds to $\nu_b$ in Equation \ref{eqn:nub}.  Assuming $\epsilon_B = 3 \times 10^{-3}$ as done for previous Galactic PWN models \citep{Tanaka2010, Tanaka2013}, $\gamma_b = 3 \times 10^5$ for this model, similar to the $10^5 - 10^6$ inferred for Galactic PWNe.  A value of $\epsilon_B \sim 10^{-6}$, similar to that inferred for SN 2015bn by \cite{Vurm2021}, would only increase $\gamma_b$ by a factor $\sim$ 10, leaving it only slightly higher than that of Galactic PWNe. 

\begin{figure}
    \centering
    \includegraphics[width=\linewidth]{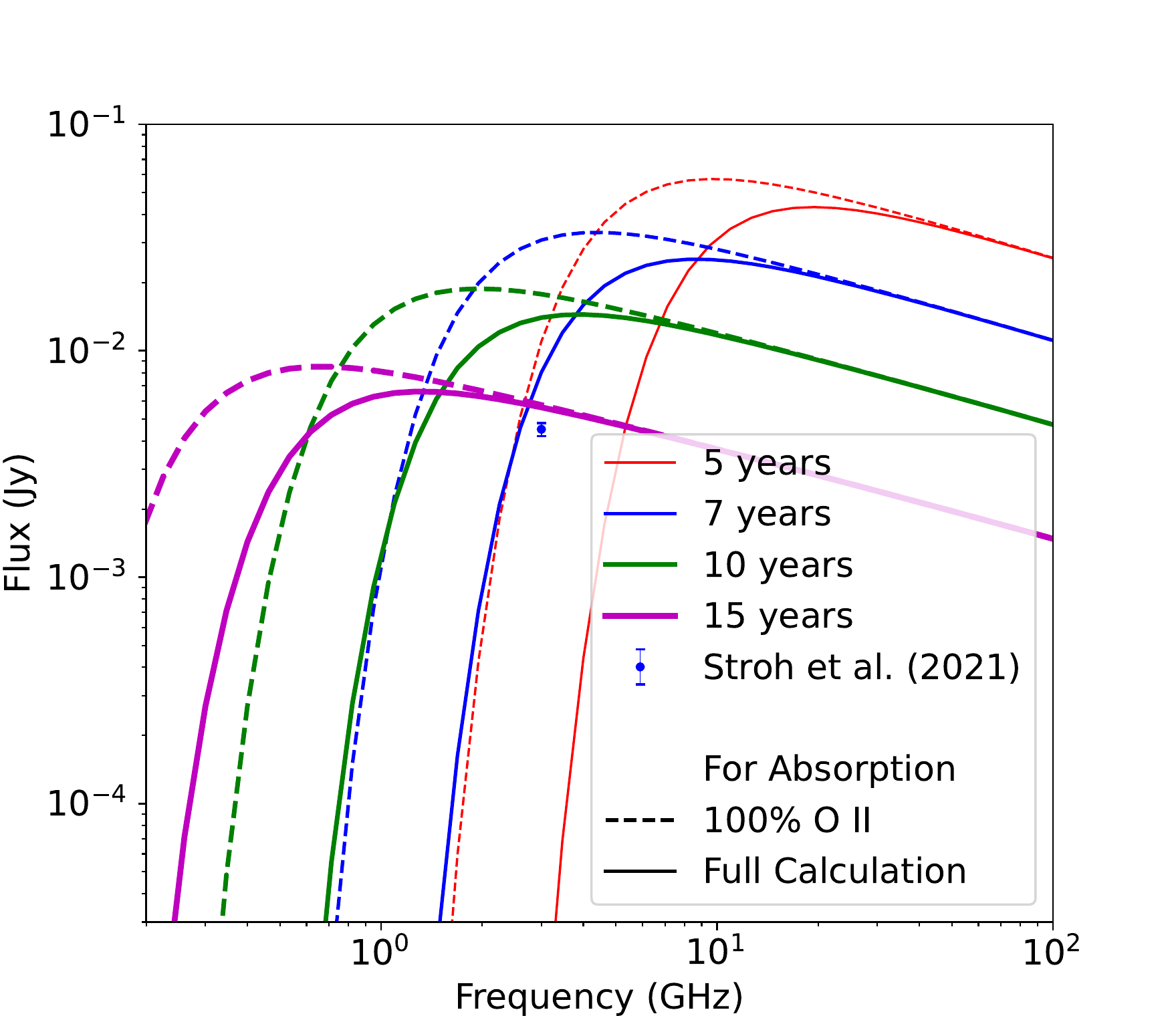}
    \caption{Predicted radio spectra for SN 2012au at 5, 7, 10, and 15 years post-explosion using the model determined by nebular spectral and light curve modelling, as well as the observed data at 7 years from \cite{Stroh2021}.  The calculation of free-free absorption is done both by assuming a completely singly ionized ejecta and electron temperature of 10$^4$, as done in previous works, as well as using the output of our nebular spectral calculations.
    The emission is extremely bright and long-lasting, with a free-free absorption break that decreases in frequency with time, roughly consistent with observations, and should be easily detectable by instruments like VLA for decades.}
    \label{fig:radio}
\end{figure}

By assuming spectral indices of $q_1 = 1.5$ and $q_2 = 2.5$ for a nebula with total power \citep{Murase2015, Murase2021}

\begin{equation}
\nu F_\nu \propto \epsilon_e L_\text{\rm PWN}
\begin{cases}
(\nu/\nu_b)^{(2-q_1)/2} & (\nu \leq \nu_b), \\
(\nu/\nu_b)^{(2-q_2)/2} & (\nu_b \leq \nu),
\end{cases}
\label{eqn:synspecfc}
\end{equation}
as done in \cite{Omand2018, Law2019, Eftekhari2021}, these parameters can be used to estimate radio emission from the PWN and compare with the observation from \cite{Stroh2021} at 7 years post-explosion.  Using the model presented in \cite{Murase2015, Murase2016}, which calculates PWN emission by self-consistently calculating synchrotron emission and self absorption, pair cascades, Compton and inverse Compton
scattering, adiabatic cooling, and both internal and external attenuation by solving the Boltzmann equation for electron-positron pairs and photons in the PWN over all electron energies and photon frequencies, we calculate the expected radio spectra for SN 2012au at 5, 7, 10, and 15 years post-explosion, which is shown alongside the observed data in Figure \ref{fig:radio}. 
This model predicts bright, long-lasting, emission with a free-free absorption break that decreases in frequency with time; this emission is roughly consistent with observations and should be easily detectable by instruments like VLA for several decades.

The low energy break in the spectra is caused by free-free absorption, which has an optical depth of

\begin{equation}
   \tau_{\rm ff} = 8.4 \times 10^{-28} \left(\frac{T_{\rm e}}{10^4 \text{ K}}\right)^{-1.35} \left(\frac{\nu}{10 \text{ GHz}}\right)^{-2.1} \int dr n_e n_i \bar{Z}^2 ,
   \label{eqn:tauff}
\end{equation}

where $T_{\rm e}$ is the electron temperature, $n_e$ and $n_i$ are the electron and ion number densities, respectively, and $\bar{Z}$ is the effective charge of the ejecta.  This is normally estimated by assuming the ejecta is 100$\%$ O II with $T_{\rm e} = 10^4$ \citep[e.g.][]{Omand2018, Law2019}, but the nebular spectral model at 6 years shows the ejecta 
has an electron fraction of $\sim$ 0.8 and an electron temperature of $\sim$ 6000 K, which we assume are constant in time for these models and use in our calculation.  These values are likely still uncertain due to model systematics, but using them to calculate absorption causes the break frequency to increase by around a factor of 2 with respect to the simple ejecta assumptions, with a small decrease in peak luminosity.  This effect should become more pronounced at very late times, so the interpretation of radio SLSNe, such as PTF10hgi \citep{Eftekhari2019, Mondal2020, Eftekhari2021, Hatsukade2021}, need to account for these deviations, and future models will have to constrain these values more strongly to be able to make predictions and aid future observational studies.  Figure \ref{fig:radio} also shows that these more detailed calculations are necessary for consistency with the previously observed data for SN 2012au \citep{Stroh2021}. 

\subsection{Mixing and Clumping} \label{sec:mixclmp}

Determining the extent of mixing in SN 2012au is very important to properly model the nebular spectra. Looking back at the results with the realistic composition, many models predicted strong optical calcium and sulphur lines, which were either barely detectable or undetectable.  The fully mixed model used here is the most extreme case of mixing, and likely unphysical \citep[e.g.,][]{Jerkstrand2017hb}.  Mixing in magnetar-driven supernovae is mostly caused by Rayleigh-Taylor instabilities from the pulsar wind pushing on the ejecta, which will disrupt the initial shell structure and cause some mixing between regions.  In simulations involving SLSN-like parameters, this usually create two major zones: an outer region with mostly carbon and oxygen, and an inner region consisting of heavier elements \citep{Chen2020, Suzuki2021}.  With more Ic-BL-like parameters (faster energy injection), these two zones will mix radially
, but still be in different angular zones, meaning that creating an angle averaged profile will miss the zonal separation (macroscopic vs microscopic mixing).  The extent of mixing within each zone will depend on the characteristic size of the Rayleigh-Taylor instabilities, and will likely cause small regions within each zone that will retain the shell composition.  However, for a less energetic magnetar, such of the $P = 15$ ms one inferred through light curve modelling in Section \ref{sec:imp}, Rayleigh-Taylor instabilities may not develop on a large scale, and the shell structure of the progenitor may be mostly retained \citep[See Figure 5 from ][for example]{Suzuki2021}.  By assuming Ic-BL-like mixing for this study instead of a shell structure consistent with a less energetic magnetar, we may have severely overestimated the extent of mixing in the ejecta: the generally unsatisfactory fits of our fully mixed models to SN 2012au strengthens the argument that complete microscopic mixing does not occur.

Understanding clumping of the ejecta (variation in densities over small or intermediate length scales in the ejecta), particularly in the inner regions most affected by hydrodynamic instabilities, is also necessary to properly model the ejecta.  Numerical simulations of supernovae with strong central engines show that the unburnt C/O ashes reside in mostly high density clumps, while the heavier elements reside in lower density gas \citep{Suzuki2021}.  \cite{Pandey2021} also find a large imaging polarization value, indicative of ejecta asymmetry.  \cite{Jerkstrand2017} found that a strongly clumped O/Mg zone with a filling factor of $\lesssim$ 0.01 was needed to reproduce the nebular spectrum of SN 2015bn; and \cite{Dessart2019} found that clumping is essential to trigger ejecta recombination and produce O I, Ca II, and Fe II lines, and that reproducing most SLSNe-I nebular spectra requires clumping.  The multi-zone models from Section \ref{sec:multiz} show that resolving the high-density inner region is important for understanding the ionization structure of the ejecta, and can affect line luminosities by a factor of a few; modelling this region with a low filling factor will allow photons to reach further without being absorbed, but they will be less able to penetrate the clumps, leading to possible small-scale variations in the ionization structure of these clumps.  This might have a large affect on the modelled spectrum, and should be investigated in future studies.

Deciding how to model these effects will be important for more detailed models.  The extent of these mixing and clumping effects depends largely on both the magnetar rotational energy and spin-down time, which both depend on the initial magnetar spin period, which the nebular spectrum can not directly probe.  Parameter estimation using the early light curve can help with this, but models of magnetar driven supernovae are generally not reliable for highly kinetic supernovae due to the physical coupling between pulsar luminosity and ejecta kinetic energy, something which is not included in some models \citep[e.g][]{Nicholl2017}; 
adapting models for magnetar-driven kilonovae \citep{Yu2013, Metzger2019, Sarin2022} could be a viable strategy, since the physics of these systems is largely similar.  Inferred parameters from these models would allow both a qualitative characterization of multidimensionality in the supernova as well as restricting the possible parameter space needed for nebular spectral synthesis calculations.

\section{Summary} \label{sec:conc}

We presented a suite of late-time (1-6y) spectral simulations of SN ejecta powered by an inner PWN. To achieve this, we implemented improved treatment of photoionization of key elements in the \texttt{SUMO} spectral synthesis code, and extended the code to allow arbitrary radiative energy injection at an inner boundary.

Over a large grid of 1-zone models, we study the behaviour of the SN physical state and line emission as PWN luminosity ($L_{\rm PWN}$), injection SED temperature ($T_{\rm PWN}$), ejecta mass ($M_{\rm ej}$), and composition (pure O or realistic) vary. We discuss the resulting emission in the context of the observed behaviour of SN 2012au, a strong candidate for a PWN-powered SN.  

We find that:

\begin{itemize}

    \item Oxygen ionization and ejecta temperature generally increase as the ejecta mass decreases and engine luminosity increases.  The temperature range is around $1,600-6,300$ K for pure oxygen models at 6 years, $400-6,300$ K for mixed composition models at 6 years, and $\sim$ $4,000-15,000$ K for models with both compositions at 1 year.  The dominant ion can be O I, O II or O III depending on the combination of engine luminosity and ejecta mass.
    
    \item Low ejecta mass models, at high PWN power, obtain runaway ionization \citep{Jerkstrand2017} for O I and, in extreme cases, also O II, causing a sharp decrease in their ion fraction over a small change in the parameter space, and quenching of [O I] $\lambda \lambda$ 6300, 6364 (and [O II] $\lambda \lambda$ 7320, 7325).

     \item As the characteristic photon energy of the PWN SED increases, the ejecta becomes less ionized due to the lower number of ionizing photons. However, the ejecta temperature increases as the characteristic photon energy of the PWN SED increases.
    
    
    \item Several pure-oxygen models are able to reproduce the observed features of SN 2012au at 6 years (which shows strong [O I], [O II] and [O III] emission) to within a factor of $\sim$ 2. These models have $L_{\rm PWN}$ $\sim \left(1-5\right) \times 10^{39}$ erg s$^{-1}$ and $M_{\rm ej}$ $\sim 1.5-6$ $M_{\odot}$.  As $T_{\rm PWN}$ increases, the best-fit models have lower $M_{\rm ej}$ and $L_{\rm PWN}$. 
    
    \item At 1 year, pure oxygen models with high $T_{\rm PWN}$ have overluminous [O II], [O III], and [O I] $\lambda$ 5577 emission compared to observations, but models with lower $T_{\rm PWN}$ can reproduce the oxygen features reasonably well. However, these models have a lower ejecta mass than at 6 years, a lower engine luminosity than one would expect for vacuum dipole spin down, and a lower injection SED temperature than one would expect for the typical evolution of a PWN SED (See Equation \ref{eqn:nub}). These best-fit parameters are likely unphysical.
    
    \item Realistic Ibc mixed composition models at 6 years require a stronger engine luminosity compared to the pure oxygen models to reproduce the same oxygen line luminosities.  These models cool strongly through IR radiation, particularly the [Ne II] 12.8 $\mu$m line, and  have lower temperatures.  Some models also exhibit strong [Ca II] and [S III] features in the optical band.  We find no satisfactory simultaneous fit for O, Ca and S lines to SN 2012au, which indicates that a fully mixed ejecta is not suitable.
    
    \item Realistic Ibc mixed composition models at 1 year show strong [O II]/[Ca II] and [O III] emission over most of the parameter space, with only low engine luminosity, low injection SED temperature models able to somewhat reproduce the spectrum, similar to the pure oxygen models. The realistic models tended to reproduce the observed oxygen/calcium lines more accurately over a larger parameter space compared to the pure oxygen models, but improved treatment of inner shell processes and mixing/clumping will be needed to reproduce all the spectral features with a high degree of accuracy.
    
    \item Multi-zone models show that in 1D stratification gradients develop both for ionization and temperature, with the innermost ejecta being the most highly ionized and hottest. 
     Oxygen line luminosities can vary by a factor of a few depending on the stratification, which provides useful information for to what level of agreement 1-zone models can be assessed against observations.
    
    \item We demonstrate how results and constraints from nebular modelling can be connected into modelling of the diffusion phase and the radio emission.  Light curves consistent with the best-fitting pure-oxygen nebular spectral models at 6 years give an initial magnetar spin period of $\sim$ 15 ms, and radio spectra predicted from these parameters have fluxes of $\sim$ 10 mJy at 10 years post-explosion with a slow decline rate, meaning the remnant should be detectable for decades.  These predictions are consistent with previous observations.
    
    
\end{itemize}

\begin{acknowledgements}
The author would like to thank the anonymous referee for their helpful comments.  The authors would also like to thank Akihiro Suzuki, Dan Milisavljevic, Mike Barlow, Maria Niculescu-Duvaz, Claes Fransson, Eliot Ayache, Bart van Baal, and Quentin Pognan for their helpful discussions. 
This project has received funding from the European Research Council (ERC) under the
European Union’s Horizon 2020
Research and Innovation Programme (ERC Starting Grant No. [803189], PI: A. Jerkstrand) The computations were enabled by resources provided by the Swedish National Infrastructure for Computing (SNIC) at the PDC Center for High Performance Computing, KTH Royal Institute of Technology, partially funded by the Swedish Research Council through grant agreement no. 2018-05973.
\end{acknowledgements}

\bibliographystyle{aa} 
\bibliography{aanda} 

%
%
%

\begin{appendix}

\section{Code Updates} \label{app:code}

\subsection{Photon Injection from the Inner Boundary}

A module external to \texttt{SUMO} was developed which can generate a pulsar wind nebula spectrum to be used as an input to \texttt{SUMO}.  This model can currently support blackbody spectra (with free parameters $L_{\rm tot}$ and $T_{\rm PWN}$), power law spectra (with free parameters $L_{\rm tot}$ and $q$, the spectral index), and broken power law spectra (with free parameters $L_{\rm tot}$, $\lambda_{\rm break}$, $q_1$, and $q_2$).  The user also defines the wavelength resolution (which does not have to match the wavelength resolution of \texttt{SUMO}) and the minimum and maximum wavelengths. 
The user is also free to generate their own spectrum from external programs, since the spectrum data files feature only one line with the number of bins and total luminosity, followed by two columns specifying the wavelength $\lambda$ in angstrom and the spectral luminosities $L_\lambda$ in erg s$^{-1}$ cm$^{-1}$.

To run a \texttt{SUMO} simulation using a central engine, the user needs to turn on the central engine flag in the runfile, as well as specify the filename for the engine spectrum.  \texttt{SUMO} will then interpolate over that spectrum to create photon packets that are injected at the inner boundary of the innermost zone.  These packets then propagate through the ejecta like using the normal Monte Carlo routine.

\subsection{Atomic Data Updates}

The recombination rate and photoionization cross sections for some ions in the atomic data set used by \texttt{SUMO} had to be updated for this study.  16 ions had their recombination rates either added to the dataset, as they had never been relevant in previous simulations, or updated due to the higher temperatures sometimes encountered in the solutions for engine-driven supernovae.  The list of ions with added/updated recombination rates, and the source where the rates were taken from, is given in Table \ref{tbl:recrates}.  

\begin{figure*}
\centering\begin{tabular}{cccc}
\includegraphics[width=.23\linewidth]{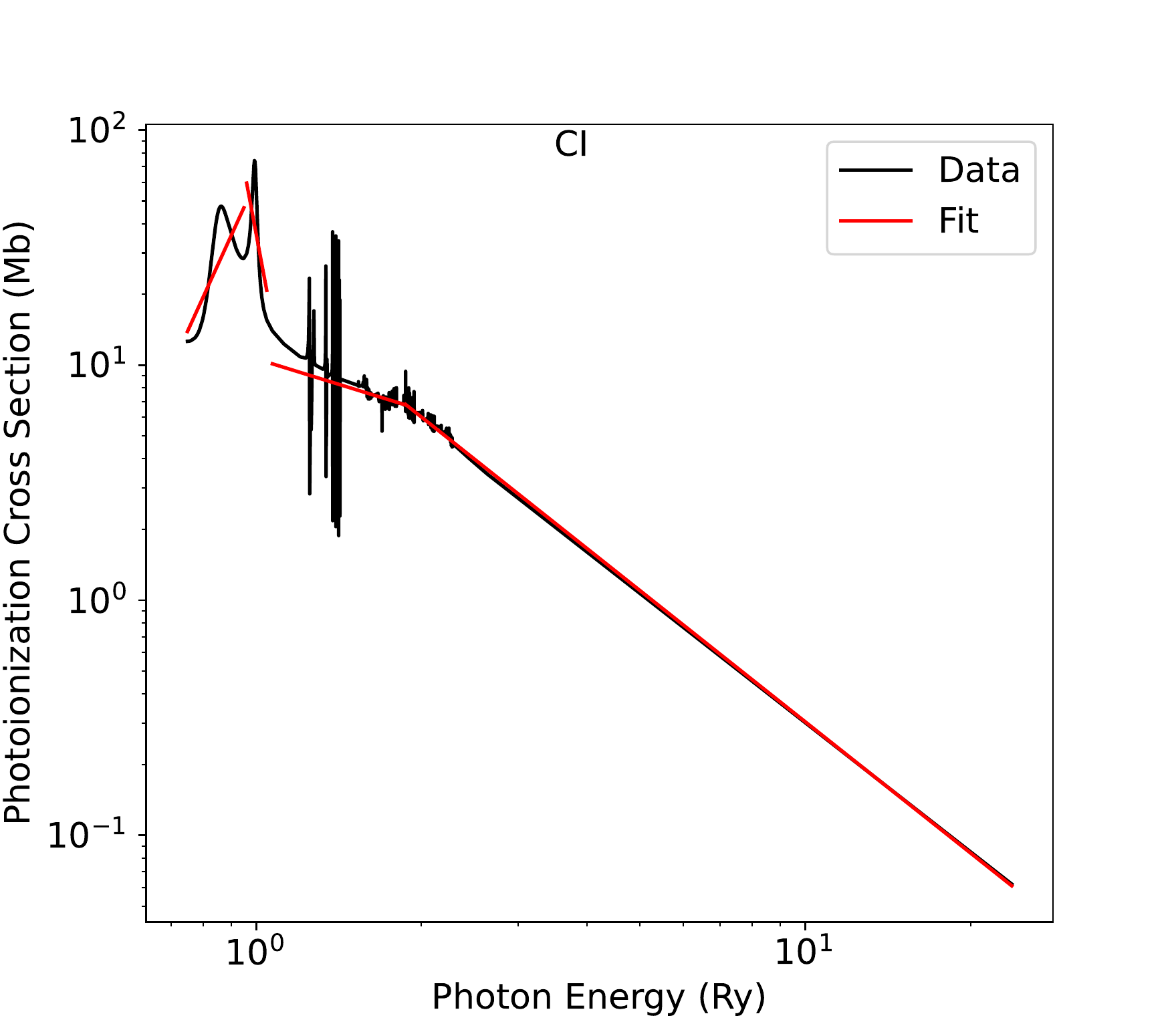}&
\includegraphics[width=.23\linewidth]{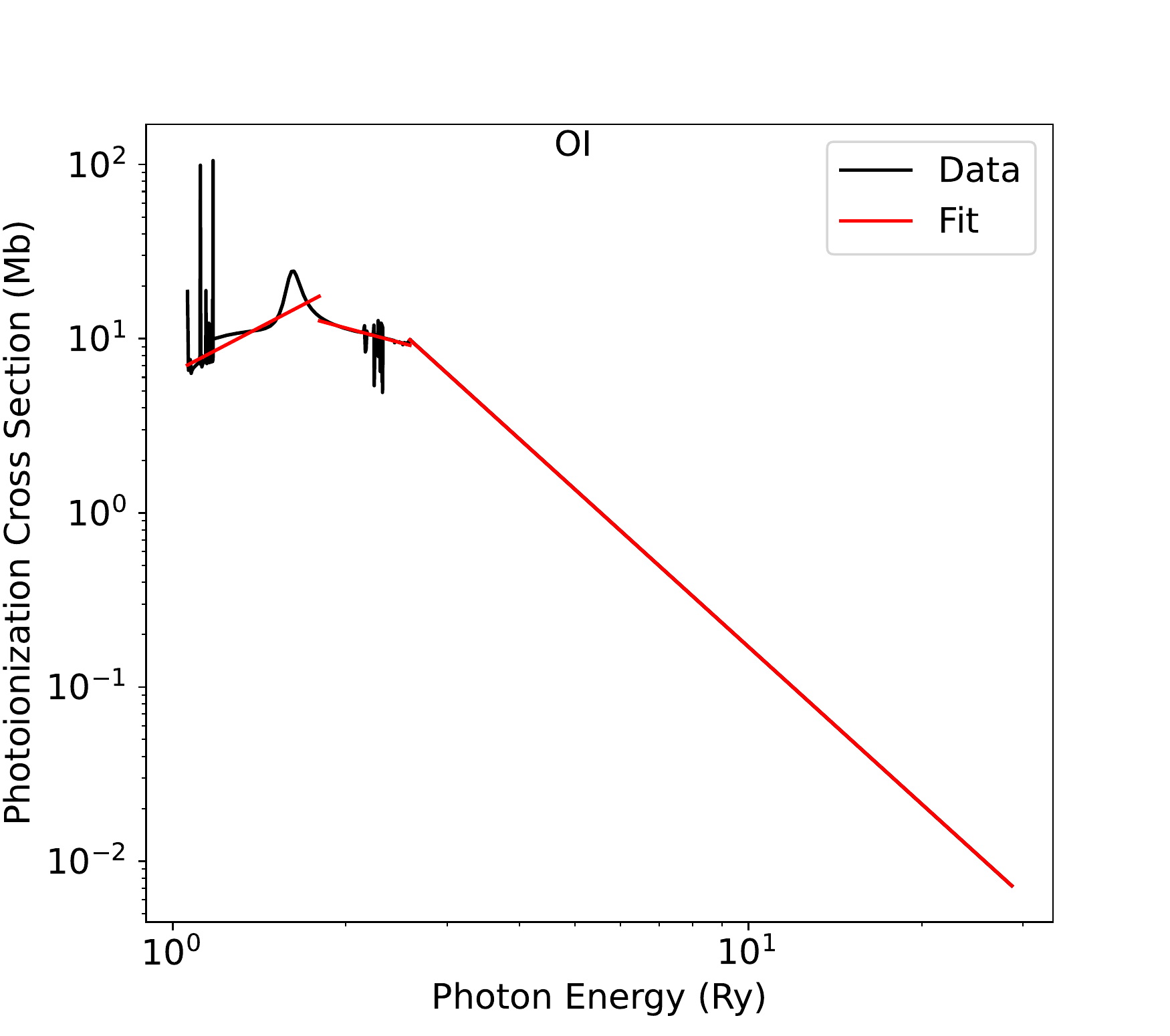}&
\includegraphics[width=.23\linewidth]{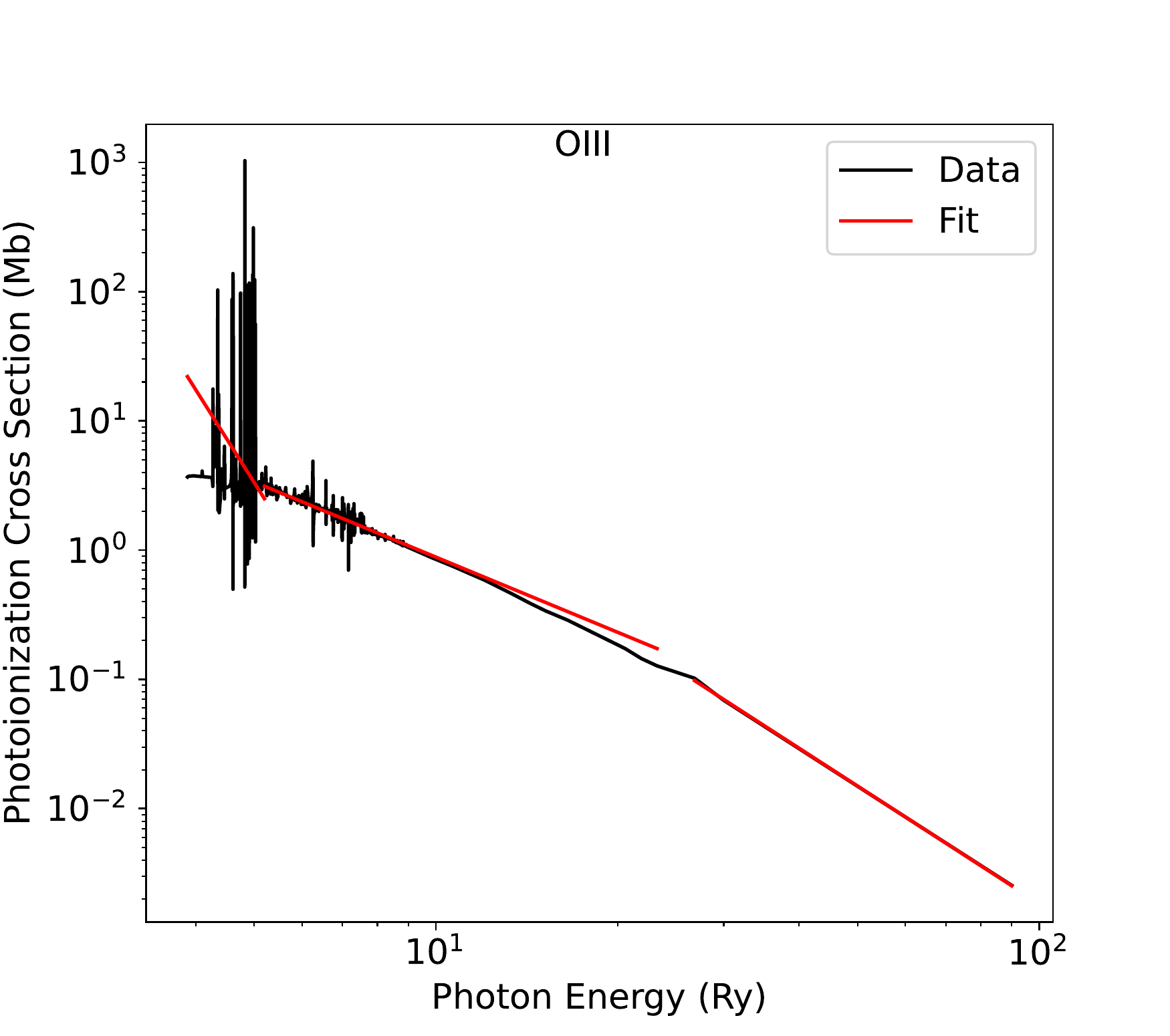}& 
\includegraphics[width=.23\linewidth]{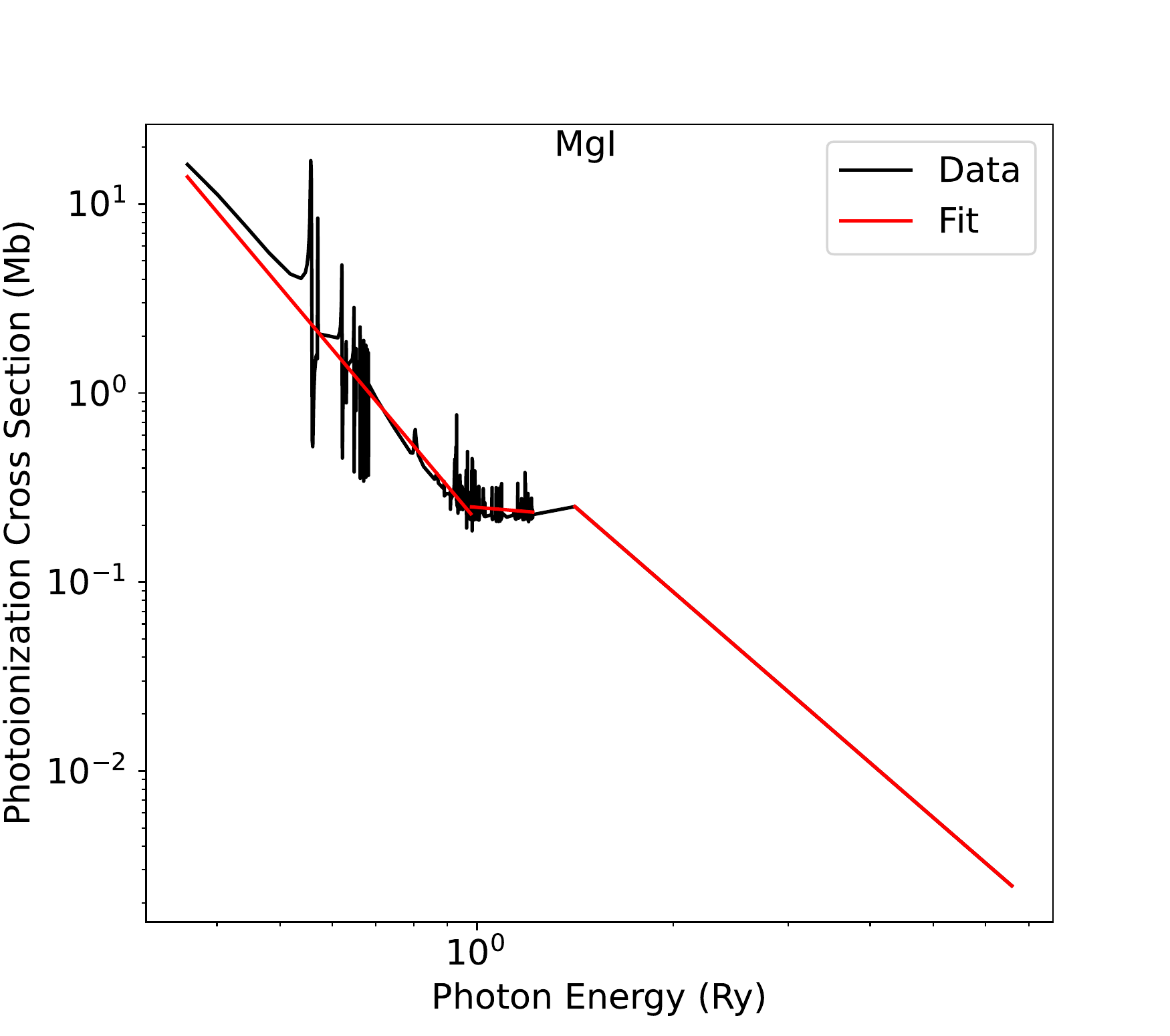} \\
\includegraphics[width=.23\linewidth]{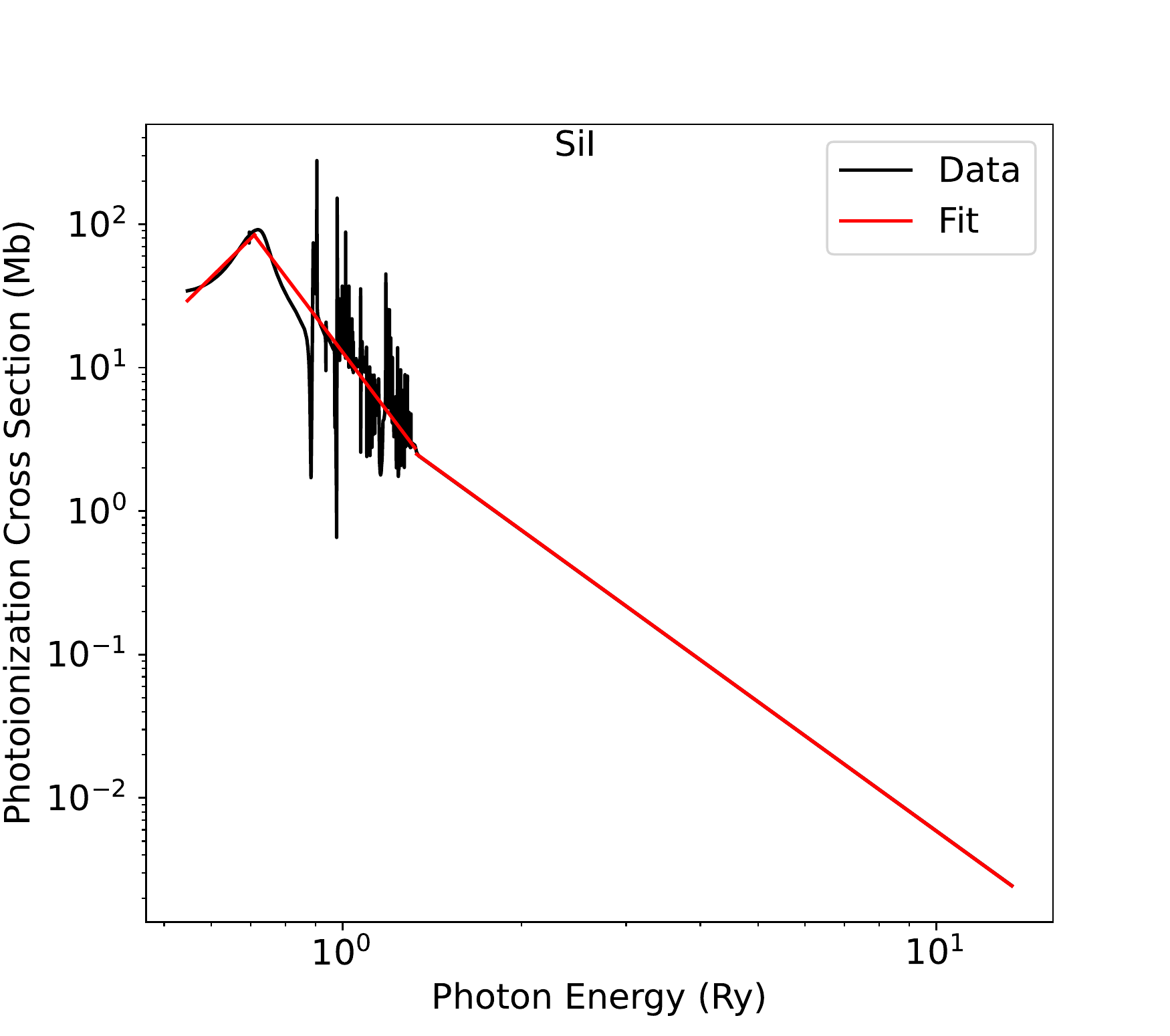}&
\includegraphics[width=.23\linewidth]{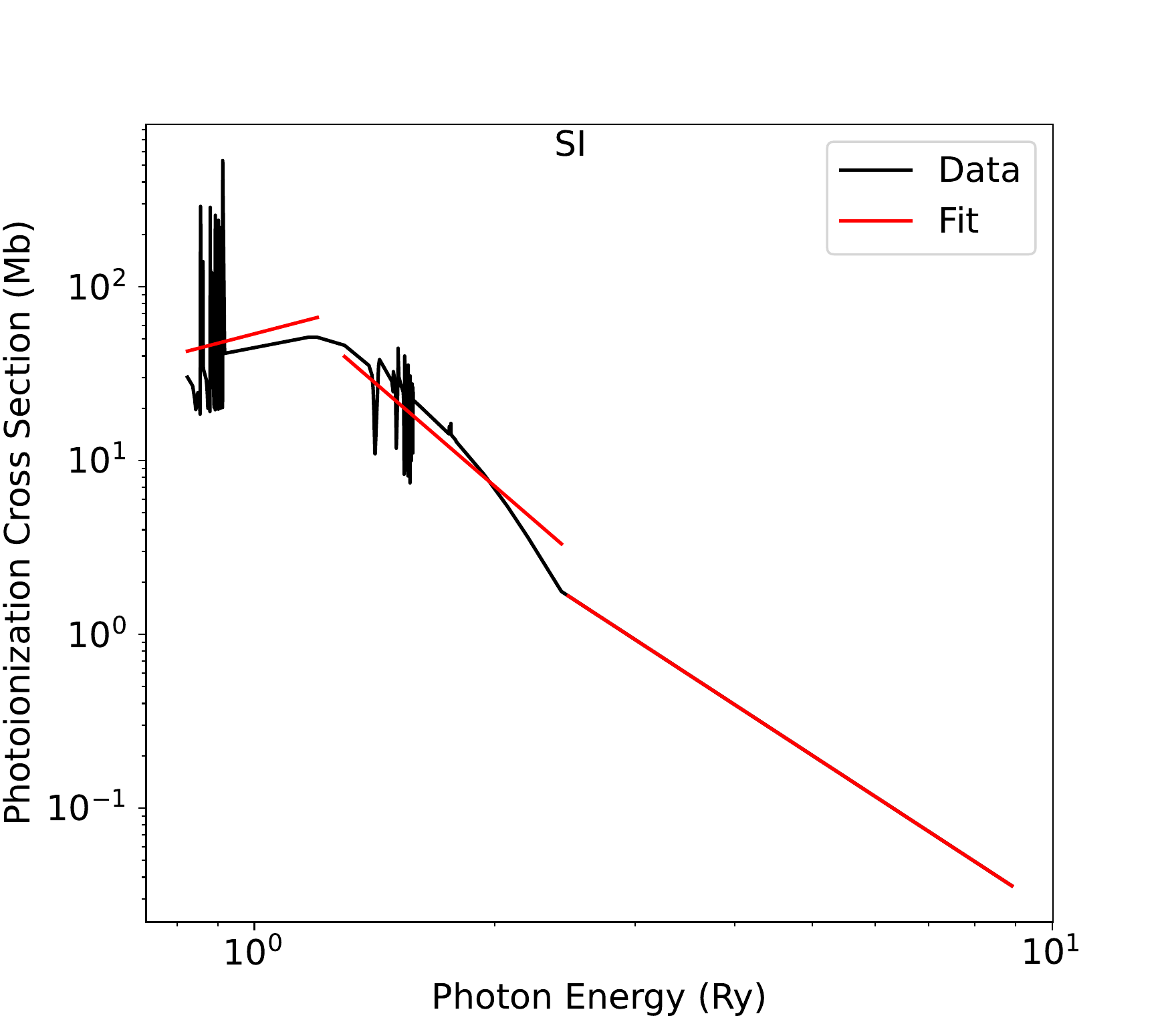}&
\includegraphics[width=.23\linewidth]{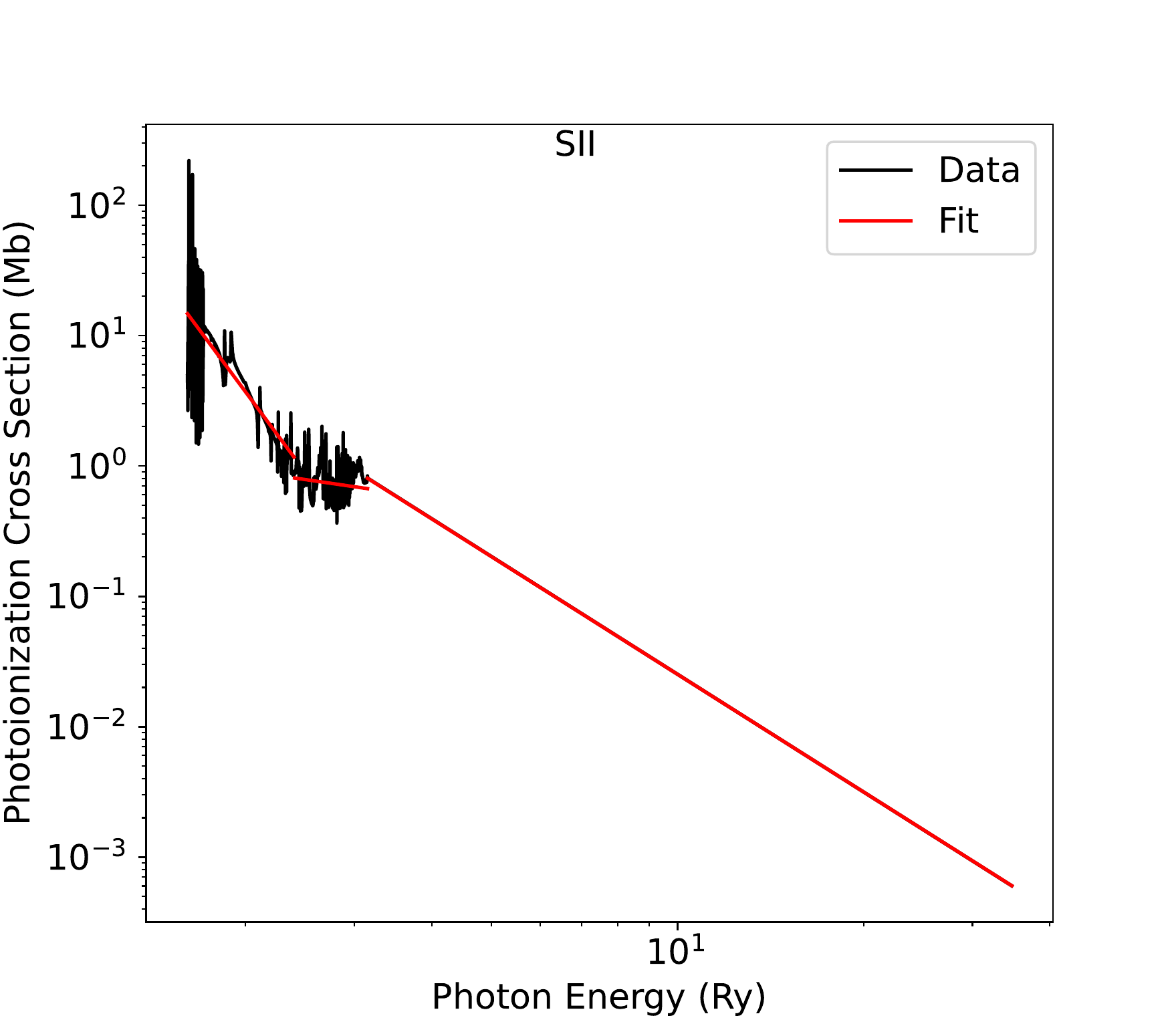}&
\includegraphics[width=.23\linewidth]{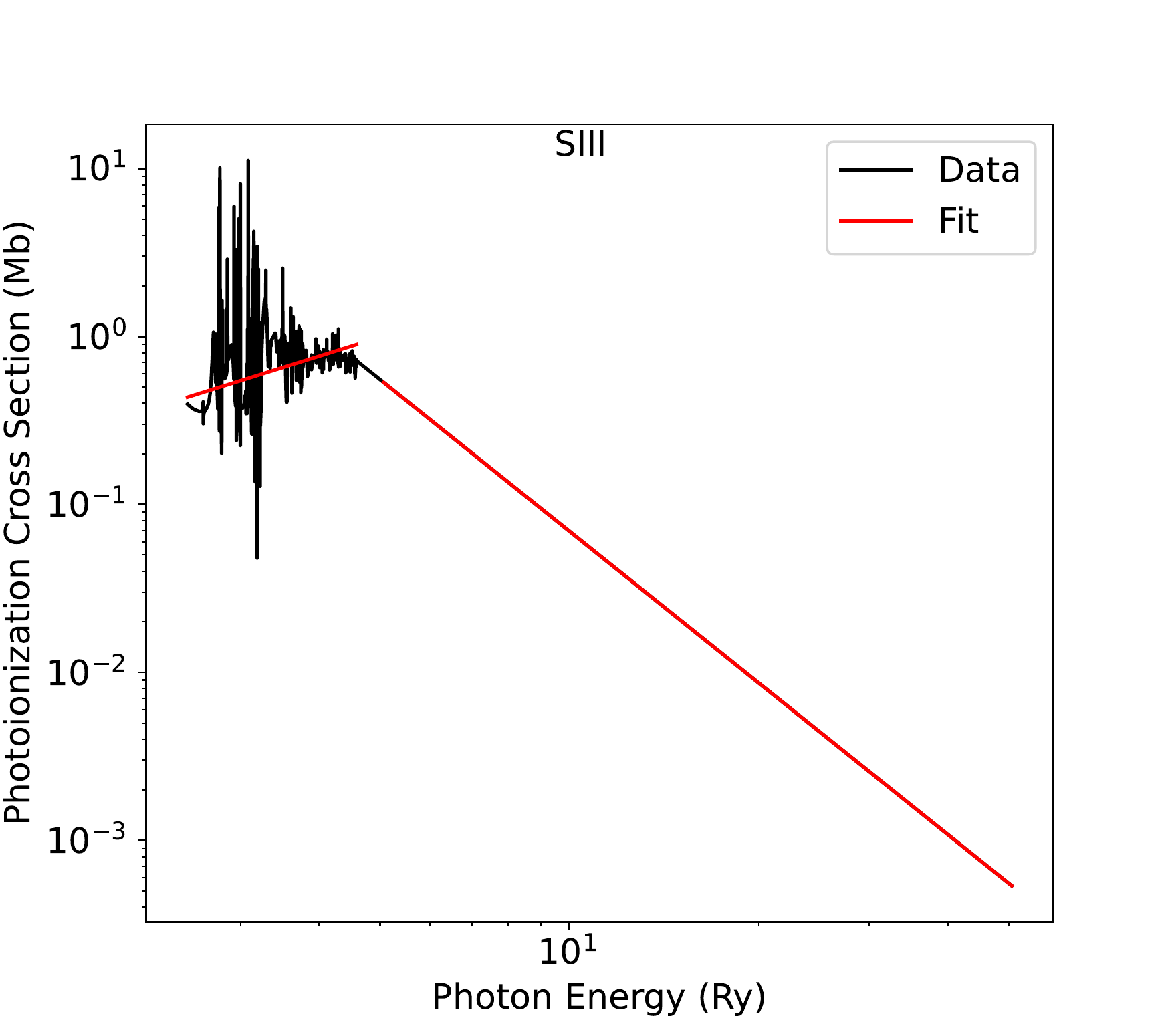}\\ 
\includegraphics[width=.23\linewidth]{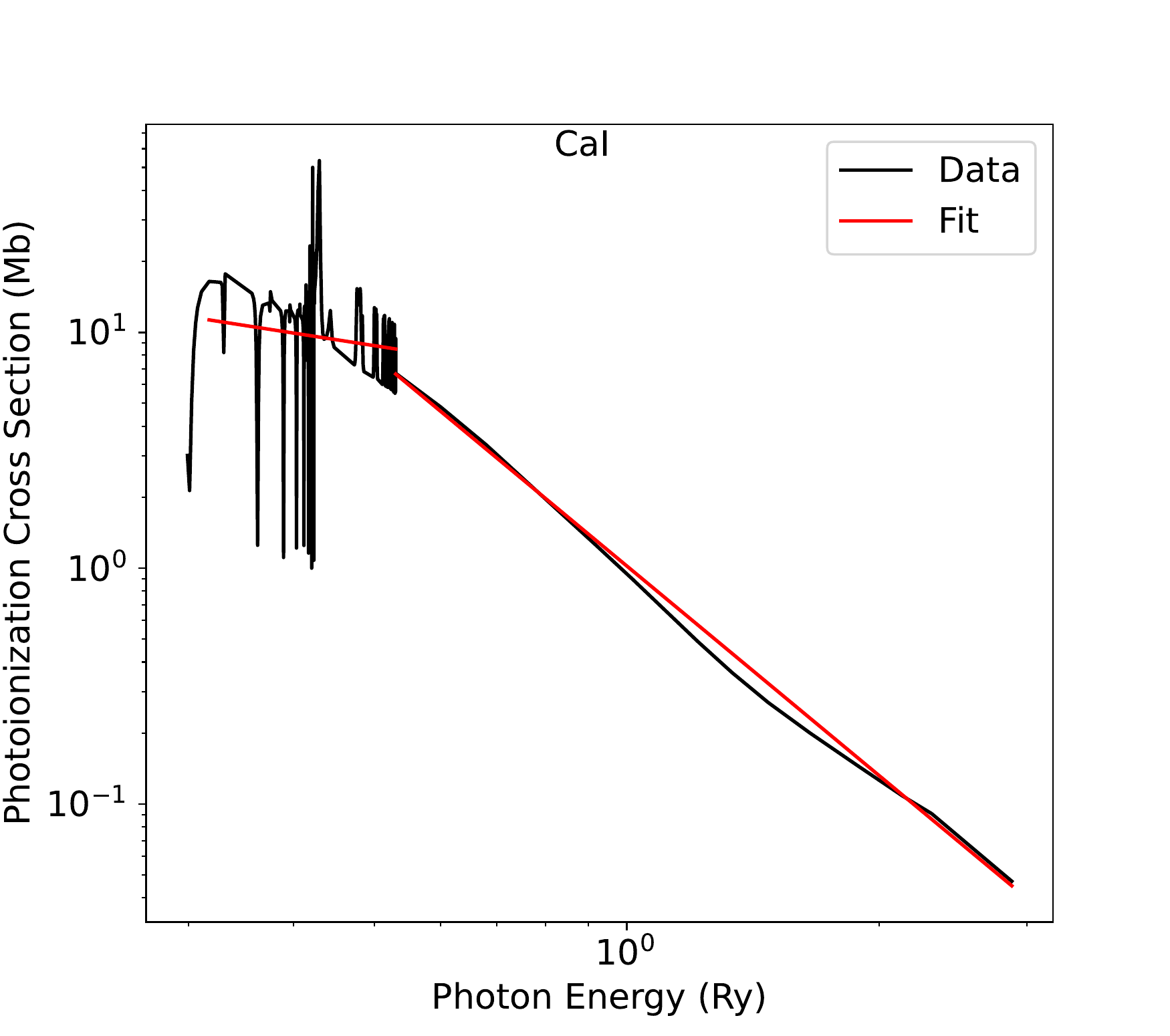}&
\includegraphics[width=.23\linewidth]{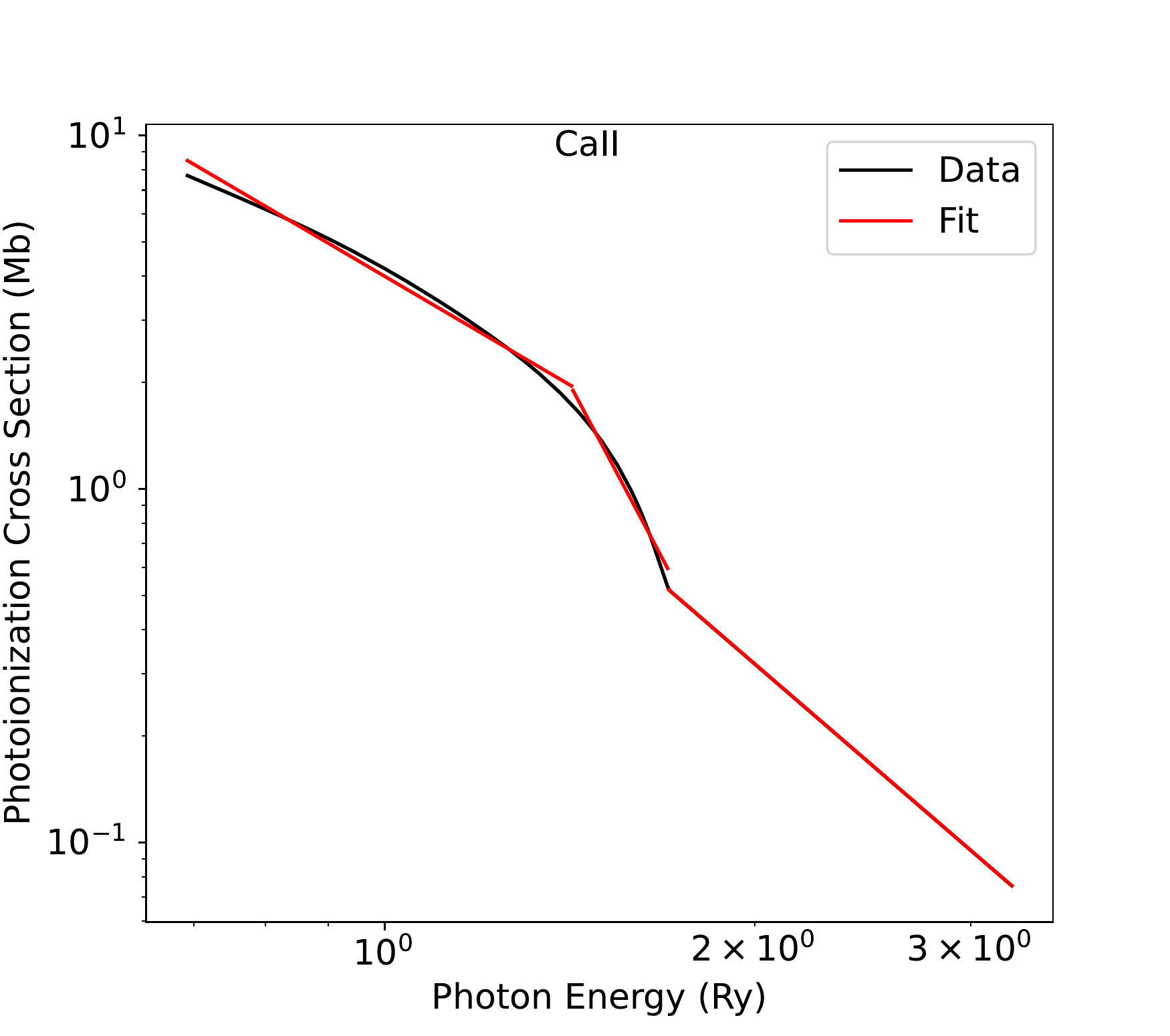}&
\includegraphics[width=.23\linewidth]{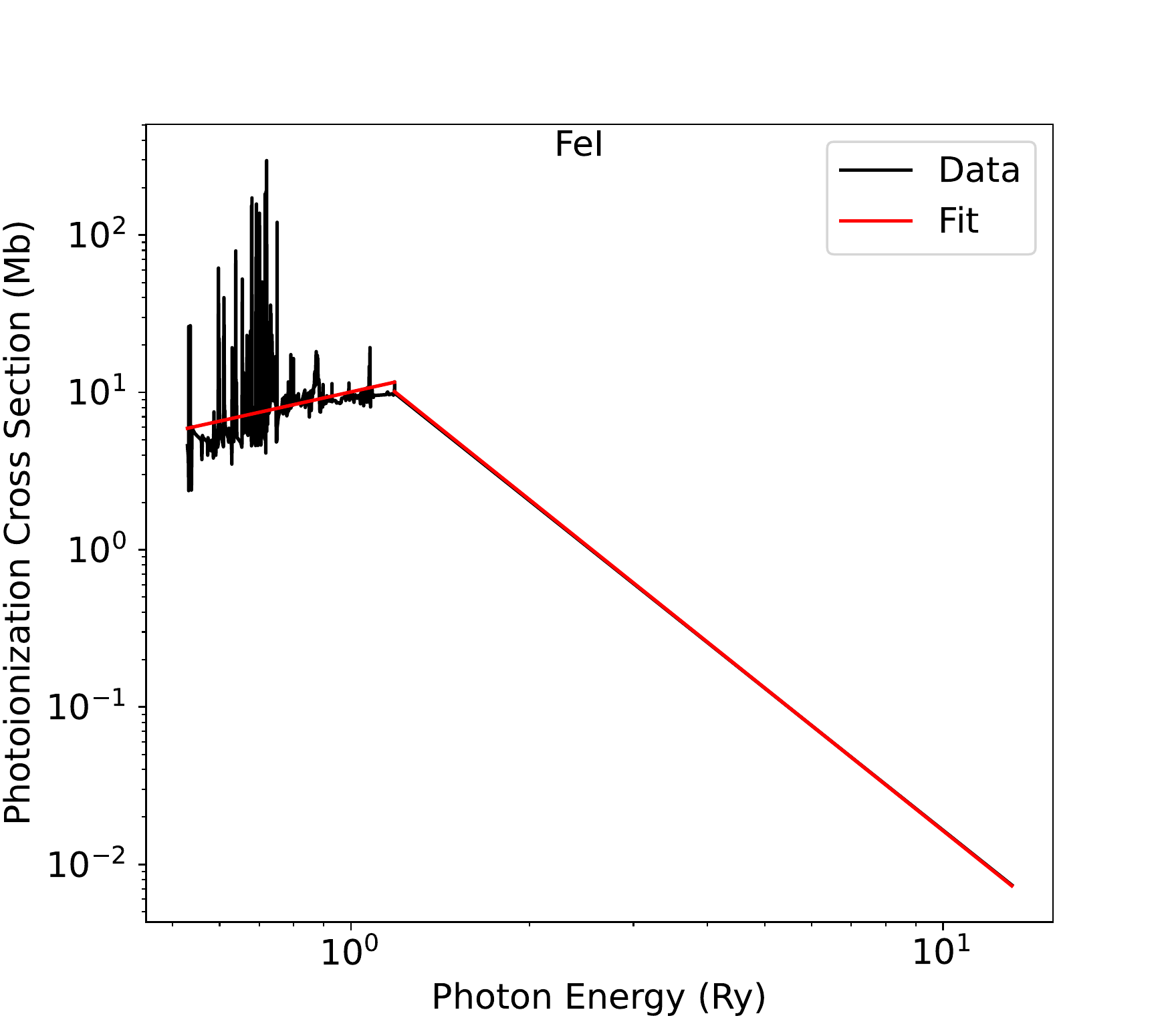}&
\includegraphics[width=.23\linewidth]{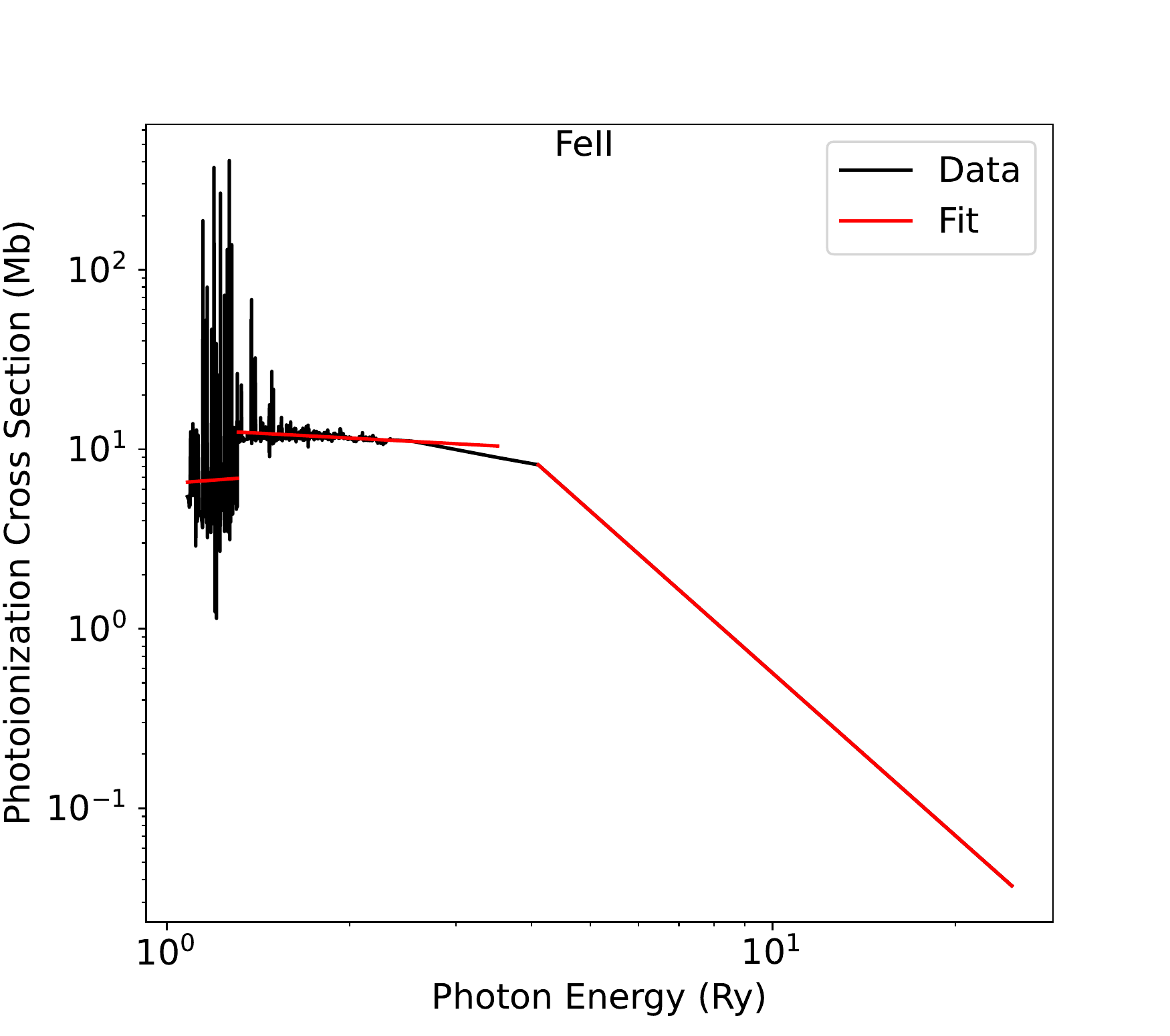} \\
\end{tabular} \\
\centering\begin{tabular}{cc}
\includegraphics[width=.23\linewidth]{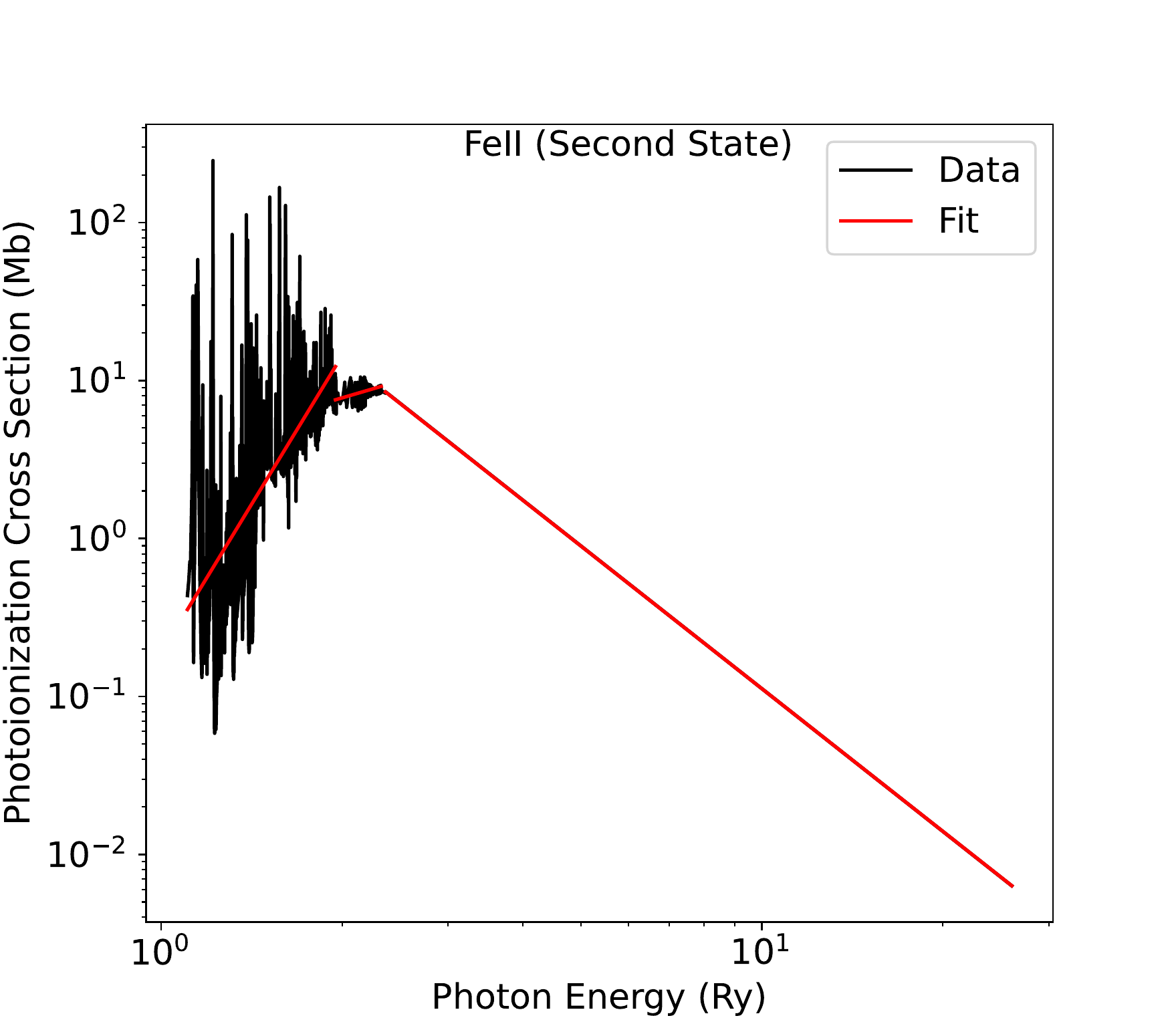}& 
\includegraphics[width=.23\linewidth]{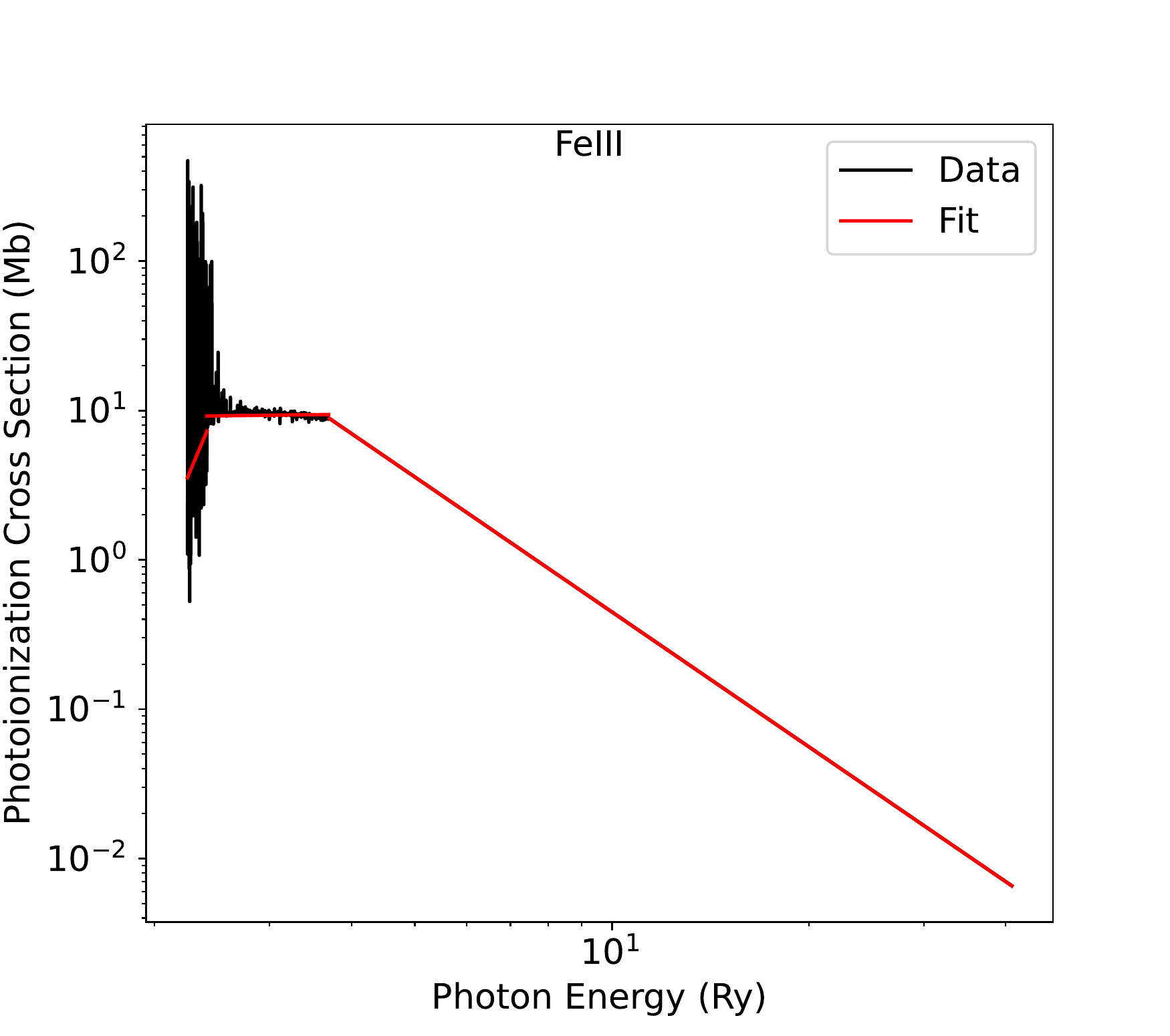}\\
\end{tabular}
\caption{Updated fits for the photoionization cross sections from the first excited state of 13 ions (and second excited state for Fe II).  Parameters for the fits are listed in Table \ref{tbl:picsfits}.}%
\label{fig:pics}
\end{figure*}

13 ions, all those that have excited states with energy lower than 2.72 eV (0.20 Ry) above the ground state (except for Fe, which had a cutoff of 1.36 eV) had their photoionization cross sections from those states updated; this cutoff was chosen because any higher energy states will be $< 5\%$ filled in thermal equilibrium at 10 000 K. 
Fe II was the only ion with multiple excited states below this cutoff.  The cross sections $\sigma$ are fit with multiple functions of the form

\begin{equation}
    \sigma = \sigma_0(E_\gamma/E_{\rm i})^\beta
    \label{eqn:pics}
\end{equation}
where $E_\gamma$ is the photon energy, $E_{\rm i}$ is the ionization energy, and $\sigma_o$ and $\beta$ are free parameters.  Since a single power law does not fit the full cross section for any ion, the cross section was fit over different energy regimes using different power laws, with the regimes transitioning at $\lambda_{\rm cut}$ . 
Lists of the ions updated and fit parameters is given in Tables \ref{tbl:pics} and \ref{tbl:picsfits} and the fits are shown in Figure \ref{fig:pics}.

By using power law fits we ignore the rapid variations of the cross sections that can occur due to resonances. In a homologously expanding ejecta, isotropic Doppler shifting smears the internal specific intensity out over frequency, so such a treatment is expected to be acceptable.

\begin{table}
\centering
\begin{tabular}{c c}
   Ion & Reference \\ \hline
   C I & \cite{Nahar1995} \\
   C II & \cite{Nahar1995} \\  
   C III & \cite{Nahar1997a} \\
   C IV & \cite{Nahar2000a} \\
   O I & \cite{Nahar1999} \\
   O II & \cite{Nahar1999} \\
   O III & \cite{Nahar1996a} \\  
   O IV & \cite{Nahar1999} \\
   Si I & \cite{Nahar2000b} \\
   Si II & \cite{Nahar1996a} \\  
   S II & \cite{Nahar1996a} \\
   S III & \cite{Nahar2000b} \\
   Fe I & \cite{Nahar1997b} \\  
   Fe II & \cite{Nahar1997c} \\
   Fe III & \cite{Nahar1996b} \\  
   Fe IV & \cite{Nahar1998a} \\
\end{tabular}
\caption{Ions with updated total recombination rates.}
\label{tbl:recrates}
\end{table}

\begin{table}
\centering
\begin{tabular}{c c c c}
   Ion & Reference & Excited State & Number of \\ 
    &  & Name & Power Laws \\ \hline 
   C I & \cite{Nahar1991} & 2p$^2$($^1$D) & 4 \\
   O I & \cite{Nahar1998b} & 2p$^4$($^1$D) & 3 \\
   O III & \cite{Nahar1994a} & 2p$^2$($^1$D) & 3 \\
   Mg I & \cite{Cunto1992} & 3p($^3$P$^{\rm o}$) & 3 \\
   Si I & \cite{Nahar1993} & 3p$^2$($^1$D) & 3 \\
   S I & \cite{Cunto1992} & 3p$^4$($^1$D) & 3 \\
   S II & \cite{Nahar1995} & 3p$^3$($^2$D$^{\rm o}$) & 3 \\
   S III & \cite{Nahar2000b} & 3p$^2$($^1$D) & 2 \\
   Ca I & \cite{Cunto1992} & 4s4p($^3$P$^{\rm o}$) & 2 \\
   Ca II & \cite{Cunto1992} & 3d($^2$D) & 3 \\
   Fe I & \cite{Bautista1997} & 3d$^7$4s(a$^5$F) & 2 \\
   Fe II & \cite{Nahar1994b} & 3d$^7$(a$^4$F) & 3 \\
    &  & 3d$^6$4s(a$^4$D) & 3 \\
   Fe III & \cite{Nahar1996c} & 3d$^6$(a$^4$P) & 3 \\
\end{tabular}
\caption{Ions with updated photoionization rates.  Fe II has two low-energy excited states that were updated.  The fitted parameters are given in Table \ref{tbl:picsfits}.}
\label{tbl:pics}
\end{table}

\begin{table*}
\centering
\begin{tabular}{c|cc|c|cc|c|cc|c|cc}
   Ions & $\sigma_0$ & $\beta$ & $\lambda_{\rm cut}$ & $\sigma_0$ & $\beta$ & $\lambda_{\rm cut}$ & $\sigma_0$ & $\beta$ & $\lambda_{\rm cut}$ & $\sigma_0$ & $\beta$  \\
    & (cm$^2$) & & ($\AA$) & (cm$^2$) & & ($\AA$) & (cm$^2$) & & ($\AA$) & (cm$^2$) & \\ \hline
    C I & 3.71E-17 & -1.85 & 487 & 1.31E-17 & -0.72 & 860 & 1.32E-15 & -12.48 & 959 & 1.39E-17 & 5.12 \\
    O I & 1.42E-16 & -3.00 & 353 & 2.02E-17 & -0.89 & 506 & 7.05E-18 & 1.73 & & & \\
    O III & 3.38E-17 & -3.02 & 36 & 5.51E-18 & -1.93 & 175 & 2.20E-17 & -7.41 & & & \\
    Mg I & 1.51E-17 & -3.00 & 646 & 3.36E-19 & -01.1 & 930 & 1.39E-17 & -4.11 & & & \\
    Si  I & 3.60E-17 & -3.00 & 685 & 3.47E-16 & -5.48 & 1283 & 2.93E-17 & 4.07 & & & \\
    S I & 4.50E-17 & -3.00 & 376 & 2.39E-16 & -3.96 & 759 & 4.26E-17 & 1.19 & & & \\
    S II & 6.01E-18 & -3.00 & 289 & 1.06E-18 & -0.69 & 380 & 1.47E-17 & -6.38 & & & \\
    S III & 4.60E-18 & -3.00 & 199 & 4.35E-19 & 1.17 & & & & & &  \\
    Ca I & 3.60E-17 & -2.95 & 1719 & 1.17E-17 & -0.55 & & & & & &  \\
    Ca II & 7.76E-18 & -3.00 & 536 & 2.13E-16 & -6.54 & 642 & 8.49E-18 & -2.04 & & & \\
    Fe I & 1.16E-16 & -3.01 & 769 & 5.91E-18 & 0.84 & & & & & & \\
    Fe II & 4.47E-16 & -3.00 & 222 & 1.29E-17 & -0.18 & 696 & 6.56E-18 & 0.25 & & & \\
        & 8.24E-17 & -3.00 & 386 & 4.01E-18 & 1.12 & 467 & 3.55E-19 & 6.24 & & & \\
    Fe III & 3.96E-17 & -3.00 & 246 & 9.16E-18 & 0.05 & 379 & 3.56E-18 & 10.44 & & & \\
\end{tabular}
\caption{Fitted parameters for the updated photoionization cross sections from highest to lowest energy. The fits are shown in Figure \ref{fig:pics}.}
\label{tbl:picsfits}
\end{table*}

\end{appendix}  

\end{document}